\pdfoutput=1
\documentclass[titlepage,oneside,a4paper,11pt]{book}
\usepackage[nottoc]{tocbibind}
\usepackage[margin=10pt,font=small,labelfont=bf]{caption}
\usepackage{microtype,amsmath,amssymb,amsfonts,amscd,setspace,palatino,mathpazo,graphicx,fancyhdr}
\usepackage[multiple,stable]{footmisc}
\usepackage[amsmath,amsthm,thmmarks]{ntheorem}
\allowdisplaybreaks[4]
\usepackage[noconfig]{refstyle}
\usepackage[square,comma,numbers,compress]{natbib}
\usepackage[hyperfootnotes=false,linktocpage=true,linkbordercolor={.1 .1 1},%
	urlbordercolor={.1 .1 1},citebordercolor={.1 .1 1},menubordercolor={.1 .1 1}]{hyperref}
\hypersetup{
	pdftitle={M-theory Calabi-Yau Quantum Mechanics},
	pdfauthor={Alexander Simon Haupt},
	pdfsubject={Theoretical Physics},
	pdfkeywords={M-theory, string theory, Calabi-Yau, five-folds, Quantum Mechanics, one dimension, sigma models, superspace},
	bookmarksnumbered}
\usepackage{geometry}
\newcommand\abstractname{Abstract}
\newenvironment{abstract}{%
        \null\vfil\vspace{-5cm}\small 
        \begin{center}%
          {\bfseries\huge \abstractname\vspace{-.5em}}%
        \end{center}%
        \quotation}
      {\if@twocolumn\else\endquotation\fi}
      
\def\addsymbol #1: #2#3{$#1$ \> \parbox{5.1in}{#2 \dotfill \pageref{#3}}\\} 
\def\newnot#1{\label{#1}} 

\let\eqref=\relax
\newref{eq}{name={eq.~},Name={Eq.~},names={eqs.~},Names={Eqs.~},rngtxt={-},refcmd=(\ref{#1})}
\newref{f}{name={footnote~},Name={Footnote~},names={footnotes~},Names={Footnotes~}}
\newref{app}{name={appendix~},Name={Appendix~},names={appendixes~},Names={Appendixes~}}
\newref{tab}{name={table~},Name={Table~},names={tables~},Names={Tables~}}
\newref{ch}{name={chapter~},Name={Chapter~},names={chapters~},Names={Chapters~}}
\newref{sec}{name={section~},Name={Section~},names={sections~},Names={Sections~}}
\newref{fig}{name={figure~},Name={Figure~},names={figures~},Names={Figures~}}
\numberwithin{equation}{section}

\newcommand{\be}{\begin{equation}}
\newcommand{\ee}{\end{equation}}
\newcommand{\bea}{\begin{equation}\begin{aligned}}	
\newcommand{\eea}{\end{aligned}\end{equation}}		
\newcommand{\UA}{{\underline{A}}}					
\newcommand{\UB}{{\underline{B}}}					
\newcommand{\UZ}{{\underline{0}}}
\newcommand{\Uth}{{\underline{\theta}}}
\newcommand{\BUth}{{\bar{\underline{\theta}}}}
\newcommand{\ul}[1]{{\underline{#1}}}
\newcommand{\cc}{\text{c.c.}}

\newcommand{\field}[1]{\mathbb{#1}}
\newcommand{\oneon}[1]{\frac{1}{#1}}
\newcommand{\manifold}{\mathcal{M}}
\newcommand{\nfold}[1]{\mathrm{CY}_{#1}}
\newcommand{\lagr}{\mathcal{L}}
\newcommand{\hamil}{\mathcal{H}}
\newcommand{\cs}{\mathcal{J}}
\newcommand{\curvtwofrm}{\mathcal{R}}
\newcommand{\susyno}{\mathcal{N}}
\newcommand{\sderiv}{\mathcal{D}}
\newcommand{\tr}{\mathrm{tr}}
\newcommand{\R}{\field{R}}
\newcommand{\C}{\field{C}}
\newcommand{\Z}{\field{Z}}
\newcommand{\CP}{\field{P}}
\newcommand{\M}{\field{M}}
\newcommand{\al}{\alpha}
\newcommand{\transp}{{\rm T}}
\newcommand{\iddots}{\mathinner{\mkern2mu\raise1pt\hbox{.}\mkern2mu \raise4pt\hbox{.}\mkern2mu\raise7pt\hbox{.}\mkern1mu}}
\DeclareMathOperator{\diag}{diag}
\DeclareMathOperator{\GL}{GL}
\DeclareMathOperator{\SU}{SU}
\DeclareMathOperator{\SO}{SO}
\DeclareMathOperator{\U}{U}

\providecommand{\id}{\leavevmode\hbox{\small$\mathrm{1}$\kern-3.8pt\normalsize$\mathrm{1}$}}

\begin{document}

\setlength{\baselineskip}{16.5pt plus 1pt minus 0.1pt} 

\pagestyle{empty}
%
%
%
\begin{center}
  {\sc{University of London}}\\[1.0cm]
  \centerline{\includegraphics[height=5cm]{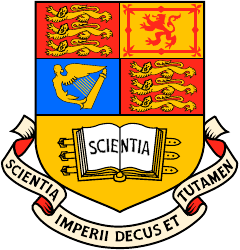}}	
  \vspace{1.0cm}
  Imperial College of
  Science, Technology and Medicine\\ The
  Blackett Laboratory\\ Theoretical Physics Group\\[1.5cm]
  
  {\Huge \sc{M-theory Calabi-Yau}}\\[0.3cm]
  {\Huge \sc{Quantum Mechanics}}\\[2cm]

  {\Large{Alexander Simon Haupt}}\\[2cm]

  Thesis submitted for the Degree of\\
  Doctor of Philosophy\\
  of the University of London\\
  and the Diploma of Membership\\
  of Imperial College\\[.8cm]
  \vfill
November 2009
\end{center}

\cleardoublepage
\pagestyle{plain}
\chapter*{Declaration}
The work presented in this thesis was carried out in the Theoretical Physics Group at Imperial College London between October 2005 and July 2009 under the supervision of Professor Kellogg Stelle. Except where otherwise stated, the work is original and has not been submitted before for a degree of this or any other university.\\
\vspace*{1cm}
\begin{flushright}
Alexander Haupt\\
19 November 2009
\end{flushright}

\cleardoublepage
%
%
%
\vspace*{2cm}
\begin{abstract}\vspace*{1.5cm}\noindent
This thesis explores an exotic class of M-theory compactifications in which the compact manifold is taken to be a Calabi-Yau five-fold -- that is, a 10-dimensional Ricci-flat K\"ahler manifold. In this way, all spatial dimensions of M-theory are compactified and the resulting effective theory is a one-dimensional $\susyno=2$ super-mechanics model that exhibits peculiar features of one-dimensional supersymmetry, such as the appearance of fermion-only super-multiplets. The latter necessitates reducing also the fermionic sector of M-theory, which is not normally included in the compactification literature and is thus presented, together with the required technology, in detail.

The one-dimensional effective theory is most elegantly described in superspace and therefore, a detailed account of one-di\-men\-sional $\susyno=2$ superspace is provided. This includes developing the theory of fermionic multiplets and the study of cross-couplings between $2a$ and $2b$ multiplets, as well as an in-depth presentation of curved one-dimensional $\susyno=2$ superspace.

Another important aspect is the inclusion of flux. We study its consistency conditions, its relation to supersymmetry and the way it gives rise to a potential in the one-dimensional effective action. It is also explained how the supersymmetry-preserving part of the potential can be obtained from a Gukov-type superpotential.

The main motivation of this compactification scenario is rooted in the realm of cosmology. Viewed as a cosmological model, its virtue is a democratic treatment of spatial dimensions. As opposed to the artificial $3+7$ split in most string compactifications, the early universe starts out with all spatial dimensions compact and small in our approach. One then seeks for dynamical ways in which three out of the ten spatial dimensions grow large at late times. Possible realisations of this idea are discussed both at the classical and at the quantum level.

Finally, preliminary work on Calabi-Yau five-fold compactifications of F-theory and the resulting two-dimensional string-like actions is presented.
\end{abstract}

\tableofcontents
\cleardoublepage
\chapter*{List of Symbols\hfill} \addcontentsline{toc}{chapter}{List of Symbols}\label{LoS}
%
%
%
%
\begin{tabbing}
\emph{Symbol}~~~~~~~~~\=\parbox{5.1in}{\emph{Meaning}\hfill \emph{Defined on p.}}\\
%
%
\addsymbol \manifold,\manifold^D: {$D$-dimensional space-time manifold with $(-+\ldots +)$ signature}{symbol:mfld}
\addsymbol X: {Calabi-Yau five-fold (or $n$-fold, more generally)}{symbol:CY5}
\addsymbol \R: {real line -- time direction in Calabi-Yau five-fold compactifications}{symbol:R}
\addsymbol \M^D: {$D$-dimensional Minkowski space with $(-+\ldots +)$ signature}{symbol:MD}
\addsymbol \CP^n: {complex projective space of complex dimension $n$}{symbol:Pn}
\addsymbol T^n: {$n$-torus}{symbol:Tn}
\addsymbol S^n: {$n$-sphere}{symbol:Sn}
\addsymbol \Sigma: {\chref{msp}: Euclidean 3-manifold; \chref{fcy5}: string-worldsheet}{symbol:Sigma}
%
%
\addsymbol M,N,\ldots: {$D$-dimensional space-time indices (range: $0,1,\ldots,(D-1)$)}{symbol:11ind}
\addsymbol m,n,\ldots: {$D=10$ Euclidean indices (range: $1,\ldots,10$)}{symbol:10indEucl}
\addsymbol \mu,\nu,\ldots: {$D=10$ Euclidean holomorphic indices (range: $1,\ldots,5$)}{symbol:10indEuclhol}
\addsymbol \bar\mu,\bar\nu,\ldots: {$D=10$ Euclidean anti-holomorphic indices (range: $\bar{1},\ldots,\bar{5}$)}{symbol:10indEuclahol}
\addsymbol A,B,\ldots: {one-dimensional $\susyno=2$ superspace indices (range: $0,\theta,\bar\theta$)}{symbol:ind_sspace}
\addsymbol i,j,\ldots: {real $(1,1)$-moduli space indices (range: $1,\ldots,h^{1,1}(X)$)}{symbol:ind_11}
\addsymbol p,q,\ldots: {complex $(2,1)$-moduli space indices (range: $1,\ldots,h^{2,1}(X)$)}{symbol:ind_21}
\addsymbol x,y,\ldots: {complex $(1,3)$-moduli space indices (range: $1,\ldots,h^{1,3}(X)$)}{symbol:ind_13}
\addsymbol e,f,\ldots: {real $(2,2)$-moduli space indices (range: $1,\ldots,h^{2,2}(X)$)}{symbol:ind_22}
\addsymbol a,b,\ldots: {complex $(1,4)$-moduli space indices (range: $1,\ldots,h^{1,4}(X)$)}{symbol:ind_14}
\addsymbol \mathcal{P,Q},\ldots: {real 3-form moduli space indices (range: $1,\ldots,b^3(X)$)}{symbol:ind_real3}
\addsymbol \mathcal{X,Y},\ldots: {real 4-form moduli space indices (range: $1,\ldots,b^4(X)$)}{symbol:ind_real4}
\addsymbol \mathcal{\hat{X},\hat{Y}},\ldots: {real $\hat{4}$-form indices ($\hat{4}=(1,3)+(3,1)$; range: $1,\ldots,2h^{1,3}(X)$)}{symbol:ind_hat4}
\addsymbol \mathcal{A,B},\ldots: {real 5-form indices (range: $1,\ldots,b^5(X)$)}{symbol:ind_real5}
%
%
\addsymbol \{\omega_i\}: {real basis of $H^{1,1}(X)$}{symbol:11_basis}
\addsymbol \{\nu_p\}: {basis of $H^{2,1}(X)$}{symbol:21_basis}
\addsymbol \{\varpi_x\}: {basis of $H^{1,3}(X)$}{symbol:13_basis}
\addsymbol \{\sigma_e\}: {real basis of $H^{2,2}(X)$}{symbol:22_basis}
\addsymbol \{\chi_a\}: {real basis of $H^{1,4}(X)$}{symbol:14_basis}
\addsymbol \{ N_{\cal{P}} \}: {basis of $H^3(X)$}{symbol:3form_basis}
\addsymbol \{ O_{\cal{X}} \}: {basis of $H^4(X)$}{symbol:4form_basis}
\addsymbol \{ O_{\hat{\cal{X}}} \}: {basis of $H^{2,2}(X) \subset H^4(X)$}{symbol:hat4form_basis}
\addsymbol \{ O_{\tilde{\cal{X}}} \}: {basis of $H^{1,3}(X)\oplus H^{3,1}(X) \subset H^4(X)$}{symbol:tilde4form_basis}
\addsymbol \{\hat{\omega}^i\}: {real basis of dual $(4,4)$-forms ($\int \omega_i \wedge \hat{\omega}^j = \delta_i {}^j$)}{symbol:dual8forms}
%
%
\addsymbol N: {lapse function of one-dimensional gravity}{symbol:lapse}
\addsymbol t^i: {K\"ahler moduli fields}{symbol:moduli_11}
\addsymbol z^a: {complex structure moduli fields}{symbol:moduli_14}
\addsymbol \xi^p: {$(2,1)$-moduli fields}{symbol:moduli_21}
\addsymbol X^\mathcal{P}: {real 3-form moduli fields}{symbol:moduli_3form}
\addsymbol \psi_0: {lapsino; superpartner of $N$ in one-dimensional supergravity}{symbol:lapsino}
\addsymbol \psi^i: {$(1,1)$-fermion; superpartner of $t^i$}{symbol:fermion_11}
\addsymbol \kappa^a: {$(1,4)$-fermion; superpartner of $z^a$}{symbol:fermion_14}
\addsymbol \Lambda^\mathcal{P}: {3-form fermion; superpartner of $X^\mathcal{P}$}{symbol:fermion_3form}
\addsymbol \Upsilon^{\hat{\cal{X}}}: {$\hat{4}$-form fermion}{symbol:fermion_hat4form}
%
%
\addsymbol (\cdot)^\ast,\overline{(\cdot)}: {complex conjugation}{symbol:complex_conjug}
\addsymbol \tilde{(\cdot)}: {Dual Lefschetz operator: $\Lambda\omega = (-1)^q \tilde\omega$}{symbol:tmap}
\addsymbol k_e^i: {matrix representation of $(2,2)\to(1,1)$ version of $\Lambda$}{symbol:kei}
\addsymbol \omega_0: {primitive differential form (that is, $\tilde{\omega}_0=0$)}{symb:prim_form}
\addsymbol [\omega]: 
	{cohomological equivalence class: $[\omega]\equiv\{\omega^\prime \,|\, \omega^\prime = \omega + d\eta\}$}{symbol:cohomclass}
\addsymbol \mathfrak{A},\mathfrak{B}: {linear maps $H^3(X) \leftrightarrow H^{2,1}(X)\oplus H^{1,2}(X)$}{symbols:3_21_maps}
\addsymbol \mathfrak{C},\ldots,\mathfrak{F}: 
	{linear maps $H^4(X) \leftrightarrow H^{3,1}(X)\oplus H^{2,2}(X)\oplus H^{1,3}(X)$}{symbols:4_cplx_maps}
%
%
\addsymbol \bar\Psi: {$D=11$ Dirac conjugation $\bar\Psi \equiv i \Psi^\dag \Gamma^\UZ$}{symbol:dirac_conjug}
\addsymbol \eta,\eta^\star: {covariantly constant spinors (of opposite chirality) on the five-fold $X$}{symbol:eta}
%
%
\addsymbol \#S: {cardinality of the set $S$ (number of elements contained in $S$)}{symbol:cardinality}
\addsymbol \lfloor x\rfloor: {floor function: the largest integer less than or equal to $x$}{symbol:floor}
\addsymbol \lceil x\rceil: {ceiling function: the smallest integer greater than or equal to $x$}{symbol:ceil}
%
%
\addsymbol [i_1 \ldots i_n]: {index anti-symmetrisation with ``weight one''}{symbol:ind_asymm}
\addsymbol (i_1 \ldots i_n): {index symmetrisation with ``weight one''}{symbol:ind_symm}
\addsymbol [i_1 \ldots i_n\}: {graded index anti-symmetrisation with ``weight one''}{symbol:ind_graded_asymm}
\addsymbol | i_1\ldots i_n |: {indices excluded from (anti)-symmetrisation}{symbol:ind_no_asymm}
\addsymbol \delta^{i_1 \ldots i_n}: {multi-index Kronecker symbol: $=1$ if $i_1=\ldots=i_n$ and zero otherwise}{symbol:multind_kronecker}
\addsymbol \delta^{i_1 \ldots i_n}_{j_1 \ldots j_n}:
	{generalised anti-symmetric Kronecker symbol: 
	 $\delta^{i_1 \ldots i_n}_{j_1 \ldots j_n}\equiv\delta^{[i_1}_{[j_1} \cdots \delta^{i_n]}_{j_n]}$}{symbol:asymm_kronecker}
%
%
\addsymbol [X,Y\}: {graded commutator: $[X,Y\} \equiv XY - (-1)^{XY} YX$}{symbol:graded_comm}
\addsymbol X|: {lowest component of the superfield $X$ (short for $X |_{\theta=\bar\theta=0}$)}{symbol:sspace_lowest_comp}
%
%
\addsymbol \beta: {M-theory expansion parameter: $\beta \equiv \left(\kappa_{11}^2/(2\pi^2)\right)^{2/3} / (2\pi )^2$}{symbol:beta}
\addsymbol \beta_1: {rescaled M-theory expansion parameter: $\beta_1 \equiv (2\pi)^4\beta/v^{4/5}$}{symbol:beta1}
\addsymbol \gamma: {intermediate expansion parameter: $\gamma\equiv\sqrt{\beta}$}{symbol:gamma}
\addsymbol \gamma_1: {rescaled intermediate expansion parameter: $\gamma_1\equiv\sqrt{\beta_1}$}{symbol:gamma1}
%
\end{tabbing}\label{LoSend}

\cleardoublepage
\thispagestyle{empty}
\vspace*{\stretch{1}}
\begin{quotation}
``If I have seen a little further it is by standing on the shoulders of Giants.''
\begin{flushright}
--- \textsc{Sir Isaac Newton} in a letter to Robert Hooke dated February 5, 1676
\end{flushright}
\end{quotation}
\vspace*{\stretch{3}}

\pagestyle{fancy}
\fancyhead{}
\fancyfoot{}
\lhead{\leftmark}
\cfoot{\thepage}

%
%

\chapter{Introduction}\chlabel{intro}

\section{Preliminary Remarks}

Around four billion years ago, in a chain of random events totally irrelevant for the universe as a whole, on some planet in some arm of a spiral galaxy, some molecules made up of mainly carbon, hydrogen, oxygen, nitrogen, sulphur and phosphorous (so-called biogenic elements) collided and formed prebiotic structures. This idea of ``abiogenesis'' (see, for example, ref.~\cite{DysonOOL}) is but one out of a number of possible scenarios of how life could have started out on our planet. During the following few billion years, DNA and conglomerates thereof as well as other organic compounds (such as peptides and proteins) had evolved out of this ``prebiotic soup'' into a species of intelligent life known as homo sapiens. At some point, humans began to wonder how the world they live in works, came into being and will ultimately end up. It is perhaps this innate curiosity about the history, inner working and fate of the universe and its contents that distinguishes us from most other living organisms on this planet. Science strives to guide, formalise and optimise the accumulation of knowledge that results from this insatiable curiosity.

These ultimate questions mentioned in the previous paragraph find their manifestation, for example, in contemporary theoretical physics in the form of the more than half a century old puzzle of how to reconcile quantum theory with the theory of general relativity. It is a curious fact by itself that the two central pillars of modern physics and arguably two of the most remarkable achievements of the human intellect are notoriously incompatible. One of several theoretical ideas potentially capable of shedding light on this puzzle is ``string theory'' (reviewed, for example, in refs.~\cite{GSW,Ortin:2004ms,Becker:2007zj}), although it was originally devised in an entirely different context.

In the late 1960's, it grew out of an attempt to model the strong interactions in hadronic physics. At the end, quantum chromodynamics took up its place as model of the strong nuclear force. String theory fell out of favour at that time not least because it possessed some unwanted features such as the need for extra dimensions, the occurrence of tachyons, the lack of fermions and the existence of massless spin-2 particles. However, the latter was turned into a success by interpreting the unwanted massless spin-2 particles as gravitons and -- banning the theory from the hadronic scale of $10^{-15} \,m$ down to the Planck scale of $10^{-35} \,m$ -- string theory was henceforth regarded as a candidate for a theory of quantum gravity and possibly a unified theory of nature.

In the 1970's, string theory received further support from an \emph{a priori} completely independent development, namely the discovery of supersymmetry. As far as string theory was concerned, this new type of symmetry allowed the consistent introduction of fermions while, at the same time, removing the bothersome tachyon from the spectrum. Later, in the mid-1980's, five distinct consistent, anomaly-free superstring theories defined in 10 dimensions were found and the theory achieved widespread acceptance.

Albeit a vital ingredient of modern string theory, supersymmetry is an independent theoretical framework completely in its own right. The central idea behind supersymmetry is to introduce a symmetry interchanging bosons and fermions, thereby extending the usual concept of symmetries of quantum field theories based on Lie groups. In particular, Haag, \L opusza\'nski and Sohnius showed in 1975~\cite{Haag:1974qh} that supersymmetry offers the only possible way to evade the famous Coleman-Mandula no-go theorem~\cite{Coleman:1967ad}. Equipping quantum field theories with this symmetry implies truly remarkable properties, most importantly improved ultraviolet behaviour.\footnote{For a recent review, see, for example, the article~\cite{Wess:2009cy} by the late Julius Wess -- one of the founders of supersymmetry.} On the downside, however, a supersymmetric theory requires an equal number of fermions and bosons in each multiplet and this is, for example, not how the standard model is organised. In order to make the standard model supersymmetric each of its elementary particles must be accompanied by a superpartner of opposite spin statistics but with the same mass. Since these superpartners have not been seen at the currently accessible energy scales, it is clear that supersymmetry must be broken if it is to be a symmetry of nature.

It is possible to marry supersymmetry with the theory of general relativity and the result is known as supergravity. Just as general relativity can be obtained by gauging the Poincar\'{e} group of space-time symmetries, supergravity can be viewed as the gauged version of theories with global supersymmetry. In other words, supersymmetry is realised as a local symmetry in supergravity theories. Supergravity theories can be defined in various dimensions (see, for example, refs.~\cite{Tanii:1998px,deWit:1997sz}, for an introduction). Assuming Lorentzian signature, eleven is the highest number of dimensions\footnote{At least formally, it is possible to go to higher dimensions by allowing the space-time manifold to have more than one time direction. The physical interpretation of such theories becomes problematic, however. We will come back to this point later in \chref{fcy5}.} in which a supergravity theory can be formulated consistently~\cite{Nahm:1977tg} -- that is, such that the theory does not contain elementary particles with spin greater than two. Furthermore, supergravity in $D=11$ leads to a unique theory, which was first constructed by Cremmer, Julia and Scherk in 1978~\cite{Cremmer:1978km}. Most of the numerous supergravity theories that exist in lower dimensions can be obtained from $D=11$ supergravity by a Kaluza-Klein reduction on some compact manifold.

For more than two decades, string theory and 11-dimensional supergravity have evolved largely alongside each other without a great deal of interaction. In fact, the theoretical physics community has been rather polarised between the two ideas and there has been a long-standing question of whether and how the two are related. Viewed in this light, an intriguing fact hinting towards a possible relation is the proximity of dimensions in which the two theories are defined: string theory in 10 versus supergravity in 11 dimensions. An answer was finally proposed by Witten and others during a period of intense activity in 1995~\cite{Witten:1995ex,Townsend:1995kk}. The result was a new over-arching framework called ``M-theory'' whose very name let alone its foundations, fundamental principles and definitions remain largely elusive.

\section{Overview and Motivation}\seclabel{why_mcy5}

In order to make contact with four-dimensional physics, the old idea by Kaluza and Klein from the 1920's of how to treat extra dimensions was revived in string-/M-theory from the outset. By assuming that the extra dimensions are compact (hence the name compactification) and tiny compared to our macroscopic four dimensions, one arrives at a lower-dimensional effective description of a higher-dimensional theory. In this approach, the properties of the lower-dimensional effective theory are determined by the topology and geometry of the compact manifold that represents the extra dimensions (and, of course, by the form of the higher-dimensional theory itself). 

In 1985, Candelas, Horowitz, Strominger and Witten~\cite{Candelas:1985en} made use of precisely this observation by asking how a six-dimensional compact manifold must look like in order to obtain a physically interesting four-dimensional theory from compactifying 10-dimensional string theory. The answer was Calabi-Yau three-folds. Since then, string-/M-theory has been compactified on many different kinds of manifolds (and even non-manifolds) to many different numbers of dimensions, thereby unveiling a series of powerful and intriguing properties of the theory, such as dualities and hidden symmetries. Calabi-Yau three-folds~\cite{Candelas:1985en}, four-folds~\cite{Brunner:1996pk,Becker:1996gj,Haack:2002tu} and two-folds (that is, $K3$)~\cite{Hull:1994ys,Witten:1995ex} have played a central r\^ole in this context.

In the scheme of string-/M-theory compactifications, Calabi-Yau five-folds have so far largely been treated as ``orphans'' and the aim of this thesis is to remedy this shortcoming.\footnote{It is important to remark that Calabi-Yau five-folds have already appeared before in the physics literature, for example, in ref.~\cite{Kumar:1996zx} where subclasses of those manifolds feature in the discussion of certain vacuum constructions of F-theory and thirteen dimensional S-theory leading to supersymmetric two-dimensional $\susyno=(1,1)$ and three-dimensional $\susyno=2$ theories, respectively, and then again more detailed later in ref.~\cite{Curio:1998bv} in a similar but more general context. Moreover, M-theory backgrounds based on Calabi-Yau five-folds and their corrections induced by higher-order curvature terms have been considered in ref.~\cite{Lu:2004ng}.} Since Calabi-Yau five-folds are ten-dimensional, only M-theory (and F-theory) can be dimensionally reduced on these spaces. The resulting effective theory is one-dimensional (two-dimensional, if one starts from F-theory) and has a residual local $\susyno=2$ supersymmetry. It is interesting to note that $D=1$ and $D=11$ are, respectively, the lowest and highest dimension in which supergravity theories can be defined (although, supergravity is of course non-dynamical in $D=1$) and are thus special from this viewpoint. Via Calabi-Yau five-fold reductions, these two ``extreme'' versions of supergravity are connected.

When the legendary English mountaineer George Leigh Mallory was asked by journalists in 1923 why he wanted to climb Mount Everest -- the highest mountain on earth --, he famously replied, ``because it's there.'' While this is a perfectly legitimate motivation for climbing a mountain, it is not sufficient for motivating the study of a scientific problem. We will therefore now discuss the motivations for studying Calabi-Yau five-fold reductions.

The ultimate vision behind this type of compactification is to potentially provide a different angle on the ``dimensionality problem'' of string theory. As described above, in most attempts to cure string theory's requirement of more than four dimensions, an \emph{ad hoc} split between the observed four extended space-time dimensions and an additional number of compact extra dimensions is introduced by hand. It would be conceptually more desirable to treat at least all the spatial dimensions on equal footings in the early universe. In such a ``democratic'' picture, one assumes that the early universe started out with all spatial dimensions compact and tiny. One may then study how three large spatial dimensions emerge at later times by considering the dynamics on the Calabi-Yau five-fold moduli space, which is governed by the actions derived in this thesis.

To this end, we will begin with a dimensional reduction of 11-dimensional supergravity on Calabi-Yau five-folds. The result is a one-dimensional effective action in the form of a non-linear sigma-model coupled to one-dimensional $\susyno=2$ supergravity. We will write this action in curved one-dimensional $\susyno=2$ superspace to make the supersymmetry manifest. In the next step, we will consider higher-order quantum corrections to the 11-dimensional action and their effects on the one-dimensional effective theory. At this point, we will also allow non-zero internal background flux to be present, which gives rise to a potential in the one-dimensional effective theory, thereby lifting parts of the otherwise totally flat moduli space. The question of moduli space dynamics is then examined both at the classical and at the quantum, with the latter being approached via a mini-superspace quantisation procedure. Due to the complexity of the dynamical equations, we will have to restrict our present analysis to the simplest cases which can be worked out explicitly, but do unfortunately not lead to realistic physical behaviour. 

Besides the potential cosmological applications, the one-dimensional effective theory possesses some peculiar features, which turn it into an interesting object to study in its own right. Even though the computational complexity is generally smaller than in higher dimensions, one-dimensional theories can have conceptual subtleties. Indeed, many concepts of ordinary field theory are put to the extreme in one dimension. The r\^{o}le of gravity, degree of freedom counting and the possibility of supersymmetric theories with a different number of on-shell bosonic and fermionic degrees of freedom are examples where subtleties arise. Some of these features have been proposed as explanations of puzzling properties of four-dimensional theories (see, for example, ref.~\cite{Witten:1995rz}).

Finally, we will take a first look at F-theory compactifications on Calabi-Yau five-folds to two dimensions. Here, we will restrict to a particular proposal for a 12-dimensional Lagrangian and its reduction to two dimensions. We will find that the resulting two-dimensional action allows an interpretation as a bosonic string moving in the moduli space of Calabi-Yau five-folds. Studying the supersymmetric completion of the two-dimensional theory may potentially shed some light on the structure of the elusive fermionic side of the 12-dimensional theory. Further motivation comes from the question of whether Calabi-Yau five-fold reductions favour a particular space-time signature and whether there is a relation to four dimensions.

\section{About this thesis}

The remainder of the thesis is organised as follows. In \chrangeref{mthy}{superspace}, some of the necessary background knowledge is developed to make the thesis self-contained and to put it into a wider perspective. These chapters do not contain new results but rather represent short reviews of some of the existing literature. We begin with \chref{mthy}, where 11-dimensional supergravity and relevant aspects of M-theory are briefly reviewed. This also serves to fix and explain our notation. 

\Chref{cy5} concentrates on the mathematics of Calabi-Yau manifolds. After presenting the definitions, general properties and the way they originally entered the physics literature, we specialise on five-folds. Most of the mathematical results on three- and four-folds carry over to the case of five-folds. However, there are also a few differences, which will be pointed out in due course. At the end of \chref{cy5}, we present constructions of explicit examples of Calabi-Yau five-folds. They are highly relevant as case studies in the application to M- and F-theory reductions later.

In \chref{superspace}, we develop flat and curved one-dimensional $\susyno=2$ superspace to the level of generality required for our purposes. This also proves useful as a check that our superspace conventions are self-consistent, which in turn is important since superspace methods are employed to infer results about supersymmetry later.

Most of the original research reported on in this thesis has been pursued in collaboration with Andre Lukas and Kellogg Stelle and has been published in ref.~\cite{Haupt:2008nu}. The findings are presented in \chref{mcy5} and at the beginning of \chref{msp}. The introductory \chrangeref{mthy}{superspace} are to a certain degree based on review sections of ref.~\cite{Haupt:2008nu}, but represent extended versions thereof.

\Chref{mcy5} sets out to provide a holistic study of the reduction of M-theory on Calabi-Yau five-folds. After analysing the consistency conditions arising from Calabi-Yau five-fold backgrounds (slightly extending the considerations in ref.~\cite{Haupt:2008nu}), we perform the actual reduction of both the bosonic and the fermionic side of 11-dimensional supergravity to obtain the one-dimensional effective action. We then use the results of \chref{superspace} to write the action in superspace. In the last part of \chref{mcy5}, we consider the implications of the presence of non-zero internal background flux for the one-dimensional theory. We find a scalar potential, which can be obtained from a Gukov-type superpotential under certain circumstances.

In \chref{msp}, we take a first look at the physical applications of our one-dimensional effective theory. We begin by studying some of its classical solutions. After introducing quantum geometrodynamics and mini-superspace quantisation, we present the author's unpublished work on some of the quantum aspects of the one-dimensional theory obtained from Calabi-Yau five-fold reductions. We quantise the resulting model and compare it to the mini-superspace quantisation of general relativity. In the last part of \chref{msp}, we solve the Wheeler-DeWitt equation for the simplest class of Calabi-Yau five-fold reductions. This should merely be regarded as an illustrative example as the solutions display forbidding physical behaviour. 

The technology of Calabi-Yau five-fold reductions is applied to a different setting in \chref{fcy5}. We go up in dimensions from 11 to 12 and reduce a 12-dimensional action, conjectured to be related to F-theory, on a Calabi-Yau five-fold to two-dimensions. This is also based on some of the author's currently unpublished work in progress. After summarising the proposal for a 12-dimensional action, we perform the dimensional reduction in the same spirit as \chref{mcy5}. The resulting two-dimensional action allows an interpretation as a bosonic string moving in the moduli space of Calabi-Yau five-folds, which has some interesting implications for the quantum consistency of the model in relation to topological properties of the corresponding Calabi-Yau five-fold. A central element of the 12-dimensional action is its connection with 11-dimensional supergravity. We therefore devote the last part of \chref{fcy5} to the question of how this connection manifests itself at the level of the lower-dimensional theories resulting from Calabi-Yau five-fold reductions.

We conclude the thesis in \chref{conclusions} with a summary of the main results and a list of possible future directions. After \chref{conclusions}, the reader will find two appendices. \Appref{conventions} contains an in-depth explanation of our conventions and notation. A particular emphasis is placed on index-free differential form language, which is used extensively throughout the other chapters. Finally, \appref{long_calcs} comprises a collection of some longer calculations and proofs, which were omitted in the main text to avoid distraction.
%
%

\chapter{\texorpdfstring{$D=11$}{D=11} Basics}\chlabel{mthy}

\section{11-dimensional supergravity \`{a} la Cremmer-Julia-Scherk}\seclabel{sugra11}

The field content of $D=11$ supergravity \`{a} la Cremmer-Julia-Scherk (CJS)~\cite{Cremmer:1978km} is remarkably simple compared to most of the supergravity theories in lower dimensions. It comprises the metric\footnote{Our conventions for indices, metric signature, symbols and other choices are summarised in \appref{conventions}\newnot{symbol:11ind} and in the list of symbols on pages~\pageref{LoS}-\pageref{LoSend}. Here, $M,N,\ldots=0,1,\ldots,10$ are $D=11$ curved space-time indices.} $g_{MN}$ (spin 2) and an Abelian 3-form gauge field $A$ (spin 1) on the bosonic side and the gravitino $\Psi_M$ (spin $3/2$) on the fermionic side. In $D$ dimensions, the metric, a $p$-form gauge field and the gravitino contain $\oneon{2} D (D-1)$, $\left( \begin{smallmatrix} D-1 \\ p \\ \end{smallmatrix} \right)$ and $(D-1) f$ off-shell degrees of freedom, respectively. In the same order, the counting of on-shell degrees of freedom is $\oneon{2} D (D-3)$, $\left( \begin{smallmatrix} D-2 \\ p \\ \end{smallmatrix} \right)$ and $\oneon{2} (D-3) f$, respectively. The quantity $f$ in the formul\ae\ for the degrees of freedom of the gravitino represents the dimension of the smallest irreducible spinor representation of $\SO(D-1,1)$, for example $f=32$ for $D=11$. Derivations of the counting formul\ae\ can be found, for example, in ref.~\cite{deWit:1997sz}. In 11-dimensional supergravity, the metric, 3-form and gravitino comprise 44, 84 and 128 independent on-shell degrees of freedom, respectively. There is an equal number of bosonic ($44+84$) and fermionic (128) degrees of freedom as required by supersymmetry.

Using only the fields introduced above, requiring diffeomorphism invariance, local Lorentz invariance, local supersymmetry, Abelian gauge invariance $\delta A = d\Lambda$ and stopping at the two derivative level, leads to the unique 11-dimensional classical CJS action~\cite{Cremmer:1978km}
\begin{align}
	S_{\rm CJS} &= S_{\rm CJS,B} + S_{\rm CJS,F} \; , \eqlabel{S_CJS} \\
	S_{\rm CJS,B} &= \oneon{2\kappa^2_{11}} 
	  \int_\manifold \left\{ R \ast 1 - \oneon{2} G\wedge\ast G - \oneon{6} G\wedge G \wedge A \right\} , \eqlabel{S_CJS_B} \\ \begin{split}
	S_{\rm CJS,F} &= - \oneon{2\kappa^2_{11}} \int_\manifold d^{11}x \sqrt{-g} \left\{ \vphantom{\oneon{96}}
	  \bar{\Psi}_M \Gamma^{MNP} D_N \left(\frac{\omega+\hat{\omega}}{2}\right) \Psi_P \right. \\ &\quad\;+ \left.
	  \oneon{192} \left( \bar{\Psi}_M \Gamma^{MNPQRS} \Psi_S + 12 \bar{\Psi}^N \Gamma^{PQ} \Psi^R \right) (G_{NPQR} +\hat{G}_{NPQR}) 
	  \right\} , \end{split} \eqlabel{S_CJS_F}
\end{align}
where $\kappa_{11}$ is the 11-dimensional gravitational constant, $G=dA$ is the 4-form field strength of the 3-form gauge field $A$, $R$ is the Ricci scalar of the 11-dimensional metric $g_{MN}$, ${\cal M}$\newnot{symbol:mfld} is the space-time manifold and $g\equiv\det{g_{MN}}$. We use index-free notation where possible. Our conventions for differential forms are summarised in \appref{forms}. In local coordinates, the bosonic action~\eqref*{S_CJS_B} becomes
\be
	S_{\rm CJS,B} = \oneon{2\kappa^2_{11}} \int_\manifold d^{11}x \left\{ \sqrt{-g} R - \frac{\sqrt{-g}}{48} G^2 + \oneon{(3!4!)^2} GGA \right\} ,
\ee
where $G^2 \equiv G_{MNPQ} G^{MNPQ}$ and $GGA \equiv G_{M_1 \ldots M_4} G_{M_5 \ldots M_8} A_{M_9 \ldots M_{11}} \varepsilon^{M_1 \ldots M_{11}}$. 

The gravitino $\Psi_M$ is an 11-dimensional Majorana spinor,\footnote{Our conventions for spinors are summarised in \appref{spinorconv}.} the Dirac conjugate $\bar{\Psi}_M$ is given by $\bar{\Psi}_M = i \Psi_M^\dagger \Gamma^\UZ$ and the Majorana representation is chosen for the gamma matrices $\Gamma^M$. Note that the gamma matrices appearing in \eqref*{S_CJS_F} are curved gamma matrices, related to the flat space gamma matrices by a contraction with the inverse vielbein $\Gamma^M = e_{\underline{N}}^M \Gamma^{\underline{N}}$, where underlined indices denote tangent space (local Lorentz) indices. The covariant derivative $D_N (\omega)$ is defined by
\be\eqlabel{covder}
	D_N (\omega) \Psi_P = \left(\partial_N + 
		\oneon{4} {\omega_N}^{\underline{Q}\underline{R}} \Gamma_{\underline{Q}\underline{R}} \right) \Psi_P \; , 
\ee
with spin connection ${\omega_N}^{\underline{Q}\underline{R}}$. The supercovariant tensors $\hat\omega$ and $\hat{G}$ are given by
\begin{align}
	\hat{\omega}_{MNP} &= \omega_{MNP} - \oneon{4} \bar{\Psi}^Q \Gamma_{MNPQR} \Psi^R \; , \eqlabel{scov_omega} \\
	\hat{G}_{MNPQ} &= G_{MNPQ} - 6 \bar{\Psi}_{\left[M\right.} \Gamma_{NP} \Psi_{\left. Q\right]} \; . \eqlabel{scov_G}
\end{align}
They are responsible for introducing terms quartic in the gravitino into the fermionic action~\eqref*{S_CJS_F}, which are necessary for supersymmetry to work. The particular form in \eqrangeref{scov_omega}{scov_G} renders them supercovariant, which means their supersymmetry variations do not contain derivatives of the infinitesimal supersymmetry parameter $\epsilon^{(11)}$. The structure of the 4-fermi interactions is rather complicated and hence, in most of what follows, these terms are ignored. It is therefore appropriate to re-write the fermionic action~\eqref*{S_CJS_F} with the 4-fermi terms separated out
\begin{multline}
	S_{\rm CJS,F} = - \oneon{2\kappa^2_{11}} \int_\manifold d^{11}x \sqrt{-g} \left\{ \vphantom{\oneon{96}}
		\bar{\Psi}_M \Gamma^{MNP} D_N(\omega) \Psi_P \right. \\ + \left.
		\oneon{96} \left( \bar{\Psi}_M \Gamma^{MNPQRS} \Psi_S + 12 \bar{\Psi}^N \Gamma^{PQ} \Psi^R \right) G_{NPQR}  
		\right\} + S_{\rm CJS,F,4} \; , \eqlabel{S_CJS_F1}
\end{multline}
where just for completeness, we state
\begin{multline}
	S_{\rm CJS,F,4} = \oneon{2\kappa^2_{11}} \oneon{32} \int_\manifold d^{11}x \sqrt{-g} \left\{
		\vphantom{\Psi^{\left. R\right]}}
		\left(\bar{\Psi}_M \Gamma^{MNP}\Gamma^{QR} \Psi_P \vphantom{\Psi^{\left. R\right]}} \right) 
		\left(\bar{\Psi}^S \Gamma_{NQRST} \Psi^T \vphantom{\Psi^{\left. R\right]}} \right) \right. \\ \left. + 
		\left(\bar{\Psi}_M \Gamma^{MNPQRS} \Psi_S  + 
		 12 \bar{\Psi}^{\left[N\right.} \Gamma^{PQ} \Psi^{\left. R\right]} \right) 
		\left(\bar{\Psi}_N \Gamma_{PQ} \Psi_R \vphantom{\Psi^{\left. R\right]}} \right) \right\} .
\end{multline}
Contributions from the 4-fermi terms in $S_{\rm CJS,F,4}$ to other equations are henceforth abbreviated to $(\text{fermi})^4$ (or $(\text{fermi})^3$ in case of supersymmetry transformations) and not considered explicitly.

Varying the action~\eqref*{S_CJS} yields the equations of motion (for a step-by-step derivation, we refer to ref.~\cite{Miemiec:2005ry})
\begin{gather}
	R_{MN} = \oneon{12} G_{M M_2\ldots M_4}{G_N}^{M_2\ldots M_4} - 
	 	\oneon{144} g_{MN} G_{M_1\ldots M_4}G^{M_1\ldots M_4} \; , \eqlabel{sugra11_geom} \\
	d \ast G + \oneon{2} G\wedge G = 0 \; , \eqlabel{sugra11_Geom_class} \\
	\Gamma^{MNP} \hat{D}_N \Psi_P = 0 \; , \eqlabel{sugra11_gravitino_eom}
\end{gather}
where $R_{MN}$ is the Ricci tensor of the metric $g_{MN}$ and the supercovariant derivative $\hat{D}_N$ is defined as
\be
	\hat{D}_M = D_M (\hat{\omega}) + \oneon{288} \left(\Gamma_M {}^{NPQR} - 8 \delta_M^N \Gamma^{PQR} \right) \hat{G}_{NPQR} \; .
\ee
In \eqref{sugra11_geom,sugra11_Geom_class}, the contributions from the fermionic action~\eqref*{S_CJS_F} are missing. However, we do not need these contributions for we will use the field equations solely to study supergravity backgrounds with vanishing fermions. Ignoring the 4-fermi terms in \eqref{S_CJS_F}, one may write \eqref{sugra11_gravitino_eom} as
\be\eqlabel{sugra11_gravitino_eom1}
	\Gamma^{MNP} D_N(\omega) \Psi_P + \oneon{96} \left( \Gamma^{MNPQRS} \Psi_S 
			+ 12 g^{MN} \Gamma^{PQ} \Psi^R \right) G_{NPQR} + (\text{fermi})^3 = 0 \; .
\ee
In local coordinates, the 3-form equation of motion~\eqref*{sugra11_Geom_class} reads
\be\eqlabel{sugra11_Geom_class_loc_coords}
	\nabla_M G^{MNPQ} + \oneon{2 (4!)^2} G_{M_1\ldots M_4}G_{M_5\ldots M_8}\epsilon^{M_1\ldots M_8 NPQ} = 0 \; .
\ee

The classical CJS action~\eqref*{S_CJS} is invariant under local supersymmetry transformations of the form
\begin{align}
	\delta_{\epsilon^{(11)}} g_{MN} &= 2 \bar{\epsilon}^{(11)} \Gamma_{\left(M\right.} \Psi_{\left.N\right)} \; , \eqlabel{sugra11_susy_transf_g} \\
	\delta_{\epsilon^{(11)}} A_{MNP} &= -3 \bar{\epsilon}^{(11)} \Gamma_{\left[MN\right.} \Psi_{\left.P\right]} \; , \eqlabel{sugra11_susy_transf_A} \\
	\delta_{\epsilon^{(11)}} \Psi_M &= 2\hat{D}_M \epsilon^{(11)} \; , \eqlabel{sugra11_susy_transf_gravitino}
\end{align}
which are parameterised by an 11-dimensional anti-commuting Majorana spinor $\epsilon^{(11)}$. Again, ignoring the 4-fermi terms in \eqref{S_CJS_F}, one may write \eqref{sugra11_susy_transf_gravitino} as
\be\eqlabel{sugra11_susy_transf_gravitino1}
	\delta_{\epsilon^{(11)}} \Psi_M  = 
	   2 D_M (\omega) \epsilon^{(11)} + \oneon{144} (\Gamma_M {}^{NPQR} - 8 \delta_M^N \Gamma^{PQR}) \epsilon^{(11)} G_{NPQR} + 
	   (\text{fermi})^3 \; .
\ee
The actual proof that the action~\eqref*{S_CJS} is invariant under the transformations~\eqrangeref*{sugra11_susy_transf_g}{sugra11_susy_transf_gravitino} is quite long and in parts tedious. On the other hand, we believe it is instructive to go through it, particularly in view of the necessary manipulations of fermionic terms, which are similar to the calculations in \secref{ferm_red} on the fermionic reduction. Moreover, we will later use the supersymmetry of the action~\eqref*{S_CJS} to infer key facts about the kind of supersymmetry realised in the one-dimensional models obtained by dimensional reduction and therefore, it is very important to ascertain ourselves that there are no mistakes in the equations written above. However, presenting the proof here would disrupt the flow of this short summary, which is why we instead opted for showing the proof in the appendix and the interested reader is kindly referred to \appref{proof_CJS_action_susy} for the detailed proof.

\section{M-theory snippets}\seclabel{sugra_mthy_corr}

Dimensionally reducing\footnote{For a general introduction to Kaluza-Klein reductions, we refer to refs.~\cite{Duff:1986hr,Becker:2007zj}.} the 11-dimensional CJS action~\eqref*{S_CJS} on a circle leads to the so-called type IIA supergravity in ten dimensions~\cite{Giani:1984wc,Campbell:1984zc,Huq:1983im}, whose bosonic action (in the ``string frame'') reads
\begin{multline}\eqlabel{sugra_IIA}
	S_{\text{IIA}} = \frac{1}{2\kappa_{10}^2}\int \left\{
	e^{-2\phi} \left[ R\ast 1 + 4\, d\phi\wedge\ast d\phi - \oneon{2}H\wedge\ast H \right] \right. \\ \left. 
	- \oneon{2} F\wedge\ast F - \oneon{2} G\wedge\ast G
	- \oneon{2} G\wedge G\wedge B \right\} ,
\end{multline}
with metric $g$, dilaton $\phi$, NS-NS 2-form $B$, R-R 1-form $C$, R-R 3-form $A$ and corresponding field strengths $F= dC$, $H = dB$ and $G = dA + H\wedge C$. In order to arrive at~\eqref*{sugra_IIA}, the following compactification ansatz has been made for the 11-dimensional bosonic fields
\begin{gather}
	d\hat{s}^2 = e^{-\frac{2}{3}\phi} ds^2 +e^{\frac{4}{3}\phi}(dx^{10}+C_m dx^m)^2 \; , \eqlabel{sugra_11_10_metric_ansatz} \\
	\hat{A}_{mn\,10} = B_{mn} \; , \qquad \hat{A}_{mnp} = A_{mnp} \; , \eqlabel{sugra_11_10_3form_ansatz}
\end{gather}
where, for distinction, hatted objects denote 11-dimensional fields and $m,n,\ldots = 0,\ldots,9$ are 10-dimensional curved indices. In addition, one learns a relation between the radius $R$ of the compactified 11th dimension and the gravitational constants $\kappa_{10}$ and $\kappa_{11}$
\be\eqlabel{sugra_11_10_kappa}
	\kappa_{11}^2 = 2\pi R \, \kappa_{10}^2 \; .
\ee

Type IIA supergravity is the low-energy effective theory of the 10-dimensional type IIA superstring~\cite{GSW,Ortin:2004ms,Becker:2007zj} with parameters identified as follows
\be\eqlabel{sugra_11_10_kappa10_ls_gs}
	2 \kappa_{10}^2 = (2\pi)^7 l_s^8 g_s^2 \; ,
\ee
where $l_s = \sqrt{\al^\prime}$ is the string length and $g_s = e^{\left\langle\phi\right\rangle}$ is the type IIA superstring coupling constant. From~\eqref*{sugra_11_10_metric_ansatz} it follows that a distance measured in 10-dimensional string units is equal to $g_s^{1/3}$ times the same distance measured in 11-dimensional Planck units. The fundamental length scales in 10 and 11 dimensions are thus related by
\be\eqlabel{sugra_11_10_lp_ls_gs}
	l_p = g_s^{1/3} l_s \; ,
\ee
where $l_p$ is the 11-dimensional Planck length $2\kappa_{11}^2 = \oneon{2\pi} (2\pi l_p)^9$. Plugging \eqref{sugra_11_10_kappa10_ls_gs,sugra_11_10_lp_ls_gs} into \eqref{sugra_11_10_kappa} yields
\be
	R = g_s^{2/3} l_p = g_s l_s \; ,
\ee
suggesting that, in the strong coupling limit, $g_s \rightarrow \infty$, of the type IIA superstring, an 11th dimension grows large and the full non-perturbative theory becomes 11-dimensio\-nal~\cite{Witten:1995ex,Townsend:1995kk}. The conjectured 11-dimensional theory -- dubbed M-theory -- is characterized by reproducing 10-dimensional type IIA superstring theory upon dimensional reduction on a circle and having 11-dimensional CJS supergravity as its low-energy limit.\footnote{This is by no means all the evidence there is for M-theory. The conjecture is also supported by arguments involving various branes and dualities. The presentation in this section is streamlined to the context and applications discussed in this thesis. More complete reviews of M-theory can be found, for example, in refs.~\cite{Duff:1996aw,Townsend:1996xj,Sezgin:1997qf,Argurio:1998cp,Becker:2007zj}.}

In its r\^{o}le as the low-energy effective theory of M-theory, the CJS action~\eqref*{S_CJS} receives an infinite series of higher-order derivative corrections which are organised by integer powers of the quantity\newnot{symbol:beta}
\be
	\beta =\frac{1}{(2\pi )^2}\left(\frac{\kappa_{11}^2}{2\pi^2}\right)^{2/3} \; ,
\ee
and can be written schematically in a perturbative expansion as
\be\eqlabel{S11_pert_series}
	S_{11} = S_{\rm CJS} + \sum_{i=1}^\infty \beta^i \, S_{11,(i)} \; .
\ee
In this thesis, we work at most up to order $\mathcal{O}(\beta^1)$, at which the famous $R^4$ terms appear
\be\eqlabel{S11}
	S_{11} = S_{\rm CJS} + \beta\, S_{11,(1)} + \ldots = S_{\rm CJS} + \beta\, S_{11,\rm GS} + \beta\, S_{11,R^4} + \ldots \; .
\ee
The two distinct terms $S_{11,\rm GS}$ and $S_{11,R^4}$ present at $\mathcal{O}(\beta^1)$ are discussed below.

The first correction term we consider is the so-called Green-Schwarz term~\cite{Duff:1995wd}
\be\eqlabel{sugra11_gs}
	S_{11,\rm GS} = - \frac{(2\pi )^4}{2\kappa_{11}^2} \int_\manifold A \wedge X_8 ,
\ee
where $X_8$ is a quartic polynomial in the curvature 2-form $\curvtwofrm^\ul{M} {}_\ul{N} \equiv \oneon{2} R^\ul{M} {}_{\ul{N}PQ} dx^P \wedge dx^Q$. It can be conveniently expressed in terms of the first and second Pontrjagin classes~\cite{Nakahara:1990th}, $p_1(\manifold )$ and $p_2(\manifold )$, of $\manifold$ as
\bea\eqlabel{def_x8}
	X_8 &= \oneon{48} \left(\left(\frac{p_1}{2}\right)^2 - p_2\right), \\
	p_1(\manifold)  &= - \oneon{2} \left(\oneon{2\pi}\right)^2 \tr \, \curvtwofrm^2 \; , \\
	p_2(\manifold)  &=   \oneon{8} \left(\oneon{2\pi}\right)^4 \left((\tr\,\curvtwofrm^2)^2 - 2\,\tr\,\curvtwofrm^4\right) ,
\eea
thereby making its topological nature manifest. For more details on Pontrjagin classes and $X_8$, we refer to \appref{cp_classes}. The existence of~\eqref*{sugra11_gs} is inferred from M5-brane worldvolume anomaly cancellation~\cite{Duff:1995wd}. Signs and pre-factors are determined by supersymmetry and the anomaly cancellation condition on the five-brane world volume~\cite{Bilal:2003es,deAlwis:1996ez,deAlwis:1996hr}. In view of later chapters, it is important to notice that the relative sign between $S_{11,\rm GS}$ and the Chern-Simons term, $GGA$, is positive in our conventions. The Green-Schwarz term~\eqref*{sugra11_gs} leads to a correction to the equation of motion~\eqref*{sugra11_Geom_class} for $A$, which now reads
\be\eqlabel{sugra11_3form_eom}
	d \ast G =- \oneon{2} G\wedge G - (2\pi)^4 \beta X_8 \; .
\ee
We note that the exactness of $d\ast G$ implies that the eight-form $\frac{1}{2}G\wedge G+(2\pi )^4\beta X_8$ must be cohomologically trivial on $\manifold$. This integrability condition will become important later when discussing Calabi-Yau five-fold compactifications.

The second term appearing at order $\mathcal{O}(\beta^1)$ comes from uplifting a known 10-di\-men\-sio\-nal counterterm of the type IIA superstring,
\be
	S_{{\rm IIA},R^4} \sim \int d^{10} x \sqrt{-g} \, t_8^{m_1\dots m_8} t_8^{n_1\dots n_8} R_{m_1m_2n_1n_2}\dots R_{m_7m_8n_7n_8} \; ,
\ee
to 11 dimensions~\cite{Green:1997di,Green:1997as,Russo:1997mk}. The famous rank eight tensor $t_8$ has been defined in ref.~\cite{Schwarz:1982jn}. In our convention, the uplift to 11 dimensions produces
\be\eqlabel{R4}
	S_{11,R^4} = \oneon{2\kappa_{11}^2}\frac{1}{9\cdot 2^{11}}\int_{\cal M} d^{11} x \sqrt{-g}\, 
	t_8^{M_1\dots M_8} t_8^{N_1\dots N_8} R_{M_1M_2N_1N_2}\dots R_{M_7M_8N_7N_8} \; ,
\ee
where the 11-dimensional $t_8$ corresponds to the 10-dimensional $t_8$ with the index range extended by one. Since $t_8$ contains a term proportional to the 8-index $\varepsilon$-symbol, \eqref*{R4} contains a piece proportional to the product of two 8-index $\varepsilon$-symbols, which should be regarded as a formal expression that is to be replaced by products of Kron\-ecker-$\delta$s according to the fundamental $\varepsilon$-symbol identity~\eqref*{epsilon_identity1}. Supersymmetry in 11 dimensions relates $S_{11,R^4}$ to $S_{11,\rm GS}$ and therefore provides a way of checking the sign and pre-factor in~\eqref*{R4}.

\section[Branes in 11-dimensional supergravity]{Branes in 11-dimensional supergravity\footnote{This short section is specifically tailored to include -- out of this vast subject -- only what will be needed in later chapters of this thesis. For a review of the subject, we refer to, for example, ref.~\cite{Stelle:1998xg}.}}

As part of the $D=11$ super-Poincar\'e algebra acting on the objects defined in \secref{sugra11}, the most general anti-commutation relations amongst the 32-component Majorana supercharges $Q_\alpha$, allowing central charges, reads as follows~\cite{vanHolten:1982mx,Townsend:1995gp}
\be
	\{ Q_\alpha, Q_\beta \} = (\Gamma^M C)_{\alpha\beta} P_M +
		\oneon{2!} (\Gamma^{MN} C)_{\alpha\beta} Z_{MN} +
		\oneon{5!} (\Gamma^{MNPQR} C)_{\alpha\beta} Z_{MNPQR} \; ,
\ee
where $\alpha,\beta,\ldots=1,\ldots,32$ are 11-dimensional spinor indices and $C$ is the charge conjugation matrix defined in~\eqref*{gamma11_D_and_C}. The 528 components of the left hand side are split into $528=11+55+462$ on the right hand side. The tensors $Z_{MN}$ and $Z_{MNPQR}$ are called central charges for they commute with all the other generators of the super-Poincar\'{e} algebra. These 2-form and 5-form charges are \emph{inter alia}\footnote{The other objects arising from $Z_{MN}$ and $Z_{MNPQR}$ are the so-called MW- and KK6-branes corresponding to waves and Kaluza-Klein monopoles, respectively. The associated supergravity solutions are purely gravitational, that is $A = \Psi_M = 0$. In addition, there is an M9-brane, which couples to a non-dynamical 10-form. For more details, see, for example, ref.~\cite{Hull:1997kt}.} carried by 2-branes and 5-branes, respectively. To highlight the M-theory context, these are denoted M2- and M5-branes.

These branes have indeed been found as classical $1/2$-BPS solutions of the classical CJS action~\eqref*{S_CJS}~\cite{Duff:1990xz,Gueven:1992hh}. The fundamental ``electric'' M2-brane solution~\cite{Duff:1990xz} takes the form
\begin{equation}\eqlabel{soliton_m2_brane}
\begin{aligned}
	ds^2 &= H^{-\frac{2}{3}}\eta_{\al\beta}d\sigma^\al d\sigma^\beta + H^{\frac{1}{3}}(dr^2+r^2 d\Omega_7^2) \; , \\
	A &= H^{-1} d\sigma^0\wedge d\sigma^1\wedge d\sigma^2 \; , \qquad \Psi_M = 0 \; ,
\end{aligned}
\end{equation}
where $d\Omega_7$ is the volume element of the 7-dimensional unit sphere $S^7$\newnot{symbol:Sn}, $H \equiv 1+\frac{k_3}{r^6}$ with integration constant $k_3>0$ and $\sigma^\al$, $\al,\beta,\ldots=0,1,2$, are the coordinates on the M2-brane world-volume. Near the M2-brane, which is located at $r=0$, the geometry is $AdS_4\times S^7$ breaking $\SO(10,1)$ to $\SO(2,1)\times\SO(8)$. Far away ($r\rightarrow\infty$) from the M2-brane, the space-time is approximated by 11-dimensional Minkowski-space $\M^{11}$\newnot{symbol:MD} with full $\SO(10,1)$.

For completeness, we also state the solitonic ``magnetic'' M5-brane solution~\cite{Gueven:1992hh}
\begin{equation}\eqlabel{soliton_m5_brane}
\begin{aligned}
	ds^2 &= H^{-\frac{1}{3}}\eta_{\al\beta}d\sigma^\al d\sigma^\beta + H^{\frac{2}{3}}(dr^2+r^2 d\Omega_4^2) \; , \\
	\ast A &= H^{-1} d\sigma^0\wedge\cdots\wedge d\sigma^5 \; , \qquad \Psi_M = 0 \; ,
\end{aligned}
\end{equation}
where $d\Omega_4$ is the volume element of the 4-dimensional unit sphere $S^4$, $H \equiv 1+\frac{k_6}{r^3}$ with integration constant $k_6>0$ and $\sigma^\al$, $\al,\beta,\ldots=0,\ldots,5$, are the coordinates on the M5-brane world-volume. This solution interpolates between $AdS_7\times S^4$ near the M5-brane at $r=0$ and $\M^{11}$ far away, $r\rightarrow\infty$, with isometries $\SO(5,1)\times\SO(5)$ and $\SO(10,1)$, respectively.

The classical action for the M2-brane is given by
\be\eqlabel{membraneaction}
	S_3 = -T_3 \int_{{\cal M}^3} \left\{d^3\sigma\,\sqrt{-\tilde{g}}+\tilde{A}\right\} ,
\ee
where $\tilde{g}$ and $\tilde{A}$ are the pullbacks of the $11$-dimensional metric $g_{MN}$ and 3-form $A$ under the embedding $x^M = x^M(\underline{\sigma} )$ of the M2-brane world-volume ${\cal M}^3$ into space-time ${\cal M}$. The M2-brane tension $T_3$ is given by
\be\eqlabel{T3def}
	T_3 = \frac{1}{2\pi\sqrt{\beta}} \; .
\ee
In the presence of M2-branes, the action~\eqref*{membraneaction} must be added to the 11-dimensional action~\eqref*{S11}. Since the M2-brane couples to the supergravity fields, it alters their equations of motion and can, in particular, affect the integrability of \eqref{sugra11_3form_eom} making it necessary to take it into account. Re-computing the equation of motion for $A$ including the contribution from \eqref*{membraneaction} leads to
\be\eqlabel{Geom_corr}
	d \ast G =- \oneon{2} G\wedge G - (2\pi)^4 \beta X_8  -2\kappa_{11}^2 T_3\,\delta({\cal M}^3) \; .
\ee
Here, $\delta ({\cal M}^3)$ is an 8-form current associated with the M2-brane world-volume. It is characterised by the property
\be
	\int_{{\cal M}^3}w = \int_{\cal M}w\wedge\delta ({\cal M}^3)
\ee
for any 3-form $w$.
%
%

\chapter{Calabi-Yau five-folds}\chlabel{cy5}

In this chapter, we develop and present some mathematical background knowledge on Calabi-Yau five-folds necessary to perform dimensional reductions on these spaces. As such, this represents the most mathematical chapter (besides the appendices, of course). It is independent of the rest of the text and is meant to provide a brief yet stand-alone introduction to the subject of Calabi-Yau manifolds. Readers familiar with the subject may wish to skip this chapter and instead use it as a reference to look up formul\ae.

We will indicate when a result is not only valid for Calabi-Yau five-folds in particular, but more generally for Calabi-Yau $n$-folds, where $n$ is the complex dimension of the Calabi-Yau manifold. Many of the results discussed here can also be found in the mathematical literature, such as refs.~\cite{Candelas:1987is,Nakahara:1990th}.

\section{Definition and basic properties}\seclabel{cy_def}

A Calabi-Yau $n$-fold $X$\newnot{symbol:CY5}, with $n=\dim_\C X$, is defined to be a smooth, compact K\"{a}hler manifold with vanishing first Chern class $c_1 (X) = 0$. The requirements of compactness or smoothness are sometimes given up in some of the physics literature in order to study generalisations of the strict mathematical concept of a Calabi-Yau manifold. However, we restrict our attention to the previously stated narrower definition in the present thesis.

A K\"{ahler} manifold $X$ is a Hermitian manifold whose K\"{ahler} form $J$ is closed, $dJ = 0$. The K\"{a}hler form of a Hermitian manifold is defined as $J \equiv i g_{\mu\bar\nu} dz^\mu \wedge d\bar{z}^{\bar\nu}$, where $g_{\mu\bar\nu}$ is a Hermitian metric on $X$ and $z^\mu$, $\mu,\nu,\ldots = 1,\ldots,n$, are complex (holomorphic) local coordinates on $X$. Our full list of conventions used for complex manifolds can be found in \appref{complex_geometry}. (Note, however, that in order to make contact with 11 dimensions, the notation is slightly changed so that the Hermitian metric is denoted $g_{\mu\bar\nu}$ in this chapter as opposed to $G_{\mu\bar\nu}$ in the appendix.)

It was conjectured by Calabi~\cite{Calabi:1955} and later proved by Yau~\cite{Yau:1977ms} that the condition $c_1 (X) = 0$ is equivalent to the existence of a Ricci-flat metric if $X$ is K\"ahler. This explains the naming of these manifolds. The Calabi-Yau condition is mathematically equivalent to any one of the statements that the global holonomy group of $X$ is contained in $\SU(n)$, that there exists a (up to a constant) unique nowhere vanishing holomorphic $(n,0)$-form $\Omega$ and that there exists a pair of globally defined covariantly constant spinors $\eta$\newnot{symbol:eta} and $\eta^\star$ of opposite (same) chirality for $n$ odd (even).

Henceforth, the Ricci-flat metric is denoted by $g_{\mu\bar\nu}$. The Ricci-flatness of $g_{\mu\bar\nu}$ together with the existence and properties of $\Omega$ imply that $\Omega$ is harmonic and covariantly constant with respect to $g_{\mu\bar\nu}$. It is thus locally of the form
\be
	\Omega = \oneon{n!} f(z) \epsilon_{\mu_1\ldots\mu_n} dz^{\mu_1} \wedge \cdots \wedge dz^{\mu_n} \; ,
\ee
with a local holomorphic function $f(z)$ that is non-zero in its respective coordinate patch. As a consequence, $\Omega$ is intimately related to the invariant volume element $\Omega_X$ defined in~\eqref*{volume_form}
\be\eqlabel{omegaX_hol_50_form}
	\Omega_X = \ast 1 = \frac{J^n}{n!} = \frac{i^n (-1)^{\frac{n(n-1)}{2}}}{||\Omega||^2} \Omega \wedge \bar{\Omega} \; . 
\ee
where we have introduced $||\Omega||^2 \equiv \Omega_{\mu_1\ldots\mu_n} \bar{\Omega}^{\mu_1\ldots\mu_n} / n!$. We remark that locally, $||\Omega||^2 = |f(z)|^2$, which is useful in some computations.

The existence of the covariantly constant spinor $\eta$ implies that $J$ and $\Omega$ can be expressed as spinor bilinears
\be
	J_{\mu\bar{\nu}}= i \eta^\dagger \gamma_{\mu\bar{\nu}} \eta\; ,\qquad
	\Omega_{\mu_1 \ldots \mu_n}=||\Omega || \eta^\dagger \gamma_{\mu_1 \ldots \mu_n} \eta^\star \;  , \eqlabel{JOdef} 
\ee
where $\gamma^\mu$ are the curved gamma matrices on $X$ in local holomorphic coordinates and the index on $\gamma^\mu$ is raised and lowered using the Ricci-flat metric $g_{\mu\bar\nu}$. (Our spinor and gamma matrix conventions on Calabi-Yau five-folds are summarised in \appref{spinorconv}.) Throughout this thesis, $\eta$ is assumed to be normalised to unity, that is $\eta^\dagger \eta = 1$. The two expressions in~\eqref*{JOdef} define an $\SU(n)$-structure on $X$ and are the only two non-vanishing non-trivial spinor bilinears on $X$.

In this thesis, dimensional reductions on Calabi-Yau manifolds (specifically, five-folds) are considered. In order to avoid non-generic cases of reductions, we require Calabi-Yau manifolds to have a sufficiently large global holonomy group ${\rm Hol}(X) \subseteq \SU(n)$ such that they allow only one out of $2^{n-1}$ supersymmetries. This excludes, for example, spaces such as $2n$-tori $T^{2n}$\newnot{symbol:Tn} and direct products of the form $\nfold{n-m}\times\nfold{m}$ and $\nfold{n-m}\times T^{2m}$, since all of them preserve a larger amount of supersymmetry.

\section{Why Calabi-Yau?}

Before examining the mathematical properties of Calabi-Yau manifolds more thoroughly, we first of all stop for a brief detour into why Calabi-Yau manifolds play such an important r\^ole in string-/M-theory.

Calabi-Yau manifolds, specifically Calabi-Yau three-folds, first entered the phy\-sics literature as a result of investigating~\cite{Candelas:1985en} how to obtain 4-dimensional effective actions with a minimal amount of supersymmetry from the 10-dimensional string theories. The original line of thought of ref.~\cite{Candelas:1985en} is briefly summarised in this section.

In order to be left with a minimal amount of supersymmetry, that is $\susyno=1$, in 4 dimensions, it is a good idea to start with a theory that has minimal $\susyno=1$ supersymmetry in 10 dimensions, which makes the heterotic string a natural candidate. The massless bosonic modes of the heterotic string are described by the effective action
\be\eqlabel{sugra_het}
	S_{\text{het}} = \frac{1}{2\kappa_{10}^2}\int e^{-2\phi} \left\{
		R\ast 1 + 4\, d\phi\wedge\ast d\phi - \oneon{2} \tilde{H}\wedge\ast \tilde{H} - \frac{\al^\prime}{4} \mathrm{Tr}\, (F\wedge\ast F) \right\} ,
\ee
with metric $g$, dilaton $\phi$, NS-NS 2-form $B$, non-Abelian $E_8 \times E_8$ Yang-Mills 1-form gauge potential $A$ and corresponding field strengths $F= dA$ and $\tilde{H} = dB + \text{CS-terms}$.

The following three conditions are imposed in order to obtain a phenomenologically interesting 4-dimensional model. First, the 10-dimensional space-time manifold $\manifold$ is factorised such that $\manifold = \manifold^4 \times X$ with $\manifold^4$ being a 4-dimensional maximally symmetric Lorentzian manifold and $X$ being a 6-dimensional compact manifold. Being maximally symmetric, $\manifold^4$ must correspond to either Minkowski space $\M^4$, de Sitter space $dS_4$ or anti-de Sitter space $AdS_4$. The second condition is that, 4 out of the 16 local supersymmetries are left unbroken after the dimensional reduction thereby leaving local $\susyno=1$ supersymmetry in 4 dimensions. The third and final condition is that $d\phi$ and $H=dB$ can be consistently set to zero on $\manifold$.

The second of these three conditions places the strongest constraints. First, it implies the vanishing of the 4-dimensional Ricci-scalar $R^{(4)} = 0$, forcing $\manifold^4$ to be 4-dimensional Minkowski space $\manifold^4 = \M^4$. Two further necessary conditions for unbroken $\susyno=1$ supersymmetry are a constant dilaton, which is already catered for by imposing the third condition, and the existence of a covariantly constant spinor $\eta$ on $X$. From the covariant constancy $D_m \eta = 0$ of $\eta$, an important integrability condition follows
\be
	[D_m, D_n] \, \eta = \oneon{4} R^{(6)}_{mnpq} \gamma^{pq} \eta = 0 \; ,
\ee
where $R^{(6)}_{mnpq}$ is the Riemann tensor of $X$. After contracting with $\gamma^n$, this implies $R^{(6)}_{mn} = 0$, that is $X$ must be Ricci-flat. The covariantly constant spinor $\eta$ can be used to construct
\be\eqlabel{J_spin_bilin_CY3}
	J_m {}^n = - i \bar\eta \gamma_m {}^n \eta \; ,
\ee
which can be shown to satisfy $\nabla_m J_{np} = 0$ and $J_m {}^n J_n {}^p = - \delta_m {}^p$. That means, $X$ is K\"ahler with K\"ahler form $J$ as in~\eqref*{J_spin_bilin_CY3} and Hermitian metric $g^{(6)}_{mn}$ satisfying $J_m {}^p J_n {}^q g^{(6)}_{pq} = g^{(6)}_{mn}$. By Calabi's theorem, $X$ must therefore be a Calabi-Yau three-fold, or in other words ${\rm Hol}(X) \subseteq \SU(3)$, in order to satisfy the conditions listed above.

To complete the argument, the third condition also needs to be checked. It can in fact be satisfied by embedding the spin connection $\omega_m {}^{\underline{p}\underline{q}}$ of $X$ into the $E_8\times E_8$ gauge group, which is sufficiently large to permit such an embedding. This leaves an $E_6\times E_8$ gauge part and therefore amounts to a breaking scheme $E_8\times E_8 \rightarrow \SU(3)\times E_6\times E_8$. The gauge group can be broken down further by other means. Finally, we mention that the number $\# f$ of massless fermion generations is related to the Euler number $\eta(X)$ of $X$ by $\# f = |\eta(X)|/2$, which is an example of the important class of physical quantities that are determined by purely topological properties of the internal manifold.

\section{Calabi-Yau topology}\seclabel{cy_topology}

For complex manifolds, the $(n+1)^2$ Hodge numbers $h^{p,q}(X)$, defined as the complex dimension of the $(p,q)$-th $\bar\partial$-cohomology group (see \appref{complex_geometry}), are conventionally arranged into the so-called Hodge diamond
\begin{equation}
\begin{array}{*{20}c}
   {} & {} & {} & {} & {} & {h^{0,0}} & {} & {} & {} & {} & {}  \\
   {} & {} & {} & {} & {h^{1,0}} & {} & {h^{0,1}} & {} & {} & {} & {}  \\
   {} & {} & {} &  {\mathinner{\mkern2mu\raise1pt\hbox{.}\mkern2mu
 \raise4pt\hbox{.}\mkern2mu\raise7pt\hbox{.}\mkern1mu}}  & {} & {h^{1,1}} & {} &  \ddots  & {} & {} & {}  \\
   {} & {} & {h^{i,0}} & {} &  {\mathinner{\mkern2mu\raise1pt\hbox{.}\mkern2mu
 \raise4pt\hbox{.}\mkern2mu\raise7pt\hbox{.}\mkern1mu}}  & {} &  \ddots  & {} & {h^{0,i}} & {} & {}  \\
   {} &  {\mathinner{\mkern2mu\raise1pt\hbox{.}\mkern2mu
 \raise4pt\hbox{.}\mkern2mu\raise7pt\hbox{.}\mkern1mu}}  & {} &  \ddots  & {} & {} & {} &  {\mathinner{\mkern2mu\raise1pt\hbox{.}\mkern2mu
 \raise4pt\hbox{.}\mkern2mu\raise7pt\hbox{.}\mkern1mu}}  & {} &  \ddots  & {}  \\
   {h^{n,0}} & {} &  \vdots  & {} & {} &  \vdots  & {} & {} &  \vdots  & {} & {h^{0,n}}  \\
   {} &  \ddots  & {} &  {\mathinner{\mkern2mu\raise1pt\hbox{.}\mkern2mu
 \raise4pt\hbox{.}\mkern2mu\raise7pt\hbox{.}\mkern1mu}}  & {} & {} & {} &  \ddots  & {} &  {\mathinner{\mkern2mu\raise1pt\hbox{.}\mkern2mu
 \raise4pt\hbox{.}\mkern2mu\raise7pt\hbox{.}\mkern1mu}}  & {}  \\
   {} & {} & {h^{n,n - i}} & {} &  \ddots  & {} &  {\mathinner{\mkern2mu\raise1pt\hbox{.}\mkern2mu
 \raise4pt\hbox{.}\mkern2mu\raise7pt\hbox{.}\mkern1mu}}  & {} & {h^{n - i,n}} & {} & {}  \\
   {} & {} & {} &  \ddots  & {} & {h^{n - 1,n - 1}} & {} &  {\mathinner{\mkern2mu\raise1pt\hbox{.}\mkern2mu
 \raise4pt\hbox{.}\mkern2mu\raise7pt\hbox{.}\mkern1mu}}  & {} & {} & {}  \\
   {} & {} & {} & {} & {h^{n,n - 1}} & {} & {h^{n - 1,n}} & {} & {} & {} & {}  \\
   {} & {} & {} & {} & {} & {h^{n,n}} & {} & {} & {} & {} & {}  \\
 \end{array} 
\end{equation}
In specialising consecutively to Hermitian, K\"ahler and Calabi-Yau manifolds not all of the $(n+1)^2$ Hodge numbers remain independent. Due to \eqref{cplx_lapl_props}, $h^{p,q}(X) = h^{n-p,n-q}(X)$ on a Hermitian manifold. This reduces the number of independent Hodge numbers to $\lceil (n+1)^2 /2 \rceil$\newnot{symbol:ceil}, that is $(n+1)^2 /2$ for $n$ odd and $(n+1)^2 /2 + 1/2$ for $n$ even. For K\"ahler manifolds, also \eqref{kahler_laplacian} holds. Together with the previous result, this implies $h^{p,q}(X) = h^{q,p}(X)$ for K\"ahler manifolds and the number of independent Hodge numbers is reduced to $[n(n+4)+3]/4$ for $n$ odd and $[n(n+4)+4]/4$ for $n$ even. A Calabi-Yau manifold with the restriction on the holonomy group described in the last paragraph of \secref{cy_def} does not possess any continuous isometries, which translates into $h^{0,1}(X)=h^{1,0}(X)=0$. In addition, the existence of the unique holomorphic $(n,0)$-form $\Omega$ implies $h^{n,0}(X)=h^{0,n}(X)=h^{0,0}(X)=h^{n,n}(X)=1$ as well as another duality of the form $h^{p,0}(X)=h^{0,n-p}(X)$ since the contraction of a harmonic $(p,0)$-form with $\bar\Omega$ yields a harmonic $(0,n-p)$-form. This reduces the Hodge numbers by a further $(n+1)$. Thus, for $n$ odd there are $(n^2-1)/4$ and for $n$ even there are $n^2/4$ \emph{a priori} independent Hodge numbers and the top left quadrant of the Hodge diamond for Calabi-Yau manifolds becomes

\begin{equation}
\begin{array}{*{20}c}
   {} & {} & {} & {} & {} & {} & 1 \\
   {} & {} & {} & {} & {} & 0 & {} \\
   {} & {} & {} & {} & \iddots & {} & h^{1,1} \\
   {} & {} & {} & \iddots & {} & \iddots & {} \\
   {} & {} & \iddots & {} & \iddots & {} & h^{2,2} \\
   {} & 0 & {} & \iddots & {} & \iddots & \vdots \\
   1 & {} & h^{n-1,1} & {} & h^{n-2,2} & \cdots & \!\!\cdot\raisebox{2pt}{:} \\
\end{array} 
\end{equation}

\noindent Only the top left quadrant is shown here, since the Hodge diamond for Calabi-Yau manifolds is symmetric about the vertical and horizontal axes. In particular, the Hodge diamond of Calabi-Yau five-folds reads

\begin{equation}\eqlabel{cy5_hodgediamond}
\begin{array}{*{20}c}
   {} & {} & {} & {} & {} & 1 & {} & {} & {} & {} & {}  \\
   {} & {} & {} & {} & 0 & {} & 0 & {} & {} & {} & {}  \\
   {} & {} & {} & 0 & {} & {h^{1,1}} & {} & 0 & {} & {} & {}  \\
   {} & {} & 0 & {} & {h^{2,1}} & {} & {h^{2,1}} & {} & 0 & {} & {}  \\
   {} & 0 & {} & {h^{3,1}} & {} & {h^{2,2}} & {} & {h^{3,1}} & {} & 0 & {}  \\
   1 & {} & {h^{4,1}} & {} & {h^{3,2}} & {} & {h^{3,2}} & {} & {h^{4,1}} & {} & 1  \\
   {} & 0 & {} & {h^{3,1}} & {} & {h^{2,2}} & {} & {h^{3,1}} & {} & 0 & {}  \\
   {} & {} & 0 & {} & {h^{2,1}} & {} & {h^{2,1}} & {} & 0 & {} & {}  \\
   {} & {} & {} & 0 & {} & {h^{1,1}} & {} & 0 & {} & {} & {}  \\
   {} & {} & {} & {} & 0 & {} & 0 & {} & {} & {} & {}  \\
   {} & {} & {} & {} & {} & 1 & {} & {} & {} & {} & {}  \\
 \end{array} 
\end{equation}

\noindent with \emph{a priori} six independent Hodge numbers. In fact, there is one more relation

\be\eqlabel{Hodgecons}
	11 h^{1,1} - 10 h^{2,1} + 10 h^{3,1} - 11 h^{4,1} - h^{2,2} + h^{3,2} = 0 \; ,
\ee

\noindent which can be derived by an index theorem calculation using the Calabi-Yau condition $c_1(X)=0$. Hence, Calabi-Yau five-folds are characterised by five rather than six independent Hodge numbers. An analogous relation also holds for Calabi-Yau four-folds~\cite{Klemm:1996ts}.

Other relevant topological invariants of Calabi-Yau manifolds, apart from the Hodge numbers $h^{p,q}(X)$ and the Euler number
\be\eqlabel{def_euler_no}
	\eta (X) 	\equiv 	\sum_{i=0}^{2n}(-1)^i b^i(X) 
			= 		\sum_{i=0}^{2n}(-1)^i \sum_{p+q=i} h^{p,q}(X) \; , 
\ee
are the Chern classes $c_i(X)$, the intersection numbers $d_{{i_1}\dots {i_n}}$ of $n$ 2-cycles and various other intersection numbers, which will be introduced later.

\section{Calabi-Yau five-fold complex geometry}\seclabel{cy5_geom}

Henceforth, we concentrate on the case $n=5$, that is Calabi-Yau five-folds. Our conventions for complex differential forms and some general results from complex geometry are summarised in \appref{complex_geometry}.

By straightforward but in part somewhat tedious component calculations, the Hodge-star of $(p,q)$-forms can be expressed in terms of the K\"ahler form $J$ and the map $\tilde{(\cdot)}$ defined in~\eqref*{def_tilde_map}. In general, on a Hermitian manifold of complex dimension five, one finds
\begin{equation}\eqlabel{hodge_special}
\begin{aligned}
	&(0,1): \ast \zeta 	= \frac{i}{4!} J^4 \wedge \zeta , \qquad & 
	&(1,1): \ast \omega	= -\oneon{3!} J^3 \wedge \omega - \frac{i}{4!} J^4 \wedge \tilde{\omega} , \\
	&(2,1): \ast \nu 		= \frac{i}{2} J^2 \wedge \nu + \oneon{3!} J^3 \wedge \tilde{\nu} , \qquad & 
	&(3,1): \ast \varpi	= - J \wedge \varpi - \frac{i}{2} J^2 \wedge \tilde{\varpi} , \\
	&(4,1): \ast \chi 	= i \chi + J \wedge \tilde{\chi} , \qquad & 
	&(2,2): \ast \sigma	= J \wedge \sigma - \frac{i}{2} J^2 \wedge \tilde{\sigma} + \oneon{12} J^3 \wedge \tilde{\tilde{\sigma}}, \\
	&(3,2): \ast \omega = -i \omega - J \wedge \tilde{\omega} - \frac{i}{12} J^2 \wedge \tilde{\tilde{\omega}}. & & 
\end{aligned}
\end{equation}
Note the pattern for the $(p,1)$-forms: the second term in the $(p,1)$-expression is minus the first term in the $(p-1,1)$-expression, with the forms being replaced according to $\tilde{\omega}^{(p,1)} \rightarrow \omega^{(p-1,1)}$. 

The above equations simplify for harmonic $(p,q)$-forms on a Calabi-Yau five-fold $X$. First, we recall that Calabi-Yau five-folds have vanishing Hodge numbers $h^{p,0}(X)=h^{0,p}(X)$ for $p=1,\ldots,4$. This means non-zero harmonic $(p,0)$- and $(0,p)$- forms do not exist for $p=1,\ldots,4$ and consequently $\tilde{\omega}^{(p,1)} = \tilde{\omega}^{(1,p)} = 0$ for harmonic $(p,1)$- and $(1,p)$-forms with $p>1$. Moreover, a harmonic $(0,0)$-form is a constant and hence, $\tilde{\omega} = {\rm const.}$ for a harmonic $(1,1)$-form $\omega$. The value of the constant is determined in \eqref{tilde_11}. Combining these facts with the formul\ae~\eqref*{hodge_special}, one finds for the Hodge-star of harmonic $(p,q)$-forms on a Calabi-Yau five-fold $X$
\begin{align}
	&(1,1):\;	\ast \omega	= -\oneon{3!} J^3 \wedge \omega - \frac{i}{4!} \tilde{\omega} J^4, & \nonumber \\
	&(2,1):\; 	\ast \nu 		= \frac{i}{2} J^2 \wedge \nu , \qquad & \nonumber \\
	&(3,1):\; 	\ast \varpi		= - J \wedge \varpi , \qquad & 
	&(2,2):\; 	\ast \sigma	= J \wedge \sigma - \frac{i}{2} J^2 \wedge \tilde{\sigma} + \oneon{12}\tilde{\tilde{\sigma}}J^3, \nonumber \\
	&(4,1):\; 	\ast \chi 		= i \chi , &
	&(3,2):\; 	\ast \phi 		= -i \phi - J \wedge\tilde{\phi}\; , \eqlabel{dualforms}
\end{align}
where we should keep in mind that $\tilde{\omega}$ and $\tilde{\tilde{\sigma}}$ are constants, $\tilde{\sigma}$ is a harmonic $(1,1)$-form and $\tilde{\phi}$ is a harmonic $(2,1)$-form. Using the $\kappa$-map defined in~\eqref*{def_kappa_map} and the result~\eqref*{tilde_11}, we can write the Hodge-star of a $(1,1)$-form $\omega$ as
\be\eqlabel{dualforms11}
	\ast \omega = -\oneon{3!} J^3 \wedge \omega + \frac{5}{4!} \frac{\kappa(\omega,J,\ldots,J)}{\kappa(J,\ldots,J)} J^4 \; .
\ee
A further useful relation for a Hodge-star is
\be\eqlabel{sigmaJJ}
	\ast (\sigma\wedge J^2) = \tilde{\tilde{\sigma}} J - 2 i \tilde{\sigma} \; ,
\ee
where $\sigma$ is a $(2,2)$-form. In the next section, we will use this relation to explicitly compute $\tilde{\sigma}$ and $\tilde{\tilde{\sigma}}$, but before that we briefly turn to the covariantly constant spinor $\eta$ on $X$.

As mentioned in \secref{cy_def}, on a Calabi-Yau five-fold $X$ we have a spinor $\eta$, unique up to normalisation, which is invariant under the holonomy group ${\rm Hol}(X)$. This means $\eta$ is covariantly constant with respect to the Levi-Civita connection associated to the Ricci-flat metric $g_{\mu\bar\nu}$. Our conventions are such that $\eta$ has positive and $\eta^\star$ negative chirality, that is
\be\eqlabel{eta_chirality}
	\gamma^{(11)} \eta = \eta \; , \qquad \gamma^{(11)} \eta^\star = - \eta^\star \; ,
\ee
where the 10-dimensional chirality operator $\gamma^{(11)}$ is defined in~\eqref*{gamma11}. Moreover, we normalize $\eta$ such that
\be\eqlabel{etanorm}
	\eta^\dagger\eta =1 \; .
\ee
A consequence of~\eqref*{eta_chirality} is that
\be\eqlabel{etaTeta_is_zero}
	\eta^\transp \eta = 0 \; ,
\ee
which follows by inserting $(\gamma^{(11)})^2 = \id$ and using $(\gamma^{(11)})^\transp = - \gamma^{(11)}$. The spinor $\eta$ satisfies a very important annihilation condition
\be\eqlabel{eta_annihil}
	\gamma^{\bar{\mu}}\eta =0 \; ,
\ee
which follows from~\eqref{JOdef}, the definition of $J$ and the Clifford algebra~\eqref*{cy5_hol_cliff_alg} by direct calculation~\cite{Candelas:1987is}
\begin{gather}
	i g_{\mu\bar{\nu}} 	= J_{\mu\bar{\nu}} = i \eta^\dagger \gamma_{\mu\bar{\nu}} \eta
					= i \eta^\dagger \left(- \gamma_{\bar{\nu}} \gamma_\mu + g_{\mu\bar{\nu}} \right) \eta \\
	\Rightarrow \quad 0 = \eta^\dagger \gamma_{\bar{\nu}} \gamma_\mu \eta = (\gamma_\nu \eta)^\dagger (\gamma_\mu \eta) \; .
\end{gather}
The case $\mu = \nu$ then implies \eqref{eta_annihil}. The spinor $\eta$ is thus annihilated by half of the gamma matrices. The $\gamma^{\mu}$ with holomorphic (or unbarred) indices may be viewed as fermionic creation operators producing spinors on $X$ when acting on the Clifford vacuum $\eta$. A general spinor $\psi$ on $X$ can be decomposed according to~\cite{Fre:1995bc}
\be\eqlabel{cy_spinor_decomposition}
	\psi = \omega^{(0,0)} \eta + \omega^{(1,0)}_{\mu_1} \gamma^{\mu_1} \eta + \ldots 
	     + \omega^{(n,0)}_{\mu_1\ldots\mu_n} \gamma^{\mu_1\ldots\mu_n} \eta \; ,
\ee
where the coefficient functions $\omega^{(p,0)}_{\mu_1\ldots\mu_p}$ are $(p,0)$-forms written in component notation. For $\psi$ to be a zero mode of the Dirac operator $\displaystyle{\not} \nabla \equiv \gamma^\mu \nabla_\mu + \gamma^{\bar\mu} \nabla_{\bar\mu}$, the forms $\omega^{(p,0)}$ need to be harmonic. Finally, it can be shown that $\eta$ satisfies the Fierz identity~\cite{Naito:1986cr,Miemiec:2005ry}
\be\eqlabel{cy5_fierz_id}
	\eta^\star \eta^T = - \oneon{32} g_{\mu\bar\nu} \gamma^{\mu\bar\nu} \; ,
\ee
where \eqref{eta_annihil} has been used to simplify the right hand side. The Fierz identity will be useful in the chapter on dimensional reduction.

\section{Calabi-Yau five-fold deformations and moduli spaces}\seclabel{cy5_moduli_spaces}

For Calabi-Yau three-folds, the moduli space of Ricci-flat metrics is (locally) a direct product of a K\"ahler and a complex structure moduli space associated to harmonic $(1,1)$- and $(1,2)$-forms, respectively~\cite{Candelas:1990pi}. More generally, a Ricci-flat metric on $X$ is uniquely specified by fixing the two distinct forms $J$ and $\Omega$. In other words, the moduli space factors (at least locally and assuming $h^{2,0}(X)=0$) into the space of K\"ahler ($\delta J$) and complex structure ($\delta \Omega$) deformations, which are associated with harmonic $(1,1)$-forms and harmonic $(1,n-1)$-forms, respectively. This is not true for the only Calabi-Yau two-fold called $K3$, since  $h^{2,0}(K3)=1$~\cite{Aspinwall:1996mn}.

Thus, the metric deformations on a Calabi-Yau five-fold $X$ are described by K\"ahler moduli of type $(1,1)$ and complex structure moduli of type $(1,4)$. As we discussed in \secref{cy_topology}, there are many more harmonic forms on $X$. While those harmonic forms are unrelated to metric deformations, they nonetheless play a r\^{o}le in M-theory compactifications and are hence also studied in this section.

To begin, it is useful to introduce sets of harmonic basis forms for these cohomologies as follows
\begin{align}
	&H^{1,1}(X): \quad  \{\omega_i\}_{i=1,\ldots,h^{1,1}(X)} , \\
	&H^{2,1}(X): \quad  \{\nu_p\}_{p=1,\ldots,h^{2,1}(X)} , \\
	&H^{1,3}(X): \quad  \{\varpi_x\}_{x=1,\ldots,h^{1,3}(X)} , \\
	&H^{2,2}(X): \quad  \{\sigma_e\}_{e=1,\ldots,h^{2,2}(X)} , \\
	&H^{1,4}(X): \quad  \{\chi_a\}_{a=1,\ldots,h^{1,4}(X)} , 
\end{align}\newnot{symbol:ind_11}\newnot{symbol:ind_21}\newnot{symbol:ind_13}\newnot{symbol:ind_22}\newnot{symbol:ind_14}\newnot{symbol:11_basis}\newnot{symbol:21_basis}\newnot{symbol:13_basis}\newnot{symbol:22_basis}\newnot{symbol:14_basis}
with $\omega_i$ and $\sigma_e$ being real and all other forms being complex. These forms can be used to construct various intersection numbers
\be\eqlabel{def_int_nos}
\begin{array}{lllllll}
	d_{i_1\ldots i_5} &=& \int_X \omega_{i_1} \wedge \cdots \wedge \omega_{i_5} ,&\quad&
	d_{p\bar{q}ij} &=& \int_X \nu_p \wedge \bar{\nu}_{\bar{q}} \wedge \omega_i \wedge \omega_j , \\
	d_{eijk}&=&\int_X\sigma_e\wedge \omega_i\wedge \omega_j\wedge \omega_k ,&\quad&
	d_{p\bar{q}e}&=&\int_X \nu_p \wedge \bar{\nu}_{\bar{q}} \wedge \sigma_e,\\
	d_{efi}&=&\int_X\sigma_e\wedge\sigma_f\wedge\omega_i,&\quad&
	d_{x\bar{y}i} &=& \int_X \varpi_x \wedge \bar{\varpi}_{\bar{y}} \wedge \omega_i .
\end{array}
\ee
which will play a r\^{o}le later on. The term \emph{intersection number} is a slight misnomer in this context, as all of these integrals, except $d_{i_1\ldots i_5}$, in general depend on the complex structure (due to the use of complex $(p,q)$-forms) and thus do not represent topological invariants.

We begin with the metric moduli. The basic requirement is that a variation 
\be
	g_{mn} \rightarrow g^\prime_{mn} = g_{mn} + \delta g_{mn}
\ee
of the metric leaves the Ricci tensor zero at linear order in $\delta g_{mn}$ so as to stay in the realm of Calabi-Yau five-folds. Here, $m,n,\ldots = 1,\ldots,10$ are real indices on $X$. Complexifying this expression, reveals that the $(1,1)$ part of $\delta g_{mn}$ can be expanded in terms of harmonic $(1,1)$-forms, while the $(2,0)$ and $(0,2)$ parts can be expressed in terms of harmonic $(1,4)$-forms. Explicitly, one has
\be\eqlabel{deltag}
	\delta g_{\mu\bar{\nu}}=-iw_{i,\mu\bar{\nu}}\delta t^i\; ,\quad 
	\delta g_{\mu\nu}=-\frac{2}{4!||\Omega ||^2}{\Omega_{\mu}}^{\bar{\rho}_1\ldots\bar{\rho}_4}\chi_{a,\nu\bar{\rho}_1\ldots\bar{\rho}_4}\delta z^a\; ,
\ee
with the variations $\delta t^i$, $i,j,\ldots=1,\ldots,h^{1,1}(X)$, and $\delta z^a$, $a,b,\ldots=1,\ldots,h^{1,4}(X)$, being elements of the K\"ahler and complex structure moduli space, respectively. The standard moduli space metric on the space of metric deformations is defined by
\be\eqlabel{wp}
	{\cal G}(\delta g_{mp} ,\widetilde{\delta g}_{nq}) \equiv 
		\oneon{4V}\int_X d^{10}x\,\sqrt{g}\,g^{mn}g^{pq}\delta g_{mp}\widetilde{\delta g}_{nq} \; .
\ee
Here, $V$ denotes the volume of $X$ defined in~\eqref*{def_volume}. This metric splits into a K\"ahler and a complex structure part which can be worked out separately. 

Let us first discuss the K\"ahler deformations. A straightforward calculation, inserting the first \eqref{deltag}, shows that
\be
	{\cal G}^{(1,1)}_{ij}(\underline{t}) = \frac{1}{2V}\int_X\omega_i\wedge\ast\omega_j \; .
\ee
Using~\eqref*{dualforms11}, this can be written in terms of topological integrals which involve $J$ and the forms $\omega_i$. Then, defining the K\"ahler moduli $t^i$ by
\be\eqlabel{J11_exp}
	J = t^i \omega_i \; ,
\ee
one finds
\be\eqlabel{metric11}
	{\cal G}^{(1,1)}_{ij}(\underline{t})=-10\frac{\kappa_{ij}}{\kappa}+\frac{25}{2}\frac{\kappa_i\kappa_j}{\kappa^2} \; ,
\ee
where we have introduced a shorthand notation for the $\kappa$-map defined in~\eqref*{def_kappa_map} with recurring arguments
\begin{align}
	\kappa 		&\equiv \kappa(J,\ldots,J) = 5!\, V=d_{i_1\dots i_5}t^{i_1}\dots t^{i_5} \; , \eqlabel{def_kappa0} \\
	\kappa_i		&\equiv \kappa(\omega_i,J,\ldots,J) = d_{ii_2\dots i_5}t^{i_2}\dots t^{i_5} \; , \eqlabel{def_kappa1} \\
	\kappa_{ij}	&\equiv \kappa(\omega_i,\omega_j,J,J,J) = d_{iji_1i_2i_3}t^{i_1}t^{i_2}t^{i_3} \; , \eqlabel{def_kappa2} \\
				& \qquad\qquad\quad\vdots \nonumber \\
	\kappa_{i_1 \ldots i_5} &\equiv \kappa(\omega_{i_1},\ldots,\omega_{i_5}) = d_{i_1 \ldots i_5} \; . \eqlabel{def_kappa5}
\end{align}
The intersection number $d_{i_1 \ldots i_5}$ has been defined in~\eqref*{def_int_nos}. Note that from \eqref{J11_exp}, one has $\kappa = \kappa_i t^i = \kappa_{ij} t^i t^j$ and so on. With this notation, for example
\be\eqlabel{omegat}
	\tilde{\omega}_i=5i\frac{\kappa_i}{\kappa} \; .
\ee
It is easy to check that the above moduli space metric~\eqref*{metric11} can be obtained from a ``K\"ahler potential'' $K^{(1,1)}$ as
\be
	{\cal G}^{(1,1)}_{ij} = \partial_i\partial_j K^{(1,1)} \; ,
\ee
where $K^{(1,1)} \equiv - (\ln\kappa)/2$. We can use the moduli space metric to define lower index moduli $t_i$ via $t_i \equiv {\cal G}^{(1,1)}_{ij}t^j$. From the explicit form~\eqref*{metric11} of the metric, it is easy to verify the useful relation
\be\eqlabel{tli}
	t_i = \frac{5\kappa_i}{2\kappa} = - \frac{i}{2} \tilde{\omega}_i \; .
\ee
A further useful observation is related to ``metrics'' of the form
\be\eqlabel{Gtilde}
	\tilde{\cal G}_{ij} \equiv {\cal G}^{(1,1)}_{ij}+c\frac{\kappa_i\kappa_j}{\kappa^2}
\ee
for any real number $c$. A short calculation, using \eqref{tli} and $\kappa_it^i=\kappa$ repeatedly, shows that
\be\eqlabel{GtGti}
   \tilde{\cal G}_{ij}\left({\cal G}^{(1,1)jk}+\tilde{c}\frac{\kappa^j\kappa^k}{\kappa^2}\right) = 
   	\delta_i^k+\left(c+\tilde{c}+\frac{2}{5}c\tilde{c}\right)\frac{\kappa_i\kappa^k}{\kappa^2} \; ,
\ee
where $\tilde{c}$ is an arbitrary real number. Here, the standard moduli space metric ${\cal G}^{(1,1)}_{ij}$ and its inverse ${\cal G}^{(1,1)ij}$ have been used to lower and raise indices. The above relation shows that for all $c\neq -5/2$, the metric~\eqref*{Gtilde} is invertible and that its inverse is given by
\be\eqlabel{Gti}
	\tilde{\cal G}^{ij}={\cal G}^{(1,1)jk}+\tilde{c}\frac{\kappa^j\kappa^k}{\kappa^2}\; ,\qquad \tilde{c}=-\frac{5c}{5+2c} \; .
\ee
These relations will be helpful in \secref{action1d_corr}. 

In summary, the K\"ahler moduli space for Calabi-Yau five-folds can be treated in complete analogy with the one for Calabi-Yau three-folds. The main difference is that the moduli space metric is now governed by a quintic pre-potential $\kappa$ instead of a cubic one for three-folds.

We now move on to the complex structure moduli. Evaluating the standard moduli space metric~\eqref*{wp} for the $(2,0)$ variation of the metric in~\eqref*{deltag}, one finds
\be
	{\cal G}^{(1,4)}_{a\bar{b}} = \oneon{V||\Omega ||^2}\int_X\chi_a\wedge\ast\bar{\chi}_{\bar{b}} \; .
\ee
Using the result in~\eqref*{dualforms} for the Hodge dual of $(4,1)$-forms together with \eqref{omegaX_hol_50_form} then leads to the standard result
\be\eqlabel{metric41}
	{\cal G}_{a\bar{b}}^{(1,4)} (\underline{z}, \underline{\bar{z}}) = 
		\frac{\int_X \chi_a \wedge \bar{\chi}_{\bar{b}}}{\int_X \Omega\wedge\bar\Omega} \; .
\ee
Under an infinitesimal variation of the complex structure, the holomorphic and anti-holomorphic differentials $dz^\mu$ and $d\bar{z}^{\bar\mu}$ mix linearly with each other, $\frac{\partial}{\partial z^a} dz^\mu = \sigma_{a\nu} {}^\mu dz^\nu + \rho_{a\bar\nu} {}^\mu d\bar{z}^{\bar\nu}$ (Note that the complex structure moduli $z^a$ should not be confused with the local holomorphic coordinates $z^\mu$ on $X$. They are distinguished by the different types of indices). Therefore, a $(p,q)$-form mixes only with $(p\pm 1, q\mp 1)$-forms under an infinitesimal variation of the complex structure~\cite{Strominger:1990pd}
\be
	\frac{\partial}{\partial z^a} \omega^{(p,q)} = \eta^{(p,q)} + \eta^{(p-1,q+1)} + \eta^{(p+1,q-1)} \; .
\ee
Closedness of a differential form is preserved under an infinitesimal variation of the complex structure, since the exterior derivative $d$ is independent of the complex structure moduli $z^a$, that is $[d, \frac{\partial}{\partial z^a}] = 0$. For the case of the holomorphic $(5,0)$-form $\Omega$, this implies Kodaira's relation
\be\eqlabel{kodairas_relation}
	\frac{\partial}{\partial z^a} \bar\Omega = k_a \bar\Omega + \chi_a \; ,
\ee
where, as before, $\chi_a \in H^{1,4} (X)$ and $k_a$ may depend on the complex structure moduli $z^a$ but not on the coordinates $z^\mu$ of $X$. This is exactly analogous to the case of Calabi-Yau three-folds~\cite{Candelas:1990pi,Hubsch:1992,Fre:1995bc} and Calabi-Yau four-folds~\cite{Haack:2002tu}. Kodaira's relation implies, via direct differentiation, that the moduli space metric~\eqref*{metric41} can be obtained from the K\"ahler potential $K^{(1,4)}$ as
\be\eqlabel{K41}
	{\cal G}^{(1,4)}_{a\bar{b}}=\partial_a\partial_{\bar{b}} K^{(1,4)} \; ,
\ee
where $\partial_a \equiv \partial/\partial z^a$, $\partial_{\bar{b}} \equiv \partial/\partial \bar{z}^{\bar{b}}$ and $K^{(1,4)} \equiv \ln \left[i \int_X \Omega \wedge \bar{\Omega}\right]$. The K\"ahler potential $K^{(1,4)}$ also serves to determine the function $k_a$ in Kodaira's relation, namely $k_a = - \partial_a K^{(1,4)}$. In order to express $K^{(1,4)}$ more explicitly in terms of moduli, we introduce a symplectic basis $(A^{\cal A}, B_{\cal B})$ of 5-cycles and a dual basis $(\alpha_{\cal A},\beta^{\cal B})$ of 5-forms satisfying
\be\eqlabel{sympbasis}
	\int_{A^{\cal B}} \alpha_{\cal A} = \int_X \alpha_{\cal A} \wedge \beta^{\cal B} = \delta_{\cal A}^{\cal B} \; , \qquad
	\int_{B_{\cal A}} \beta^{\cal B} = \int_X \beta^{\cal B} \wedge \alpha_{\cal A} = -\delta_{\cal A}^{\cal B} \; .
\ee
Then, the period integrals are defined in the usual way as
\be\eqlabel{periods}
	{\cal Z}^{\cal A} \equiv \int_{A^{\cal A}} \Omega \; , \qquad\qquad \mathcal{G}_{\cal A} \equiv \int_{B_{\cal A}} \Omega \; .
\ee
and the periods ${\cal G}_{\cal A}$ can be shown to be functions of ${\cal Z}^{\cal A}$, just as in the three-fold case. In the dual basis $(\alpha_{\cal A},\beta^{\cal B})$, the holomorphic $(5,0)$-form $\Omega$ can then be expanded as $\Omega = {\cal Z}^{\cal A} \alpha_{\cal A} - \mathcal{G}_{\cal A} \beta^{\cal A}$ and inserting this into the expression~\eqref*{K41} for the K\"ahler potential yields
\be
	K^{(1,4)} = \ln \left[i (\mathcal{G}_{\cal A} \bar{\cal Z}^{\cal A} - {\cal Z}^{\cal A} \bar{\mathcal{G}}_{\cal A})\right] .
\ee
By virtue of Kodaira's relation~\eqref*{kodairas_relation}, $\int_X\Omega\wedge\frac{\partial\Omega}{\partial {\cal Z}^{\cal A}}=0$ which immediately leads to ${\cal G}_{\cal A}=\frac{1}{2}\frac{\partial}{\partial {\cal Z}^{\cal A}}({\cal G}_{\cal B}{\cal Z}^{\cal B})$. Hence, the periods ${\cal G}_{\cal A}$ can be obtained as derivatives
\be
	{\cal G}_{\cal A} = \frac{\partial {\cal G}}{\partial{\cal  Z}^{\cal A}}
\ee
of a pre-potential ${\cal G}$ which is homogeneous of degree two in the projective coordinates ${\cal Z}^{\cal A}$. This is formally very similar to the Calabi-Yau three-fold case. However, an important difference is that the 5-forms here contain not only $(5,0)$, $(0,5)$, $(4,1)$ and $(1,4)$ pieces but also $(3,2)$ and $(2,3)$ parts. That is, $\mathcal{A,B},\ldots = 0,1,\ldots, h^{1,4}+h^{2,3}$\newnot{symbol:ind_real5}. As a consequence, the periods ${\cal Z}^{\cal A}$ do not simply serve as projective coordinates on the complex structure moduli space, though they can in principle be computed as functions of the $z^a$. However, their vast redundancy renders them much less useful as compared to the three-fold case.

In the expression~\eqref*{dualforms} for the Hodge-star of a harmonic $(2,2)$-form $\sigma_e$, the contractions $\tilde{\sigma}_e$ and $\tilde{\tilde{\sigma}}_e$ appear. We are now in a position to explicitly compute these contractions. First, we note that the harmonicity of $\sigma_e$ implies that $\tilde{\sigma}_e$ is a harmonic $(1,1)$-form (see \eqref{dtildeomega1,dtildeomega2}) and can therefore be expanded in terms of the basis of harmonic $(1,1)$-forms $\omega_i$
\be
	\tilde{\sigma}_e = i\, k_e^i\omega_i \; ,
\ee
with some coefficients $k_e^i$\newnot{symbol:kei}, which generally depend on the K\"ahler moduli $t^i$.  Applying one more contraction and using \eqref{omegat}, we learn that
\be 
	\tilde{\tilde{\sigma}}_e = -\frac{5}{\kappa}k_e^i\kappa_i \; .
\ee
We can determine all the contractions of harmonic $(2,2)$-forms if we are able to compute the coefficients $k_e^i$. This can be accomplished by multiplying \eqref{sigmaJJ} with $\omega_j$ and integrating over the Calabi-Yau five-fold $X$. This results in
\be\eqlabel{kei}
	k_e^i = \frac{1}{4V}\left({\cal G}^{(1,1)ij}-\frac{25}{6}\frac{\kappa^i\kappa^j}{\kappa^2}\right) d_{ejkl}t^kt^l \; ,
\ee
where ${\cal G}^{(1,1)ij}$ is the inverse of ${\cal G}^{(1,1)}_{ij}$.

\subsection{Real vs. complex forms}\seclabel{real_form_formalism}

For later chapters, it will be useful to know how to treat the $(2,1)$- and $(1,3)$-forms as real 3- and 4-forms,\footnote{All differential forms occurring in this subsection are henceforth implicitly assumed to be harmonic.} respectively, and to know how they are related to each other. This is advantageous since the real forms are topological and in particular do not depend on the complex structure moduli. 

Real harmonic 3-forms are naturally locked to 3-cycles and are thus topologically invariant. The fact that $h^{3,0}(X)=0$ for Calabi-Yau five-folds ensures that a real 3-form is exclusively made up of a $(2,1)$- and a $(1,2)$-piece. However, the way in which a particular 3-form is split into $(2,1)$- and $(1,2)$-parts evidently depends on the choice of complex structure. We can parametrically represent this fact by introducing complex structure dependent linear maps $\mathfrak{A}$\newnot{symbols:3_21_maps} and $\mathfrak{B}$ from real 3-forms to complex $(2,1)$-forms and vice versa.

For fixed bases, the linear maps have a matrix representation according to
\begin{align}
	\nu_p &= \mathfrak{A}_p {}^{\cal{Q}} N_{\cal{Q}} \qquad \qquad \qquad 
	\text{(and: $\bar{\nu}_{\bar{p}} = \bar{\mathfrak{A}}_{\bar{p}} {}^{\cal{Q}} N_{\cal{Q}}$)} \; ,\eqlabel{21_3_rel1}\\
	N_{\cal{P}} &= \mathfrak{B}_{\cal{P}} {}^q \nu_q + \bar{\mathfrak{B}}_{\cal{P}} {}^{\bar{q}} \bar{\nu}_{\bar{q}} \; ,\eqlabel{21_3_rel2}
\end{align}
where $\{ N_{\cal{P}} \}_{\mathcal{P}=1,\ldots,b^{3}(X)}$\newnot{symbol:ind_real3}\newnot{symbol:3form_basis} is a real basis of $H^{3} (X)$ and $\{ \nu_p \}_{p=1,\ldots,h^{2,1}(X)}$ is a basis of $H^{2,1} (X)$. To avoid confusion with symbols defined elsewhere, we use Fraktur font letters to denote maps translating between real and complex forms and calligraphic letters for real form indices. Note that $\mathfrak{A}_p {}^{\cal{Q}}$ and $\mathfrak{B}_{\cal{P}} {}^q$ are complex and have dependence $\mathfrak{A}_p {}^{\cal{Q}} = \mathfrak{A}_p {}^{\cal{Q}} (\underline{z},\bar{\underline{z}})$, $\mathfrak{B}_{\cal{P}} {}^q = \mathfrak{B}_{\cal{P}} {}^q (\underline{z},\bar{\underline{z}})$, where $z^a$ and $\bar{z}^{\bar{a}}$ are the complex structure moduli of the Calabi-Yau five-fold. The equations above have two faces, for they can either be written in local real 10-dimensional coordinates or in local holomorphic coordinates. For example, \eqref{21_3_rel2} in real coordinates is
\be
	N_{\mathcal{P},m_1 m_2 m_3} 	= \mathfrak{B}_{\cal{P}} {}^q \nu_{q,m_1 m_2 m_3} 
								+ \bar{\mathfrak{B}}_{\cal{P}} {}^{\bar{q}} \bar{\nu}_{\bar{q},m_1 m_2 m_3} \; ,
\ee
whereas in local holomorphic coordinates it reads
\be\eqlabel{21_3_cplx_coords}
	N_{\mathcal{P},\mu_1 \mu_2 \bar{\nu}} = \mathfrak{B}_{\cal{P}} {}^q \nu_{q,\mu_1 \mu_2 \bar{\nu}} \; , \qquad \text{(and $\cc$)} \; ,
\ee
where forms with unnatural index types are to be translated manually using \eqref{real_complex_coords}. Inserting \eqref{21_3_rel1} into \eqref{21_3_rel2} and vice versa, we learn relations between the $\mathfrak{A}$ and $\mathfrak{B}$ maps
\begin{align}
	& \mathfrak{A}_p {}^{\cal{Q}} \mathfrak{B}_{\cal{Q}} {}^q = \delta_p {}^q &\text{(and $\cc$)} \; , \eqlabel{AB_rel1} \\
	& \mathfrak{A}_p {}^{\cal{Q}} \bar{\mathfrak{B}}_{\cal{Q}} {}^{\bar{q}} = 0 &\text{(and $\cc$)} \; , \eqlabel{AB_rel2} \\
	& \mathfrak{B}_{\cal{P}} {}^q \mathfrak{A}_q {}^{\cal{Q}} + \bar{\mathfrak{B}}_{\cal{P}} {}^{\bar{q}} \bar{\mathfrak{A}}_{\bar{q}} {}^{\cal{Q}} = 
		\delta_{\cal{P}} {}^{\cal{Q}} \; . &\eqlabel{AB_rel3} 
\end{align}
For the complex structure dependence, one finds
\begin{align}
	\partial_a N_{\cal{P}} &= 0 \; , &
	\partial_a \nu_p &= 
		\mathfrak{A}_p {}^{\cal{Q}} {}_{,a} \mathfrak{B}_{\cal{Q}} {}^q \nu_q + 
		\mathfrak{A}_p {}^{\cal{Q}} {}_{,a} \bar{\mathfrak{B}}_{\cal{Q}} {}^{\bar{q}} \bar{\nu}_{\bar{q}} \; , \\
	\partial_{\bar{a}} N_{\cal{P}} &= 0 \; , &
	\partial_{\bar{a}} \nu_p &= 
		\mathfrak{A}_p {}^{\cal{Q}} {}_{,\bar{a}} \mathfrak{B}_{\cal{Q}} {}^q \nu_q + 
		\mathfrak{A}_p {}^{\cal{Q}} {}_{,\bar{a}} \bar{\mathfrak{B}}_{\cal{Q}} {}^{\bar{q}} \bar{\nu}_{\bar{q}} \; .
\end{align}
Using \eqrangeref{21_3_rel1}{21_3_rel2} and \eqref{dualforms}, one can compute the Hodge star of the real 3-form $N_{\cal{P}}$
\be\eqlabel{3form_hodgestar}
	\ast N_{\cal{P}} = \oneon{2} \cs_{\cal{P}} {}^{\cal{Q}} N_{\cal{Q}} \wedge J^2 \; , 
\ee
where $\cs_{\cal{P}} {}^{\cal{Q}} \equiv i(\mathfrak{B}_{\cal{P}} {}^q \mathfrak{A}_q {}^{\cal{Q}} - \bar{\mathfrak{B}}_{\cal{P}} {}^{\bar{q}} \bar{\mathfrak{A}}_{\bar{q}} {}^{\cal{Q}})$. The linear map $\cs$ provides a complex structure on the \emph{moduli space} of real 3-forms induced by the complex structure of the Calabi-Yau five-fold itself. It satisfies (cf. \appref{complex_geometry})
\be\eqlabel{3form_cplx_struct_rel}
	\cs_{\cal{P}} {}^{\cal{Q}} \cs_{\cal{Q}} {}^{\cal{R}} = - \delta_{\cal{P}} {}^{\cal{R}} \; , \qquad\qquad
	(\cs_{\cal{P}} {}^{\cal{Q}})^\ast = \cs_{\cal{P}} {}^{\cal{Q}} \; , \qquad\qquad
	\tr \, \cs = 0 \; .
\ee
Using the complex structure $\cs$, we define projection operators
\be\eqlabel{3form_projectors_def}
	P_{\pm} {}_{\cal{P}} {}^{\cal{Q}} \equiv \oneon{2} (\id \mp i \cs)_{\cal{P}} {}^{\cal{Q}}
\ee
satisfying
\be\eqlabel{3form_projectors_rel}
	P_{\pm} {}_{\cal{P}} {}^{\cal{Q}} P_{\pm} {}_{\cal{Q}} {}^{\cal{R}} = P_{\pm} {}_{\cal{P}} {}^{\cal{R}} \; , \qquad
	P_{+} {}_{\cal{P}} {}^{\cal{Q}} P_{-} {}_{\cal{Q}} {}^{\cal{R}} = P_{-} {}_{\cal{P}} {}^{\cal{Q}} P_{+} {}_{\cal{Q}} {}^{\cal{R}} = 0 \; , \qquad
	(P_{\pm} {}_{\cal{P}} {}^{\cal{Q}})^\ast = P_{\mp} {}_{\cal{P}} {}^{\cal{Q}} \; .
\ee
In terms of the $\mathfrak{A}$ and $\mathfrak{B}$ maps, they are explicitly given by
\be
	P_{+} {}_{\cal{P}} {}^{\cal{Q}} = \mathfrak{B}_{\cal{P}} {}^q \mathfrak{A}_q {}^{\cal{Q}} \; , \qquad\qquad
	P_{-} {}_{\cal{P}} {}^{\cal{Q}} = \bar{\mathfrak{B}}_{\cal{P}} {}^{\bar{q}} \bar{\mathfrak{A}}_{\bar{q}} {}^{\cal{Q}} \; .
\ee
The standard metric on the moduli space of real 3-forms is
\be
	{\cal G}^{(3)}_{\cal{PQ}} = \int_X N_{\cal{P}} \wedge \ast N_{\cal{Q}} \; .
\ee
Using the expression for the Hodge-star~\eqref*{3form_hodgestar}, we can rewrite this so as to make the dependence on the moduli more explicit

\be
	{\cal G}^{(3)}_{\cal{PQ}} (\underline{t},\underline{z},\underline{\bar{z}}) = 
		\oneon{2} \cs_{\cal{\left(P\right.}} {}^{\cal{R}} d_{\left.\mathcal{Q}\right)\mathcal{R}ij} t^i t^j \; ,
\ee

\vspace{\parskip}\noindent where we have defined a new intersection number $d_{\mathcal{P}\mathcal{Q}ij} \equiv \int_X N_{\cal{P}} \wedge N_{\cal{Q}} \wedge \omega_i \wedge \omega_j$, which is purely topological. Note that $d_{\mathcal{P}\mathcal{Q}ij} = - d_{\mathcal{Q}\mathcal{P}ij}$. The metric anti-commutes with the complex structure

\be
	\cs_{\cal{P}} {}^{\cal{Q}} \mathcal{G}^{(3)}_{\cal{QR}} + \mathcal{G}^{(3)}_{\cal{PQ}} \cs_{\cal{R}} {}^{\cal{Q}} = 0 \; , 
\ee

\vspace{\parskip}\noindent which, in fact, becomes a Hermiticity condition on the metric $\mathcal{G}^{(3)}$

\be\eqlabel{G3_hermitian}
	\mathcal{G}^{(3)}_{\cal{PQ}} = \cs_{\cal{P}} {}^{\cal{R}} \cs_{\cal{Q}} {}^{\cal{S}} \mathcal{G}^{(3)}_{\cal{RS}} \; .
\ee

\vspace{\parskip}\noindent Thus, the 3-form moduli space is a Hermitian manifold with $\mathcal{G}^{(3)}$ being a Hermitian metric.

A real 4-form, which is topologically invariant, can be decomposed into the sum of $(1,3)$-, $(3,1)$- and $(2,2)$-forms using the complex structure of the Calabi-Yau five-fold $X$. In the same spirit as for the 3-forms, we introduce linear maps $\mathfrak{C}$, $\mathfrak{D}$, $\mathfrak{E}$ and $\mathfrak{F}$ to translate between real 4-forms and their $(1,3)$-, $(3,1)$- and $(2,2)$-pieces\newnot{symbols:4_cplx_maps}
\begin{align}
	\varpi_x &= \mathfrak{C}_x {}^{\cal{X}} O_{\cal{X}} \qquad \qquad \qquad 
	\text{(and: $\bar{\varpi}_{\bar{x}} = \bar{\mathfrak{C}}_{\bar{x}} {}^{\cal{X}} O_{\cal{X}}$)} \; ,\eqlabel{31_4_rel1} \\
	\sigma_e &= \mathfrak{E}_e {}^{\cal{X}} O_{\cal{X}} \; , \eqlabel{31_4_rel2} \\
	O_{\cal{X}} &= \mathfrak{D}_{\cal{X}} {}^x \varpi_x + \bar{\mathfrak{D}}_{\cal{X}} {}^{\bar{x}} \bar{\varpi}_{\bar{x}} 
			    + \mathfrak{F}_{\cal{X}} {}^e \sigma_e \; , \eqlabel{31_4_rel3} 
\end{align}

\vspace{\parskip}\noindent where $\{ \varpi_x \}$ is a basis of $H^{1,3} (X)$, whereas $\{ \sigma_e \}$ and $\{ O_{\cal{X}} \}$\newnot{symbol:ind_real4}\newnot{symbol:4form_basis} are real bases of $H^{2,2} (X)$ and $H^{4} (X)$, respectively. Unlike $\mathfrak{C}$ and $\mathfrak{D}$, $\mathfrak{E}$ and $\mathfrak{F}$ are real. All linear maps $\mathfrak{C}$, $\mathfrak{D}$, $\mathfrak{E}$ and $\mathfrak{F}$ \emph{a priori} depend on the complex structure moduli $z^a$ and $\bar{z}^{\bar{a}}$. By consecutively inserting \eqrangeref{31_4_rel1}{31_4_rel3} into each other, we learn relations among the linear maps
\begin{align}
	&\mathfrak{C}_x {}^{\cal{X}} \mathfrak{D}_{\cal{X}} {}^y = \delta_x {}^y \; , \qquad
	\bar{\mathfrak{C}}_{\bar{x}} {}^{\cal{X}} \bar{\mathfrak{D}}_{\cal{X}} {}^{\bar{y}} = \delta_{\bar{x}} {}^{\bar{y}} \; , \qquad
	\mathfrak{E}_e {}^{\cal{X}} \mathfrak{F}_{\cal{X}} {}^f = \delta_e {}^f \; , \eqlabel{CD_rel1} \\
	&\mathfrak{C}_x {}^{\cal{X}} \bar{\mathfrak{D}}_{\cal{X}} {}^{\bar{y}} = 
	\mathfrak{C}_x {}^{\cal{X}} \mathfrak{F}_{\cal{X}} {}^e =
	\mathfrak{E}_e {}^{\cal{X}} \mathfrak{D}_{\cal{X}} {}^x = 0 \; , \qquad\qquad\qquad \text{(and $\cc$)} \; , \eqlabel{CD_rel2} \\
	&\mathfrak{D}_{\cal{X}} {}^x \mathfrak{C}_x {}^{\cal{Y}} + 
	\bar{\mathfrak{D}}_{\cal{X}} {}^{\bar{x}} \bar{\mathfrak{C}}_{\bar{x}} {}^{\cal{Y}} +
	\mathfrak{F}_{\cal{X}} {}^e \mathfrak{E}_e {}^{\cal{Y}} = \delta_{\cal{X}} {}^{\cal{Y}} \; . \eqlabel{CD_rel3}
\end{align}

The wedge product of two harmonic $(1,1)$-forms is a harmonic $(2,2)$-form. For the purpose of this thesis, we will restrict attention to the case where all $(2,2)$-forms are obtained by wedging together two $(1,1)$-forms, that is we require\footnote{In the Calabi-Yau four-fold literature, the right hand side of \eqref{22_vertical} is often referred to as the vertical part, denoted $H^{2,2}_V$, of $H^{2,2}$ (see, for example, ref.~\cite{Haack:2002tu}). The total space $H^{2,2}$ is given by $H^{2,2} = H^{2,2}_V \oplus H^{2,2}_H$, where $H^{2,2}_H$ comprises all $(2,2)$-forms that can \emph{not} be obtained by the product of two $(1,1)$-forms. In this terminology, we are considering Calabi-Yau five-folds $X$ for which $H^{2,2} (X) = H^{2,2}_V (X)$ and $H^{2,2}_H (X) = 0$.}
\be\eqlabel{22_vertical}
	H^{2,2}(X) = H^{1,1}(X) \wedge H^{1,1}(X) \; .
\ee
All explicit examples of Calabi-Yau five-folds presented in this thesis satisfy \eqref{22_vertical}. The significance of this restriction is that, since the $(1,1)$-forms (being naturally locked to 2-cycles) are independent of the complex structure, so are the $(2,2)$-forms if they are entirely generated by the square of $(1,1)$-forms. This implies that $\sigma_e$, $\mathfrak{E}_e {}^{\cal{X}}$ and $\mathfrak{F}_{\cal{X}} {}^e$ are all independent of the complex structure moduli (and of any other moduli, in fact). Since the left hand side and the last term on the right hand side of \eqref{31_4_rel3} are independent of the complex structure, the same must be true for the \emph{sum} of the first two terms on the right hand side. This observation allows us to treat the $(1,3)$ and $(3,1)$ part together in a complex structure independent way.

Let us now choose the basis $\{ O_{\cal{X}} \}$ such that the first $2 \, h^{1,3}(X)$ indices lie in the $(1,3)+(3,1)$ directions and the remaining indices lie in the $(2,2)$ direction, that is we divide the index range $\mathcal{X}=(\hat{\mathcal{X}},\tilde{\mathcal{X}})$, where $\hat{\mathcal{X}}=1,\ldots,2h^{1,3}(X)$\newnot{symbol:ind_hat4} and $\tilde{\mathcal{X}}=1,\ldots,h^{2,2}(X)$. This rearrangement is also independent of the complex structure. \Eqrangeref{31_4_rel1}{31_4_rel3} then become
\begin{align}
	\varpi_x &= \mathfrak{C}_x {}^{\hat{\cal{X}}} O_{\hat{\cal{X}}} \qquad \qquad \qquad 
	\text{(and: $\bar{\varpi}_{\bar{x}} = \bar{\mathfrak{C}}_{\bar{x}} {}^{\hat{\cal{X}}} O_{\hat{\cal{X}}}$)} \; ,\eqlabel{31_4_new_rel1} \\
	\sigma_e &= \mathfrak{E}_e {}^{\tilde{\cal{X}}} O_{\tilde{\cal{X}}} \; , \eqlabel{31_4_new_rel2} \\
	O_{\hat{\cal{X}}} &= \mathfrak{D}_{\hat{\cal{X}}} {}^x \varpi_x + \bar{\mathfrak{D}}_{\hat{\cal{X}}} {}^{\bar{x}} \bar{\varpi}_{\bar{x}}, \qquad
	O_{\tilde{\cal{X}}} = \mathfrak{F}_{\tilde{\cal{X}}} {}^e \sigma_e \; , \eqlabel{31_4_new_rel3} 
\end{align}
where $O_{\hat{\cal{X}}}$\newnot{symbol:hat4form_basis}, $O_{\tilde{\cal{X}}}$\newnot{symbol:tilde4form_basis}, $\mathfrak{F}_{\tilde{\cal{X}}} {}^e$, $\mathfrak{E}_e {}^{\tilde{\cal{X}}}$ and $\sigma_e$ are independent of the complex structure moduli $z^a$, whereas all other objects are dependent on them. Instead of \eqrangeref{CD_rel1}{CD_rel3} we have
\begin{align}
	&\mathfrak{C}_x {}^{\hat{\cal{X}}} \mathfrak{D}_{\hat{\cal{X}}} {}^y = \delta_x {}^y \; , \qquad
	\bar{\mathfrak{C}}_{\bar{x}} {}^{\hat{\cal{X}}} \bar{\mathfrak{D}}_{\hat{\cal{X}}} {}^{\bar{y}} = \delta_{\bar{x}} {}^{\bar{y}} \; , \qquad
	\mathfrak{E}_e {}^{\tilde{\cal{X}}} \mathfrak{F}_{\tilde{\cal{X}}} {}^f = \delta_e {}^f \; , \eqlabel{CD_new_rel1} \\
	&\mathfrak{C}_x {}^{\hat{\cal{X}}} \bar{\mathfrak{D}}_{\hat{\cal{X}}} {}^{\bar{y}} = 0 \; , 
		\qquad\qquad\qquad \text{(and $\cc$)} \; , \eqlabel{CD_new_rel2} \\
	&\mathfrak{D}_{\hat{\cal{X}}} {}^x \mathfrak{C}_x {}^{\hat{\cal{Y}}} + 
	\bar{\mathfrak{D}}_{\hat{\cal{X}}} {}^{\bar{x}} \bar{\mathfrak{C}}_{\bar{x}} {}^{\hat{\cal{Y}}} = \delta_{\hat{\cal{X}}} {}^{\hat{\cal{Y}}} \; , \qquad
	\mathfrak{F}_{\tilde{\cal{X}}} {}^e \mathfrak{E}_e {}^{\tilde{\cal{Y}}} = \delta_{\tilde{\cal{X}}} {}^{\tilde{\cal{Y}}} \; . \eqlabel{CD_new_rel3}
\end{align}
The relations between $\mathfrak{C}_x {}^{\hat{\cal{X}}}$, $\mathfrak{D}_{\hat{\cal{X}}} {}^y$, $O_{\hat{\cal{X}}}$ and $\varpi_x$ are very similar to the relations between $\mathfrak{A}_p {}^{\cal{P}}$, $\mathfrak{B}_{\cal{P}} {}^q$, $N_{\cal{P}}$ and $\nu_p$ for the 3-form case discussed above. The complex structure dependence in the $(1,3)$-sector is parametrised by $\mathfrak{C}_x {}^{\hat{\cal{X}}}$ and $\mathfrak{D}_{\hat{\cal{X}}} {}^y$
\begin{align}
	\partial_a O_{\hat{\cal{X}}} &= 0 \; , &
	\partial_a \varpi_x &= 
		\mathfrak{C}_x {}^{\hat{\cal{Y}}} {}_{,a} \mathfrak{D}_{\hat{\cal{Y}}} {}^y \varpi_y + 
		\mathfrak{C}_x {}^{\hat{\cal{Y}}} {}_{,a} \bar{\mathfrak{D}}_{\hat{\cal{Y}}} {}^{\bar{y}} \bar{\varpi}_{\bar{y}} \; , \\
	\partial_{\bar{a}} O_{\hat{\cal{X}}} &= 0 \; , &
	\partial_{\bar{a}} \varpi_x &= 
		\mathfrak{C}_x {}^{\hat{\cal{Y}}} {}_{,\bar{a}} \mathfrak{D}_{\hat{\cal{Y}}} {}^y \varpi_y + 
		\mathfrak{C}_x {}^{\hat{\cal{Y}}} {}_{,\bar{a}} \bar{\mathfrak{D}}_{\hat{\cal{Y}}} {}^{\bar{y}} \bar{\varpi}_{\bar{y}} \; .
\end{align}
Using \eqref{31_4_new_rel1,31_4_new_rel3,dualforms}, one can compute the Hodge star of the real 4-form $O_{\hat{\cal{X}}}$
\be\eqlabel{4form_hodgestar}
	\ast O_{\hat{\cal{X}}} = - O_{\hat{\cal{X}}} \wedge J \; .
\ee
Whenever we use the forms $O_{\hat{\cal{X}}}$ to describe $(1,3)$- and $(3,1)$-forms we will refer to it as the $\hat{4}$-form formulation. The standard metric on the moduli space of real $\hat{4}$-forms is given by
\be\eqlabel{4form_metric}
	{\cal G}^{(\hat{4})}_{\cal{\hat{X}\hat{Y}}} = \int_X O_{\hat{\cal{X}}} \wedge \ast O_{\hat{\cal{Y}}} \; .
\ee
Using the expression for the Hodge star \eqref*{4form_hodgestar}, we can rewrite this so as to make the dependence on the moduli more explicit
\be
	{\cal G}^{(\hat{4})}_{\cal{\hat{X}\hat{Y}}} (\underline{t}) = 
		- d_{\mathcal{\hat{X}}\mathcal{\hat{Y}}i} t^i \; ,
\ee
where we have defined a new intersection number $d_{\mathcal{\hat{X}}\mathcal{\hat{Y}}i} \equiv \int_X O_{\hat{\cal{X}}} \wedge O_{\hat{\cal{Y}}} \wedge \omega_i$, which is purely topological. Note that $d_{\mathcal{\hat{X}}\mathcal{\hat{Y}}i} = d_{\mathcal{\hat{Y}}\mathcal{\hat{X}}i}$.

Similarly to the 3-form case, there is a complex structure $\cs_{\hat{\cal{X}}} {}^{\hat{\cal{Y}}}$ on the $\hat{4}$-form moduli space inherited from the complex structure of the Calabi-Yau five-fold and given by
\be
	\cs_{\hat{\cal{X}}} {}^{\hat{\cal{Y}}} \equiv i (\mathfrak{D}_{\hat{\cal{X}}} {}^x \mathfrak{C}_x {}^{\hat{\cal{Y}}} - 
	\bar{\mathfrak{D}}_{\hat{\cal{X}}} {}^{\bar{x}} \bar{\mathfrak{C}}_{\bar{x}} {}^{\hat{\cal{Y}}}) \; .
\ee
It satisfies relations~\eqref*{3form_cplx_struct_rel} with indices adjusted appropriately. The projection operators are
\be\eqlabel{4form_projectors_def}
	P_{\pm} {}_{\hat{\cal{X}}} {}^{\hat{\cal{Y}}} \equiv \oneon{2} (\id \mp i \cs)_{\hat{\cal{X}}} {}^{\hat{\cal{Y}}} \; ,
\ee
which satisfy \eqref{3form_projectors_rel} and are explicitly given by
\be
	P_{+} {}_{\hat{\cal{X}}} {}^{\hat{\cal{Y}}} = \mathfrak{D}_{\hat{\cal{X}}} {}^y \mathfrak{C}_y {}^{\hat{\cal{Y}}} \; , \qquad\qquad
	P_{-} {}_{\hat{\cal{X}}} {}^{\hat{\cal{Y}}} = \bar{\mathfrak{D}}_{\hat{\cal{X}}} {}^{\bar{y}} \bar{\mathfrak{C}}_{\bar{y}} {}^{\hat{\cal{Y}}} \; .
\ee
Note, however, that unlike in the 3-form case, the standard $\hat{4}$-form metric~\eqref*{4form_metric} is not Hermitian with respect to the complex structure $\cs_{\hat{\cal{X}}} {}^{\hat{\cal{Y}}}$.

\section{Examples of Calabi-Yau five-folds}\seclabel{cy5_examples}

When studying a new type of mathematical entity, it is of utmost importance to assure oneself that one is not dealing with an empty set. The simplest way to achieve this is to find some explicit examples, in this case of Calabi-Yau five-folds. Of course, this only establishes mathematical existence. In addition, M-theory places some further restrictions on potential compactification manifolds. In a second step, one thus needs to check whether the constructed examples provide suitable backgrounds for performing M-theory reductions. This is the matter of \secref{topol_constraint, flux_cy5_examples}. Here, however, we start first of all by summarising some well-known mathematical constructions of Calabi-Yau manifolds, specialised to the case of five-folds. These constructions have first been accomplished for Calabi-Yau three-folds~\cite{Hubsch:1992,Candelas:1987kf,Candelas:1987du,Green:1986ck,Candelas:1985en} and four-folds~\cite{Klemm:1996ts,Brunner:1996pk,Brunner:1996bu,Gopakumar:1996mu}.

\subsection{Complete intersections in products of projective spaces}\seclabel{cicy5}

Arguably, the simplest constructions of explicit Calabi-Yau manifolds are the so-called complete intersection Calabi-Yau manifolds (CICYs). In order to define them, we first need the notion of a complex projective space $\CP^n$\newnot{symbol:Pn} of complex dimension $n$, which is the space of complex lines in $\C^{n+1}$
\be
	\CP^n \equiv \{ z^\mu \in \C^{n+1} \backslash \{0\} \} / \{ z^\mu \sim \lambda z^\mu , \, \lambda \in \C^\ast \} \; ,
\ee
where $\C^\ast \equiv \C \backslash \{0\}$ and $z^\mu$ are homogeneous coordinates on $\CP^n$. The complex projective space $\CP^n$ is a K\"ahler manifold and can also be written as $\CP^n = S^{2n+1} / \U(1)$. The Hodge numbers of $\CP^n$ are $h^{p,q} (\CP^n) = 1$ for $p=q$ ($p,q=0,\ldots,n$) and zero otherwise. The zero locus of a holomorphic polynomial $p(z^\mu)$ of homogeneous degree $q$,
\be
	p(z^\mu) = 0 \; ,
\ee
defines a complex co-dimension one hypersurface in $\CP^n$. We recall that a polynomial $p(z^\mu)$ is called homogeneous of degree $q$ if it satisfies $p(\lambda z^\mu) = \lambda^q p(z^\mu)$. The hypersurface inherits both the complex structure and the K\"ahlerity property from the embedding space.

CICYs are embedded in an ambient space ${\cal A}$ consisting of one or more complex projective spaces of various dimensions
\be\eqlabel{def_CICY_amb_space}
	\mathcal{A} \equiv \bigotimes_{r=1}^m \CP^{n_r} \; ,
\ee
which is also K\"ahler and $\dim_\C \mathcal{A} = \sum_{r=1}^m n_r$. Each $\CP^{n_r}$ in~\eqref*{def_CICY_amb_space} has its own closed K\"ahler form $J_r$  and we choose a normalisation of the form
\be\eqlabel{Pnorm}
	\int_{\CP^{n_r}} J_r^{n_r} = 1 \; .
\ee
A CICY manifold $X$ is defined as the intersection of $K$ hypersurfaces in the ambient space ${\cal A}$ such that the resulting space is a smooth complex co-dimension $K$ manifold and such that each hypersurface is given by the zero locus of a holomorphic polynomial $p_\al (z^{\mu_{(1)}}_{(1)}, \ldots , z^{\mu_{(m)}}_{(m)}) = 0$, $\al = 1,\ldots,K$, of homogeneous degree $q_\al^r$ in the coordinates $z^{\mu_{(r)}}_{(r)}$ of the $r$-th factor $\CP^{n_r}$ in ${\cal A}$.

A family of CICYs is specified by the non-negative integers $n_r$ and $q_\al^r$, where $r=1,\ldots,m$ and $\al = 1,\ldots,K$. This is conventionally arranged into a configuration matrix $[{\bf n}|{\bf q}]$
\be\eqlabel{conf2}
 [{\bf n}|{\bf q}] \equiv \left[\begin{array}{c}n_1\\\vdots\\n_m\end{array}\right|\left.\begin{array}{ccc}q^1_1&\dots&q^1_K\\
 \vdots&\ddots&\vdots\\
 q^m_1&\hdots&q^m_K\\\end{array}\right] .
\ee
The most famous CICY is undoubtedly the Calabi-Yau three-fold $[4|5]$, known as the quintic in $\CP^4$ since it is given by the zero locus of a holomorphic polynomial of homogeneous degree 5 in $\CP^4$. Here, however, we are interested in Calabi-Yau five-folds, which means we restrict to configuration matrices that satisfy
\be\eqlabel{dimcons}
	K=\sum_{r=1}^mn_r-5 \; .
\ee
In that case, the analogue of the quintic in $\CP^4$ is the septic in $\mathbb{P}^6$ denoted $[6|7]$, which is defined by the zero locus of a holomorphic polynomial of homogeneous degree 7 in $\CP^6$. In order to get acquainted with the notation of configuration matrices consider the example (which is the 5-dimensional cousin of the famous 3-dimensional Tian-Yau manifold~\cite{Tian1985})
\be\eqlabel{TY5}
	\left[\begin{array}{c}4\\4\end{array}\right|\left.\begin{array}{ccc}4&0&1\\0&4&1\end{array}\right] ,
\ee
which corresponds to three polynomials of the form
\begin{gather}
	f_{\mu_{(1)} \nu_{(1)} \rho_{(1)} \sigma_{(1)}} z_{(1)}^{\mu_{(1)}} z_{(1)}^{\nu_{(1)}} z_{(1)}^{\rho_{(1)}} z_{(1)}^{\sigma_{(1)}} = 0 \; , \\
	g_{\mu_{(2)} \nu_{(2)} \rho_{(2)} \sigma_{(2)}} z_{(2)}^{\mu_{(2)}} z_{(2)}^{\nu_{(2)}} z_{(2)}^{\rho_{(2)}} z_{(2)}^{\sigma_{(2)}} = 0 \; , \\
	h_{\mu_{(1)} \nu_{(2)}} z_{(1)}^{\mu_{(1)}} z_{(2)}^{\nu_{(2)}} = 0 \; ,
\end{gather}
embedded in $\CP^4 \times \CP^4$ and with unspecified coefficient tuples $f$, $g$ and $h$. The further analysis will not depend on the precise choice of coefficients in the polynomials. We will therefore not distinguish between the class defined by $[{\bf n}|{\bf q}]$ and its representatives $X\in [{\bf n}|{\bf q}]$.

By definition, CICYs are smooth complex manifolds. They are also compact, since they are submanifolds of ${\cal A}$ and each factor $\CP^{n_r}= S^{2n_r + 1} / \U(1)$ is compact, and they are K\"ahler, which follows by restricting the closed K\"ahler forms $J_r$ to the hypersurface in the K\"ahler manifold $\CP^{n_r}$. However, we have not yet checked, whether the so obtained manifolds are also Calabi-Yau. A rather amazing mathematical result is that for CICYs the Calabi-Yau condition, $c_1(X)=0$, translates into an algebraic constraint on the entries $n_r$ and $q_\al^r$ of the configuration matrix $[{\bf n}|{\bf q}]$. In general, the $i$-th Chern class $c_i([{\bf n}|{\bf q}])$ of a CICY $[{\bf n}|{\bf q}]$ only depends on the non-negative integers $n_r$ and $q_\al^r$ and the K\"ahler forms $J_r$ of the ambient $\CP^{n_r}$s. The details of the derivation of this result are presented in \appref{CICY_chern_classes}. Here, we only quote the final result. For CICY five-folds, the first five Chern classes are relevant. We start with the first Chern class $c_1([{\bf n}|{\bf q}])$
\be\eqlabel{c1}
	c_1([\mathbf{n}|\mathbf{q}])=\sum_{r=1}^m\left[n_r+1-\sum_{\alpha=1}^Kq^r_\alpha\right]J_r \; ,
\ee
which must be set to zero in order for $[{\bf n}|{\bf q}]$ to be a Calabi-Yau manifold. Hence, the Calabi-Yau condition $c_1([{\bf n}|{\bf q}])=0$ turns into the simple algebraic constraints
\vspace{\parskip}
\be\eqlabel{cycondition}
	\sum_{\al=1}^Kq^r_\al = n_r+1
\ee
\vspace{\parskip}\noindent 
for all $r = 1,\ldots, m$. This means each row in the $\mathbf{q}$-part of the configuration matrix has to sum up to the complex dimension of the associated projective space plus~1. The higher Chern classes of $[\mathbf{n}|\mathbf{q}]$ are given by
\vspace{\parskip}
\begin{align}
	c_2([\mathbf{n}|\mathbf{q}])&= c_2^{rs} J_r J_s = 
		\oneon{2} \left[-(n_r+1)\delta^{rs} + \sum_{\al=1}^K q_\al^r q_\al^s \right] J_r J_s \; , \eqlabel{c2} \\
	c_3([\mathbf{n}|\mathbf{q}])&= c_3^{rst} J_r J_s J_t = 
		\oneon{3} \left[(n_r+1)\delta^{rst} - \sum_{\al=1}^K q_\al^r q_\al^s q_\al^t \right] J_r J_s J_t \; , \eqlabel{c3} \\
	c_4 ([\mathbf{n}|\mathbf{q}]) &= c_4^{rstu} J_r J_s J_t J_u = \oneon{4}
	\left[ -(n_r+1)\delta^{rstu} + \sum_{\al=1}^K q_\al^r q_\al^s q_\al^t q_\al^u + 2 c_2^{rs} c_2^{tu} \right] J_r J_s J_t J_u \; , \eqlabel{c4}\\
	c_5 ([\mathbf{n}|\mathbf{q}]) &= c_5^{r_1 \ldots r_5} J_{r_1} \cdots J_{r_5} \nonumber \\ &= \oneon{5}
	\left[ (n_r+1)\delta^{r_1 \ldots r_5} - \sum_{\al=1}^K q_\al^{r_1} \cdots q_\al^{r_5} 
		+ 5 c_3^{\left(r_1 r_2 r_3 \right.} c_2^{\left. r_4 r_5 \right)} \right] J_{r_1} \cdots J_{r_5} \; , \eqlabel{c5}
\end{align}

\vspace{\parskip}\noindent where $c_1([\mathbf{n}|\mathbf{q}]) = 0$ has been used to simplify the expressions, the $\wedge$-product has been omitted for brevity and the multi-index Kronecker-$\delta$ is $\delta^{r_1 \ldots r_i} = \prod_{k=1}^{i-1} \delta^{r_k r_{k+1}}$ (in other words, $\delta^{r_1 \ldots r_i} = 1$ if $r_1 = \ldots = r_i$ and zero otherwise)\newnot{symbol:multind_kronecker}. The fifth Chern class $c_5(X)$ is related to the Euler number defined in~\eqref*{def_euler_no} by a variant of the Gauss-Bonnet formula
\be\eqlabel{Euler_c5}
	\eta (X) = \int_X c_5(X) \; .
\ee
Of particular importance for M-theory reductions is the value of the fourth Chern class $c_4(X)$. The expression~\eqref*{c4} for $c_4 ([\mathbf{n}|\mathbf{q}])$ can be further re-written by introducing the dual $(4,4)$-forms $\hat{J}^r$\newnot{symbol:dual8forms} via
\be\eqlabel{def_dual_44_forms}
	\int_X J_r \wedge \hat{J}^s = \delta_r^s \; ,
\ee
and expanding $c_4 ([\mathbf{n}|\mathbf{q}]) = \hat{c}_{4r} \hat{J}^r$. From \eqref{c4,def_dual_44_forms} and the definition~\eqref*{def_kappa_map} of the $\kappa$-map, it follows that
\be\eqlabel{intersec}
	\hat{c}_{4r} = c_4^{stuv} \kappa(J_r,J_s,J_t,J_u,J_v) \; .
\ee
The expression $\kappa(J_r,J_s,J_t,J_u,J_v)$ can be evaluated by means of the general theorem for the pull back in terms of the K\"ahler class (see, for example, ref.~\cite{Hubsch:1992})
\be\eqlabel{mudef}
	\int_{[\mathbf{n}|\mathbf{q}]} \omega = \int_{\cal A} \omega \wedge\mu \; ,\qquad 
	\mu \equiv \bigwedge_{\al=1}^K\left(\sum_{r=1}^mq^r_\al J_r\right) ,
\ee
which converts integration of a closed $(5,5)$-form $\omega$ over $[\mathbf{n}|\mathbf{q}]$ into an integration over the ambient space ${\cal A}$. Taking into account the normalisation~\eqref*{Pnorm}, we find $c_4 = 720 \hat{J}^1 + 720 \hat{J}^2$ and $\eta = - 4128$ for the example~\eqref*{TY5}.

The Hodge numbers for CICY five-folds that satisfy $q^r_\alpha>0$ for all $r$ and $\alpha$ can be determined straightforwardly by making use of two results from algebraic topology, namely the Lefschetz hyperplane theorem (see, for example, ref.~\cite{Hubsch:1992})
\be
	H^{p,q}(X) \simeq H^{p,q}({\cal A}) \qquad \text{for $p+q\neq 5$} 
\ee
and the K\"unneth formula (see, for example, ref.~\cite{Hubsch:1992})
\be
	H^n (X\times Y) \simeq \bigoplus_{i+j=n} H^i(X)\otimes H^j(Y) \; .
\ee
Out of the six independent Hodge numbers $h^{1,1}(X)$, $h^{1,2}(X)$, $h^{1,3}(X)$, $h^{2,2}(X)$, $h^{1,4}(X)$ and $h^{2,3}(X)$, all but the last two can be directly calculated in this way and one finds
\begin{align}
	h^{1,1}([\mathbf{n}|\mathbf{q}])&= h^{1,1}({\cal A})=m\\
	h^{1,2}([\mathbf{n}|\mathbf{q}])&= 0\\
	h^{1,3}([\mathbf{n}|\mathbf{q}])&= 0\\
	h^{2,2}([\mathbf{n}|\mathbf{q}])&= h^{2,2}({\cal A})=\frac{m(m-1)}{2}+\#\{r | n_r\geq 2\}\; ,
\end{align} 
where $\# S$\newnot{symbol:cardinality} denotes the cardinality of the set $S$. Note that the first equation implies that the $J_r$ inherited from the ambient $\CP^{n_r}$s form a basis of $H^{1,1}([\mathbf{n}|\mathbf{q}])$. The remaining two Hodge numbers $h^{1,4}([\mathbf{n}|\mathbf{q}])$ and $h^{2,3}([\mathbf{n}|\mathbf{q}])$ are fixed using \eqref{c5,Euler_c5}, the definition~\eqref*{def_euler_no} and the constraint~\eqref*{Hodgecons}. If some of the $q^r_\alpha$ are zero, more subtle variants of the above arguments may be applicable or in other cases, more complicated techniques, such as spectral sequence methods~\cite{Green:1987cr}, may need to be invoked to determine the Hodge numbers.

\begin{table}[t]\begin{center}
\begin{tabular}{|l|l|l|l|l|l|}
\hline
$[n|q_1\dots q_K]$&$c_2/J^2$&$c_4/J^4$&$\eta$&$h^{1,4}$&$h^{2,3}$\\\hline\hline
$[6|7]$&$21$&$819$&$-39984$&$1667$&$18327$\\\hline
$[7|6\,2]$&$16$&$454$&$-32544$&$1357$&$14917$\\\hline
$[7|5\,3]$&$13$&$259$&$-19440$&$811$&$8911$\\\hline
$[7|4\,4]$&$12$&$198$&$-14208$&$593$&$6513$\\\hline
$[8|5\,2\,2]$&$12$&$234$&$-23280$&$971$&$10671$\\\hline
$[8|4\,3\,2]$&$10$&$136$&$-13392$&$559$&$6139$\\\hline
$[8|3\,3\,3]$&$9$&$99$&$-9720$&$406$&$4456$\\\hline
$[9|4\,2\,2\,2]$&$9$&$114$&$-14592$&$609$&$6689$\\\hline
$[9|3\,3\,2\,2]$&$8$&$78$&$-9648$&$403$&$4423$\\\hline
$[10|3\,2\,2\,2\,2]$&$7$&$58$&$-8832$&$369$&$4049$\\\hline
$[11|2\,2\,2\,2\,2\,2]$&$6$&$39$&$-6912$&$289$&$3169$\\\hline
\end{tabular}
\caption{The 11 CICY five-folds that can be defined in a single projective space. The Hodge numbers are $h^{1,2}(X)=h^{1,3}(X)=0$ and $h^{1,1}(X)=h^{2,2}(X)=1$ for all manifolds.}
\tablabel{tab:cicy1}\end{center}\end{table}
CICY five-folds have so far not been classified systematically. In contrast, CICY three-folds have been classified and 7868 distinct configurations have been found~\cite{Candelas:1987kf}. It is beyond the scope of this thesis to attempt a full classification of CICY five-folds. Instead, we are content with establishing the mere existence of configurations that lead to viable M-theory compactifications. In \tabref{tab:cicy1}, we list all the CICY five-folds that can be defined in a single projective space. Since
\be
	[n|q_1\,\dots \,q_{K-1}\, 1] \simeq [n-1|q_1\,\dots\, q_{K-1}] \; ,
\ee
one can restrict to $q_\al > 1$ for all $\al$, without loss of generality. One then finds 11 distinct configurations. For comparison, 5 distinct CICY three-folds in a single projective space exist. 
\begin{table}[t]\begin{center}
\begin{tabular}{|l|l|l|l|l|l|l|l|}
\hline
$[\mathbf{n}|\mathbf{q}]$&$c_2$&$c_4$&$\eta$&$h^{1,1}$&$h^{2,2}$&$h^{1,4}$&$h^{2,3}$\\\hline\hline
$\left[\begin{array}{l}1\\5\end{array}\right|\left.\begin{array}{l}2\\6\end{array}\right]$&$\begin{array}{l}12J_1J_2+\\15J_2^2\end{array}$&$\begin{array}{l}2610\hat{J}^1+\\4542\hat{J}^2\end{array}$&$-32280$&$2$&$2$&$1347$&$14797$\\\hline
$\left[\begin{array}{l}2\\4\end{array}\right|\left.\begin{array}{l}3\\5\end{array}\right]$&$\begin{array}{l}3J_1^2+\\15J_1J_2+\\10J_2^2\end{array}$&$\begin{array}{l}3240\hat{J}^1+\\3975\hat{J}^2\end{array}$&$-29400$&$2$&$3$&$1227$&$13478$\\\hline
$\left[\begin{array}{l}3\\3\end{array}\right|\left.\begin{array}{l}4\\4\end{array}\right]$&$\begin{array}{l}6J_1^2+\\16J_1J_2+\\6J_2^2\end{array}$&$\begin{array}{l}3600\hat{J}^1+\\3600\hat{J}^2\end{array}$&$-28608$&$2$&$3$&$1194$&$13115$\\\hline
$\left[\begin{array}{l}1\\2\\3\end{array}\right|\left.\begin{array}{l}2\\3\\4\end{array}\right]$&$\begin{array}{l}3J_2^2+\\12J_2J_3+\\6J_3^2+\\6J_1J_2+\\8J_1J_3\end{array}$&$\begin{array}{l}84\hat{J}^1+\\114\hat{J}^2+\\130\hat{J}^3\end{array}$&$-24480$&$3$&$5$&$1023$&$11225$\\\hline
\end{tabular}
\caption{Examples of CICY five-folds defined in a product of projective spaces. The Hodge numbers are $h^{1,2}(X)=h^{1,3}(X)=0$ for all manifolds.}
\tablabel{tab:cicy2}\end{center}\end{table}
\Tabref{tab:cicy2} lists some of the properties of four examples of CICY five-folds defined in a product of multiple projective spaces. In \secref{topol_constraint}, we examine whether these examples are well-suited as compactification manifolds for M-theory.

\subsection{Torus quotients}\seclabel{torus_quotients}

For reasons that will become clear later, it is desirable to have examples of Calabi-Yau five-folds $X$ with vanishing fourth Chern class $c_4(X)=0$. Amongst the CICY five-folds with ``small'' configuration matrices, there are no such examples, given the additional restriction that the global holonomy group ${\rm Hol}(X) \subseteq \SU(5)$ be sufficiently large to allow only one out of 16 supersymmetries, as stated in the definition in \secref{cy_def}. We thus need to turn to another way of constructing Calabi-Yau manifolds to establish a simple example with $c_4(X)=0$.

A good starting point is the 10-torus $T^{10} = T^2 \times \cdots \times T^2$ thought of as five 2-tori. This is a complex dimensional manifold with $\dim_\C T^{10} = 5$ and holomorphic coordinates $z^\mu$, $\mu=1,\ldots,5$ with identifications $z^\mu\sim z^\mu+1$ and $z^\mu\sim z^\mu +i$. The $\mu$-th coordinate $z^\mu$ is the single holomorphic coordinate in the $\mu$-th 2-torus. The 10-torus is K\"ahler and Ricci-flat with respect to the canonical metric $\delta_{\mu\bar\nu}$. However, it does not break any supersymmetry and is therefore not a Calabi-Yau five-fold in the strict sense defined in \secref{cy_def}. On the other hand, it has the desirable feature of vanishing Chern classes $c_i(T^{10})=0$ for $i=1,\ldots,5$.

Consider the symmetry $\mathbb{Z}_2^4$ acting on the holomorphic coordinates $z^\mu$ of $T^{10}$ and being generated by
\begin{align}
 \gamma_1(z^1,\ldots ,z^5) &= (-z^1+1/2,-z^2+i/2,z^3+1/2,z^4,z^5)\\
 \gamma_2(z^1,\dots ,z^5) &= (z^1,-z^2+1/2,-z^3+i/2,z^4+1/2,z^5)\\
 \gamma_3(z^1,\ldots ,z^5) &= (z^1,z^2,-z^3+1/2,-z^4+i/2,z^5+1/2)\\
 \gamma_4(z^1\ldots ,z^5) &= (z^1+1/2,z^2,z^3,-z^4+1/2,-z^5+i/2)\; .
\end{align}
The 16 elements of this $\mathbb{Z}_2^4$-group are fixed point free. Therefore, $\mathbb{Z}_2^4$ is freely acting and the quotient space $X=(T^2)^5/\mathbb{Z}_2^4$ constitutes a smooth manifold with the inherited properties of being a Ricci-flat K\"ahler manifold of complex dimension five and with vanishing Chern classes. Despite the smallness of the holonomy group ${\rm Hol}(X) = \mathbb{Z}_2^4$, it is nonetheless sufficient to preserve only one out of 16 supersymmetries and therefore $X$ constitutes a Calabi-Yau five-fold in the strict sense defined in \secref{cy_def}. The Hodge numbers $h^{p,q}(X)$ of $X$ are obtained by counting the number of $\mathbb{Z}_2^4$-invariant $(p,q)$-differentials $dz^{\mu_1}\wedge\dots\wedge dz^{\mu_p}\wedge d\bar{z}^{\nu_1}\wedge\dots\wedge d\bar{z}^{\nu_q}$. One finds
\begin{align}
	&h^{1,1}(X)=5\; ,& &h^{1,2}(X)=0\; ,& &h^{1,3}(X)=0\; , \eqlabel{torus_quotient_hpq1} \\ 
	&h^{2,2}(X)=10\; ,& &h^{1,4}(X)=5\; ,& &h^{2,3}(X)=10\; . \eqlabel{torus_quotient_hpq2}
\end{align}
An interesting, but as yet open, question is whether explicit Calabi-Yau five-folds can be found that have both full $\SU(5)$-holonomy and vanishing fourth Chern class $c_4(X)=0$.
%
%

\chapter{\texorpdfstring{$\susyno=2$}{N=2} Supertime}\chlabel{superspace}

In this chapter, we review and develop one-dimensional $\susyno=2$ supersymmetry to the level necessary to describe the effective actions arising from Calabi-Yau five-fold reductions of M-theory. One-dimensional supersymmetry has previously been studied, for example, in refs.~\cite{Howe:1989vn,Coles:1990hr,vanHolten:1995qt,Machin:2002zn} and notably in the context of black hole moduli spaces~\cite{Gibbons:1997iy}. However, in order to be able to describe the effective actions that will arise in the next chapter in superspace language, some generalisations and extensions of the one-dimensional $\susyno=2$ formulations discussed in the literature are required.

These extensions include couplings between $2a$ and $2b$ multiplets -- the two main irreducible multiplets in one-dimensional $\susyno=2$ supersymmetry -- as well as actions and component versions for fermionic multiplets, which are multiplets whose lowest component is a fermion. By means of these fermionic multiplets, a peculiar feature of one-dimensional supersymmetry will manifest itself, namely the possible mismatch of the number of on-shell bosonic and fermionic degrees of freedom, which is in stark contrast to higher dimensional supersymmetry where the matching of these numbers is a cornerstone feature. Even though gravity in one dimension is non-dynamical, it leads to constraints which cannot be ignored. This means we also need to consider one-dimensional {\em local} supersymmetry. Finally, we study some superpotentials in preparation of the flux compactification performed at the end of the next chapter.

All those features have not been fully worked out in the literature. We therefore provide a systematic exposition of one-dimensional $\susyno=2$ global and local supersymmetry in this chapter. Before starting, however, we take a step back and briefly summarise the merits of superspace formulations of supersymmetric theories in order to motivate our efforts.

\section{Superspace \texorpdfstring{$1 \times 1$}{1x1}}\seclabel{sspace101}

As mentioned in \chref{intro}, supersymmetry is a symmetry interchanging bosons and fermions. Supersymmetric theories possess many desirable and promising features from a theoretical and phenomenological perspective. In particular, Haag, \L opusza\'nski and Sohnius showed in 1975~\cite{Haag:1974qh} that supersymmetry offers the only possible way to evade the famous Coleman-Mandula no-go theorem~\cite{Coleman:1967ad}.

However, as we saw via the 11-dimensional example in \chref{mthy} (and \appref{proof_CJS_action_susy}), supersymmetry is also a rather complicated symmetry and checking \emph{ad hoc} whether an action is invariant under a given set of supersymmetry transformations, let alone determining the correct transformations, is a highly non-trivial task. It would therefore be very helpful to devise a mechanism such that actions become manifestly supersymmetric. This was achieved by the invention of superspace by Salam and Strathdee in 1974~\cite{Salam:1974yz,Salam:1974za,Salam:1974jj}. If a given action can be written in superspace, this constitutes a proof of supersymmetry for any consistent superspace action is automatically guaranteed to be supersymmetric by the very construction of superspace. Moreover, the supersymmetry of actions derived from superspace expressions is realised off-shell, that is without any reference to the equations of motion. This is achieved through incorporating additional so-called auxiliary fields, which are governed by purely algebraic equations of motion. Upon elimination of the auxiliary fields, on-shell supersymmetry is regained. Another virtue of superspace is its usefulness as a machinery to easily generate new supersymmetric theories by building new actions according to the rules of superspace.

Unfortunately, however, the superspace methods described in this section break down when attempting to describe supersymmetries generated by 8 or more real supercharges, although more sophisticated techniques may be applicable instead. An example is harmonic superspace~\cite{Galperin:2001uw}, which involves infinite sets of auxiliary fields, to describe supersymmetries with 8 to 12 real supercharges. Fortunately, one-dimensional $\susyno=2$ supersymmetry -- the case relevant for this thesis -- is generated by only two real supercharges and can therefore be described using the techniques introduced below.

The prize for having manifestly realised off-shell supersymmetry is an enlarged space-time manifold $\manifold^{D|\susyno}$ parametrised not only by ordinary (commuting) space-time coordinates $x^\mu$ with $\mu=0,\ldots,(D-1)$, but in addition also containing anti-commuting directions\footnote{This section aims to provide a brief schematical overview of some of the ideas and concepts behind superspace and is by no means complete or rigourous. In particular, the precise details of the equations presented below often depend on the space-time dimension $D$ and to some extend also on $\susyno$. For a more complete and rigourous exposition, we refer to, for example, refs.~\cite{Gates:1983nr,West:1990tg,Wess:1992cp} and ref.~\cite{DeWitt:1992cy} for more on the mathematical underpinnings.} $\theta^{\al i}$ with $\al=1,\ldots,f$ and $i=1,\ldots,\susyno$. Here, $f$ is the dimension of the spinor representation, and $\susyno$ counts the number of independently realised supersymmetries. The conjugate spinors $\bar\theta^{\dot\al i} = \theta^{\al i} {}^\dag$ are in general also present. The anti-commuting nature of these extra coordinates is expressed by
\be\eqlabel{graded_coords}
	[ x^\mu, x^\nu ] = 0 \; , \qquad
	[ x^\mu, \theta^{\al i} ] = 0 \; , \qquad
	\{ \theta^{\al i}, \theta^{\beta j} \} = 0 \; , \qquad
	\{ \theta^{\al i}, \bar\theta^{\dot\beta j} \} = 0 \; ,
\ee
valid for all index choices, which shows that $\manifold^{D|\susyno}$ looks locally like a $\Z_2$-graded vector space with even subspace $\M^D$ and odd subspace $\R^{f \susyno}$ ($\simeq\C^{f \susyno/2}$ for $f \susyno$ even). The components of the $\Z_2$-graded coordinates $\theta_\al^i$ are called Grassmann numbers. The associated derivatives $\partial_{\al i} \equiv \partial / \partial \theta^{\al i}$ and $\bar\partial_{\dot\al i} \equiv \partial / \partial \bar\theta^{\dot\al i}$ satisfy
\begin{gather}
	\partial_{\al i} \theta^{\beta j} = \delta_\al^\beta \delta_i^j \; , \qquad
	\bar\partial_{\dot\al i} \bar\theta^{\dot\beta j} = \delta_{\dot\al}^{\dot\beta} \delta_i^j \; , \qquad
	\bar\partial_{\dot\al i} \theta^{\beta j} = \partial_{\al i} \bar\theta^{\dot\beta j} = 0 \; , \\
	\{ \partial_{\al i}, \partial_{\beta j} \} = 0 \; , \qquad
	\{ \bar\partial_{\dot\al i}, \bar\partial_{\dot\beta j} \} = 0 \; , \qquad
	\{ \partial_{\al i}, \bar\partial_{\dot\beta j} \} = 0 \; . \qquad
\end{gather}
Integration over Grassmann numbers (known as Berezin integration) is formally equivalent to differentiation $\int \leftrightarrow \partial$ (see \appref{spinorconv}) and hence
\begin{gather}
	\int d^n \theta = \int d^n \bar\theta = \int d^{2n} \theta = 0 \; , \\
	\int d^n \theta \, \theta^n = \int d^n \bar\theta \, \bar\theta^n = \int d^{2n} \theta \, \theta^n\bar\theta^n = 1 \; ,
\end{gather}
with $n\equiv f \susyno/2$ (assuming the total number of real supercharges $f \susyno$ is even) and $d^{2n} \theta \equiv d^n \theta d^n \bar\theta$.

Superfields $\Phi(x, \theta, \bar\theta)$ are general functions on superspace. They can be understood as a set of ordinary fields $\phi(x^\mu)$ by Taylor expanding the superfield $\Phi$ in $\theta^{\al i}$ and $\bar\theta^{\dot\al i}$. The Taylor series terminates after a finite number of terms due to~\eqref*{graded_coords}. The product of two superfields is itself a superfield. Supercovariant derivatives can be defined as\footnote{By ignoring the position of the extension index $i$, we are also ignoring the possibility of a non-trivial automorphism group -- called $R$-symmetry -- acting on the indices $i,j,\ldots=1,\ldots,\susyno$.}
\be
	D^i_\al \equiv \partial_{\al i} + i (\gamma^\mu \bar\theta^i)_{\al} \partial_\mu \; , \qquad
	\bar{D}^i_{\dot\al} \equiv - \bar\partial_{\dot\al i} - i (\gamma^\mu \theta^i)_{\dot\al} \partial_\mu \; ,
\ee
with the important property that the supercovariant derivative $D^i_\al \Phi$ of a superfield $\Phi$ is itself a superfield. The derivative $D^i_\al$ is intimately linked to the generators of infinitesimal supersymmetry transformations $Q^i_\al$ given by
\be\eqlabel{gen_def_Q}
	Q^i_\al \equiv \partial_{\al i} - i (\gamma^\mu \bar\theta^i)_{\al} \partial_\mu \; , \qquad
	\bar{Q}^i_{\dot\al} \equiv - \bar\partial_{\dot\al i} + i (\gamma^\mu \theta^i)_{\dot\al} \partial_\mu \; .
\ee
An infinitesimal supersymmetry transformation becomes $\delta_{\epsilon^{\al i},{\rm tot.}} = i \epsilon^{\al i} Q_\al^i + i \bar{\epsilon}^{\dot\al i} \bar{Q}_{\dot\al}^i$. Together, $Q^i_\al$ and $D^i_\al$ satisfy
\begin{gather}
	\{ Q^i_\al, \bar{Q}^j_{\dot\beta} \} = 2 i \delta^{ij} (\gamma^\mu C)_{\al\dot\beta} \partial_\mu \; , \qquad
	\{ Q^i_\al, Q^j_\beta \} = \{ \bar{Q}^i_{\dot\al}, \bar{Q}^j_{\dot\beta} \} = 0 \; , \eqlabel{sspace_gen_QD_prop1} \\
	\{ D^i_\al, \bar{D}^j_{\dot\beta} \} = - 2 i \delta^{ij} (\gamma^\mu C)_{\al\dot\beta} \partial_\mu \; , \qquad
	\{ D^i_\al, D^j_\beta \} = \{ \bar{D}^i_{\dot\al}, \bar{D}^j_{\dot\beta} \} = 0 \; , \eqlabel{sspace_gen_QD_prop2} \\
	\{ Q^i_\al, D^j_\beta \} = \{ Q^i_\al, \bar{D}^j_{\dot\beta} \} = \{ \bar{Q}^i_{\dot\al}, D^j_\beta \} =
	\{ \bar{Q}^i_{\dot\al}, \bar{D}^j_{\dot\beta} \} = 0 \; . \eqlabel{sspace_gen_QD_prop3}
\end{gather}
For cases with an even number of real supercharges $f \susyno$, a chiral superfield $\Psi$ can be defined by requiring $\bar{D}^i_{\dot\al} \Psi = 0$.

The power and usefulness of superspace is encapsulated in the following result which can be checked using the expressions given above: A general superfield $\Phi$ integrated over full superspace $\int d^D x d^{2n} \theta \, \Phi$ is invariant under the supersymmetry generated by~\eqref*{gen_def_Q}. Also, a chiral superfield $\Psi$ integrated over half-superspace $\int d^D x d^n \theta \, \Psi$ is supersymmetric in the same sense. The most general manifestly supersymmetric real action of a single general superfield $\Phi$ and a single chiral superfield $\Psi$ can be written as
\be
	\int d^D x d^{2n} \theta \, f(\Phi, D_\al^i\Phi) + \left(\int d^D x d^n \theta \, W(\Psi) + \text{h.c.} \right) ,
\ee
with some real-valued function $f$ and holomorphic function $W$ and with the supplementary condition that the integrand does not depend explicitly on the coordinates. Frequently encountered specific examples include mass terms $f = \Phi^\dag \Phi$, kinetic terms $f=\overline{(D^i\Phi)} \cdot D^i \Phi$ and sigma-models $f=g_{a\bar{b}}(\Phi) \overline{(D^i\Phi^b)} \cdot D^i \Phi^a$.

So far, we dealt with flat superspace, which describes global supersymmetry. In order to write locally supersymmetric theories -- that is, supergravity theories -- in superspace, we need to consider curved superspace. The subject of supergeometry deals with describing the geometry of curved superspace. In supergeometry, the coordinates $x^\mu$, $\theta^{\al i}$ and $\bar\theta^{\dot\al i}$ are assembled into a supercoordinate $z^A = (x^\mu, \theta^{\al i}, \bar\theta^{\dot\al i})$ satisfying a graded-commutation relation\newnot{symbol:graded_comm}
\be
	[ z^A, z^B \} \equiv z^A z^B - (-1)^{AB} z^B z^A = 0 \; ,
\ee
where, in the exponent of $(-1)^{AB}$, one sets $A=0$ ($A=1$) for $z^A$ Grassmann even (odd) and similarly for $B$. The associated derivatives are $\partial_A \equiv \partial / \partial z^A$. Supervectors $V^A (z)$ are superfields carrying a superindex $A$ and they transform as $\delta V^A = - \xi^B (\partial_B V^A) - (\partial_B \xi^A) V^B$ under infinitesimal super-general coordinate transformations (SGCTs) $z^A \rightarrow z^\prime {}^A = z^A - \xi^A$. SGCTs contain both ordinary diffeomorphisms and local supersymmetry transformations. Higher-rank supertensors are defined accordingly. One also introduces superforms $\Omega$
\be
	\Omega = \oneon{p!} dz^{A_p} \wedge \cdots \wedge dz^{A_1} \Omega_{A_1 \ldots A_p} \; ,
\ee
with the help of a graded-wedge product $dz^A \wedge dz^B = - (-1)^{AB} dz^B \wedge dz^A$. The exterior derivative $d = dz^A \partial_A$ acts as
\be
	d\Omega = \oneon{p!} dz^{A_p} \wedge \cdots \wedge dz^{A_1} \wedge dz^B \partial_B \Omega_{A_1 \ldots A_p} \; ,
\ee
whereby the Leibniz rule becomes $d(\Omega_1 \wedge \Omega_2) = \Omega_1 \wedge d\Omega_2 + (-1)^{\deg\Omega_2} (d\Omega_1) \wedge \Omega_2$.

Many other laws of ordinary Riemannian geometry carry over to supergeometry with powers of $(-1)$ appropriately inserted to reflect the $\Z_2$-grading. The formul\ae\ given here thus resemble those of \appref{forms}. Since the abundant occurrence of the prefix \emph{super-} leads to awkward notation, we henceforth drop this prefix where no confusion is possible. In particular, objects carrying indices $A,B,\ldots$ are understood to be superobjects. 

The fundamental dynamical objects of supergravity formulated in curved superspace are the Cartan variables, that is the vielbein $E_A {}^\UB$ and the spin-connection $\Omega_{A\UB} {}^{\ul{C}}$. It is important to emphasise that the tangent space indices $\UA,\UB,\ldots$ are taken to be valued in the bosonic Lorentz group $\SO(D-1,1)$ only and not in the full super-Lorentz group $\SO(D-1,1|\susyno)$ as one might have na\"ively guessed. This is in order to avoid representation-mixing transformations that would obscure the physical interpretation.

The vielbein $E_A {}^\UB$ and its inverse $E_\UA {}^B$ translate between curved and tangent-space indices 
\be
	V_A = E_A {}^\UB V_\UB \; , \qquad\qquad 
	V_\UA = E_\UA {}^B V_B
\ee
and they are related via
\be\eqlabel{sspace_curved_supervielbein_inverse_rel}
	E_A {}^\UB E_\UB {}^C = \delta_A {}^C \; , \qquad\qquad 
	E_\UA {}^B E_B {}^{\underline{C}} = \delta_\UA {}^{\underline{C}} \; .
\ee
The vielbein $E_A {}^\UB$ transforms as a co-vector under SGCTs
\be\eqlabel{sspace_curved_inf_sgct_supervielbein}
	\delta E_A {}^\UB = \xi^C (\partial_C E_A {}^\UB) + (\partial_A \xi^C) E_C {}^\UB \; .
\ee
With the help of the spin-connection 1-form $\Omega = dz^A \Omega_A {}^{\ul{rs}} M_{\ul{rs}}$, where $M_{\ul{rs}}$ are generators of the bosonic Lorentz group $\SO(D-1,1)$, covariant derivatives $\sderiv_A$ are introduced as
\be\eqlabel{sspace_gen_def_curved_sderiv}
	\sderiv \equiv dz^A \sderiv_A = d + \Omega\wedge \; , \qquad \sderiv_A = \partial_A + \Omega_A {}^{\ul{rs}} M_{\ul{rs}} \; ,
\ee
with tangent space version
\be\eqlabel{sspace_gen_def_flat_sderiv}
	\sderiv 	= E^\UA \sderiv_\UA \; , \qquad \sderiv_\UA = E_\UA {}^B \sderiv_B 
			= E_\UA {}^B \partial_B + E_\UA {}^B \Omega_B {}^{\ul{rs}} M_{\ul{rs}} \; ,
\ee
in terms of the vielbein 1-form $E^\UA = dz^B E_B {}^\UA$. By comparison with~\eqref*{sspace_gen_QD_prop2}, we see that even flat superspace -- despite having vanishing curvature -- has non-zero torsion. Further geometric objects are introduced via the graded-commutator of two covariant derivatives
\be\eqlabel{sspace_graded_comm_of_cov_derivs}
	[ \sderiv_\UA , \sderiv_\UB \} = - T_{\UA\UB} {}^{\ul{C}} \sderiv_{\ul{C}} - R_{\UA\UB} {}^{\ul{rs}} M_{\ul{rs}} \; ,
\ee
with torsion 2-form $T^\UA = \oneon{2} E^\UB \wedge E^\ul{C} T_{\ul{C}\UB} {}^\UA = \sderiv E^\UA$ and curvature 2-form $R = \oneon{2} E^\UB \wedge E^\ul{C} R_{\ul{C}\UB} {}^{\ul{rs}} M_{\ul{rs}} = \sderiv\Omega$. The first and second Bianchi identities (BIs) follow from \eqref{sspace_graded_comm_of_cov_derivs} as integrability conditions
\be
	[ [ \sderiv_\UA , \sderiv_\UB \} , \sderiv_\ul{C} \} = 0
\ee
and are explicitly given by
\begin{align}
	\sderiv T^\UA &= E^\UA \wedge R \; , \eqlabel{sspace_gen_BI1} \\
	\sderiv R &= 0 \; , \eqlabel{sspace_gen_BI2}
\end{align}
respectively.

Expressions and rules of flat superspace carry over to curved superspace by covariantisation, which amounts to promoting derivatives $d \rightarrow \sderiv$ and measures $d^D x d^{2n} \theta \rightarrow d^D x d^{2n} \theta \, \mathcal{E}$ to their respective covariant analogues. The determinant of the vielbein $\mathcal{E} \equiv {\rm sdet} E_A {}^\UB$ precisely cancels off the Jacobian from the transformation of $d^D x d^{2n} \theta$ under SGCTs.

As before, by Taylor expanding in the Grassmann odd coordinates $\theta^{\al i}$ and $\bar\theta^{\dot\al i}$, component actions of supergravity theories can be obtained. However, as it stands, the current formalism is not well-suited for describing physical systems efficiently. This is because the vielbein and spin-connection contain a very large number of \emph{a priori} independent components and therefore are vastly redundant descriptions of the underlying fundamental degrees of freedom. The over-counted degrees of freedom can be removed in a two-step process. First, the gauge freedom is removed by fixing the gauge. The residual freedom is eliminated by imposing constraints on various components of the torsion. It must then however be checked whether the torsion constraints are consistent and in particular, it must be checked whether the BIs~\eqrangeref*{sspace_gen_BI1}{sspace_gen_BI2} are still satisfied.

The precise details of this procedure are rather sensitive to the choice of $D$ and $\susyno$. Rather than dwelling further on the abstract general case, we will discuss the aforementioned techniques by means of the two explicit examples $\R^{1|2}$ and $\manifold^{1|2}$. These are the two cases relevant for this thesis. In the next section, we specialise to the flat case $\R^{1|2}$ before turning to the general curved case $\manifold^{1|2}$ in the last section of this chapter.

\section{Flat \texorpdfstring{$\susyno=2$}{N=2} supertime\texorpdfstring{: $\R^{1|2}$}{}}\seclabel{sspace_flat}

One-dimensional $\susyno=2$ superspace~\cite{Coles:1990hr} -- or $\susyno=2$ \emph{supertime} for short -- is most easily obtained by dimensional reduction from two dimensions, which has attracted a lot of attention in view of formulating superstring actions in superspace~\cite{Howe:1977xz,Martinec:1983um}. In two dimensions, there are Majorana, Weyl and Majorana-Weyl spinors and hence the same amount of supersymmetry can be realized by different choices of spinorial representation for the supercharges (see, for example, ref.~\cite{West:1990tg}). For $\susyno=2$, the two options are $(1,1)$ and $(2,0)$ supersymmetry.

Upon reduction to one dimension, these two choices of two-dimensional $\susyno=2$ supersymmetry lead to two different irreducible $\susyno=2$ supertime multiplets, referred to as $2a$ (descending from two-dimensional $(1,1)$ supersymmetry) and $2b$ (descending from two-dimensional $(2,0)$ supersymmetry) multiplets. These two irreducible representations play a central r\^{o}le in $\susyno=2$ supertime. 

Off-shell, the $2a$ multiplet contains a real scalar as its lowest component plus a complex fermion and a real scalar auxiliary field, while the $2b$ multiplet contains a complex scalar as its lowest component, accompanied by a complex fermion. The $2b$ multiplet does not contain an auxiliary field.  Other off-shell multiplets, not obtained from a standard toroidal reduction, are the fermionic $2a$ and $2b$ multiplets and the non-linear multiplet~\cite{vanHolten:1995qt}. From those we will only need and discuss in detail the fermionic $2b$ multiplet. It has a complex fermion as its lowest component which is balanced by a complex scalar at the next level.

\subsection{\texorpdfstring{$\R^{1|2}$}{Flat N=2, D=1} supergeometry}

To describe flat $\susyno=2$ supertime $\R^{1|2}$, we just need to suitably restrict the index ranges of the formul\ae\ presented in \secref{sspace101}. Specifically, $\R^{1|2}$ is parametrised by coordinates $z^A = (x^0=\tau,\theta,\bar\theta)$, where $\theta$ is a complex anti-commuting one-dimensional spinor and the superindices $A,B,\ldots$\newnot{symbol:ind_sspace} run over the values $0$, $\theta$ and $\bar{\theta}$. In analogy to~\eqref*{gen_def_Q}, the supercharges are
\be\eqlabel{sspace_flat_superchrg_defn}
	Q = \partial_\theta - \frac{i}{2} \bar\theta \partial_0 \; , \qquad\qquad
	\bar{Q} = - \partial_{\bar\theta} + \frac{i}{2} \theta \partial_0 \; ,
\ee
where $\partial_\theta = \partial / \partial \theta$, $\partial_{\bar\theta} = \partial / \partial \bar\theta = - \left(\partial_\theta\right)^\ast$, $\partial_0 = \partial / \partial x^0 = \partial / \partial\tau$. Using the conventions for one-dimensional spinors summarised in \appref{spinorconv}, it is easy to verify that $Q$ and $\bar{Q}$ satisfy the algebra
\be	
	\{Q, \bar{Q} \} = i \partial_0 = H , \quad \{ Q,Q \} = 0, \quad \{ \bar{Q},\bar{Q} \} = 0 \; .
\ee
Supersymmetry transformations of ${\cal N}=2$ supertime are parameterised by a complex one-di\-men\-sio\-nal spinor $\epsilon$ and act as
\be\eqlabel{susytransform}
	\delta_\epsilon = i \epsilon Q\; ,\qquad\qquad \delta_{\bar{\epsilon}} = i \bar{\epsilon}\bar{Q} \; .
\ee
This choice ensures that the total supersymmetry variation $\delta_{\epsilon,\text{tot.}} = \delta_\epsilon + \delta_{\bar{\epsilon}}$ is real. The associated covariant derivatives $D$ and $\bar{D}$ anti-commute with the supercharges, that is $\{D,Q\} = \{D,\bar{Q}\} = \{\bar{D},Q\} = \{\bar{D},\bar{Q}\} = 0$, and are explicitly given by
\be\eqlabel{sspace_flat_cov_deriv_defn}
	D = \partial_\theta + \frac{i}{2} \bar\theta \partial_0 \; , \qquad\qquad \bar{D} = - \partial_{\bar\theta} - \frac{i}{2} \theta \partial_0 \; .
\ee
They satisfy the anti-commutation relations
\be\eqlabel{sspace_flat_cov_deriv_comm}
	\{D,\bar{D}\} = - i \partial_0 = - H, \quad \{D,D\} = 0,\quad \{\bar{D},\bar{D}\} = 0 \; .
\ee
Although not really required for the global case, it is useful for comparison with local supersymmetry later on to develop the geometry of flat supertime. By comparing~\eqref*{sspace_flat_cov_deriv_defn} with the general definition~\eqref*{sspace_gen_def_flat_sderiv}, one can straightforwardly work out the components of the vielbein $E_A {}^\UB$ (note that $\sderiv_0=\partial_0$, $\sderiv_\theta = D$ and $\sderiv_{\bar{\theta}}=\bar{D}$). One finds (cf. ref.~\cite{Howe:1989vn} eq. (3.4))
\begin{align}
	&E_0 {}^{\underline{0}} = 1 , 				&\quad &  E_0 {}^\Uth = 0 , 			&\quad &  E_0 {}^\BUth = 0 , \nonumber \\
	&E_\theta {}^\UZ = -\frac{i}{2} \bar\theta , 	&\quad &  E_\theta {}^\Uth = 1 , 		&\quad & E_\theta {}^\BUth = 0 , \nonumber \\
	&E_{\bar{\theta}} {}^\UZ = -\frac{i}{2} \theta , 	&\quad &  E_{\bar{\theta}} {}^\Uth = 0 ,	&\quad &  E_{\bar{\theta}} {}^\BUth = -1 . \eqlabel{sspace_flat_seinbein}
\end{align}
The torsion $T^\UA$ and curvature $R$ are computed from~\eqref*{sspace_graded_comm_of_cov_derivs}. Note that in $D=1$, the Lorentz group $\SO(0,1) = \SO(1)$ is trivial. That means the single Lorentz generator $M_{\ul{rs}} = M_{\ul{11}} = - M_{\ul{11}} = 0$ and consequently also the curvature 2-form $R$ vanishes identically. There is, however, still a global $\U(1)\simeq\SO(2)$ $R$-symmetry present, which acts by rotating the complex anti-commuting coordinate $\theta \rightarrow e^{-i\alpha} \theta$ by some angle $\alpha$. For the torsion of $\R^{1|2}$, one finds (cf. ref.~\cite{Howe:1989vn} eq. (3.7))
\be\eqlabel{sspace_flat_torsion_value}
	T_{\Uth\BUth} {}^{\underline{0}} = i
\ee
as the only non-vanishing component. Finally, the determinant of the flat vielbein~\eqref*{sspace_flat_seinbein} is given by
\be
	\mathrm{sdet} E_A{}^{\underline{B}} = -1\; .
\ee

Next, we introduce superfields. Their component field content can be worked out by Taylor expanding in $\theta$ and $\bar{\theta}$. Since $\theta^2=\bar{\theta}^2=0$, only the terms proportional to $\theta$, $\bar{\theta}$ and $\theta\bar{\theta}$ arise, in addition to the lowest, $\theta$-independent component. Different types of irreducible superfields can be obtained by imposing constraints on this general superfield. We now discuss these various types in turn.

\subsection{\texorpdfstring{$2a$}{2a} multiplets}

A $2a$ superfield $\phi(\tau,\theta ,\bar{\theta})$ is defined to be a real superfield. It is constrained by the reality condition $\phi =\phi^\dagger$. A short calculation shows that the most general component expansion consistent with this constraint is
\be\eqlabel{sspace_flat_2a_comp}
	\phi = \varphi + i \theta\psi + i\bar\theta \bar\psi + \oneon{2} \theta\bar\theta f\; ,
\ee
where $\varphi$ and $f$ are real scalars and $\psi$ is a complex fermion. The highest component $f$ will turn out to be an auxiliary field so that a $2a$ superfield contains one real physical scalar field. From \eqref{sspace_flat_superchrg_defn,susytransform}, the supersymmetry transformations of these components are given by
\begin{align}
 &\delta_\epsilon \varphi = - \epsilon \psi, \quad \delta_\epsilon \psi = 0, \quad 
					\delta_\epsilon \bar\psi = \frac{i}{2} \epsilon \dot{\varphi} - \oneon{2} \epsilon f , \quad
					\delta_\epsilon f = - i \epsilon \dot{\psi}, \eqlabel{sspace_flat_transf_2a} \\
 &\delta_{\bar\epsilon} \varphi = \bar\epsilon \bar\psi, \quad \delta_{\bar\epsilon} \bar\psi = 0, \quad
                                             \delta_{\bar\epsilon} \psi = - \frac{i}{2} \bar\epsilon\dot{\varphi} - \oneon{2} \bar\epsilon f,\quad
					\delta_{\bar\epsilon} f = - i\bar\epsilon \dot{\bar\psi} \; . \eqlabel{sspace_flat_transf_2a_bar}
\end{align}
For a set, $\{\phi^i\}$, of $2a$ superfields the most general non-linear sigma model\footnote{For an introduction to supersymmetric non-linear sigma models in one and two dimensions, see, for example, ref.~\cite{Machin:2002zn}.} can be written in superspace as~\cite{Coles:1990hr,Gibbons:1997iy,Hull:1999ng}
\begin{multline}\eqlabel{nlsm_2a}
S_{2a} = \oneon{4} \int d\tau d^2\theta \left\{ (G(\underline{\phi})+B(\underline{\phi}))_{ij} D \phi^i \bar{D} \phi^j + L_{ij}(\underline{\phi}) D \phi^i D \phi^j
	\right. \\ \left.	+ M_{ij}(\underline{\phi}) \bar{D} \phi^i \bar{D} \phi^j + {\cal W}(\underline{\phi}) \right\} ,
\end{multline}
where $G_{ij}$ is symmetric, $B_{ij}$, $L_{ij}$, $M_{ij}$ are anti-symmetric and ${\cal W}$ is an arbitrary function of $\phi^i$. The component version of ${\cal W}(\underline{\phi})$ is obtained by a Taylor expansion about $\varphi^i$: 
\be
	{\cal W}(\underline{\phi}) = {\cal W}(\underline{\varphi}) + i\theta\psi^i {\cal W}_{,i}(\underline{\varphi})
		+ i\bar\theta\bar\psi^i {\cal W}_{,i}(\underline{\varphi}) 
		+ \oneon{2} \theta\bar\theta ({\cal W}_{,i}(\underline{\varphi}) f^i + 2 {\cal W}_{,ij}(\underline{\varphi}) \psi^i \bar\psi^j ) .
\ee
The $,i$ notation denotes the ordinary derivative with respect to $\varphi^i$. From this and the other formul\ae\ given in this section, it is straightforward to work out the component action of this superspace action. Here, we will not present the most general result but focus on the first and last term in~\eqref*{nlsm_2a}. One finds
\begin{align}
 S_{2a} &= \oneon{4} \int d\tau d^2\theta \left\{ G_{ij}(\underline{\phi}) D \phi^i \bar{D} \phi^j + {\cal W}(\underline{\phi}) \right\}\\
 &= \oneon{4}\int d\tau\,\left\{ \oneon{4} G_{ij}(\underline{\varphi}) \dot{\varphi}^i \dot{\varphi}^j
			- \frac{i}{2} G_{ij}(\underline{\varphi}) (\psi^i \dot{\bar\psi}^j - \dot{\psi}^i\bar{\psi}^j)
			+ \oneon{4}G_{ij}(\underline{\varphi}) f^i f^j\right. \nonumber\\
			&- \oneon{2}G_{ij,k}(\underline{\varphi}) (\psi^i\bar{\psi}^j f^k - \psi^k\bar{\psi}^j f^i - \psi^i\bar{\psi}^k f^j)
			+ \frac{i}{2} G_{ij,k}(\underline{\varphi}) (\psi^k\bar{\psi}^i + \bar{\psi}^k\psi^i) \dot{\varphi}^j \\
			&\left.- G_{ij,kl}(\underline{\varphi}) \psi^i\bar{\psi}^j\psi^k\bar{\psi}^l-\frac{1}{2}{\cal W}_{,i}(\underline{\varphi})f^i- {\cal W}_{,ij}(\underline{\varphi})\psi^i\bar{\psi}^j\right\}\nonumber .
\end{align}
Apart from the standard kinetic terms we have Pauli terms (coupling two fermions and the time derivative of a scalar), Yukawa couplings and four-fermi terms. We also see that the highest components $f^i$ are indeed auxiliary fields. The $f^i$ equation of motion can be solved explicitly and leads to
\be
	f^i =  G^{ij}{\cal W}_{,j} + \ldots \; ,
\ee
where $G^{ij}$ is the inverse of $G_{ij}$.  The dots indicate fermion bilinear terms which we have not written down explicitly. Using this solution to integrating out the $f^i$ produces additional four-fermi terms and  the scalar potential
\be
	S_{2a,{\rm pot}} = - \oneon{8} \int d\tau\, {\cal U}\; ,\qquad {\cal U} = \frac{1}{2}G^{ij}{\cal W}_{,i}{\cal W}_{,j}\; .
\ee 

\subsection{\texorpdfstring{$2b$}{2b} multiplets}

The other major type of multiplet is the $2b$ multiplet $Z(\tau,\theta,\bar{\theta})$ which is defined by the chirality constraint $\bar{D}Z=0$. Working out its most general component expansion, one finds
\be\eqlabel{sspace_flat_sfield_2b}
	Z = z + \theta\kappa + \frac{i}{2} \theta\bar\theta \dot{z} \; ,
\ee
where $z$ is a complex scalar and $\kappa$ is a complex fermion. We note that, unlike for the $2a$ multiplet, the highest component is not an independent field but simply $\dot{z}$. Hence, a $2b$ multiplet contains a complex physical scalar field and no auxiliary field. This difference in physical bosonic field content in comparison with the $2a$ multiplet will be quite useful when it comes to identifying which supermultiplets arise from an M-theory reduction.

\Eqref{sspace_flat_superchrg_defn,susytransform} lead to the following component supersymmetry transformations
\begin{align}
&\delta_\epsilon z = i\epsilon \kappa, \quad\delta_\epsilon \bar{z} = 0, \quad\delta_\epsilon \kappa = 0, \quad
					\delta_\epsilon \bar\kappa = \epsilon \dot{\bar{z}}, \eqlabel{sspace_flat_transf_2b} \\
&\delta_{\bar\epsilon} \bar{z} = i\bar\epsilon \bar\kappa, \quad\delta_{\bar\epsilon} z = 0,\quad
                                              \delta_{\bar\epsilon} \bar\kappa = 0,\quad\delta_{\bar\epsilon} \kappa = \bar\epsilon \dot{z}\; .
                                              \eqlabel{sspace_flat_transf_2b_bar}
\end{align}
A general non-linear sigma model for a set, $\{ Z^a\}$, of $2b$ multiplets has the form~\cite{Coles:1990hr,Gibbons:1997iy,Hull:1999ng}
\be\eqlabel{nlsm_2b}
  S_{2b} = \oneon{4} \int d\tau d^2\theta \left\{ G_{a{\bar{b}}}(\underline{Z},\underline{\bar{Z}}) DZ^a \bar{D}\bar{Z}^{\bar{b}} 
		+ \left[\oneon{2} B_{ab}(\underline{Z},\underline{\bar{Z}}) DZ^a DZ^b + \cc\right] +
		  F(\underline{Z},\bar{\underline{Z}})\right\} ,
\ee
where $G_{a\bar{b}}$ is hermitian, $B_{ab}$ is anti-symmetric and $F$ is an arbitrary real function. The component version of $F(\underline{Z},\bar{\underline{Z}})$ is obtained by a Taylor expansion about $z^a$ and $\bar{z}^{\bar{a}}$
\begin{multline}
	F(\underline{Z},\bar{\underline{Z}}) = 
	F(\underline{z},\bar{\underline{z}})
	+ \theta \kappa^a F_{,a}(\underline{z},\bar{\underline{z}})
	- \bar\theta \bar{\kappa}^{\bar{a}} F_{,\bar{a}}(\underline{z},\bar{\underline{z}}) \\ 
	+ \oneon{2} \theta\bar\theta \left\{ i F_{,a}(\underline{z},\bar{\underline{z}}) \dot{z}^a 
		- i F_{,\bar{a}}(\underline{z},\bar{\underline{z}}) \dot{\bar{z}}^{\bar{a}}
		+ 2 F_{,a\bar{b}}(\underline{z},\bar{\underline{z}}) \kappa^a \bar{\kappa}^{\bar{b}} \right\} .
\end{multline}
The component form of the action~\eqref*{nlsm_2b} can again be worked out straightforwardly from the above formal\ae\ but we will not pursue this here. 

Instead, we focus on a slightly different superspace action which is better adapted to what we will need in \chref{mcy5}. First, we drop the term proportional to $B_{ab}$, which does not arise from M-theory. Secondly, we introduce a slight generalisation in that we allow the sigma model metric $G_{a\bar{b}}$ to also depend on $2a$ superfields $\phi^i$, in addition to the $2b$ superfields $Z^a$ and their complex conjugates. A multi-variable Taylor expansion of a function $G(\underline{\phi},\underline{Z},\underline{\bar{Z}})$ depending on $2a$ as well as $2b$ superfields yields the following component form
\bea
	G(\underline{\phi},\underline{Z},\underline{\bar{Z}}) = 
	  G(\underline{\varphi},\underline{z},\underline{\bar{z}})
	+ \theta [i\psi^i G_{,i}(\underline{\varphi},\underline{z},\underline{\bar{z}}) 
		+ \kappa^a G_{,a}(\underline{\varphi},\underline{z},\underline{\bar{z}})]
	+ \bar\theta [i\bar\psi^i G_{,i}(\underline{\varphi},\underline{z},\underline{\bar{z}}) 
		\\ - \bar{\kappa}^{\bar{a}} G_{,\bar{a}}(\underline{\varphi},\underline{z},\underline{\bar{z}})] 
	+ \theta\bar\theta \left[
		   \oneon{2} G_{,i}(\underline{\varphi},\underline{z},\underline{\bar{z}}) f^i
		+ G_{,ij}(\underline{\varphi},\underline{z},\underline{\bar{z}}) \psi^i \bar\psi^j \right.\\ \left.
		+ i G_{,ia}(\underline{\varphi},\underline{z},\underline{\bar{z}}) \bar\psi^i \kappa^a
		+ i G_{,i\bar{a}}(\underline{\varphi},\underline{z},\underline{\bar{z}}) \psi^i \bar{\kappa}^{\bar{a}} 
		+ G_{,a\bar{b}}(\underline{\varphi},\underline{z},\underline{\bar{z}}) \kappa^a \bar{\kappa}^{\bar{b}} \right.\\ \left.
		+ \frac{i}{2} G_{,a}(\underline{\varphi},\underline{z},\underline{\bar{z}}) \dot{z}^a
		- \frac{i}{2} G_{,\bar{a}}(\underline{\varphi},\underline{z},\underline{\bar{z}}) \dot{\bar{z}}^{\bar{a}}
		\right] .
\eea
The relevant action is
\bea
 S_{2b} &= \frac{1}{4}\int d\tau\,d^2\theta\,\left\{G_{a{\bar{b}}}(\underline{\phi},\underline{Z},\underline{\bar{Z}}) DZ^a \bar{D}\bar{Z}^{\bar{b}} +  F(\underline{Z},\bar{\underline{Z}})\right\} \\
 &= \frac{1}{4}\int d\tau\,\left\{ 
 	G_{a\bar{b}}(\underline{\varphi}, \underline{z}, \underline{\bar{z}}) \dot{z}^a \dot{\bar{z}}^{\bar{b}}
	- \frac{i}{2} G_{a\bar{b}}(\underline{\varphi}, \underline{z}, \underline{\bar{z}}) 
		(\kappa^a \dot{\bar{\kappa}}^{\bar{b}} - \dot{\kappa}^a \bar{\kappa}^{\bar{b}}) \right. \\  & \left.
	- \frac{i}{2} G_{a\bar{b}, c}(\underline{\varphi}, \underline{z}, \underline{\bar{z}}) 
		(\kappa^a \bar{\kappa}^{\bar{b}} \dot{z}^c - 2\kappa^c \bar{\kappa}^{\bar{b}} \dot{z}^a ) 
	+ \frac{i}{2} G_{a\bar{b}, \bar{c}}(\underline{\varphi}, \underline{z}, \underline{\bar{z}}) 
		(\kappa^a \bar{\kappa}^{\bar{b}} \dot{\bar{z}}^{\bar{c}} + 2\kappa^a \bar{\kappa}^{\bar{c}} \dot{\bar{z}}^{\bar{b}} )   \right. \\  & \left.
	- G_{a\bar{b}, c\bar{d}}(\underline{\varphi}, \underline{z}, \underline{\bar{z}}) 
		\kappa^a \bar{\kappa}^{\bar{b}} \kappa^c \bar{\kappa}^{\bar{d}}
	- \oneon{2} G_{a\bar{b}, i}(\underline{\varphi}, \underline{z}, \underline{\bar{z}}) \kappa^a \bar{\kappa}^{\bar{b}} f^i
	- G_{a\bar{b}, ij}(\underline{\varphi}, \underline{z}, \underline{\bar{z}}) \kappa^a \bar{\kappa}^{\bar{b}} \psi^i \bar{\psi}^j   \right. \\  & \left.
	- i G_{a\bar{b}, ic}(\underline{\varphi}, \underline{z}, \underline{\bar{z}}) \kappa^a \bar{\kappa}^{\bar{b}} \bar{\psi}^i \kappa^c 
	- i G_{a\bar{b}, i\bar{c}}(\underline{\varphi}, \underline{z}, \underline{\bar{z}}) \kappa^a \bar{\kappa}^{\bar{b}} \psi^i \bar{\kappa}^{\bar{c}}
	- G_{a\bar{b}, i}(\underline{\varphi}, \underline{z}, \underline{\bar{z}}) 
		\psi^i \bar{\kappa}^{\bar{b}} \dot{z}^a  \right. \\  & \left.
	+ G_{a\bar{b}, i}(\underline{\varphi}, \underline{z}, \underline{\bar{z}}) 
		\bar{\psi}^i \kappa^a \dot{\bar{z}}^{\bar{b}}
	- \frac{i}{2} (F_{,a}\dot{z}^a-F_{,\bar{b}}\dot{\bar{z}}^{\bar{b}})-F_{,a\bar{b}}\kappa^a\bar{\kappa}^{\bar{b}}\right\} .
\eea
Note that the function $F$ gives rise to a Chern-Simons type term (and fermion mass terms) but not to a scalar potential.

\subsection{Fermionic multiplets}

The $2a$ and $2b$ superfields introduced above are bosonic superfields in the sense that their lowest components are bosons. However, for both types of multiplets there also exists a fermionic version, satisfying the same constraint as their bosonic counterparts but starting off with a fermion as the lowest component. Here, we will only consider fermionic $2b$ superfields. The details for fermionic $2a$ superfields can be worked out analogously.

Fermionic $2b$ superfields $R(\tau ,\theta,\bar{\theta})$ have a spinorial lowest component and are defined by the chirality constraint $\bar{D}R=0$. Their general component expansion reads
\be\eqlabel{2bfexp}
	R = \rho + \theta h + \frac{i}{2} \theta\bar\theta \dot{\rho} \; ,
\ee
where $\rho$ is a complex fermion and $h$ is a complex scalar. For its component supersymmetry transformations, one finds
\begin{align}
	&\delta_\epsilon \rho = i\epsilon h, \quad\delta_\epsilon \bar\rho = 0, \quad\delta_\epsilon h = 0, \quad
						\delta_\epsilon \bar{h} = - \epsilon \dot{\bar{\rho}}, \eqlabel{sspace_flat_transf_2bf} \\
	&\delta_{\bar\epsilon} \bar\rho = -i\bar\epsilon \bar{h}, \quad\delta_{\bar\epsilon} \rho = 0, \quad
	                                             \delta_{\bar\epsilon} \bar{h} = 0.\quad\delta_{\bar\epsilon} h = \bar\epsilon \dot{\rho}\; .
           \eqlabel{sspace_flat_transf_2bf_bar}
\end{align}
A set, $\{R^x\}$, of fermionic $2b$ superfields can be used to build non-linear sigma models where only fermions are propagating. A class of such models is given by
\bea
	S_{2b,{\rm F}} &= \oneon{4} \int d\tau\, d^2\theta\, G_{x\bar{y}}(\underline{\phi}) R^x \bar{R}^{\bar{y}} \\
	                        &=\frac{1}{4}\int d\tau \, \left\{
		   \frac{i}{2} G_{x\bar{y}}(\underline{\varphi}) ( \rho^x \dot{\bar{\rho}}^{\bar{y}} - \dot{\rho}^x \bar{\rho}^{\bar{y}} ) 
		- G_{x\bar{y}}(\underline{\varphi}) h^x \bar{h}^{\bar{y}}
		- \oneon{2} G_{x\bar{y},i}(\underline{\varphi}) \rho^x \bar{\rho}^{\bar{y}} f^i  \right. \\ & \left. \qquad\qquad\qquad
		- i G_{x\bar{y},i}(\underline{\varphi}) ( \psi^i \rho^x \bar{h}^{\bar{y}} + \bar{\psi}^i \bar{\rho}^{\bar{y}} h^x ) 
		- G_{x\bar{y},ij}(\underline{\varphi}) \rho^x \bar{\rho}^{\bar{y}} \psi^i \bar{\psi}^j  \vphantom{\frac{i}{2}} \right\} .
\eea
Here, we have allowed the sigma model metric to depend on a set, $\{\phi^i\}$, of $2a$ moduli -- a situation which will arise from M-theory reductions.  Note that the bosons $h^x$ are indeed auxiliary fields and only the fermions $\rho^x$ have kinetic terms.

\section{Curved \texorpdfstring{$\susyno=2$}{N=2} supertime\texorpdfstring{: $\manifold^{1|2}$}{}}\seclabel{sspace_curved}

The goal of this section is to develop curved $\susyno=2$ supertime to an extent that will allow us to write down actions over this superspace and compare their component expansion with the results from \chref{mcy5}. We will use the formul\ae\ of this section to write the one-dimensional effective action obtained from dimensionally reducing M-theory on Calabi-Yau five-folds in curved $\susyno=2$ supertime thereby making the residual supersymmetry manifest. 

The well-known case of $\susyno=1$ in four dimensions~\cite{Wess:1977fn,Grimm:1977kp,Wess:1978bu,Stelle:1978ye,Ferrara:1978em,Sohnius:1981tp} and supergravity theories in two-dimensions~\cite{Howe:1978ia,Ertl:2001sj} will guide us in constructing curved $\susyno=2$ supertime. 

The on-shell one-dimensional $\susyno=2$ supergravity multiplet comprises the lapse function (or ``einbein'') $N$, which is a real scalar, and the ``lapsino'' $\psi_0$, which is a one-component complex spinor. In all expressions provided in this section, flat superspace (and thus the equations of the previous section) can be recovered by gauge fixing the supergravity fields to $N=1$ and $\psi_0 = 0$. From a more geometric viewpoint, $\manifold^{1|2}$ looks locally like $\R^{1|2}$.

\subsection{Superspace formulation of one-dimensional \texorpdfstring{$\susyno=2$}{N=2} supergravity}

As in the previous section, we start by introducing super-coordinates $z^A = (x^0=\tau,\theta,\bar\theta)$. We recall the rather subtle r\^{o}le of the local Lorentz indices in one dimension: These indices are taken to be valued in the trivial Lorentz group $\SO(0,1) = \SO(1)$. Since there is no Lie algebra for the trivial group, there are no Lorentz generators in one dimension and the local Lorentz indices $\UA, \UB, \ldots$ are taken to not transform under any local group action but should merely be thought of as labels. They label the two different representations of $\mathrm{Spin}(1)$, namely for $\UA=\UZ$ the ``vector'' representation, which is nothing but the real numbers in one dimension, and for $\UA=\Uth$ the spinor representation which are real Grassmann numbers. In addition, the fact that we want to realise $\susyno$-extended supersymmetry (with $\susyno > 1$) means we need in principle another index, say $i,j=1,\ldots,\susyno$ on the $\UA=\Uth$ components to label the $\susyno$-extendedness of the spinorial components (see \secref{sspace101}). Here, $\susyno=2$ and hence $\UA,\UB,\ldots = \UZ, \Uth_{1}, \Uth_{2}$. For ease of notation, we combine the two $\Uth_i$ into a combination of one complex index $\Uth = \Uth_{1} + i \Uth_{2}$ (and similarly $\BUth = \Uth_{1} - i \Uth_{2}$) thereby suppressing the additional $\susyno$-extension index $i$. After this step, the local Lorentz indices $\UA, \UB, \ldots$ range over $\UZ$, $\Uth$ and $\BUth$. Note that this coincides precisely with the notation used for curved indices except for the additional underline added for distinction.

As in the flat case, there is, of course, a global $\U(1)\simeq\SO(2)$ $R$-symmetry present, which rotates the complex anti-commuting coordinate $\theta \rightarrow e^{-i\alpha} \theta$ by some angle $\alpha$. This $\U(1)$ $R$-symmetry can be gauged as was shown in ref.~\cite{Howe:1989vn}. The tangent space group then becomes $\U(1)$ and tangent vectors $V^\UA$ transform as $\delta V^\UA = V^\UB L_\UB {}^\UA$, where $L_\Uth {}^\Uth = -i \alpha$, $L_\BUth {}^\BUth = i\alpha$ and all others zero. In the one-dimensional component action this leads to additional terms containing the new $\U(1)$ gauge field. Since we are, in this thesis, focussing on one-dimensional actions obtained from reductions on Calabi-Yau five-folds $X$, we anticipate that no such terms will arise in our lower dimensional actions. This is a consequence of $h^{1,0}(X) = 0$ (there being no continuous isometries on a Calabi-Yau five-folds $X$), which precludes the occurrence of massless Kaluza-Klein vectors in the lower dimensional action. Thus, we shall ignore the possibility of a gauged $\U(1)$ R-symmetry. The choice of anti-commuting coordinate $\theta$ may in this context be viewed as a gauge fixing, which we shall make from the outset. Note, though, that the equations of this section nonetheless respect the global $\U(1)$ $R$-symmetry.

In summary, even though the local Lorentz indices can take on three different values, we will restrict to the case in which no group is acting on them. Objects carrying an anti-symmetrised combination of two or more local Lorentz indices vanish identically, since the Lorentz generator in each representation of $\mathrm{Spin}(1)$ is zero and there are, by construction, no representation-mixing Lorentz transformations. This immediately implies $\Omega = 0$ and $R = 0$, which profoundly simplifies the further discussion. The fact that the curvature two-form $R$ vanishes identically for our ``curved'' superspace formulation of one-dimensional $\susyno=2$ supergravity may be regarded as another manifestation that this theory is in fact non-dynamical.

Since the on-shell supergravity multiplet contains only one real scalar, we take the geometrical supertime tensors to be $2a$ superfields, which means they comprise four component fields when expanded out in powers of $\theta$ and $\bar{\theta}$ (see~\eqref{sspace_flat_2a_comp}). The supervielbein $E_A {}^\UB$, in general, consists of a set of $3\times 3=9$ $2a$ superfields, which totals to $9 \times 4 = 36$ component (that is, off-shell) fields, and is expanded as 
\be\eqlabel{sspace_curved_supervielbein_gen_expansion}
	E_A {}^\UB = E_{A(0)} {}^\UB + i \theta E_{A(1)} {}^\UB + i \bar\theta E_{A(\bar{1})} {}^\UB + \oneon{2} \theta\bar\theta E_{A(2)} {}^\UB\; .
\ee
This is a large number of apparently independent fields given that on-shell, we just have three, namely $N$, $\psi_0$ and $\bar{\psi}_0$. In order to not obscure the physical content and to formulate supertime theories in the most efficient way, it is important to find a formulation with the minimum number of component fields. 

This can be achieved by imposing covariant constraints on the supervielbein and by gauging away some components using the SGCTs~\eqref*{sspace_curved_inf_sgct_supervielbein}. The infinitesimal parameters $\xi^A$ in~\eqref*{sspace_curved_inf_sgct_supervielbein} comprise a set of three four-component $2a$ superfields (that is, 12 component fields in total). The lowest component of $\xi^0 | = \zeta$ is the infinitesimal parameter of worldline reparametrizations, whereas the lowest components of the spinorial parameters $\xi^\theta | = i\epsilon$ and $\xi^{\bar\theta} | = i\bar\epsilon$ correspond to the infinitesimal local $\susyno=2$ supersymmetry parameters. The notation $\phi |$\newnot{symbol:sspace_lowest_comp} is a shorthand for $\phi |_{\theta=\bar\theta = 0}$, that is denoting the lowest component of the superfield $\phi$.

An infinitesimal local $\susyno=2$ supersymmetry transformation with parameters $\epsilon$ and $\bar\epsilon$ on a general superfield $\phi$ can be written by means of the supercharges $Q$ and $\bar{Q}$ as
\be\eqlabel{sspace_curved_gen_sfield_transf}
	\delta_\epsilon \phi = i \epsilon Q \phi, \qquad\qquad \delta_{\bar\epsilon} \phi = i \bar\epsilon \bar{Q} \phi .
\ee
If we use the following general component expansion for $\phi$
\be\eqlabel{sspace_curved_gen_cov_sfield_expansion}
	\phi = \phi | + \theta (\sderiv\phi |) - \bar\theta (\bar\sderiv\phi |) + \oneon{2} \theta\bar\theta ( [\sderiv, \bar\sderiv]\phi | ) ,
\ee
then the components of $\phi$ transform as
\bea\eqlabel{sspace_curved_gen_sfield_comp_transf}
	\delta_\epsilon (\phi |) = i \epsilon Q \phi | , \qquad\quad
	\delta_\epsilon (\sderiv \phi |) &= i \epsilon Q \sderiv \phi | , \qquad\quad
	\delta_\epsilon (\bar\sderiv \phi |) = i \epsilon Q \bar\sderiv \phi | , \\
	\delta_\epsilon ( [\sderiv, \bar\sderiv]\phi | ) &= i \epsilon Q [\sderiv, \bar\sderiv] \phi | .
\eea
Both~\eqref*{sspace_curved_gen_cov_sfield_expansion,sspace_curved_gen_sfield_comp_transf} are manifestly super-covariant expressions since we used the tangentized covariant super-derivative defined in~\eqref*{sspace_gen_def_flat_sderiv} for building them. For $\susyno=2$ supertime, $\Omega=0$ and thus $\sderiv_\UA$ is just $\sderiv_\UA =E_\UA {}^B \partial_B$. Note that, similarly to $D$ and $\bar{D}$ in the flat case, the tangentized, spinorial super-covariant derivatives are abbreviated as $\sderiv \equiv \sderiv_\Uth = E_\Uth {}^A \partial_A$ and $\bar\sderiv \equiv \sderiv_\BUth = E_\BUth {}^A \partial_A$, which are not to be confused with the covariant exterior derivative 1-form $\sderiv = dz^A \sderiv_A$ introduced in \secref{sspace101}. 

From the general fact that $Q | = \sderiv | = \partial_\theta$, it follows that one may replace $Q$s by $\sderiv$s everywhere in~\eqref*{sspace_curved_gen_sfield_comp_transf} and hence knowing the component expansion of $\sderiv$ is enough for working out the entire component version of \eqref*{sspace_curved_gen_sfield_transf}, namely
\bea\eqlabel{sspace_curved_gen_sfield_comp_transf_final}
	\delta_\epsilon (\phi |) = i \epsilon \sderiv \phi | , \qquad\quad
	\delta_\epsilon (\sderiv \phi |) &= i \epsilon \sderiv^2 \phi | = 0 , \qquad\quad
	\delta_\epsilon (\bar\sderiv \phi |) = i \epsilon \sderiv \bar\sderiv \phi | , \\
	\delta_\epsilon ( [\sderiv, \bar\sderiv]\phi | ) &= - i \epsilon \sderiv \bar\sderiv \sderiv \phi | ,
\eea
and similarly for the $\bar\epsilon$-transformations. In the second and fourth equation in~\eqref*{sspace_curved_gen_sfield_comp_transf_final}, we used the property $\sderiv^2 = 0$.

Now continuing our quest for finding the minimal formulation of off-shell $\susyno=2$, $D=1$ supergravity, we have here opted for the analogue of the Wess-Zumino gauge in $D=4$ and the way to formulate it in the present case will be explained in the following. Since we have three physical components in the supergravity multiplet, we shall use $9=12-3$ components out of $\xi^A$ to gauge fix 9 out of the 36 components of $E_A {}^\UB$, namely
\begin{align}
	&E_\theta {}^\UZ | = E_{\bar{\theta}} {}^\UZ | =
	E_\theta {}^\BUth | = E_{\bar{\theta}} {}^\Uth | =
	\sderiv E_{\bar{\theta}} {}^\Uth | = 
	\sderiv E_{\bar{\theta}} {}^\BUth | = 0 , \eqlabel{sspace_curved_gauge_fix1}\\
	&\bar\sderiv E_\theta {}^\UZ | = \frac{i}{2} , \quad
	E_\theta {}^\Uth | = 1, \quad E_{\bar{\theta}} {}^\BUth | = -1 . \eqlabel{sspace_curved_gauge_fix2}
\end{align}
The three remaining parameters in $\xi^A$ act on the three physical fields $N$, $\psi_0$ and $\bar\psi_0$, which we choose to identify in the following way
\be\eqlabel{sspace_curved_gauge_fix3}
	E_0 {}^\UZ | = N , \qquad
	E_0 {}^\Uth | = \psi_0 , \qquad
	E_0 {}^\BUth | = - \bar\psi_0 .
\ee

We will now discuss our choice of covariant constraints. Usually, they are imposed on certain components of the tangentized supertorsion $T_{\underline{A}\underline{B}} {}^{\underline{C}}$. Our aim is to find a combination of constraints that yield the minimum number of fields in the $\theta$-expansion of the supervielbein $E_A {}^\UB$. The main idea is to take the system of constraints from $\susyno=1$, $D=4$ and restrict the index ranges appropriately. In doing so, we obtain the following torsion constraints
\begin{align}
	 T_{\Uth\BUth} {}^\UZ &= i , \qquad &
	T_{\Uth\BUth} {}^\Uth &= 0 , \qquad &
	&\text{(conventional constraints)}\eqlabel{sspace_curved_conv_constr} , \\
	 T_{\BUth\BUth} {}^\UZ &= 0 , \qquad &
	T_{\BUth\BUth} {}^\Uth &= 0 , \qquad &
	&\text{(representation preserving constraints)}\eqlabel{sspace_curved_rep_pres_constr} , \\
	& &T_{\Uth\Uth} {}^\Uth &= 0 , \qquad &
	&\text{(``type 3'' constraint)}\eqlabel{sspace_curved_type3_constr} , 
\end{align}
and their complex conjugates, of course. We are equating superfields to superfields here and hence each of the above relations is manifestly (super-)covariant. The first line is the analogue of the conventional constraints in $\susyno=1$, $D=4$. In the absence of $R$, \eqref{sspace_graded_comm_of_cov_derivs} becomes
\be
	[ \sderiv_\UA , \sderiv_\UB \} = - T_{\UA\UB} {}^{\underline{C}} \sderiv_{\underline{C}} \; .
\ee
The conventional constraints stem from imposing (cf. \eqref{sspace_flat_cov_deriv_comm})
\be
	\{ \sderiv, \bar\sderiv \} = - i \sderiv_\UZ \; ,
\ee
which guarantees that the \emph{tangentized} covariant super-derivatives of curved superspace, $\sderiv$ and $\bar\sderiv$, satisfy the \emph{flat} algebra. A $2b$ superfield $Z$ satisfies $\bar\sderiv Z = 0$ by definition. The representation preserving constraints listed in~\eqref*{sspace_curved_rep_pres_constr} follow from the corresponding integrability condition
\be
	\{ \bar\sderiv, \bar\sderiv \} Z = 0
\ee
for all $2b$ superfields $Z$. For the constraint in~\eqref*{sspace_curved_type3_constr}, we do not have a direct motivation from a one-dimensional viewpoint, so we impose it purely by analogy to the conformal constraint of $\susyno=1$ in $D=4$.

Now, we need to examine the BIs~\eqrangeref*{sspace_gen_BI1}{sspace_gen_BI2}. Specializing to $\susyno=2$ supertime, the second BI identically vanishes due to $R = 0$ and the first BI becomes\newnot{symbol:ind_graded_asymm}
\be\eqlabel{sspace_curved_1d_BI}
	d T^\UA = 0  \qquad\Leftrightarrow\qquad
	\sderiv_{\left[\UA\right.} T_{\left.\UB\underline{C}\right\}} {}^{\underline{D}} + 
	T_{\left[\UA\UB\right.} {}^{\underline{E}} T_{\left.|\underline{E}|\underline{C}\right\}} {}^{\underline{D}} = 0 \; .
\ee
In the presence of constraints, consistency requires that the BIs sill be obeyed and this needs to be checked by explicit calculation. In this respect, the BIs become ``contentful'' (rather than being genuine identities) when constraints are present and then the BIs must be \emph{imposed}. For the case at hand, one learns from the BI~\eqref*{sspace_curved_1d_BI} that all remaining torsion components which are not already fixed by the constraints~\eqrangeref*{sspace_curved_conv_constr}{sspace_curved_type3_constr} must be zero.

From the choice of gauge fixing~\eqrangeref*{sspace_curved_gauge_fix1}{sspace_curved_gauge_fix3} and torsion constraints~\eqrangeref*{sspace_curved_conv_constr}{sspace_curved_type3_constr} and the imposition of the BI~\eqref*{sspace_curved_1d_BI}, all 36 components in the supervielbein expansion~\eqref*{sspace_curved_supervielbein_gen_expansion} are fixed uniquely to
\vspace{\parskip}
\begin{align}
	&E_0 {}^\UZ = N + i \theta \bar\psi_0 + i \bar\theta \psi_0 , \eqlabel{sspace_curved_seinbein1} \\
	&E_0 {}^\Uth = \psi_0 , \qquad
	E_0 {}^\BUth = - \bar\psi_0 , \eqlabel{sspace_curved_seinbein2} \\
	&E_\theta {}^\UZ = -\frac{i}{2} \bar\theta , \qquad
	E_{\bar{\theta}} {}^\UZ = -\frac{i}{2} \theta , \eqlabel{sspace_curved_seinbein3} \\
	&E_\theta {}^\Uth = 1 , \quad
	E_\theta {}^\BUth = 0 , \quad
	E_{\bar{\theta}} {}^\Uth = 0 , \quad
	E_{\bar{\theta}} {}^\BUth = -1 \eqlabel{sspace_curved_seinbein4}.
\end{align}
\vspace{\parskip}\noindent 
Note that the minimal set of fields of off-shell pure $\susyno=2$, $D=1$ supergravity does not comprise any auxiliary fields. From \eqref{sspace_curved_supervielbein_inverse_rel} we compute the component expansion of the inverse supervielbein
\vspace{\parskip}
\begin{align}
	&E_\UZ {}^0 = N^{-1} - \frac{i}{2} \theta N^{-2} \bar\psi_0 - \frac{i}{2} \bar\theta N^{-2} \psi_0
		- \oneon{2} \theta\bar\theta N^{-3} \psi_0 \bar\psi_0 , \\
	&E_\UZ {}^\theta = - N^{-1} \psi_0 - \frac{i}{2} \theta N^{-2} \psi_0 \bar\psi_0 , \qquad
	E_\UZ {}^{\bar\theta} = - N^{-1} \bar\psi_0 + \frac{i}{2} \bar\theta N^{-2} \psi_0 \bar\psi_0 , \\
	&E_\Uth {}^0 = \frac{i}{2} \bar\theta N^{-1} - \oneon{4} \theta\bar\theta N^{-2} \bar\psi_0 , \quad\quad
	E_\BUth {}^0 = - \frac{i}{2} \theta N^{-1} - \oneon{4} \theta\bar\theta N^{-2} \psi_0 , \\
	&E_\Uth {}^\theta = 1 - \frac{i}{2} \bar\theta N^{-1} \psi_0 - \oneon{4} \theta\bar\theta N^{-2} \psi_0 \bar\psi_0 ,
		\qquad\qquad
	E_\Uth {}^{\bar\theta} = -\frac{i}{2} \bar\theta N^{-1} \bar\psi_0 , \\
	&E_\BUth {}^{\bar\theta} = -1 + \frac{i}{2} \theta N^{-1} \bar\psi_0 + \oneon{4} \theta\bar\theta N^{-2} \psi_0 \bar\psi_0 , 
		\qquad\qquad
	E_\BUth {}^\theta = \frac{i}{2} \theta N^{-1} \psi_0 .
\end{align}
\vspace{\parskip}\noindent 
Since $\sderiv_\UA = E_\UA {}^B \partial_B$, the above expressions directly allow us to write down the component expansion of the tangentized, spinorial super-covariant derivative
\begin{equation}\eqlabel{sspace_curved_sderiv_comp_exp}
	\sderiv = \left( 1 - \frac{i}{2}N^{-1}\bar\theta\psi_0 - \oneon{4}N^{-2}\theta\bar\theta\psi_0\bar\psi_0 \right) \partial_\theta
		+ \left( \frac{i}{2}N^{-1}\bar\theta - \oneon{4}N^{-2}\theta\bar\theta\bar\psi_0 \right) \partial_0 
		- \frac{i}{2}N^{-1}\bar\theta\bar\psi_0 \partial_{\bar\theta} ,
\end{equation}
and similarly for the complex conjugate version $\bar\sderiv$. By comparing the component expansion of~\eqref*{sspace_curved_inf_sgct_supervielbein} with \eqrangeref{sspace_curved_seinbein1}{sspace_curved_seinbein4}, we learn how the supergravity fields transform under local $\susyno=2$ supersymmetry
\be
	\delta_\epsilon N = - \epsilon\bar\psi_0 , \quad
	\delta_{\bar\epsilon} N = \bar\epsilon\psi_0 , \qquad 
	\delta_\epsilon\psi_0 = i\dot{\epsilon} , \quad
	\delta_\epsilon\bar\psi_0 = 0 , \quad
	\delta_{\bar\epsilon}\psi_0 = 0 , \quad
	\delta_{\bar\epsilon}\bar\psi_0 = -i\dot{\bar\epsilon}.
\ee
In order to build curved superspace actions that are manifestly invariant under local $\susyno=2$ supersymmetry, we need an expression for the super-determinant $\mathcal{E}$ of the vielbein defined, in general, as
\be
	\mathcal{E} \equiv \mathrm{sdet} E_A {}^\UB = (\det E_a {}^{\underline{b}})
		( \det [E_\alpha {}^{\underline{\beta}} 
		- E_\alpha {}^{\underline{c}} (E_d {}^{\underline{c}})^{-1} E_d {}^{\underline\beta}] )^{-1} ,
\ee
where $a,b,\ldots$ and $\alpha, \beta, \ldots$ denote vector and spinor indices, respectively. Specializing to $\susyno=2$ supertime and inserting \eqrangeref{sspace_curved_seinbein1}{sspace_curved_seinbein4}, one finds
\be
	\mathcal{E} = - N - \frac{i}{2} \theta\bar\psi_0 - \frac{i}{2} \bar\theta\psi_0 \; .
\ee
Since there is no $\theta\bar\theta$-component in this expression, it follows that the canonical action of pure supergravity vanishes as expected
\be
	S_{\text{pure sugra}} = \int d\tau d^2\theta \, \mathcal{E} = 0 \; .
\ee
Also, as an additional consistency check, one may verify that $\mathcal{E}$ is covariantly constant, in the sense that
\be
	\int d\tau d^2\theta \, \sderiv \mathcal{E} = \int d\tau d^2\theta \, \bar\sderiv \mathcal{E} = \text{(total derivative)} = 0 \; .
\ee
This allows us to use the partial-integration rule for superspace.

\subsection{\texorpdfstring{$2a$}{2a} multiplets}

In analogy to the flat superspace case, we will now present the different irreducible multiplets. We begin with the $2a$ multiplet defined by the constraint $\phi=\phi^\dagger$. The general solution to this constraint leads to the component expansion
\be
	\phi = \varphi + i \theta\psi + i\bar\theta \bar\psi + \oneon{2} \theta\bar\theta f \; , \eqlabel{sspace_curved_sfield_2a} 
\ee
where the component fields are labelled as in \eqref{sspace_flat_2a_comp}. This can also be written in a manifestly super-covariant fashion as
\be
	\phi = \phi | + \theta (\sderiv\phi |) - \bar\theta (\bar\sderiv\phi |) + \oneon{2} \theta\bar\theta ( [\sderiv, \bar\sderiv]\phi | ) .
\ee
For the supersymmetry transformations of the $2a$ component fields, one finds
\vspace{\parskip}
\begin{align}
	        \eqlabel{sspace_curved_transf_2a} \begin{aligned} &
		\delta_\epsilon \varphi = - \epsilon \psi, \quad 
		\delta_\epsilon \psi = 0, \quad 
		\delta_\epsilon \bar\psi = 	\frac{i}{2} N^{-1} \epsilon \dot{\varphi} - \oneon{2} \epsilon f 
							+ \oneon{2} N^{-1} \epsilon (\psi_0\psi + \bar{\psi}_0\bar\psi), \\& 
		\delta_\epsilon f = 	- i N^{-1} \epsilon \dot{\psi} + \frac{i}{2} N^{-2} \epsilon\bar{\psi}_0 \dot{\varphi} 
						+ \oneon{2} N^{-1} \epsilon\bar{\psi}_0 f - \oneon{2} N^{-2} \epsilon\psi\psi_0\bar{\psi}_0, \end{aligned} \\
	         \eqlabel{sspace_curved_transf_2a_bar}  \begin{aligned} &
		\delta_{\bar\epsilon} \varphi = \bar\epsilon \bar\psi, \quad 
		\delta_{\bar\epsilon} \psi = 	- \frac{i}{2} N^{-1} \bar\epsilon \dot{\varphi} - \oneon{2} \bar\epsilon f 
								- \oneon{2} N^{-1} \bar\epsilon (\psi_0\psi + \bar{\psi}_0\bar\psi), \quad 
		\delta_{\bar\epsilon} \bar\psi = 0, \\& 
		\delta_{\bar\epsilon} f = 	- i N^{-1} \bar\epsilon \dot{\bar\psi} + \frac{i}{2} N^{-2} \bar\epsilon\psi_0 \dot{\varphi} 
							- \oneon{2} N^{-1} \bar\epsilon\psi_0 f + \oneon{2} N^{-2} \bar\epsilon\bar\psi\psi_0\bar{\psi}_0 \; .
							\end{aligned}
\end{align}
\vspace{\parskip}\noindent
This is obtained by plugging the component expansions~\eqref*{sspace_curved_sderiv_comp_exp,sspace_curved_sfield_2a} into the general formula~\eqref*{sspace_curved_gen_sfield_comp_transf_final}. A standard kinetic term for a single $2a$ superfield $\phi$ and its associated component action are given by
\bea
	S_{\rm 2a,kin} &= -\oneon{4} \int d\tau\, d^2\theta\, \mathcal{E}\, \sderiv \phi \bar\sderiv \phi = \oneon{4} \int d\tau\, \mathcal{L}_{\rm 2a,kin}, \\
	\mathcal{L}_{\rm 2a,kin} &= \oneon{4}N^{-1}\dot{\varphi}^2 - \frac{i}{2}(\psi\dot{\bar\psi} - \dot\psi\bar\psi) + \oneon{4} N f^2
			+ \frac{i}{2}N^{-1}(\psi\psi_0+\bar\psi\bar\psi_0)\dot\varphi 
			+ \oneon{2}N^{-1}\psi_0\bar\psi_0\psi\bar\psi .
\eea
In the context of M-theory five-fold compactifications, we need to consider more general actions, representing non-linear sigma models for a set of $2a$ fields $\phi^i$ which also include a (super)-potential term. The superspace and component forms for such actions read
\bea\eqlabel{sspace_curved_2a_action}
	S_{\rm 2a} &= -\oneon{4} \int d\tau\, d^2\theta\, \mathcal{E}\, \{ G_{ij}(\underline{\phi}) \sderiv \phi^i \bar\sderiv \phi^j + {\cal W}(\underline{\phi}) \} 
			    = \oneon{4} \int d\tau\, \mathcal{L}_{\rm 2a}, \\
	\mathcal{L}_{\rm 2a} 	&= \oneon{4} N^{-1} G_{ij}(\underline{\varphi}) \dot{\varphi}^i \dot{\varphi}^j
			- \frac{i}{2} G_{ij}(\underline{\varphi}) (\psi^i \dot{\bar\psi}^j - \dot{\psi}^i\bar{\psi}^j)
			+ \oneon{4} N G_{ij}(\underline{\varphi}) f^i f^j \\
			&+ \frac{i}{2} N^{-1} G_{ij}(\underline{\varphi}) (\psi^i\psi_0 + \bar{\psi}^i\bar\psi_0) \dot{\varphi}^j
			+ \oneon{2} N^{-1} G_{ij}(\underline{\varphi}) \psi_0\bar\psi_0 \psi^i \bar{\psi}^j \\
			&- \oneon{2} N G_{ij,k}(\underline{\varphi}) (\psi^i\bar{\psi}^j f^k - \psi^k\bar{\psi}^j f^i - \psi^i\bar{\psi}^k f^j)
			+ \frac{i}{2} G_{ij,k}(\underline{\varphi}) (\psi^k\bar{\psi}^i + \bar{\psi}^k\psi^i) \dot{\varphi}^j \\
			&- N G_{ij,kl}(\underline{\varphi}) \psi^i\bar{\psi}^j\psi^k\bar{\psi}^l 
			- \oneon{2} N {\cal W}_{,i}(\underline{\varphi}) f^i 
			- N {\cal W}_{,ij}(\underline{\varphi}) \psi^i \bar{\psi}^j
			- \oneon{2} {\cal W}_{,i}(\underline{\varphi}) (\psi^i \psi_0 - \bar{\psi}^i \bar{\psi}_0) ,
\eea
with a sigma model metric $G_{ij}(\underline{\phi})$ and a superpotential ${\cal W}(\underline{\phi})$ . Here, $G_{\ldots,i}$  denotes differentiation with respect to the bosonic fields $\varphi^i$. Note that the fields $f^i$ are indeed auxiliary. Solving their equations of motion leads to
\be
	f^i = G^{ij}{\cal W}_j + G^{ij} G_{kl,j} \psi^k \bar{\psi}^l - G^{ij} G_{jk,l} (\psi^k \bar{\psi}^l + \psi^l \bar{\psi}^k) \; ,
\ee
where $G^{ij}$ is the inverse of $G_{ij}$ and ${\cal W}_i = {\cal W}_{,i} = \frac{\partial{\cal W}}{\partial\varphi^i}$. Inserting this back into the component action leads, amongst other terms, to the scalar potential
\be
	S_{2a,{\rm pot}} = - \oneon{8} \int d\tau\, N\, {\cal U}\; ,\qquad {\cal U} = \oneon{2} G^{ij} {\cal W}_i {\cal W}_j\; ,
\ee
for the scalars $\varphi^i$ in the $2a$ multiplets. We will also need a slight generalization of the action~\eqref*{sspace_curved_2a_action}, namely an action for a set of $2a$ superfields $X^p$ coupling to another set of $2a$ superfields $\phi^i$ and to a set of $2b$ superfields $Z^a$ via the sigma model metric $G_{pq}(\underline{\phi},\underline{Z},\bar{\underline{Z}})$
\begin{align}
	S_{\rm 2a,gen.} &= -\oneon{4} \int d\tau\, d^2\theta\, \mathcal{E}\, \{ 
		G_{pq}(\underline{\phi},\underline{Z},\bar{\underline{Z}}) \sderiv X^p \bar\sderiv X^q\} 
			    = \oneon{4} \int d\tau\, \mathcal{L}_{\rm 2a,gen.}, \nonumber \\
	\mathcal{L}_{\rm 2a,gen.} &= \oneon{4} N^{-1} G_{pq}(\underline{\varphi},\underline{z},\bar{\underline{z}}) \dot{x}^p \dot{x}^q
			- \frac{i}{2} G_{pq}(\underline{\varphi},\underline{z},\bar{\underline{z}}) 
				(\lambda^p \dot{\bar\lambda}^q - \dot{\lambda}^p \bar{\lambda}^q)
			+ \oneon{4} N G_{pq}(\underline{\varphi},\underline{z},\bar{\underline{z}}) g^p g^q \nonumber \\
			&+ \frac{i}{2} N^{-1} G_{pq}(\underline{\varphi},\underline{z},\bar{\underline{z}}) 
				(\lambda^p\psi_0 + \bar{\lambda}^p\bar\psi_0) \dot{x}^q
			+ \oneon{2} N^{-1} G_{pq}(\underline{\varphi},\underline{z},\bar{\underline{z}}) \psi_0\bar\psi_0 \lambda^p \bar{\lambda}^q 
			\nonumber \\
			&- \oneon{2} N G_{pq,i}(\underline{\varphi},\underline{z},\bar{\underline{z}}) \left(
				(\lambda^p\bar{\lambda}^q f^i - \psi^i\bar{\lambda}^p g^q + \bar{\psi}^i \lambda^p g^q)
			- i N^{-1} (\psi^i\bar{\lambda}^p + \bar{\psi}^i\lambda^p) \dot{x}^q \right)
			\nonumber \\
			&- N G_{pq,ij}(\underline{\varphi},\underline{z},\bar{\underline{z}}) \lambda^p\bar{\lambda}^q\psi^i\bar{\psi}^j 
			- \frac{i}{2} G_{pq,a}(\underline{\varphi},\underline{z},\bar{\underline{z}}) \lambda^p \bar\lambda^q (\dot{z}^a - \psi_0 \kappa^a) 
			\nonumber \\
			&+ \frac{i}{2} G_{pq,\bar{a}}(\underline{\varphi},\underline{z},\bar{\underline{z}}) 
				\lambda^p \bar\lambda^q (\dot{\bar{z}}^{\bar{a}} + \bar\psi_0 \bar\kappa^{\bar{a}}) 
			- \frac{i}{2} N G_{pq,a}(\underline{\varphi},\underline{z},\bar{\underline{z}}) \kappa^a \bar\lambda^p g^q 
			\nonumber \\
			&- \frac{i}{2} N G_{pq,\bar{a}}(\underline{\varphi},\underline{z},\bar{\underline{z}}) \bar\kappa^{\bar{a}} \lambda^p g^q 
			+ \oneon{2} G_{pq,a}(\underline{\varphi},\underline{z},\bar{\underline{z}}) \kappa^a \bar\lambda^p \dot{x}^q
			- \oneon{2} G_{pq,\bar{a}}(\underline{\varphi},\underline{z},\bar{\underline{z}}) \bar\kappa^{\bar{a}} \lambda^p \dot{x}^q 
			\nonumber \\ 
			&- N G_{pq,a\bar{b}}(\underline{\varphi},\underline{z},\bar{\underline{z}}) \lambda^p \bar\lambda^q \kappa^a \bar\kappa^{\bar{b}}
			- i N G_{pq,ia}(\underline{\varphi},\underline{z},\bar{\underline{z}}) \lambda^p \bar\lambda^q \bar\psi^i \kappa^a 
			\nonumber \\
			&- i N G_{pq,i\bar{a}}(\underline{\varphi},\underline{z},\bar{\underline{z}}) \lambda^p \bar\lambda^q \psi^i \bar\kappa^{\bar{a}} \; .
			\eqlabel{sspace_curved_2a_action_3form}
\end{align}

\subsection{\texorpdfstring{$2b$}{2b} multiplets}

\begin{sloppypar} 
Next we turn to $2b$ multiplets. They are defined by the constraint $\bar{\cal D}Z=0$ which leads to the component expansion
\begin{equation}
  Z = z + \theta\kappa + \frac{i}{2} N^{-1} \theta\bar\theta (\dot{z} - \psi_0 \kappa) \; . \eqlabel{sspace_curved_sfield_2b}
\end{equation}
Here, $N$ and $\psi_0$ are the components of the supergravity multiplet and the other fields are labelled in analogy with the globally supersymmetric case \eqref*{sspace_flat_sfield_2b}. 
Expression~\eqref*{sspace_curved_sfield_2b} is equivalent to the manifestly super-covariant version
\be
	Z = Z | + \theta (\sderiv Z |) - \oneon{2} \theta\bar\theta ( \bar\sderiv \sderiv Z | ) .
\ee
By plugging the component expansions~\eqref*{sspace_curved_sderiv_comp_exp, sspace_curved_sfield_2b} into the general formula~\eqref*{sspace_curved_gen_sfield_comp_transf_final}, the component field supersymmetry transformations are found to be
\vspace{\parskip}
\begin{align}
                              \delta_\epsilon z = i \epsilon \kappa, \quad
					\delta_\epsilon \bar{z} = 0, \quad
					\delta_\epsilon \kappa = 0, \quad
					\delta_\epsilon \bar\kappa = N^{-1} \epsilon (\dot{\bar{z}} + \bar{\psi}_0 \bar\kappa), 
\eqlabel{sspace_curved_transf_2b}\\
	                                     \delta_{\bar\epsilon} z = 0, \quad
					\delta_{\bar\epsilon} \bar{z} = i \bar\epsilon \bar\kappa, \quad
					\delta_{\bar\epsilon} \kappa = N^{-1} \bar\epsilon (\dot{z} - \psi_0 \kappa), \quad
					\delta_{\bar\epsilon} \bar\kappa = 0. \eqlabel{sspace_curved_transf_2b_bar}
\end{align}
\vspace{\parskip}\noindent
A standard kinetic term for a single $2b$ multiplet $Z$ can be written and expanded into components as
\vspace{\parskip}
\bea
	S_{\rm 2b,kin} &= - \oneon{4} \int d\tau\, d^2\theta\, \mathcal{E}\, \sderiv Z \bar\sderiv \bar{Z} = \oneon{4} \int d\tau\, \mathcal{L}_{\rm 2b,kin}, \\
	\mathcal{L}_{\rm 2b,kin} &= N^{-1}\dot{z}\dot{\bar{z}} - \frac{i}{2}(\kappa\dot{\bar\kappa} - \dot\kappa\bar\kappa) 
			- N^{-1}(\psi_0\kappa\dot{\bar{z}} - \bar\psi_0\bar\kappa\dot{z}) + N^{-1}\psi_0\bar\psi_0\kappa\bar\kappa.
\eea
\vspace{\parskip}\noindent
The generalization to a non-linear sigma model for a set, $\{ Z^a \}$, of $2b$ multiplets is given by
\vspace{\parskip}
\bea\eqlabel{sspace_curved_2b_action}
	S_{\rm 2b} &= - \oneon{4} \int d\tau\, d^2\theta\, \mathcal{E}\, G_{a\bar{b}}(\underline{Z}, \underline{\bar{Z}}) 
		\sderiv Z^a \bar\sderiv \bar{Z}^{\bar{b}} = \oneon{4} \int d\tau\, \mathcal{L}_{\rm 2b}, \\
	\mathcal{L}_{\rm 2b} &= N^{-1} G_{a\bar{b}}(\underline{z}, \underline{\bar{z}}) \dot{z}^a \dot{\bar{z}}^{\bar{b}}
		- \frac{i}{2} G_{a\bar{b}}(\underline{z}, \underline{\bar{z}}) (\kappa^a \dot{\bar{\kappa}}^{\bar{b}} - \dot{\kappa}^a \bar{\kappa}^{\bar{b}})
		- N^{-1} G_{a\bar{b}}(\underline{z}, \underline{\bar{z}}) (\psi_0 \kappa^a \dot{\bar{z}}^{\bar{b}} - \bar\psi_0 \bar{\kappa}^{\bar{b}} \dot{z}^a) \\ &
		+ N^{-1} G_{a\bar{b}}(\underline{z}, \underline{\bar{z}}) \psi_0 \bar\psi_0 \kappa^a \bar{\kappa}^{\bar{b}}
		- \frac{i}{2} G_{a\bar{b}, c}(\underline{z}, \underline{\bar{z}}) 
			(\kappa^a \bar{\kappa}^{\bar{b}} (\dot{z}^c - 2\psi_0 \kappa^c) - 2\kappa^c \bar{\kappa}^{\bar{b}} \dot{z}^a ) \\ &
		+ \frac{i}{2} G_{a\bar{b}, \bar{c}}(\underline{z}, \underline{\bar{z}}) 
			(\kappa^a \bar{\kappa}^{\bar{b}} (\dot{\bar{z}}^{\bar{c}} + 2\bar{\psi}_0 \bar{\kappa}^{\bar{c}}) 
			- 2\kappa^a \bar{\kappa}^{\bar{c}} \dot{\bar{z}}^{\bar{b}} ) 
		- N G_{a\bar{b}, c\bar{d}}(\underline{z}, \underline{\bar{z}}) \kappa^a \bar{\kappa}^{\bar{b}} \kappa^c \bar{\kappa}^{\bar{d}} .
\eea
\vspace{\parskip}\noindent
Here, $G_{\ldots,a}$ means differentiation with respect to the bosonic fields $z^a$. In our application to M-theory, we need a variant of this action where the sigma model metric $G_{a\bar{b}}$ is also allowed to depend on a set of $2a$ multiplets $\phi^i$ in addition to $Z^a$ and $\bar{Z}^{\bar{b}}$. This leads to a coupling between $2a$ and $2b$ multiplets. The action for this case reads
\vspace{\parskip}
{\allowdisplaybreaks
\begin{align}
	S_{\rm 2b} &= - \oneon{4} \int d\tau\, d^2\theta\, \mathcal{E}\, G_{a\bar{b}}(\underline{\phi}, \underline{Z}, \underline{\bar{Z}}) 
		\sderiv Z^a \bar\sderiv \bar{Z}^{\bar{b}} = \oneon{4} \int d\tau\, \mathcal{L}_{\rm 2b}, \nonumber \\ \allowdisplaybreaks
	\mathcal{L}_{\rm 2b} &= N^{-1} G_{a\bar{b}}(\underline{\varphi}, \underline{z}, \underline{\bar{z}}) \dot{z}^a \dot{\bar{z}}^{\bar{b}}
		- \frac{i}{2} G_{a\bar{b}}(\underline{\varphi}, \underline{z}, \underline{\bar{z}}) 
			(\kappa^a \dot{\bar{\kappa}}^{\bar{b}} - \dot{\kappa}^a \bar{\kappa}^{\bar{b}})
		- N^{-1} G_{a\bar{b}}(\underline{\varphi}, \underline{z}, \underline{\bar{z}}) 
			(\psi_0 \kappa^a \dot{\bar{z}}^{\bar{b}} \nonumber \\ \allowdisplaybreaks & - \bar\psi_0 \bar{\kappa}^{\bar{b}} \dot{z}^a) 
		+ N^{-1} G_{a\bar{b}}(\underline{\varphi}, \underline{z}, \underline{\bar{z}}) \psi_0 \bar\psi_0 \kappa^a \bar{\kappa}^{\bar{b}}
		- \frac{i}{2} G_{a\bar{b}, c}(\underline{\varphi}, \underline{z}, \underline{\bar{z}}) 
			(\kappa^a \bar{\kappa}^{\bar{b}} (\dot{z}^c - 2\psi_0 \kappa^c) - 2\kappa^c \bar{\kappa}^{\bar{b}} \dot{z}^a ) \nonumber \\ \allowdisplaybreaks &
		+ \frac{i}{2} G_{a\bar{b}, \bar{c}}(\underline{\varphi}, \underline{z}, \underline{\bar{z}}) 
			(\kappa^a \bar{\kappa}^{\bar{b}} (\dot{\bar{z}}^{\bar{c}} + 2\bar{\psi}_0 \bar{\kappa}^{\bar{c}}) 
			- 2\kappa^a \bar{\kappa}^{\bar{c}} \dot{\bar{z}}^{\bar{b}} ) 
		- N G_{a\bar{b}, c\bar{d}}(\underline{\varphi}, \underline{z}, \underline{\bar{z}}) \kappa^a \bar{\kappa}^{\bar{b}} \kappa^c \bar{\kappa}^{\bar{d}} \nonumber \\ \allowdisplaybreaks &
		- \oneon{2} N G_{a\bar{b}, i}(\underline{\varphi}, \underline{z}, \underline{\bar{z}}) \kappa^a \bar{\kappa}^{\bar{b}} f^i
		- N G_{a\bar{b}, ij}(\underline{\varphi}, \underline{z}, \underline{\bar{z}}) \kappa^a \bar{\kappa}^{\bar{b}} \psi^i \bar{\psi}^j
		- i N G_{a\bar{b}, ic}(\underline{\varphi}, \underline{z}, \underline{\bar{z}}) \kappa^a \bar{\kappa}^{\bar{b}} \bar{\psi}^i \kappa^c \nonumber \\ \allowdisplaybreaks &
		- i N G_{a\bar{b}, i\bar{c}}(\underline{\varphi}, \underline{z}, \underline{\bar{z}}) \kappa^a \bar{\kappa}^{\bar{b}} \psi^i \bar{\kappa}^{\bar{c}}
		- G_{a\bar{b}, i}(\underline{\varphi}, \underline{z}, \underline{\bar{z}}) 
			\psi^i \bar{\kappa}^{\bar{b}} (\dot{z}^a - \oneon{2} \psi_0 \kappa^a) \nonumber \\ \allowdisplaybreaks &
		+ G_{a\bar{b}, i}(\underline{\varphi}, \underline{z}, \underline{\bar{z}}) 
			\bar{\psi}^i \kappa^a (\dot{\bar{z}}^{\bar{b}} + \oneon{2} \bar{\psi}_0 \bar{\kappa}^{\bar{b}}) .  \eqlabel{sspace_curved_2b_action_coupled}
\end{align}}
\vspace{\parskip}\noindent
This result can be readily specialized to $G_{a\bar{b}}(\underline{\phi}, \underline{Z}, \underline{\bar{Z}}) = f(\underline{\phi}) G_{a\bar{b}}(\underline{Z}, \underline{\bar{Z}})$, for a real function $f=f(\underline{\phi})$, which is the case relevant to our M-theory five-fold compactifications.
\end{sloppypar}

\subsection{Fermionic multiplets}

Finally, we discuss fermionic $2b$ multiplets, that is super-multiplets $R$ with a fermionic lowest component and  satisfying $\bar{\cal D}R=0$. Their component expansion is given by
\be\eqlabel{sspace_curved_sfield_2bf}
	 R = \rho + \theta h + \frac{i}{2} N^{-1} \theta\bar\theta (\dot{\rho} - \psi_0 h),
\ee
where the notation for the component fields is completely analogous to the globally supersymmetric case~\eqref*{2bfexp}. The component supersymmetry transformations follow from plugging the component expansions~\eqref*{sspace_curved_sderiv_comp_exp, sspace_curved_sfield_2bf} into the general formula~\eqref*{sspace_curved_gen_sfield_comp_transf_final} and are given by
\begin{align}
	\delta_\epsilon \rho = i \epsilon h, \quad
						\delta_\epsilon \bar\rho = 0, \quad
						\delta_\epsilon h = 0, \quad
						\delta_\epsilon \bar{h} = - N^{-1} \epsilon (\dot{\bar{\rho}} - \bar{\psi}_0 \bar{h}), \eqlabel{sspace_curved_transf_2bf} \\
		\delta_{\bar\epsilon} \rho = 0, \quad
						\delta_{\bar\epsilon} \bar\rho = - i \bar\epsilon \bar{h}, \quad
						\delta_{\bar\epsilon} h = N^{-1} \bar\epsilon (\dot{\rho} - \psi_0 h), \quad
						\delta_{\bar\epsilon} \bar{h} = 0. \eqlabel{sspace_curved_transf_2bf_bar}
\end{align}
A simple kinetic term for a single fermionic $2b$ superfield $R$ takes the form
\bea
	S_{\rm 2b-f,kin} &= - \oneon{4} \int d\tau\, d^2\theta\, \mathcal{E}\, R \bar{R} = \oneon{4} \int d\tau\, \mathcal{L}_{\rm 2b-f,kin}, \\
	\mathcal{L}_{\rm 2b-f,kin} &= \frac{i}{2} (\rho\dot{\bar{\rho}} - \dot{\rho}\bar{\rho}) - N h \bar{h} \; .
\eea
Note that the only bosonic field, $h$, in this multiplet is auxiliary and, hence, we are left with only fermionic physical degrees of freedom. This observation will be crucial for writing down a superspace version of the effective one-dimensional theories obtained from M-theory. As for the other types of multiplets, we need to generalise to a sigma model for a set, $\{R^x\}$, of fermionic $2b$ multiplets. The sigma model metric $G_{x\bar{y}}=G_{x\bar{y}}(\underline{\phi})$ should be allowed to depend on $2a$ multiplets $\phi^i$. Such an action takes the form
\bea\eqlabel{sspace_curved_2bf_action}
	S_{\rm 2b-f} 	&= - \oneon{4} \int d\tau\, d^2\theta\, \mathcal{E}\, G_{x\bar{y}}(\underline{\phi}) R^x \bar{R}^{\bar{y}} 
				  = \oneon{4} \int d\tau\, \mathcal{L}_{\rm 2b-f}, \\
	\mathcal{L}_{\rm 2b-f} &= \frac{i}{2} G_{x\bar{y}}(\underline{\varphi}) ( \rho^x \dot{\bar{\rho}}^{\bar{y}} - \dot{\rho}^x \bar{\rho}^{\bar{y}} ) 
		- N G_{x\bar{y}}(\underline{\varphi}) h^x \bar{h}^{\bar{y}}
		- i N G_{x\bar{y},i}(\underline{\varphi}) ( \psi^i \rho^x \bar{h}^{\bar{y}} + \bar{\psi}^i \bar{\rho}^{\bar{y}} h^x ) \\ &
		- \oneon{2} N G_{x\bar{y},i}(\underline{\varphi}) \rho^x \bar{\rho}^{\bar{y}} f^i
		- N G_{x\bar{y},ij}(\underline{\varphi}) \rho^x \bar{\rho}^{\bar{y}} \psi^i \bar{\psi}^j
		+ \oneon{2} G_{x\bar{y},i}(\underline{\varphi}) \rho^x \bar{\rho}^{\bar{y}} ( \psi_0 \psi^i - \bar{\psi}_0 \bar{\psi}^i ) \; .
\eea
%
%
%

\chapter{M-theory on Calabi-Yau Five-Folds}\chlabel{mcy5}

The first part of this thesis finds its culmination in the present chapter. All the knowledge developed in the preceding chapters is put to use in the dimensional reduction of M-theory on Calabi-Yau five-folds. Each facet of this reduction will receive a thorough examination.

\section{Calabi-Yau five-fold compactification of M-theory}\seclabel{topol_constraint}

Before performing the dimensional reduction of M-theory on Calabi-Yau five-folds, we need to make sure that a background configuration of the form $\manifold = \R \times X$\newnot{symbol:R}, where $X$ is a Calabi-Yau five-fold, is compatible with the 11-dimensional equations of motion up to a given order in the $\beta$-expansion.

We recall that at ${\cal O}(\beta^0)$, that is at the purely classical level, the bosonic equations of motion are
\begin{gather}
	R_{MN} = \oneon{12} G_{M M_2\ldots M_4}{G_N}^{M_2\ldots M_4} - 
	 	\oneon{144} g_{MN} G^2 \; , \eqlabel{mcy5_eom_1} \\
	d \ast G + \oneon{2} G\wedge G = 0 \; . \eqlabel{mcy5_eom_2} 
\end{gather}
We will always assume a vanishing fermion background $\Psi_M = 0$ and thus, we do neither need to take into account the fermionic contributions to the above equations nor the gravitino equation of motion. For such a background to respect supersymmetry, one must in addition demand that $\delta_{\epsilon^{(11)}} \Psi_M = 0$. This leads to the Killing spinor equation, to which we will turn later in this section. First, however, observe that to zeroth order in $\beta$, it is consistent to set $G=0$, since $R_{MN} (\R \times X) = 0$ owing to the Ricci-flatness of $X$.

At the next order, ${\cal O}(\beta^1)$, the equations of motion schematically read
\begin{gather}
	R_{MN} = \oneon{12} G_{M M_2\ldots M_4}{G_N}^{M_2\ldots M_4} - 
	 	\oneon{144} g_{MN} G^2 + \mathcal{O}(\beta^1) \; , \eqlabel{mcy5_eom_3} \\
	d \ast G =- \oneon{2} G\wedge G - (2\pi)^4 \beta X_8 \; . \eqlabel{mcy5_eom_4}
\end{gather}
The consistency of these equation for a background $\manifold = \R \times X$ was examined in ref.~\cite{Lu:2004ng}. We see that $G=0$ will in general no longer be consistent with the second equation. This is not surprising, as we expect the zeroth order background to be corrected $G=0+\beta G^{(1)} + \mathcal{O}(\beta^2)$. This choice satisfies \eqref{mcy5_eom_4}, provided $d\ast G^{(1)} \sim X_8 (\R \times X)$. \Eqref{mcy5_eom_3} then becomes $R_{MN} (\R \times X) \sim \mathcal{O}(\beta^1) \neq 0$ and thus the zeroth order Calabi-Yau metric $g^{(0)}_{mn}(\text{CY})$ (with $m,n,\ldots=1,\ldots,10$) on $X$ receives corrections at $\mathcal{O}(\beta^1)$, that is $g_{mn}(X) = g^{(0)}_{mn}(\text{CY}) + \beta g^{(1)}_{mn} + \mathcal{O}(\beta^2)$, which renders $X$ to be no longer Calabi-Yau at this order. In fact, a more careful treatment~\cite{Lu:2004ng} reveals that $X$ is still a complex manifold with $c_1(X)=0$ but it is neither K\"ahler nor Ricci-flat. In addition, a warp factor is introduced into the 11-dimensional metric endowing it with a 0-brane structure. It turns out that the original 32 supersymmetries can be maintained by allowing the supersymmetry transformations to also be modified at $\mathcal{O}(\beta^1)$.

Since we would like to consider both higher order $\beta$-corrections and honest Calabi-Yau five-fold compactifications in this thesis, we are therefore forced into a hybrid picture. Namely, we will work at $\mathcal{O}(\sqrt{\beta})$. In order to illustrate how the Calabi-Yau background prevails against the higher order corrections up to this order, it is instructive to temporarily think of $\sqrt{\beta}$ as a new and independent expansion parameter $\gamma = \sqrt{\beta}$\newnot{symbol:gamma}. Our aim then is to work to first order in $\gamma$. With the new expansion parameter, the equations of motion become
\begin{gather}
	R_{MN} = \oneon{12} G_{M M_2\ldots M_4}{G_N}^{M_2\ldots M_4} - 
	 	\oneon{144} g_{MN} G^2 + \mathcal{O}(\gamma^2) \; , \eqlabel{mcy5_eom_5} \\
	d \ast G =- \oneon{2} G\wedge G - (2\pi)^4 \gamma^2 X_8 \; . \eqlabel{mcy5_eom_6}
\end{gather}
As before, we start by assuming $G = 0 + \gamma G^{(1)} + \mathcal{O}(\gamma^2)$. The Einstein equation then reads $R_{MN} = 0 + \mathcal{O}(\gamma^2)$ and thus $X$ is still Ricci-flat at $\mathcal{O}(\gamma^1)$, that is $g_{mn}(X) = g^{(0)}_{mn}(\text{CY}) + \mathcal{O}(\gamma^2)$. As for the $G$-equation of motion, we observe that the left hand side of \eqref{mcy5_eom_6} is exact and thus, cohomologically the equation reads
\be\eqlabel{G1_2conds}
	g\wedge g + 2\, X_8(\R \times X) = 0 \; ,
\ee
where the re-scaled cohomological class $g \equiv \left[G^{(1)}/(2\pi)^2 \right]$\newnot{symbol:cohomclass} was defined for later convenience and the Hodge-$\ast$ is taken with respect to the zeroth order Ricci-flat Calabi-Yau metric $g^{(0)}_{mn}(\text{CY})$. Recall that $X_8$ is an expression in Pontrjagin classes (see~\eqref*{def_x8}), which are by definition cohomological classes although the square brackets are sometimes omitted in the literature. Hence, $[X_8] = X_8$. Note also, that $[G\wedge G] = [G]\wedge [G]$. 

In addition, the internal background 4-form flux obeys a flux quantisation condition~\cite{Witten:1996md} as a consequence of demanding single-valuedness of the path integral. This is similar to the Dirac quantisation law for magnetic flux in Maxwell theory, but with subtle differences owing to the presence of the higher-order M-theory corrections. With these subtleties, which are explained in ref.~\cite{Witten:1996md}, the quantisation law reads
\be
	g - \oneon{4} p_1(\manifold) \in H^4(\manifold,\mathbb{Z}) \; ,
\ee
where $p_1 (\manifold)$ is the first Pontrjagin class of $\manifold$. Specialising to $\manifold=\R \times X$ and using the result $p_1 (\R \times X) = - 2 c_2(X)$ explained in \appref{cp_classes}, we conclude
\be\eqlabel{gquant}
	g + \oneon{2} c_2(X) \in H^4(X,\mathbb{Z}) \; ,
\ee
which shows that $g$ is quantised in (half-)integer units if $c_2(X)$ is even (odd).

Returning to \eqref{G1_2conds}, we now need to evaluate $X_8(\R \times X)$ for a Calabi-Yau five-fold $X$. This calculation, presented in detail in \appref{cp_classes}, yields
\be\eqlabel{X8res}
	X_8(\R \times X) = - \oneon{24} c_4(X) \; ,
\ee
where $c_4(X)$ is the fourth Chern class of $X$. Together with the membrane contribution from \eqref{Geom_corr}, this leads to the following topological constraint
\be\eqlabel{intcond}
	c_4(X) - 12\, g\wedge g = 24\, W \; ,
\ee
where $W=[C]\in H_2(X,\mathbb{Z})$ represents the second integral homology class of the curve $C$, wrap\-ped by the membranes.

Finally, we need to consider the Killing spinor equation $\delta_{\epsilon^{(11)}} \Psi_M = 0$ at $\mathcal{O}(\gamma^1)$ for Calabi-Yau five-fold backgrounds. Since $g_{mn}(X) = g^{(0)}_{mn}(\text{CY}) + \mathcal{O}(\gamma^2)$, we expect the gravitino transformation~\eqref*{sugra11_susy_transf_gravitino1} to be corrected only at $\mathcal{O}(\gamma^2)$~\cite{Lu:2004ng}
\begin{multline}
	\delta_{\epsilon^{(11)}} \Psi_M  = 
	   2 (D^{(0)}_M + \gamma^2 D^{(2)}_M) \epsilon^{(11)} + \oneon{144} ((g^{(0)}_{MS}(\R\times\text{CY}) + \gamma^2 g^{(2)}_{MS})
	   	\gamma^{NPQRS} \\ - 8 \delta_M^N \gamma^{PQR}) \epsilon^{(11)} G_{NPQR} + \mathcal{O}(\gamma^2) \; ,
\end{multline}
where $g^{(0)}_{MN}(\R\times\text{CY})$ is specified by the zeroth order $\R\times X$ line element
\be\eqlabel{bg_metric}
	ds^2 = g^{(0)}_{MN}(\R\times\text{CY}) dx^M dx^N = -d\tau^2 + g^{(0)}_{mn}(\text{CY}) dx^mdx^n \; .
\ee
For brevity, we will henceforth denote the Ricci-flat metric $g^{(0)}_{mn}(\text{CY})$ on $X$ simply by $g_{mn}$. We conclude that at $\mathcal{O}(\gamma^1)$ the Killing spinor equation $\delta_{\epsilon^{(11)}} \Psi_M = 0$ follows from the classical gravitino transformation law~\eqref*{sugra11_susy_transf_gravitino1}.

The so obtained Killing spinor equation was analysed for $\R\times X$ backgrounds in ref.~\cite{Gillard:2004xq}.\footnote{The analysis in ref.~\cite{Gillard:2004xq} was carried out to zeroth order in $\gamma$. Repeating the calculation at $\mathcal{O}(\gamma^1)$ is guaranteed to yield a valid solution to this order. This is however not necessarily the most general solution at this order, since there is potentially more freedom to satisfy the uncorrected Killing spinor equation if the fields are allowed to acquire $\mathcal{O}(\gamma^1)$ corrections. This is currently under investigation.} Upon insertion of the compactification ansatz for the 11-dimensional infinitesimal supersymmetry parameter $\epsilon^{(11)}$
\be\eqlabel{KSeq_susyparam}
	\epsilon^{(11)} = \frac{i}{2} \epsilon \otimes \eta^\star - \frac{i}{2} \bar\epsilon \otimes \eta\; ,
\ee
with $\eta$ being the covariantly constant spinor on $X$ introduced in \chref{cy5}, the internal part $\delta_{\epsilon^{(11)}} \Psi_m = 0$ of the Killing spinor equation becomes
\be\eqlabel{KSeq1}
	D^{(0)}_m \eta + \frac{(2\pi)^2 \gamma}{288} (\gamma_m {}^{npqr} - 8 \delta_m^n \gamma^{pqr}) \eta \, g_{npqr} = 0\; .
\ee
Here, indices are raised and lowered using $g_{mn}$. The first term vanishes since $\eta$ is covariantly constant by assumption. The remaining second part is most easily analysed in holomorphic coordinates. The real $4$-form $g$ decomposes into a sum of $(p,q)$-forms (with p+q=4) according to
\be\eqlabel{gflux_pq_decomp}
	g = g^{(1,3)} + g^{(2,2)} + g^{(3,1)} \; ,
\ee
with $g^{(2,2)}$ real and $(g^{(1,3)})^\ast = g^{(3,1)}$. With this decomposition, the transition to holomorphic coordinates and the annihilation condition~\eqref*{eta_annihil}, one finds for the second part of~\eqref*{KSeq1}
\be
	\gamma^{\mu_1 \mu_2 \mu_3} \eta \, g^{(3,1)}_{\mu_1 \mu_2 \mu_3 \bar\nu} +
	(\gamma_{\bar\nu} {}^{\mu_1 \mu_2 \bar{\rho}_1 \bar{\rho}_2} - 4 \delta_{\bar\nu}^{\bar{\rho}_1} 
		\gamma^{\mu_1 \mu_2 \bar{\rho}_2} ) \eta \, g^{(2,2)}_{\mu_1 \mu_2 \bar{\rho}_1 \bar{\rho}_2} = 0 \; .
\ee
The gamma matrix products in the second expression can be simplified further by anti-commuting the gamma matrices. One arrives at
\be
	\gamma^{\mu_1 \mu_2 \mu_3} \eta \, g^{(3,1)}_{\mu_1 \mu_2 \mu_3 \bar\nu} +
	\gamma^\mu \eta \, (4 \tilde{g}^{(2,2)}_{\mu\bar\nu} - g_{\mu\bar\nu} \tilde{\tilde{g}}^{(2,2)} ) = 0 \; .
\ee
with $\tilde{g}^{(2,2)}_{\mu\bar\nu} = g^{(2,2)}_{\mu\rho\bar{\nu}\bar\sigma} g^{\rho\bar\sigma}$ as defined in~\eqref*{def_tilde_map}. Thus, $g^{(3,1)}$ and the expression in parenthesis must vanish separately. In fact, the expression in parenthesis implies $\tilde{g}^{(2,2)}=0$ as can be seen by taking the trace and re-inserting it into the expression. 

In summary, the preservation of $\susyno=2$ supersymmetry demands that
\be\eqlabel{gflux_KSeq_cond}
	g^{(3,1)} = g^{(1,3)} = \tilde{g}^{(2,2)} = 0 \; .
\ee

After inserting~\eqref*{KSeq_susyparam,gflux_pq_decomp}, changing to holomorphic coordinates and using properties of $\gamma^\mu$ and $\eta$, the uncompactified part $\delta_{\epsilon^{(11)}} \Psi_0 = 0$ of the Killing spinor equation translates into the following one-dimensional equation
\be
	\partial_0 \epsilon + \frac{(2\pi)^2 i \gamma}{4!} \tilde{\tilde{g}}^{(2,2)} \epsilon = 0 \; .
\ee
Imposing~\eqref*{gflux_KSeq_cond} leads to the condition $\epsilon = \text{const.}$, which is the expected result for $\R = \M^1$, in analogy to maximally symmetric Minkowski space in higher dimensions. Thus, setting $\manifold=\R\times X$ with metric~\eqref*{bg_metric}, $G=(2\pi)^2 \gamma g^{(2,2)}$ with $\tilde{g}^{(2,2)}=0$ and $\Psi_M=0$ solves the equations of motion at order $\gamma$.

It is important to note that for a given real 4-form $g$, the decomposition~\eqref*{gflux_pq_decomp} is complex structure dependent and hence, the condition~\eqref*{gflux_KSeq_cond} imposes restrictions on the allowed variations of the complex structure on $X$. In addition, the map $\tilde{(\cdot)}$ depends on the K\"ahler class, which is therefore also affected by~\eqref*{gflux_KSeq_cond}. This will be discussed in more detail in \secref{flux_comp}, but beforehand we will perform a dimensional reduction to zeroth order in $\gamma$, that is for a background $\manifold = \R \times X$ and without flux $G=0$.

\section{The bosonic reduction}\seclabel{bos_red}

In this section, we consider the Kaluza-Klein reduction of 11-dimensional supergravity, given by~\eqref*{S_CJS_B}, on a space-time manifold of the form $\manifold = \mathbb{R}\times X$, where $X$ is a Calabi-Yau five-fold. At zeroth order in $\beta$, we start with the background configuration
\be\eqlabel{bg}
	ds^2 = -d\tau^2 + g_{mn}dx^mdx^n \; ,\qquad G = 0 \; ,
\ee
where $g_{mn}=g_{mn}(x^p)$ is the Ricci-flat metric on $X$ and $m,n\ldots = 1,\dots ,10$. As we saw in the previous section, this background solves the leading order bosonic equations of motion~\eqref*{mcy5_eom_1} and \eqref*{mcy5_eom_2}.

We now need to identify the moduli of this background. As discussed in detail in \secref{cy5_moduli_spaces}, the formalism to deal with Calabi-Yau five-fold moduli spaces is largely similar to the one developed for Calabi-Yau three-folds~\cite{Candelas:1990pi}. Here, we only summarise the essential information needed for the dimensional reduction.

As for Calabi-Yau three-folds, the moduli space of Ricci-flat metrics on Calabi-Yau five-folds is (locally) a product of a K\"ahler and a complex structure moduli space respectively associated to $(1,1)$ and $(2,0)$ deformations of the metric. They can be described in terms of harmonic $(1,1)$-forms for the K\"ahler moduli space and harmonic $(1,4)$-forms for the complex structure moduli space. We begin with the K\"ahler moduli which we denote by $t^i(\tau )$\newnot{symbol:moduli_11}, where $i,j,\ldots = 1,\dots ,h^{1,1}(X)$ and $\tau$ is time. They are real scalar fields and can be defined by expanding the K\"ahler form $J$ on $X$ in terms of a basis $\{\omega_i \}$ of $H^{1,1}(X)$ as
\be
	J=t^i\omega_i \; .
\ee
The complex structure moduli are denoted by $z^a(\tau)$\newnot{symbol:moduli_14}, where $a,b,\ldots = 1,\dots , h^{1,4}(X)$, and these are, of course, complex scalar fields.

\subsection{The reduction ansatz}\seclabel{bos_red_ansatz}

After this preparation, the ansatz for the 11-dimensional metric including moduli can be written as
\be\eqlabel{gansatz}
	ds^2 = -\oneon{4}N(\tau )^2 d\tau^2 + g_{mn} (t^i,z^a,\bar{z}^{\bar{a}}) dx^m dx^n \; ,
\ee
where $N(\tau )$\newnot{symbol:lapse} is the einbein or lapse function. The lapse function can, of course, be removed by a time reparametrisation. However, its equation of motion in the one-di\-men\-sio\-nal effective theory is the usual zero-energy constraint (the equivalent of the Friedman equation in four-dimensional cosmology; see \chref{msp}). In order not to miss this constraint, we will keep $N$ explicitly in our metric ansatz.

The zero modes of the M-theory 3-form $A$ are obtained by an expansion in harmonic forms, as usual. From the Hodge diamond~\eqref*{cy5_hodgediamond} of Calabi-Yau five-folds, it is clear that only the harmonic 2- and 3-forms on $X$ are relevant in this context. For the latter we also introduce a basis $\{\nu_p\}$, where $p,q,\ldots = 1,\ldots ,h^{2,1}(X)$. The zero mode expansion for $A$ can then be written as
\be\eqlabel{Aansatz}
	A=(\xi^p(\tau )\nu_p +{\rm c.c.})+ N \mu^i(\tau )\,\omega_i\wedge d\tau \; ,
\ee
with $h^{2,1}(X)$ complex scalar fields $\xi^p$\newnot{symbol:moduli_21} and $h^{1,1}(X)$ real scalars $\mu^i$. It is clear that the latter correspond to gauge degrees of freedom since $N \mu^i(\tau )\,\omega_i\wedge d\tau = d(f^i(\tau )\,\omega_i)$ with the function $f^i$ being integrals of $N \mu^i$. Note that $N$ enters here merely to ensure worldline reparameterization covariance. Since the fields $\mu^i$ do not represent physical degrees of freedom, the one-dimensional effective action should not depend on these modes. It is, therefore, safe to ignore them in the above ansatz for $A$. Nevertheless, we will find it instructive to keep these modes for now to see explicitly how they drop out of the effective action.

Further zero-modes can arise from membranes if they are included in the compactification, such as moduli of the complex curve $C$ which they wrap and their superpartners. Here, we will not include these additional modes but rather focus on the modes from pure 11-dimensional supergravity.

While the way we have parametrised the zero modes of $A$ in \eqref*{Aansatz} appears to be the most natural one, it is not actually the most well-suited ansatz for performing the dimensional reduction. This is due to the fact that we have split a 3-form into $(2,1)$- and $(1,2)$-pieces (ignoring the gauge part for the moment) the choice of which implicitly depends on the complex structure moduli. If carried through, this leads to an unfavourable and complicated intertwining of kinetic terms of the $(2,1)$- and $(1,4)$-fields in the one-dimensional effective action (that is, terms involving products of the like $\dot{\xi}^p \dot{z}^a$ etc.), which would in turn force us into attempting lengthy field re-definitions in order to diagonalise the kinetic terms.

It would, on the other hand, be much more economic to start out with a formulation in which no such mixing of kinetic terms arises in the first place. Indeed, it is possible to circumvent, yet fully capture, this complication by using real harmonic 3-forms instead of complex $(2,1)$- and $(1,2)$-forms to parametrise the 3-form zero modes. Real harmonic 3-forms can be naturally locked to 3-cycles and thus represent topological invariants of $X$. In order to employ them in the ansatz for $A$, we first need to introduce a basis $\{ N_{\cal{P}} \}_{\mathcal{P}=1,\ldots,b^3 (X)}$ of real harmonic 3-forms on $X$. Instead of \eqref*{Aansatz}, we can then write
\be\eqlabel{Aansatz_3forms}
	A = X^{\cal{P}} (\tau) N_{\cal{P}} + N \mu^i(\tau )\,\omega_i\wedge d\tau \; ,
\ee
with $b^{3}(X) = 2 h^{2,1} (X)$ real scalar fields $X^{\cal{P}}$\newnot{symbol:moduli_3form} and $h^{1,1}(X)$ real scalars $\mu^i$. The two ans\"{a}tze for $A$ are readily related by using the linear maps $\mathfrak{A}$ and $\mathfrak{B}$ introduced in \secref{real_form_formalism}. Inserting \eqrangeref{21_3_rel1}{21_3_rel2} into \eqrangeref{Aansatz}{Aansatz_3forms}, we learn how the two formulations are related at the level of zero mode fields
\begin{align}
	\xi^p &= X^{\cal{Q}} \mathfrak{B}_{\cal{Q}} {}^p \qquad\qquad \text{(and $\cc$)} , \eqlabel{21_3_rel_xi_X_1} \\
	X^{\cal{P}} &= \xi^q \mathfrak{A}_q {}^{\cal{P}} + \bar{\xi}^{\bar{q}} \bar{\mathfrak{A}}_{\bar{q}} {}^{\cal{P}} \; . \eqlabel{21_3_rel_xi_X_2}
\end{align}
For the reasons outlined above, we henceforth adopt the 3-form formulation. At each step of the calculation, one may, of course, revert if desired to the complex $(2,1)$-form formulation using \eqrangeref{21_3_rel1}{21_3_rel2},~\eqrangeref*{21_3_rel_xi_X_1}{21_3_rel_xi_X_2} and the results of \secref{real_form_formalism}.

The $\mathfrak{A}$ and $\mathfrak{B}$ matrices turn out to be an effective way to parametrise our ignorance of the actual dependence of the $(2,1)$-forms on the complex structure moduli and it would be nice to find explicit expressions instead. However, we are not aware of a method to calculate this dependence explicitly.

\subsection{Computing the one-dimensional effective action}\seclabel{bos_red_comp}

Returning to the metric ansatz in~\eqref*{gansatz}, we now compute the 11-dimensional Ricci scalar $R$. As usual, for given values of the complex structure moduli, we introduce local holomorphic coordinates $z^\mu$ on $X$, where $\mu ,\nu ,\ldots =1,\dots ,5$, so that the metric is purely $(1,1)$, that is the components $g_{\mu\bar{\nu}}$ are the only non-vanishing ones and consequently, the 11-dimensional line element becomes
\be\eqlabel{gansatz_cplx}
	ds^2 = -\oneon{4}N(\tau )^2 d\tau^2 + 2 g_{\mu\bar\nu} (t^i,z^a,\bar{z}^{\bar{a}}) dz^\mu d\bar{z}^{\bar\nu} \; .
\ee
The factor of 2 in the second term is a consequence of our transition to holomorphic coordinates as described in \appref{complex_geometry} (in particular, see \eqref{real_complex_tensor_contractions_conversion,real_complex_ds_conversion}). The Christoffel symbol $\Gamma^L_{MN} \equiv \oneon{2} g^{LP} \left( g_{PN,M} + g_{PM,N} - g_{MN,P} \right)$ for this ansatz yields
\bea\eqlabel{bos_red_connection_comp_lapse_fnc}
	\Gamma^0_{00}			&= N^{-1} \dot{N} \; , \qquad
	&\Gamma^0_{\mu\nu} 		&= 2 N^{-2} \dot{g}_{\mu\nu} \; , \qquad
	&\Gamma^0_{\mu\bar\nu} 	&= 2 N^{-2} \dot{g}_{\mu\bar\nu} \; , \\
	\Gamma^\mu_{0\nu} 		&= \oneon{2} g^{\mu\bar\rho}\dot{g}_{\nu\bar\rho} \; , \qquad
	&\Gamma^\mu_{0\bar\nu} 	&= \oneon{2} g^{\mu\bar\rho}\dot{g}_{\bar\nu\bar\rho} \; , \qquad
	&\Gamma^\rho_{\mu\nu} 		&= g^{\rho\bar\sigma} g_{\mu\bar\sigma,\nu} = \tilde{\Gamma}^\rho_{\mu\nu} \; ,
\eea
where here and in the following the dot denotes the derivative with respect to $\tau$ and the tilde denotes purely 10-dimensional quantities on $X$. The complex conjugates of the above components are, of course, also present and all other components vanish identically. Note that even though one may choose the metric to be hermitian, this is not true for its variations precisely because they may change the complex structure and thereby the Hermiticity condition. Explicit expressions for the metric variations were stated in~\eqref*{deltag} and we will insert them in a moment.

Next, we will work out the Riemann tensor $R_{MNPQ}$. Since we are ultimately interested in the Ricci scalar $R \equiv g^{MN} R_{MN}$, we only need certain components of $R_{MNPQ}$. For our $\R\times X$ ansatz,
\be
	R = g^{00} R_{00} + 2 g^{\mu\bar\nu} R_{\mu\bar\nu} \; ,
\ee
where the Ricci tensor $R_{MN}$ is generally defined as $R_{MN} \equiv g^{PQ} R_{MPNQ}$ and in our case, it becomes
\be
	R_{00} = 2 g^{\mu\bar\nu} R_{0\mu 0\bar\nu} \; , \qquad
	R_{\mu\bar\nu} = g^{00} R_{0\mu 0\bar\nu} + 2 g^{\rho\bar\sigma} R_{\mu\left(\rho|\bar{\nu}|\bar{\sigma}\right)} \; .
\ee
We thus only need three components of the Riemann tensor, namely $R_{0\mu 0\bar\nu}$, $R_{\mu\rho\bar{\nu}\bar{\sigma}}$ and $R_{\mu\bar{\sigma}\bar{\nu}\rho}$, to determine $R$
\be
	R = 4 g^{00} g^{\mu\bar\nu} R_{0\mu 0\bar\nu} + 4 g^{\mu\bar\nu} g^{\rho\bar\sigma} R_{\mu\left(\rho|\bar{\nu}|\bar{\sigma}\right)} \; .
\ee
Using the standard formula from general relativity
\be
	R_{MNPQ}=\oneon{2}\left(g_{MQ,NP} + g_{NP,MQ} - g_{MP,NQ} - g_{NQ,MP}\right) 
		+ g_{RS} \left(\Gamma^R_{NP} \Gamma^S_{MQ} - \Gamma^R_{NQ} \Gamma^S_{MP} \right) ,
\ee
we find
\begin{align}
	R_{0\mu 0\bar\nu} &= - \oneon{2} \ddot{g}_{\mu\bar\nu} - \oneon{2} N^{-1} \dot{N} \dot{g}_{\mu\bar\nu}
		+ \oneon{2} g^{\rho\bar\sigma} \dot{g}_{\mu(\bar\sigma} \dot{g}_{\rho)\bar\nu} \; , \\
	R_{\mu\rho\bar{\nu}\bar{\sigma}} &= 2 N^{-2} \dot{g}_{\mu\left[\bar\nu\right.} \dot{g}_{\left.|\rho|\bar\sigma\right]} 
		+ \tilde{R}_{\mu\rho\bar{\nu}\bar{\sigma}} \; , \\
	R_{\mu\bar{\sigma}\bar{\nu}\rho} &= 2 N^{-2} \dot{g}_{\mu\left[\bar\nu\right.} \dot{g}_{\left.\rho\right]\bar\sigma} 
		+ \tilde{R}_{\mu\bar{\sigma}\bar{\nu}\rho} \; .
\end{align}
This leads to
\begin{multline}
	\oneon{2}N^2 R	=	   4 N^2 \frac{d}{d \tau} \left(N^{-2} g^{\mu\bar{\nu}}\dot{g}_{\mu\bar{\nu}}\right)
						+    g^{\mu\bar\rho} g^{\nu\bar\sigma} \dot{g}_{\mu\bar\sigma} \dot{g}_{\nu\bar\rho}
						+    g^{\mu\bar\rho} g^{\sigma\bar\nu} \dot{g}_{\mu\sigma} \dot{g}_{\bar\nu\bar\rho}
						\\ + 2 g^{\mu\bar\nu} \dot{g}_{\mu\bar\nu} g^{\sigma\bar\rho} \dot{g}_{\sigma\bar\rho}
						+ 4 N^{-1} \dot{N} g^{\mu\bar\nu}\dot{g}_{\mu\bar\nu} \; ,
\end{multline}
where terms containing $\tilde{R}_{\mu\bar\nu}$ have been dropped because $g_{\mu\bar\nu}$ is Ricci-flat. Into this expression, we have to insert the expansion of the metric~\eqref*{deltag} which can also be written as
\be\eqlabel{metric_mode_expansion}
	\dot{g}_{\mu\bar{\nu}} = -i \omega_{i,\mu\bar{\nu}} \dot{t}^i\; ,\quad
	\dot{g}_{\mu\nu} = - \oneon{12||\Omega||^2} \Omega_\mu {}^{\bar{\mu}_1\ldots\bar{\mu}_4} 
	  \chi_{a,\bar{\mu}_1\ldots\bar{\mu}_4\nu} \dot{z}^a\; ,\quad
	  \dot{g}_{\bar{\mu}\bar{\nu}}=\left(\dot{g}_{\mu\nu}\right)^\ast \; .
\ee
Here $\{\chi_a\}$, where $a,b,\ldots = 1,\dots ,h^{1,4}(X)$, is a basis of harmonic $(1,4)$-forms and the comma in the subscripts of the forms $\omega$ and $\chi$ separates moduli space from Calabi-Yau indices and is not to be confused with the shorthand for differentiation used elsewhere.

Further, we need the field strength $G=dA$ (keeping in mind that $d$ is the 11-di\-men\-sio\-nal exterior derivative $d = d\tau \partial_0 + \tilde{d}$) for the three-form ansatz~\eqref*{Aansatz_3forms} and its Hodge dual which are given by
\be
	G = \dot{X}^{\cal{P}} d\tau \wedge N_{\cal{P}} \; ,\qquad \ast G=-N^{-1} \dot{X}^{\cal{P}} \cs_{\cal{P}} {}^{\cal{Q}} N_{\cal{Q}}\wedge J^2 \; .
\ee
To derive the second equation we have used the result~\eqref*{3form_hodgestar} for the dual of a real 3-form on a Calabi-Yau five-fold.

Inserting the ans\"atze~\eqref*{Aansatz_3forms,gansatz_cplx} together with the last three equations into the bosonic action~\eqref*{S_CJS_B} and integrating over the Calabi-Yau five-fold, one finds the bosonic part of the one-dimensional effective action
\be\eqlabel{S1B}
	S_{\rm B,kin} = \frac{l}{2} \int d\tau N^{-1} \left\{ \oneon{4} G_{ij}^{(1,1)}(\underline{t}) \dot{t}^i \dot{t}^j  
		+ \oneon{2} G_{\mathcal{PQ}}^{(3)}(\underline{t},\underline{z},\underline{\bar{z}}) \dot{X}^{\cal{P}} \dot{X}^{\cal{Q}} 
		+ 4 V(\underline{t}) G_{a\bar{b}}^{(1,4)} (\underline{z},\underline{\bar{z}}) \dot{z}^a \dot{\bar{z}}^{\bar{b}}  \right\}
\ee
at order zero in the $\beta$ expansion. Here $l = v/\kappa_{11}^{2}$, $v$ is an arbitrary reference volume of the Calabi-Yau five-fold\footnote{Related factors of $1/v$ should be included in the definition of the moduli space metrics~\eqrangeref*{G11def}{G3def} but will be suppressed in order to avoid cluttering the notation. These factors can easily be reconstructed from dimensional analysis.} and a total derivative term has been dropped. The moduli space metrics in the $(1,1)$, $(1,4)$ and 3-form sectors are given by
\begin{align}
 G_{ij}^{(1,1)}(\underline{t}) &= 4\int_X\omega_i\wedge\ast\omega_j+8V\tilde{w}_i\tilde{w}_j \; , \eqlabel{G11def}\\
 G_{a\bar{b}}^{(1,4)}(\underline{z},\underline{\bar{z}}) &= \frac{\int_X\chi_a\wedge\bar{\chi}_{\bar{b}}}{\int_X\Omega\wedge\bar{\Omega}} \; , \\
 G_{\mathcal{PQ}}^{(3)}(\underline{t},\underline{z},\underline{\bar{z}}) &= \int_X N_{\cal{P}}\wedge\ast N_{\cal{Q}} \; , \eqlabel{G3def}
\end{align}
where $\tilde{\omega}_i= g^{\mu\bar{\nu}}\omega_{i,\mu\bar{\nu}}$.

\subsection{Physical moduli space metrics}\seclabel{bos_red_metrics}

Since $h^{1,1}(X)$ need not be even, $G_{ij}^{(1,1)}$ is a genuinely real metric that cannot be complexified in general. This is compatible with the anticipated $\susyno=2$ supersymmetry in one dimension, which only demands target spaces of sigma models to be Riemannian manifolds~\cite{Coles:1990hr}. Using the results of \secref{cy5_moduli_spaces}, these metrics can be computed as functions of the moduli. In the $(1,1)$ sector we have
\be\eqlabel{G11}
 G_{ij}^{(1,1)}(\underline{t})=8V\left[{\cal G}_{ij}^{(1,1)}(t)-25\frac{\kappa_i\kappa_j}{\kappa^2}\right]=-\frac{2}{3}\kappa_{ij}-\frac{5}{6}\frac{\kappa_i\kappa_j}{\kappa} \; ,
\ee
where $\kappa$, $\kappa_i$, $\kappa_{ij}$ etc. were defined in~\eqrangeref*{def_kappa0}{def_kappa5}.

The standard moduli space metric ${\cal G}_{ij}^{(1,1)}$, as defined in \secref{cy5_moduli_spaces}, can be obtained from the K\"ahler potential $K^{(1,1)}=-\frac{1}{2}\ln\kappa$ as ${\cal G}_{ij}^{(1,1)}=\partial_i\partial_jK^{(1,1)}$. We note that the physical sigma model metric~\eqref*{G11} differs from the standard moduli space metric ${\cal G}^{(1,1)}_{ij}$ by a term proportional to $\kappa_i\kappa_j$ and a rescaling by the volume. The latter is not really required at this stage and can be removed by a redefinition of time~$\tau$ but it will turn out to be a useful convention in the full supersymmetric version of the effective action. The additional term, however, cannot be removed, for example by a re-scaling of the fields~$t^i$. As a consequence, unlike the standard moduli space metric, the physical metric is not positive definite. Rather, $G^{(1,1)}$ has a Lorentzian signature $(-1,+1,\ldots ,+1)$. This is in contrast to, for example, M-theory compactifications on Calabi-Yau three-folds~\cite{Candelas:1990pi,Cadavid:1995bk} where the sigma model metric in the $(1,1)$ sector is identical to the standard moduli space metric and, in particular, is positive definite. In the present case, the appearance of a single negative direction is, of course, not a surprise. Our sigma model metric in the gravity sector can be though of as a ``mini-superspace'' version of the DeWitt-metric, which is well-known to have precisely one negative eigenvalue~\cite{DeWitt:1967yk} (see  \chref{msp}, for more details). Here, we see that this negative direction lies in the $(1,1)$ sector. Another difference to the Calabi-Yau three-fold case is the degree of the function $\kappa$. For Calabi-Yau three-folds, $\kappa$ is a cubic while, in the present case, it is a quintic polynomial.

We now turn to the $(1,4)$ moduli space metric $G^{(1,4)}_{a\bar{b}}$ which is, in fact, equal to the standard moduli space metric in this sector and can, hence, be expressed as
\be\eqlabel{G41}
	G^{(1,4)}_{a\bar{b}}=\partial_a\partial_{\bar{b}}K^{(1,4)}\; ,\qquad 
	K^{(1,4)}=\ln\left[i \int_X \Omega \wedge \bar\Omega \right]
\ee
in terms of the K\"ahler potential $K^{(1,4)}$. This is very similar to the three-fold case. In particular, $G^{(1,4)}_{a\bar{b}}$ is positive definite as it should be, given that the single negative direction arises in the $(1,1)$ sector.

Finally, in the 3-form sector one finds from the results in \secref{real_form_formalism} that the metric can be written as
\be\eqlabel{G3}
 	G_{\mathcal{PQ}}^{(3)}(\underline{t},\underline{z},\underline{\bar{z}}) = 
		\oneon{2} \cs_{\cal{\left(P\right.}} {}^{\cal{R}} d_{\left.\mathcal{Q}\right)\mathcal{R}ij} t^i t^j \; ,\qquad 
 	d_{\mathcal{P}\mathcal{Q}ij} = \int_X N_{\cal{P}} \wedge N_{\cal{Q}} \wedge\omega_i\wedge\omega_j\; ,
\ee
where we have introduced the intersection numbers $d_{\mathcal{P}\mathcal{Q}ij} = - d_{\mathcal{Q}\mathcal{P}ij}$, which are purely topological. The metric~\eqref*{G3} is Hermitian with respect to the complex structure $\cs$ (see \eqref{G3_hermitian}). 

This completes the definition of all objects which appear in the action~\eqref*{S1B}.

\subsection{What about \texorpdfstring{$\mu^i$}{the second term in the 3-form ansatz}?}\seclabel{what_about_mui}

We see that this action does not depend on the gauge degrees of freedom $\mu^i$ which appear in the ansatz~\eqref*{Aansatz} for the three-form $A$, as should be the case. This demonstrates $\mu^i$ independence at zeroth order in $\beta$. To see the contributions from $\mu^i$, we need to jump ahead and temporarily include the order $\beta$ terms in our consideration. At first order in $\beta$, there are three terms in the 11-dimensional theory, all of them topological, which contribute to $\mu^i$ dependent terms in one dimension. These are the Chern-Simons term $A\wedge G\wedge G$ in the bosonic CJS action~\eqref*{S_CJS_B} (if $\mathcal{O}(\gamma)$-flux is non-zero), the Green-Schwarz term~\eqref*{sugra11_gs} and the Wess-Zumino term in the membrane action~\eqref{membraneaction}. Evaluating these three terms leads to the one-dimensional contribution
\be\eqlabel{Sgauge}
	S_{\rm B,gauge}=-\frac{l\beta_1}{2}\int d\tau \, N\, \left[12\,g\wedge g+24\,W-c_4(X)\right]_i\mu^i \; ,
\ee
where $\beta_1=(2\pi)^4\beta/v^{4/5}$\newnot{symbol:beta1} is the one-dimensional version of the expansion parameter $\beta$. The notation $[\ldots ]_i$ indicates the components of the eight-form in brackets with respect to a basis $\{\hat{\omega}^i\}$ of harmonic 8-forms dual to the harmonic 2-forms $\{\omega_i\}$ or, in other words, $[\ldots ]_i = \int_X [\ldots ] \wedge \omega_i$.

Hence, at order $\beta$ the $\mu^i$ dependent terms do not automatically vanish. However, the bracket in \eqref{Sgauge} vanishes once the integrability condition~\eqref*{intcond} is imposed. Put in a different way, the equation of motion for $\mu^i$ from \eqref{Sgauge} is simply the integrability condition~\eqref*{intcond}
\be
	12\, g\wedge g + 24\, W - c_4(X) = 0 \; .
\ee
Therefore, the r\^{o}le of the gauge modes $\mu^i$ is to enforce the integrability condition at the level of the equations of motion and, once the condition is imposed, the gauge modes disappear from the action as they should. The condition~\eqref*{intcond} can, therefore, also be interpreted as an anomaly cancellation condition which has to be satisfied in order to prevent a gauge anomaly of the M-theory 3-form $A$ along the Calabi-Yau $(1,1)$ directions.

\section{The fermionic reduction}\seclabel{ferm_red}

One may ask if an explicit dimensional reduction of the fermionic part of the 11-di\-men\-sio\-nal action~\eqref*{S_CJS} is really necessary, for in many other cases, once the bosonic terms in the effective action are known the fermionic ones can be inferred from supersymmetry. In the present case, there are a number of reasons why reducing at least some of the fermionic terms might be useful. 

First of all, the structure of the bosonic action~\eqref*{S_CJS_B} points to some features of one-dimensional $\susyno=2$ supersymmetry which are not well-developed in the literature. For example, the bosonic action~\eqref*{S1B} indicates a coupling between the two main types of $\susyno=2$ supermultiplets, the $2a$ and $2b$ multiplets, which, to our knowledge, has not been worked out in the literature. 

Also, in \secref{bos_red}, we have seen that it is important to keep the lapse function as a degree of freedom in the one-dimensional theory, as it generates an important constraint. In the context of supersymmetry, the lapse is part of the  one-dimensional supergravity multiplet which one expects to generate a multiplet of constraints. Therefore, even though gravity is not dynamical in one dimension, we need to consider local one-dimensional $\susyno=2$ supersymmetry as developed in \chref{superspace}. 

At any rate, given that the relevant supersymmetry is not as well established as in some other cases, it seems appropriate to back up our results by reducing some of the 11-dimensional fermionic terms as well. We will indeed encounter a surprise, namely the appearance of purely fermionic $(1,3)$ zero modes without any matching bosons, which retrospectively justifies the additional effort of a fermionic reduction. This feature is somewhat puzzling from the point of view of higher-dimensional supersymmetry and can certainly not be clarified from the bosonic effective action alone. 

In this section, we will, therefore, reduce the terms in the 11-dimensional action quadratic in fermions. These results together with the bosonic action are sufficient to uniquely fix the one-dimensional action in superspace form and, in addition, provide us with a number of independent checks. Four-fermi terms in the one-dimensional theory are then obtained from the superspace action and we will not derive them by reduction from 11 dimensions.

\subsection{The gravitino zero modes ansatz}

We start by writing down a zero mode expansion of the 11-dimensional gravitino $\Psi_M$ on the space-time ${\cal M}=\mathbb{R}\times X$. The covariantly constant, positive chirality spinor on $X$ is denoted by $\eta$ and its negative chirality counterpart by $\eta^\star$ (for a summary of our spinor conventions see \appref{spinorconv}). The spinor $\eta$ is characterised by the annihilation conditions $\gamma^{\bar{\mu}}\eta =0$. Further, by $\omega^{(p,q)}_i$ we denote the harmonic $(p,q)$ forms on $X$. Then, following the known rules for writing down a fermionic zero mode ansatz (see \eqref{cy_spinor_decomposition} and refs.~\cite{Fre:1995bc,Roest:2004pk}), we have\newnot{symbol:lapsino}
\begin{align}
	\Psi_0 &= \psi_0(\tau) \otimes \eta^\star + \bar{\psi}_0 (\tau) \otimes \eta \eqlabel{ansatz_gravitino_0_00} , \\
	\Psi_{\bar{\mu}} &= \sum_{p, q} \zeta^{(i)}_{(p,q)} (\tau) \otimes ( \omega^{(p,q)}_{(i),\alpha_1 \ldots \alpha_{(p)} \bar{\beta}_1 \ldots \bar{\beta}_{(q-1)} \bar{\mu}} 
			\gamma^{\alpha_1 \ldots \alpha_{(p)} \bar{\beta}_1 \ldots \bar{\beta}_{(q-1)}} \eta )
	\nonumber \\&+ \sum_{p, q} \zeta^{\prime (i)}_{(p,q)} (\tau) \otimes ( \omega^{(p,q)}_{(i),\alpha_1 \ldots \alpha_{(p)} \bar{\beta}_1 \ldots \bar{\beta}_{(q-1)} \bar{\mu}} \gamma^{\alpha_1 \ldots \alpha_{(p)} \bar{\beta}_1 \ldots \bar{\beta}_{(q-1)}} \eta^\star ), 
	\eqlabel{ansatz_gravitino_internal_00} \\
	\Psi_{\mu} &= (\Psi_{\bar{\mu}})^\ast \; . 
\end{align}
Here, $\zeta_{(p,q)}^{(i)}$ and $\zeta^{\prime (i)}_{(p,q)}$ are one-dimensional complex fermions which represent the zero-modes in the $(p,q)$ sector of the Calabi-Yau five-fold and $\psi_0$ is the one-dimensional gravitino. The sums over $(p,q)$ in~\eqref*{ansatz_gravitino_internal_00} run over all non-trivial cohomology groups of the Calabi-Yau five-fold. Let us discuss the various $(p,q)$ sectors in the first sum in~\eqref*{ansatz_gravitino_internal_00} in detail. For $(p,q)=(1,4)$ the number of annihilating gamma matrices, $\gamma^{\bar{\mu}}$ exceeds the number of creating ones, $\gamma^\mu$, and, as a result, this term vanishes. Further, for all cases with $q=p+1$ the number of creation and annihilation gamma matrices is identical. Anti-commuting all $\gamma^{\bar{\mu}}$ to the right until they annihilate $\eta$ one picks up inverse metrics $g^{\mu\bar{\nu}}$ which ultimately contract the harmonic $(p,p+1)$ forms $\omega_i^{(p,p+1)}$ to harmonic $(0,1)$ forms. Since the latter do not exist on Calabi-Yau five-folds all terms with $q=p+1$ vanish. This leaves us with the cases where $p\geq q$. Among those, only the terms with $(p,q)=(2,2),(3,2)$ contain both creation and annihilation matrices. For $(p,q)=(2,2)$, anti-commuting leads to a single inverse metric which converts the harmonic $(2,2)$-forms into harmonic-$(1,1)$ forms. Therefore, the $(2,2)$-part can effectively be absorbed into the $(1,1)$ term and does not need to be written down independently. The same argument applies to the $(3,2)$-part which can be absorbed into the $(2,1)$ contribution. By the gamma matrix structure and the annihilation property of $\eta^\star$ all but the $(5,0)$ term in the second sum in~\eqref*{ansatz_gravitino_internal_00} vanish. Using the Fierz identity~\eqref*{cy5_fierz_id}, the $(5,0)$ term in the second sum can be converted into a term with the $(1,1)$ structure of the first sum and can, hence, be absorbed by the $(1,1)$ contribution. In summary, all we need to write down explicitly are the $(p,q)$ terms with $q=1$ and $p=1,2,3,4$.

For the same reason as explained in the previous subsection on the bosonic reduction, it is advantageous to use the real 3-form formulation, developed in \secref{real_form_formalism}, to capture the dynamics of the $(2,1)$-sector while avoiding off-diagonal kinetic terms mixing in time derivatives of $(4,1)$-fields.

Similarly, we will use the real $\hat{4}$-form formulation, also described in \secref{real_form_formalism}, in the $(1,3)$-sector. A general 4-form, which is always purely topological, can be decomposed into $(1,3)$, $(3,1)$ and $(2,2)$ pieces using the complex structure of the Calabi-Yau five-fold $X$. Henceforth, we will restrict our attention to Calabi-Yau five-folds whose $(2,2)$-forms are completely generated by the product of two $(1,1)$-forms. All the concrete examples of Calabi-Yau five-folds considered in this chapter are of this type. In this case, the $(2,2)$-piece of a real 4-form can be split off from the rest in a complex structure independent way and the fate of the $(2,2)$-part of the gravitino ansatz is as described in the previous paragraph. As a shorthand, we will refer to a 4-form that only comprises a $(1,3)$ and a $(3,1)$ piece as a $\hat{4}$-form and given the restriction on $h^{2,2}(X)$, this restriction is also purely topological. The $\hat{4}$-forms are thus well-suited to describe the $(1,3)+(3,1)$-sector of the reduction in a way independent of the complex structure moduli. To this end, it is convenient to choose a particular basis of real 4-forms, $\{ O_{\cal{X}} \}_{\mathcal{X}=1,\ldots,b^4(X)}$, such that the first $2\, h^{1,3}(X)$ 4-forms, denoted $\{ O_{\cal{\hat{X}}} \}_{\mathcal{\hat{X}}=1,\ldots,2h^{1,3}(X)} $, only contain $(1,3)$ and $(3,1)$-pieces and the remaining $h^{2,2}(X)$ 4-forms, denoted $\{ O_{\cal{\tilde{X}}} \}_{\mathcal{\tilde{X}}=1,\ldots,h^{2,2}(X)}$, only contain $(2,2)$-parts. This basis choice is complex structure independent for the class of manifolds under consideration. The $\hat{4}$-forms then lie in the sub-vector space spanned by $\{ O_{\cal{\hat{X}}} \}$. For a general Calabi-Yau five-fold, a more complicated intertwining of the K\"{a}hler and complex structure moduli with the $(1,3)$-fields arises leading to additional interaction terms in the one-dimensional effective action. It should be appreciated that this is a relatively mild restriction as it only affects the $(1,3)$-sector's couplings to fields of other sectors. Our analysis of all other sectors by themselves does not rest on this restriction.

After some relabelling, adopting the notation in \secref{cy5_moduli_spaces} for the harmonic forms and introducing numerical factors for later convenience, the gravitino ansatz now reads\newnot{symbol:fermion_11}\newnot{symbol:fermion_14}\newnot{symbol:fermion_3form}\newnot{symbol:fermion_hat4form}
\begin{align}
	\Psi_0 &= \psi_0(\tau) \otimes \eta^\star + \bar{\psi}_0 (\tau) \otimes \eta \eqlabel{ansatz_gravitino_0} , \\
	\Psi_{\bar{\mu}} &= 
			\psi^i(\tau) \otimes (\omega_{i,\alpha_1\bar{\mu}} \gamma^{\alpha_1} \eta) + \frac{i}{4}
			\Lambda^{\cal{P}} (\tau) \otimes (N_{\mathcal{P},\alpha_1\alpha_2\bar{\mu}} \gamma^{\alpha_1\alpha_2} \eta)\nonumber\\
				& + \oneon{4}
			\bar{\Upsilon}^{\hat{\cal{X}}}(\tau) \otimes (O_{{\hat{\mathcal{X}}},\alpha_1\ldots\alpha_3\bar{\mu}} 
				\gamma^{\alpha_1\ldots\alpha_3} \eta) - \oneon{4!}
			\bar{\kappa}^{\bar{a}}(\tau) \otimes (||\Omega ||^{-1} \bar{\chi}_{\bar{a},\alpha_1\ldots\alpha_4\bar{\mu}} 
				\gamma^{\alpha_1\ldots\alpha_4} \eta), 
       \eqlabel{ansatz_gravitino_internal} \\
	\Psi_{\mu} &= (\Psi_{\bar{\mu}})^\ast ,\eqlabel{ansatz_gravitino_1}
\end{align}
The four terms in~\eqref*{ansatz_gravitino_internal} correspond to the $(1,1)$, $(2,1)$, $(3,1)$ and $(4,1)$ sectors, respectively. The harmonic $(1,1)$-forms are denoted by $\omega_i$, where $i,j,\ldots =1,\dots ,h^{1,1}(X)$, the real 3-forms are denoted by $N_{\cal{P}}$, where $\mathcal{P,Q},\ldots =1,\dots ,b^{3}(X)$, the real $\hat{4}$-forms by $O_{\hat{\cal{X}}}$, where $\hat{\cal{X}},\hat{\cal{Y}},\ldots = 1,\dots , 2h^{1,3}(X)$ and the $(1,4)$-forms by $\chi_a$, where $a,b,\ldots =1,\dots ,h^{1,4}(X)$. In the same order, the associated zero modes, which are complex one-dimensional fermions, are denoted by $\psi^i$, $\Lambda^{\cal{P}}$, $\Upsilon^{\hat{\cal{X}}}$ and $\kappa^a$.

It is clear that the number of zero modes cannot be reduced any further and that these four types of modes are independent. Three of them, the $(1,1)$, 3-form and $(1,4)$ modes pair up with corresponding bosonic zero modes in the same sectors. The $\hat{4}$-form modes, however, have no bosonic zero mode partners and one of our tasks will be to understand how they can be incorporated into a supersymmetric one-dimensional effective theory.

Had we written the second term in~\eqref*{ansatz_gravitino_internal} in $(2,1)$-language $\Psi_{\bar{\mu}} = \ldots - 1/4 \lambda^p (\tau) \otimes (\nu_{p,\alpha_1\alpha_2\bar{\mu}} \gamma^{\alpha_1\alpha_2} \eta) + \ldots$, we would have identified a set of $h^{2,1}(X)$ complex one-di\-men\-sio\-nal fermions in this sector. From~\eqref*{ansatz_gravitino_internal} however, there appear to be $b^3 (X) = 2 h^{2,1}(X)$ complex one-dimensional fermions. This apparent factor of two discrepancy in the number of degrees of freedom is resolved by observing that a successive insertion of \eqrangeref{21_3_rel1}{21_3_cplx_coords} into the second term in~\eqref*{ansatz_gravitino_internal} leads to a constraint in the form of a projection condition on the 3-form fermions $\Lambda^{\cal{P}}$
\be\eqlabel{3ferm_constraint}
	P_+ {}_{\cal{P}} {}^{\cal{Q}} \Lambda^{\cal{P}} = \Lambda^{\cal{Q}} \; , \qquad\qquad (\text{and:} \;\;
	P_- {}_{\cal{P}} {}^{\cal{Q}} \bar{\Lambda}^{\cal{P}} = \bar{\Lambda}^{\cal{Q}} ) \; ,
\ee
where the projectors $P_\pm {}_{\cal{P}} {}^{\cal{Q}}$ were defined in~\eqref*{3form_projectors_def}. This condition, which is equivalent to $P_- {}_{\cal{P}} {}^{\cal{Q}} \Lambda^{\cal{P}}=0$, precisely halves the number of degrees of freedom so as to match the counting in $(2,1)$-language. In other words, there are $1/2 \, b^3 (X) = h^{2,1}(X)$ complex one-dimensional fermions in this sector. It can be shown that this constraint also applies to the time derivative and supersymmetry transformation of $\Lambda^{\cal{P}}$
\be\eqlabel{3ferm_constraint_dot_susy}
	P_+ {}_{\cal{P}} {}^{\cal{Q}} \dot{\Lambda}^{\cal{P}} = \dot{\Lambda}^{\cal{Q}} \; , \qquad\qquad\qquad
	P_+ {}_{\cal{P}} {}^{\cal{Q}} (\delta_\epsilon \Lambda^{\cal{P}}) = \delta_\epsilon \Lambda^{\cal{Q}} \; ,
\ee
implying in particular that the projection operators commute with both supersymmetry and time translation when acting on $\Lambda^{\cal{P}}$
\be\eqlabel{3ferm_projector_susy_commutator}
	\left[ P_\pm {}_{\cal{P}} {}^{\cal{Q}}, \partial_0 \right] \Lambda^{\cal{P}} = 0 \; , \qquad\qquad
	\left[ P_\pm {}_{\cal{P}} {}^{\cal{Q}}, \delta_\epsilon \right] \Lambda^{\cal{P}} = 0 \; .
\ee
The projection condition is thus preserved under both operations as is required by consistency. \Eqrangeref{3ferm_constraint}{3ferm_projector_susy_commutator} will play important r\^{o}les in finding the correct superspace formulation for this sector later in \secref{N=2}.

By complete analogy, we learn that the $\hat{4}$-form sector really only contains $h^{1,3}(X)$ complex one-dimensional fermions and not $2\, h^{1,3}(X)$ as is suggested by the third term in~\eqref*{ansatz_gravitino_internal}. By using \eqref{31_4_new_rel1,31_4_new_rel3} and the third term in~\eqref*{ansatz_gravitino_internal}, we infer
\be\eqlabel{4ferm_constraint}
	P_+ {}_{\hat{\cal{Y}}} {}^{\hat{\cal{X}}} \Upsilon^{\hat{\cal{Y}}} = \Upsilon ^{\hat{\cal{X}}} \; , \qquad\qquad (\text{and:} \;\;
	P_- {}_{\hat{\cal{Y}}} {}^{\hat{\cal{X}}} \bar{\Upsilon}^{\hat{\cal{Y}}} = \bar{\Upsilon}^{\hat{\cal{X}}} ) \; ,
\ee
thereby halving the number of degrees of freedom. The projection operators $P_\pm {}_{\hat{\cal{Y}}} {}^{\hat{\cal{X}}}$ were defined in~\eqref*{4form_projectors_def}. \Eqref{4ferm_constraint} implies
\begin{align}\eqlabel{4ferm_constraint_dot_susy}
	&P_+ {}_{\hat{\cal{Y}}} {}^{\hat{\cal{X}}} \dot{\Upsilon}^{\hat{\cal{Y}}} = \dot{\Upsilon}^{\hat{\cal{X}}} \; ,
	&&\left[ P_\pm {}_{\hat{\cal{Y}}} {}^{\hat{\cal{X}}}, \partial_0 \right] \Upsilon^{\hat{\cal{Y}}} = 0 \; , \\
	&P_+ {}_{\hat{\cal{Y}}} {}^{\hat{\cal{X}}} (\delta_\epsilon \Upsilon^{\hat{\cal{Y}}}) = \delta_\epsilon \Upsilon^{\hat{\cal{X}}} \; , 
	&&\left[ P_\pm {}_{\hat{\cal{Y}}} {}^{\hat{\cal{X}}}, \delta_\epsilon \right] \Upsilon^{\hat{\cal{Y}}} = 0 
		\eqlabel{4ferm_projector_susy_commutator}
\end{align}
guaranteeing the preservation of the projection condition under time translation and supersymmetry. The compatibility conditions~\eqrangeref*{4ferm_constraint_dot_susy}{4ferm_projector_susy_commutator} are, of course, required for consistency.

\begin{table}[t]\begin{center}
\begin{tabular}{|c|c|c|}
\hline
\emph{cohomology}&\emph{bosonic zero modes}&\emph{fermionic zero modes}\\\hline\hline
$H^{1,1}(X)$&$h^{1,1}(X)$ real, K\"ahler moduli&$\begin{array}{l}\text{$h^{1,1}(X)$ complex,}\\\text{from gravitino}\end{array}$\\\hline
$H^{1,2}(X)$&$h^{1,2}(X)$ complex, from 3-form&$\begin{array}{l}\text{$h^{1,2}(X)$ complex,}\\\text{from gravitino}\end{array}$\\\hline
$H^{1,3}(X)$&$-$&$\begin{array}{l}\text{$h^{1,3}(X)$ complex,}\\\text{from gravitino}\end{array}$\\\hline
$H^{2,2}(X)$&$-$&$-$\\\hline
$H^{1,4}(X)$&$h^{1,4}(X)$ complex structure moduli&$\begin{array}{l}\text{$h^{1,4}(X)$ complex,}\\\text{from gravitino}\end{array}$\\\hline
$H^{2,3}(X)$&$-$&$-$\\\hline
\end{tabular}
\caption{Summary of zero modes arising from an M-theory reduction on a Calabi-Yau five-fold $X$.}
\tablabel{zero_modes}
\end{center}\end{table}
To summarise, we list in \tabref{zero_modes} the various zero modes arising from an M-theory reduction. The bosonic zero modes from \secref{bos_red_ansatz} are also included in order to provide a precursor of how these fields pair up into supermultiplets as will be presented in \secref{N=2}.

\subsection{Performing the fermionic reduction}\seclabel{ferm_red_perform}

Before reducing the fermionic terms, we need explicit expressions for the vielbein, its time derivative and the spin connection. In particular, it should be kept in mind that the gravitino ansatz~\eqref*{ansatz_gravitino_0}--\eqref*{ansatz_gravitino_1} implicitly depends on the vielbein since the curved index gamma matrices $\gamma_\mu$ that appear have to be replaced by flat index gamma matrices $\gamma_{\underline{\mu}}$ via $\gamma_\mu=e_\mu {}^{\underline{\nu}}\gamma_{\underline{\nu}}$. 

We begin with the vielbein. From the metric ansatz~\eqref*{gansatz} with the $10$-dimensional metric taken to be purely $(1,1)$ its non-zero components are $e_0 {}^\UZ = -N/2$, $e_\mu {}^{\underline\nu} $ and $e_{\bar\mu} {}^{\underline{\bar\nu}}$, so that $g_{\mu\bar{\nu}}=e_\mu {}^{\underline{\rho}} e_{\bar{\nu}} {}^{\underline{\bar{\sigma}}}\eta_{\underline{\rho}\underline{\bar{\sigma}}}$ is the Ricci-flat metric on the Calabi-Yau five-fold. Of course, the $10$-dimensional part of the vielbein depends on the Calabi-Yau K\"ahler moduli $t^i(\tau)$ and the complex structure moduli $z^a(\tau )$ and, hence, its time-derivative is non-zero. From the time derivative~\eqref*{metric_mode_expansion} for the metric one finds
\begin{align}
	\dot{e}_\mu {}^{\underline\nu} &= - \frac{i}{2} \omega_{i,\mu} {}^\rho e_{\rho} {}^{\underline\nu} \dot{t}^i \; , \\
	\dot{e}_\mu {}^{\underline{\bar\nu}} &= - \oneon{12||\Omega||^2} \Omega_{\mu} {}^{\bar{\mu}_1\ldots\bar{\mu}_4} 
	  \chi_{a,\bar{\mu}_1\ldots\bar{\mu}_4\rho} e^{\rho\underline{\bar\nu}} \dot{z}^a \; ,
\end{align}
and similarly for the complex conjugates. From the equations above together with~\eqref*{bos_red_connection_comp_lapse_fnc} and the covariant constancy of the vielbein
\be
	\nabla_M e_N {}^\ul{P} = \partial_M e_N {}^\ul{P} - \Gamma^Q_{MN} e_Q {}^\ul{P} + \omega_M {}^\ul{P} {}_\ul{Q} e_N {}^\ul{Q} = 0 \; ,
\ee
we find expressions for the 11-dimensional spin-connection $\omega_N {}^{\underline{Q}\underline{R}}$. Its only non-zero components are given by
\begin{align}
	\omega_\mu {}^{\underline{\nu} \UZ} &= -i N^{-1} \omega_{i,\mu} {}^\rho e_\rho {}^{\underline\nu} \dot{t}^i \; , \eqlabel{spinconn1} \\
	\omega_\mu {}^{\underline{\bar\nu} \UZ} &= -\oneon{6||\Omega||^2} N^{-1} \Omega_{\mu} {}^{\bar{\mu}_1\ldots\bar{\mu}_4} 
	  \chi_{a,\bar{\mu}_1\ldots\bar{\mu}_4\rho} e^{\rho\underline{\bar\nu}} \dot{z}^a \; , \eqlabel{spinconn2}
\end{align}
plus their complex conjugates and the components $\omega_m{}^{\underline{n}\underline{p}}$ of the Calabi-Yau spin connection, computed from the $10$-dimensional vielbein $e_m {}^{\underline{n}}$.  The complex conjugates of the components listed above are, of course, also present. The components of the 11-dimensional covariant derivative $D_M$ defined in~\eqref*{covder} then become
\begin{align}
	D_0 &= \partial_0 \; , \eqlabel{covder1} \\
	D_\mu &= \tilde{D}_\mu  + \frac{i}{2} N^{-1} \omega_{i,\mu\bar\nu} \dot{t}^i \gamma^{\bar\nu} \Gamma^\UZ
		+ \oneon{12||\Omega||^2} N^{-1} \Omega_{\mu} {}^{\bar{\mu}_1\ldots\bar{\mu}_4} 
	 		\chi_{a,\bar{\mu}_1\ldots\bar{\mu}_4\nu} \dot{z}^a \gamma^\nu \Gamma^\UZ \; , \eqlabel{covder2} \\
	D_{\bar\mu} &= (D_\mu)^\ast \; , \eqlabel{covder3}
\end{align}
where $\tilde{D}_\mu$ is the covariant derivative on the Calabi-Yau five-fold.

Inserting the gravitino ansatz \eqrangeref*{ansatz_gravitino_0}{ansatz_gravitino_1} into the fermionic action~\eqref*{S_CJS_F1} produces a vast number of terms -- even when restricting to terms quadratic in fermions. Each of these terms contains a product of a certain number of gamma matrices sandwiched between two spinors $\eta$ or $\eta^\star$. Luckily, on a Calabi-Yau five-fold there only exist a very limited number of non-vanishing such spinor bilinears, namely $\eta^\dagger\eta$, $J_{\mu\bar{\nu}}$, $\Omega_{\mu_1\dots\mu_5}$ and their complex conjugates (see \secref{cy5_geom} for details). As a result, many terms in the reduction vanish immediately, due to their gamma matrix structure. The remaining terms can be split into two types. The first type leads to one-dimensional fermion kinetic terms and such terms originate from the 11-dimensional Rarita-Schwinger term in the action~\eqref*{S_CJS_F1}. The second type leads to one-dimensional Pauli terms, that is couplings between two fermions and the time derivative of a boson, which descend from all the remaining terms in the action~\eqref*{S_CJS_F1}, quadratic in fermions. 

After inserting the gravitino ansatz and integrating over the Calabi-Yau manifold, the Rarita-Schwinger term gives rise to the following fermion kinetic terms\footnote{For an illustrative example of how the fermionic reduction is performed explicitly, the reader is referred to \appref{ferm_red_11_example}. There, we provide a step-by-step explanation of how the kinetic term for the $(1,1)$-fermions (that is, the first term in~\eqref*{s11f}) is calculated from 11 dimensions.}
\begin{multline}
	S_{\rm F,kin} = - \frac{l}{2} \int d\tau \frac{i}{2} \left\{ G^{(1,1)}_{ij} (\underline{t}) ( \psi^i \dot{\bar{\psi}}^j - \dot{\psi}^i \bar\psi^j ) 
			+ G_{\mathcal{PQ}}^{(3)}(\underline{t},\underline{z},\underline{\bar{z}}) 
			   (\Lambda^{\cal{P}} \dot{\bar{\Lambda}}^{\cal{Q}} - \dot{\Lambda}^{\cal{P}} \bar{\Lambda}^{\cal{Q}})
		\right. \\ \left. 
			+  3 G^{(\hat{4})}_{\hat{\cal{X}}\hat{\cal{Y}}}(\underline{t}) 
			    (\Upsilon^{\hat{\cal{X}}} \dot{\bar{\Upsilon}}^{\hat{\cal{Y}}} - \dot{\Upsilon}^{\hat{\cal{X}}} \bar{\Upsilon}^{\hat{\cal{Y}}})
			+ 4 V(t) G^{(1,4)}_{a\bar{b}}(\underline{z},\underline{\bar{z}}) (\kappa^a \dot{\bar{\kappa}}^{\bar{b}} - \dot{\kappa}^a \bar{\kappa}^{\bar{b}})
		\right\} .
\eqlabel{s11f}\end{multline}
Here, $G_{ij}^{(1,1)}$, $G_{\mathcal{PQ}}^{(3)}$ and $G_{a\bar{b}}^{(1,4)}$ are the moduli space metrics for the $(1,1)$, 3-form and $(1,4)$ bosons exactly as defined in the previous section (see \eqrangeref{G11def}{G3def}). Since there are no $\hat{4}$-form bosons, we have not yet encountered the metric $G^{(\hat{4})}_{\hat{\cal{X}}\hat{\cal{Y}}}$. It is given by
\be
	G^{(\hat{4})}_{\hat{\cal{X}}\hat{\cal{Y}}}(\underline{t}) = \int_X O_{\hat{\cal{X}}} \wedge\ast O_{\hat{\cal{Y}}} 
		= - d_{\hat{\mathcal{X}}\hat{\mathcal{Y}}i} t^i\; ,\qquad
	d_{\hat{\mathcal{X}}\hat{\mathcal{Y}}i} = \int_X O_{\hat{\cal{X}}} \wedge O_{\hat{\cal{Y}}} \wedge \omega_i
\ee
in terms of the intersection numbers $d_{\hat{\mathcal{X}}\hat{\mathcal{Y}}i}=d_{\hat{\mathcal{Y}}\hat{\mathcal{X}}i}$, which are purely topological for the class of five-folds we are considering. To evaluate $\ast O_{\hat{\cal{Y}}}$ in the above integral we have used the result for the Hodge dual of $\hat{4}$-forms from \eqref{4form_hodgestar}.

Reducing the other fermion bilinear terms in the 11-dimensional action~\eqref*{S_CJS_F1}, we find for the Pauli terms
\bea\eqlabel{action1_f_pauli}
	S_{\rm F,Pauli} = \frac{l}{2} \int d\tau \left\{
		  \frac{i}{2} N^{-1} G^{(1,1)}_{ij}(\underline{t}) (\psi^i\psi_0 + \bar{\psi}^i\bar\psi_0) \dot{t}^j
		+ \frac{i}{2} G^{(1,1)}_{ij,k}(\underline{t}) (\psi^k\bar{\psi}^i + \bar{\psi}^k\psi^i) \dot{t}^j   \right. \\ \left. 
			+ i N^{-1} G^{(3)}_{\mathcal{PQ}}(\underline{t},\underline{z},\bar{\underline{z}}) 
				(\Lambda^{\cal{P}}\psi_0 + \bar{\Lambda}^{\cal{P}}\bar\psi_0) \dot{X}^{\cal{Q}}
			+ i G^{(3)}_{\mathcal{PQ},i}(\underline{t},\underline{z},\bar{\underline{z}}) 
				(\psi^i\bar{\Lambda}^{\cal{P}} + \bar{\psi}^i\Lambda^{\cal{P}}) \dot{X}^{\cal{Q}} \right. \\ \left. 
			- \frac{i}{2} G^{(3)}_{\mathcal{PQ},a}(\underline{t},\underline{z},\bar{\underline{z}}) 
				\Lambda^{\cal{P}} \bar\Lambda^{\cal{Q}} \dot{z}^a
			+ G^{(3)}_{\mathcal{PQ},a}(\underline{t},\underline{z},\bar{\underline{z}}) 
				\kappa^a \bar\Lambda^{\cal{P}} \dot{X}^{\cal{Q}}	\right. \\ \left. 
			+ \frac{i}{2} G^{(3)}_{\mathcal{PQ},\bar{a}}(\underline{t},\underline{z},\bar{\underline{z}}) 
				\Lambda^{\cal{P}} \bar\Lambda^{\cal{Q}} \dot{\bar{z}}^{\bar{a}} 
			- G^{(3)}_{\mathcal{PQ},\bar{a}}(\underline{t},\underline{z},\bar{\underline{z}}) 
				\bar\kappa^{\bar{a}} \Lambda^{\cal{P}} \dot{X}^{\cal{Q}}  \right. \\ \left.  
		+ 2i V G^{(1,4)}_{a\bar{b}, c}(\underline{z}, \underline{\bar{z}}) \kappa^a \bar{\kappa}^{\bar{b}} \dot{z}^c 
		- 2i V G^{(1,4)}_{a\bar{b}, \bar{c}}(\underline{z}, \underline{\bar{z}}) \kappa^a \bar{\kappa}^{\bar{b}} \dot{\bar{z}}^{\bar{c}}     \right. \\ \left. 
		- 4 N^{-1} V G^{(1,4)}_{a\bar{b}}(\underline{z}, \underline{\bar{z}}) 
			(\psi_0 \kappa^a \dot{\bar{z}}^{\bar{b}} - \bar\psi_0 \bar{\kappa}^{\bar{b}} \dot{z}^a)  \right. \\ \left. 
		- \oneon{3!} K_i G^{(1,4)}_{a\bar{b}}(\underline{z}, \underline{\bar{z}}) 
			( \psi^i \bar{\kappa}^{\bar{b}} \dot{z}^a - \bar{\psi}^i \kappa^a \dot{\bar{z}}^{\bar{b}} )
	\right\} .
\eea
This completes the dimensional reduction of the fermionic part of the 11-dimensional action at the level of terms quadratic in fermions. Our complete result for the one-dimensional effective action in components, four-fermi terms not included, is given by the sum of the bosonic action \eqref*{S1B} and the two fermionic parts \eqref*{s11f} and \eqref*{action1_f_pauli}.

Next, we need to verify that this action is indeed invariant under one-dimensional $\susyno=2$ local supersymmetry, as it should be. In the following section, we will do this by writing down a superspace action whose associated component action coincides with our reduction result. This superspace action then also determines the four-fermion terms, which we have not explicitly computed by dimensional reduction.

\section{Migrating to supertime}\seclabel{N=2}

In this section, we will make extensive use of the results of \chref{superspace} to cast the one-dimensional effective action calculated in the two previous sections into curved $\susyno=2$ supertime.

\subsection{\texorpdfstring{$\susyno=2$}{N=2} supersymmetry transformations and multiplets}

First, however, we need to identify how the zero modes of M-theory on Calabi-Yau five-folds fall into super-multiplets of one-dimensional $\susyno=2$ supersymmetry. It is a plausible assumption that bosonic and fermionic zero modes that arise from the same sector of harmonic $(p,q)$-forms on the five-fold pair up into supermultiplets. For example, the $h^{1,1}(X)$ K\"ahler moduli $t^i$ should combine with the same number of $(1,1)$ fermions $\psi^i$. Since the K\"ahler moduli $t^i$ are real scalars the resulting $h^{1,1}(X)$ supermultiplets must be of type $2a$. In the $(1,4)$ sector, on the other hand, we have $h^{1,4}(X)$ complex scalars $z^a$ (the complex structure moduli) and the same number of complex fermions $\kappa^a$ so one expects $h^{1,4}(X)$ supermultiplets of type $2b$. The 3-form sector is somewhat more peculiar. There are $b^3(X)$ real scalars $X^{\cal{P}}$ and the same number of complex fermions $\Lambda^{\cal{P}}$ fitting nicely into $b^3(X)$ $2a$ multiplets. However, we also need to take into account the constraint \eqref*{3ferm_constraint} on the fermions, which halves their number. The result is a set of constrained $2a$ multiplets with the same number of degrees of freedom as $1/2 \, b^3(X)$ $2b$ multiplets, reminding us of their original nature. This leaves us with the $\hat{4}$-form fermions $\Upsilon^{\hat{\cal{X}}}$. They have no bosonic zero mode partners so cannot be part of either the standard $2a$ or $2b$ multiplets. The natural guess is for them to form $2\, h^{1,3}(X)$ fermionic $2b$ multiplets. As for the 3-form fermions, there is the constraint~\eqref*{4ferm_constraint}, which reduces their number by a factor of two. That is, we have $h^{1,3}(X)$ complex one-dimensional fermions in this sector. Finally, the lapse function $N$ and the component $\psi_0$ of the 11-dimensional gravitino should form the one-dimensional gravity multiplet. We now verify this assignment of supermultiplets by a reduction of the 11-dimensional supersymmetry transformations.

Our task is to reduce the 11-dimensional supersymmetry transformations~\eqrangeref*{sugra11_susy_transf_g}{sugra11_susy_transf_gravitino1} for the metric ansatz~\eqref*{gansatz}, the associated spin connection~\eqrangeref*{spinconn1}{spinconn2}, the three-form ansatz~\eqref*{Aansatz} and the gravitino ansatz~\eqref*{ansatz_gravitino_0}--\eqref*{ansatz_gravitino_1}. We denote the spinor parametrising 11-dimensional supersymmetry transformations by $\epsilon^{(11)}$ in order to distinguish it from its one-dimensional counterpart $\epsilon$. The 11-dimensional spinor can then be decomposed as (cf. \eqref{KSeq_susyparam})
\be\eqlabel{susyparam}
	\epsilon^{(11)} = \frac{i}{2} \epsilon \otimes \eta^\star - \frac{i}{2} \bar\epsilon \otimes \eta\; ,
\ee
where $\eta$ is the covariantly constant spinor on the Calabi-Yau five-fold. Inserting all this into the 11-dimensional supersymmetry transformations and collecting terms proportional to the same harmonic Calabi-Yau forms we find the supersymmetry transformations of the various zero modes. For the lapse function $N$ and the time component $\psi_0$ of the gravitino they are
\bea
	\delta_\epsilon N &= - \epsilon\bar\psi_0, &\qquad \delta_\epsilon\psi_0 &= i \dot{\epsilon}, &\qquad \delta_\epsilon \bar\psi_0 &= 0 \\
	\delta_{\bar\epsilon} N &= \bar\epsilon\psi_0, &\qquad \delta_{\bar\epsilon} \psi_0 &= 0, &\qquad \delta_{\bar\epsilon} \bar\psi_0 &= - i\dot{\bar{\epsilon}} \; .
\eqlabel{susytrans_1d_sugra}
\eea
These transformations are identical to the one for a one-dimensional $\susyno=2$ supergravity multiplet as can be seen by comparing with \secref{sspace_curved}. 

For the other zero modes, we find the supersymmetry transformations
\begin{align}
	(1,1): \;\;
	\delta_\epsilon t^i &= - \epsilon \psi^i , 
	&\delta_\epsilon \psi^i &= 0, & \delta_\epsilon \bar{\psi}^i &= \frac{i}{2} N^{-1} \epsilon \dot{t}^i + \ldots 
	\eqlabel{susytrans_1d_11} , \\
	\text{3-form}: \;\;
	\delta_\epsilon X^{\cal{P}} &= - \epsilon \Lambda^{\cal{P}} , 
	&\delta_\epsilon \Lambda^{\cal{P}} &= 0, & \delta_\epsilon \bar{\Lambda}^{\cal{P}} &= i N^{-1} \epsilon P_- {}_{\cal{Q}} {}^{\cal{P}} 
		\dot{X}^{\cal{Q}} + \ldots 
	\eqlabel{susytrans_1d_3} , \\
	\text{$\hat{4}$-form}: \;\;
	\delta_\epsilon \Upsilon^{\hat{\cal{X}}} &= 0 + \ldots,  & \delta_\epsilon \bar\Upsilon^{\hat{\cal{X}}} &= 0, && \eqlabel{susytrans_1d_4} \\
	(1,4): \;\;
	\delta_\epsilon z^a &= i \epsilon \kappa^a, & \delta_\epsilon \bar{z}^{\bar{a}} &= 0, 
	&\delta_\epsilon \kappa^a &= 0, \qquad \delta_\epsilon \bar{\kappa}^{\bar{a}} = N^{-1} \epsilon \dot{\bar{z}}^{\bar{a}} + \ldots
	\eqlabel{susytrans_1d_41},
\end{align}
and similarly for the $\bar\epsilon$-variation. The dots indicate terms cubic in fermions which we have omitted.\footnote{It may be a bit surprising that the transformations above do not seem to mix fields of different types (that is $(1,1)$, $(1,4)$, etc.) despite the plethora of cross-sector interaction terms in the action. However, this is merely an artefact due to the omission of $(\text{fermi})^3$ terms. That is, the sector-mixing terms in the transformations are all of order $(\text{fermi})^3$, which can be seen by taking the full, off-shell supersymmetry transformations of \chref{superspace} and eliminating the auxiliary fields.} To arrive at the last equation in~\eqref*{susytrans_1d_3}, we have performed a compensating transformation, making use of a local fermionic symmetry. Namely, the action~\eqref*{s11f},~\eqref*{action1_f_pauli} is invariant under
\vspace{\parskip}
\be\eqlabel{3form_symm}
	\delta \Lambda^{\cal{P}} = P_- {}_{\cal{Q}} {}^{\cal{P}} l^{\cal{Q}} \; , \qquad\qquad\qquad (\text{and: } 
	\delta \bar\Lambda^{\cal{P}} = P_+ {}_{\cal{Q}} {}^{\cal{P}} \bar{l}^{\cal{Q}} ) \; ,
\ee
\vspace{\parskip}\noindent 
for a set of local complex fermionic parameters $l^{\cal{Q}}$, while all other fields do not transform. The constraint~\eqref*{3ferm_constraint} on $\Lambda^{\cal{P}}$ may be viewed as a gauge choice with respect to this symmetry. The form of the last equation in~\eqref*{susytrans_1d_3} then guarantees the preservation of this gauge choice under a supersymmetry transformation as required by \eqref{3ferm_constraint_dot_susy}. Even though the $\hat{4}$-form fermions $\Upsilon^{\hat{\cal{X}}}$ are subject to the same kind of constraint (cf. \eqref{4ferm_constraint}), there is no associated local symmetry. This is because the proof that~\eqref*{3form_symm} is a symmetry crucially hinges on the Hermiticity of the 3-form metric (cf. \eqref{G3_hermitian}), but the $\hat{4}$-form metric is not Hermitian.

Again, comparing with the results for the supersymmetry transformations of the various one-dimensional $\susyno=2$ multiplets given in \chref{superspace}, we confirm the assignment of zero modes into supermultiplets discussed above. In particular, the transformation of the $\hat{4}$-form fermions $\Upsilon^{\hat{\cal{X}}}$ indicates that they should indeed be part of fermionic $2b$ supermultiplets. 

To summarise these results, we write down the explicit off-shell component expansion for all superfields in terms of the Calabi-Yau five-fold zero modes and appropriate auxiliary fields. Taking into account the component structure of the various supermultiplets derived in \chref{superspace}, we have
\vspace{\parskip}
\begin{align}
	\text{SUGRA} && (2a): \quad		
		&\mathcal{E} = - N - \frac{i}{2} \theta\bar\psi_0 - \frac{i}{2} \bar\theta\psi_0 , \eqlabel{superfield_E} \\
	(1,1) && (2a): \quad
		&T^i = t^i + i \theta \psi^i + i \bar\theta \bar\psi^i + \oneon{2} \theta\bar\theta f^i , \eqlabel{superfield_T} \\
	\text{3-form} && (2a): \quad
		&\mathcal{X}^{\cal{P}} = X^{\cal{P}} + i \theta \Lambda^{\cal{P}} + i \bar\theta \bar{\Lambda}^{\cal{P}} 
			+ \oneon{2} \theta\bar\theta g^{\cal{P}} , \eqlabel{superfield_X} \\
	\text{$\hat{4}$-form} && (2b)-{\rm fermionic}: \quad
		&\mathcal{R}^{\hat{\cal{X}}} = \Upsilon^{\hat{\cal{X}}} + \theta H^{\hat{\cal{X}}} 
			+ \frac{i}{2} N^{-1} \theta\bar\theta (\dot{\Upsilon}^{\hat{\cal{X}}} - \psi_0 H^{\hat{\cal{X}}}) , \eqlabel{superfield_R} \\
	(1,4) && (2b): \quad
		&Z^a = z^a + \theta \kappa^a + \frac{i}{2} N^{-1} \theta\bar\theta (\dot{z}^a - \psi_0 \kappa^a ) , \eqlabel{superfield_Z}
\end{align}
\vspace{\parskip}\noindent 
where $f^i$, $g^{\cal{P}}$ and $H^{\hat{\cal{X}}}$ are bosonic auxiliary fields. These auxiliary fields can, of course, not be obtained from the reduction (since 11-dimensional supersymmetry is realised on-shell) and have to be included by hand. Full, off-shell supersymmetry transformations for all the above components are given in \chref{superspace}.

\subsection{Supertime version of the one-dimensional effective action}

Having identified the relevant supermultiplets and their components, our next step is to write down an $\susyno=2$ superspace version of the one-dimensional effective action. For the most part, an appropriate form for the superspace action can be guessed based on the bosonic action~\eqref*{S1B}. Basically, all one has to do is to promote the bosonic fields in this action to their associated superfields, replace time derivatives by super-covariant derivatives ${\cal D}$ or $\bar{\cal D}$ and integrate over superspace.

In addition, we need to implement the constraint~\eqref*{3ferm_constraint} on the 3-form fermions $\Lambda^{\cal{P}}$ at the superspace level. The superpartner of the constraint \eqref*{3ferm_constraint} turns out to be
\be\eqlabel{3ferm_constraint_spartner}
	g^{\cal{P}} = N^{-1} \cs_{\cal{Q}} {}^{\cal{P}} \dot{X}^{\cal{Q}} + N^{-1} (\psi_0 \Lambda^{\cal{P}} - \bar\psi_0 \bar{\Lambda}^{\cal{P}}) \; .
\ee
Note that since the only object in this equation depending on the complex structure moduli is $\cs_{\cal{Q}} {}^{\cal{P}}$, it follows that $\cs_{\cal{Q}} {}^{\cal{P}} {}_{,a} \dot{X}^{\cal{Q}} = 0$.

Constraints~\eqref*{3ferm_constraint,3ferm_constraint_spartner} form a constraint multiplet and can hence be obtained from a single complex superspace equation
\be\eqlabel{3ferm_constraint_sspace}
	P_- {}_{\cal{P}} {}^{\cal{Q}} (\underline{Z},\bar{\underline{Z}}) \sderiv \mathcal{X}^{\cal{P}} = 0 \; , \qquad\qquad\qquad (\text{and $\cc$}) \; ,
\ee
where $P_- {}_{\cal{P}} {}^{\cal{Q}} (\underline{Z},\bar{\underline{Z}})$ is the superspace version of the projection operator $P_- {}_{\cal{P}} {}^{\cal{Q}}$ defined in~\eqref*{3form_projectors_def}. The superspace constraint \eqref*{3ferm_constraint_sspace} follows from a superspace action by introducing a set of $b^3 (X)$ complex fermionic Lagrange multiplier superfields $L_{\cal{P}}$
\be\eqlabel{3ferm_lagr_mult_sfield}
	L_{\cal{P}} = L^{(0)}_{\cal{P}} + \theta L^{(1)}_{\cal{P}} + \bar\theta L^{(2)}_{\cal{P}} + \oneon{2} \theta\bar\theta L^{(3)}_{\cal{P}} \; .
\ee
The action for the fermionic Lagrange multiplier superfields is then given by
\be
	-\frac{l}{2}\int d\tau d^2 \theta\, \mathcal{E} 
		\left( L_{\cal{Q}} P_- {}_{\cal{P}} {}^{\cal{Q}} (\underline{Z},\bar{\underline{Z}}) \sderiv \mathcal{X}^{\cal{P}} -  
		\bar{L}_{\cal{Q}} P_+ {}_{\cal{P}} {}^{\cal{Q}} (\underline{Z},\bar{\underline{Z}}) \bar\sderiv \bar{\mathcal{X}}^{\cal{P}}\right) \; .
\ee
This takes care of all but the fermionic multiplets in the $\hat{4}$-form sector whose superfield action has to be inferred from the fermionic component action~\eqref*{s11f}, \eqref*{action1_f_pauli}. In particular, the $\hat{4}$-form part of the superspace action should be such that the bosons $H^{\hat{\cal{X}}}$ in the fermionic multiplets are non-dynamical. As for the 3-form case, we need to implement the constraint~\eqref*{4ferm_constraint} on the $\hat{4}$-form fermions $\Upsilon^{\hat{\cal{X}}}$ at the superspace level. The superpartner of the constraint~\eqref*{4ferm_constraint} is simply
\be\eqlabel{4ferm_constraint_spartner}
	P_+ {}_{\hat{\cal{Y}}} {}^{\hat{\cal{X}}} H^{\hat{\cal{Y}}} = H^{\hat{\cal{X}}} \; , \qquad\qquad (\text{and $\cc$}) \; .
\ee
\Eqref{4ferm_constraint,4ferm_constraint_spartner} are part of a single superspace equation
\be\eqlabel{4ferm_constraint_sspace}
	P_- {}_{\hat{\cal{Y}}} {}^{\hat{\cal{X}}}  (\underline{Z},\bar{\underline{Z}}) \mathcal{R}^{\hat{\cal{Y}}} = 0 \; , \qquad\qquad (\text{and $\cc$}) \; ,
\ee
which can be obtained from a superspace action principle
\be
	-\frac{l}{2}\int d\tau d^2 \theta\, \mathcal{E} 
		\left( L_{\hat{\cal{X}}} P_- {}_{\hat{\cal{Y}}} {}^{\hat{\cal{X}}} (\underline{Z},\bar{\underline{Z}}) \mathcal{R}^{\hat{\cal{Y}}} -  
		\bar{L}_{\hat{\cal{X}}} P_+ {}_{\hat{\cal{Y}}} {}^{\hat{\cal{X}}} (\underline{Z},\bar{\underline{Z}}) \bar{\mathcal{R}}^{\hat{\cal{X}}}\right)
\ee
by means of a set of $2\, h^{1,3}(X)$ complex fermionic Lagrange multiplier superfields $L_{\hat{\cal{X}}}$, which have the same component expansion as in \eqref{3ferm_lagr_mult_sfield}. $P_\pm {}_{\hat{\cal{Y}}} {}^{\hat{\cal{X}}}  (\underline{Z},\bar{\underline{Z}})$ are the superspace versions of the projection operators $P_\pm {}_{\hat{\cal{Y}}} {}^{\hat{\cal{X}}}$ defined in~\eqref*{4form_projectors_def}.

Combining all this, the suggested superspace action is
\begin{multline}\eqlabel{superspace_action_1d}
	S_1  = - \frac{l}{2} \int d\tau\, d^2 \theta\, \mathcal{E} \left\{ 
	 	   G^{(1,1)}_{ij} (\underline{T}) \sderiv T^i \bar\sderiv T^j 
		+ G_{\mathcal{PQ}}^{(3)}(\underline{T},\underline{Z},\underline{\bar{Z}}) \sderiv \mathcal{X}^{\cal{P}} \bar\sderiv \mathcal{X}^{\cal{Q}}
		\right. \\ \left. \qquad
		- 3 G^{(\hat{4})}_{\hat{\cal{X}}\hat{\cal{Y}}} (\underline{T}) \mathcal{R}^{\hat{\cal{X}}} \bar{\mathcal{R}}^{\hat{\cal{Y}}}
		+ 4 V(\underline{T}) G^{(1,4)}_{a\bar{b}} 
			(\underline{Z},\bar{\underline{Z}}) \sderiv Z^a \bar\sderiv \bar{Z}^{\bar{b}} 
		\right. \\ \left.
		+ \left( L_{\cal{Q}} P_- {}_{\cal{P}} {}^{\cal{Q}} (\underline{Z},\bar{\underline{Z}}) \sderiv \mathcal{X}^{\cal{P}} 
		+ L_{\hat{\cal{X}}} P_- {}_{\hat{\cal{Y}}} {}^{\hat{\cal{X}}} (\underline{Z},\bar{\underline{Z}}) \mathcal{R}^{\hat{\cal{Y}}} + \cc \right) \right\} .
\end{multline}
This action can be expanded out in components using the formul\ae\ of \chref{superspace}. The result can be split into $(1,1)$, 3-form, $\hat{4}$-form and $(1,4)$ parts by writing
\be\eqlabel{superspace_action_1d_comp}
	S_1 = \frac{l}{2} \int d\tau\left\{ \mathcal{L}^{(1,1)} + \mathcal{L}^{(3)} + \mathcal{L}^{(\hat{4})} + \mathcal{L}^{(1,4)}\right\}\; .
\ee
For these four parts of the Lagrangian in~\eqref*{superspace_action_1d_comp} we find, after taking into account the constraints~\eqref*{3ferm_constraint,3ferm_constraint_spartner} and using the formul\ae\ provided in \secref{real_form_formalism}
	\bea
	\mathcal{L}^{(1,1)} &= \oneon{4} N^{-1} G^{(1,1)}_{ij}(\underline{t}) \dot{t}^i \dot{t}^j
			- \frac{i}{2} G^{(1,1)}_{ij}(\underline{t}) (\psi^i \dot{\bar\psi}^j - \dot{\psi}^i\bar{\psi}^j)
			+ \oneon{4} N G^{(1,1)}_{ij}(\underline{t}) f^i f^j \\
			&+ \frac{i}{2} N^{-1} G^{(1,1)}_{ij}(\underline{t}) (\psi^i\psi_0 + \bar{\psi}^i\bar\psi_0) \dot{t}^j
			+ \oneon{2} N^{-1} G^{(1,1)}_{ij}(\underline{t}) \psi_0\bar\psi_0 \psi^i \bar{\psi}^j \\
			&- \oneon{2} N G^{(1,1)}_{ij,k}(\underline{t}) (\psi^i\bar{\psi}^j f^k - \psi^k\bar{\psi}^j f^i - \psi^i\bar{\psi}^k f^j)
			+ \frac{i}{2} G^{(1,1)}_{ij,k}(\underline{t}) (\psi^k\bar{\psi}^i + \bar{\psi}^k\psi^i) \dot{t}^j \\
			&- N G^{(1,1)}_{ij,kl}(\underline{t}) \psi^i\bar{\psi}^j\psi^k\bar{\psi}^l , \eqlabel{L11}
	\eea
	\begin{align}
	\mathcal{L}^{(3)} &= \oneon{2} N^{-1} G^{(3)}_{\mathcal{PQ}}(\underline{t},\underline{z},\bar{\underline{z}}) \dot{X}^{\cal{P}} \dot{X}^{\cal{Q}}
			- \frac{i}{2} G^{(3)}_{\mathcal{PQ}}(\underline{t},\underline{z},\bar{\underline{z}}) 
				(\Lambda^{\cal{P}} \dot{\bar\Lambda}^{\cal{Q}} - \dot{\Lambda}^{\cal{P}} \bar{\Lambda}^{\cal{Q}}) \nonumber \\
			&+ i N^{-1} G^{(3)}_{\mathcal{PQ}}(\underline{t},\underline{z},\bar{\underline{z}}) 
				(\Lambda^{\cal{P}}\psi_0 + \bar{\Lambda}^{\cal{P}}\bar\psi_0) \dot{X}^{\cal{Q}}
			+ N^{-1} G^{(3)}_{\mathcal{PQ}}(\underline{t},\underline{z},\bar{\underline{z}}) 
				\psi_0\bar\psi_0 \Lambda^{\cal{P}} \bar{\Lambda}^{\cal{Q}} \nonumber \\
			&- \oneon{2} N G^{(3)}_{\mathcal{PQ},i}(\underline{t},\underline{z},\bar{\underline{z}}) \Lambda^{\cal{P}}\bar{\Lambda}^{\cal{Q}} f^i
			+ i G^{(3)}_{\mathcal{PQ},i}(\underline{t},\underline{z},\bar{\underline{z}}) 
				(\psi^i\bar{\Lambda}^{\cal{P}} + \bar{\psi}^i\Lambda^{\cal{P}}) \dot{X}^{\cal{Q}} \nonumber \\
			&- \oneon{2} G^{(3)}_{\mathcal{PQ},i}(\underline{t},\underline{z},\bar{\underline{z}}) \Lambda^{\cal{P}} \bar{\Lambda}^{\cal{Q}} 
				(\psi_0 \psi^i - \bar\psi_0 \bar\psi^i) 
			- N G^{(3)}_{\mathcal{PQ},ij}(\underline{t},\underline{z},\bar{\underline{z}}) \Lambda^{\cal{P}}\bar{\Lambda}^{\cal{Q}}\psi^i\bar{\psi}^j 
			\nonumber \\
			&- \frac{i}{2} G^{(3)}_{\mathcal{PQ},a}(\underline{t},\underline{z},\bar{\underline{z}}) 
				\Lambda^{\cal{P}} \bar\Lambda^{\cal{Q}} (\dot{z}^a - 2 \psi_0 \kappa^a) 
			+ \frac{i}{2} G^{(3)}_{\mathcal{PQ},\bar{a}}(\underline{t},\underline{z},\bar{\underline{z}}) 
				\Lambda^{\cal{P}} \bar\Lambda^{\cal{Q}} (\dot{\bar{z}}^{\bar{a}} + 2 \bar\psi_0 \bar\kappa^{\bar{a}}) \nonumber \\
			&+ G^{(3)}_{\mathcal{PQ},a}(\underline{t},\underline{z},\bar{\underline{z}}) 
				\kappa^a \bar\Lambda^{\cal{P}} \dot{X}^{\cal{Q}}
			- G^{(3)}_{\mathcal{PQ},\bar{a}}(\underline{t},\underline{z},\bar{\underline{z}}) 
				\bar\kappa^{\bar{a}} \Lambda^{\cal{P}} \dot{X}^{\cal{Q}}  
			- N G^{(3)}_{\mathcal{PQ},a\bar{b}}(\underline{t},\underline{z},\bar{\underline{z}}) 
				\Lambda^{\cal{P}} \bar\Lambda^{\cal{Q}} \kappa^a \bar\kappa^{\bar{b}} \nonumber \\
			&- i N G^{(3)}_{\mathcal{PQ},ia}(\underline{t},\underline{z},\bar{\underline{z}}) 
				\Lambda^{\cal{P}} \bar\Lambda^{\cal{Q}} \bar\psi^i \kappa^a 
			- i N G^{(3)}_{\mathcal{PQ},i\bar{a}}(\underline{t},\underline{z},\bar{\underline{z}}) 
				\Lambda^{\cal{P}} \bar\Lambda^{\cal{Q}} \psi^i \bar\kappa^{\bar{a}}  ,
			\eqlabel{L3}
	\end{align}
	\bea
	\mathcal{L}^{(\hat{4})} &= - \frac{3i}{2} G^{(\hat{4})}_{\hat{\cal{X}}\hat{\cal{Y}}}(\underline{t}) 
			( \Upsilon^{\hat{\cal{X}}} \dot{\bar{\Upsilon}}^{\hat{\cal{Y}}} - \dot{\Upsilon}^{\hat{\cal{X}}} \bar{\Upsilon}^{\hat{\cal{Y}}} ) 
		+ 3 N G^{(\hat{4})}_{\hat{\cal{X}}\hat{\cal{Y}}}(\underline{t}) H^{\hat{\cal{X}}} \bar{H}^{\hat{\cal{Y}}} \\ &
		+ 3 i N G^{(\hat{4})}_{\hat{\cal{X}}\hat{\cal{Y}},i}(\underline{t}) 
			(\psi^i \Upsilon^{\hat{\cal{X}}} \bar{H}^{\hat{\cal{Y}}} + \bar{\psi}^i \bar{\Upsilon}^{\hat{\cal{Y}}} H^{\hat{\cal{X}}} ) 
		+ \frac{3}{2} N G^{(\hat{4})}_{\hat{\cal{X}}\hat{\cal{Y}},i}(\underline{t}) \Upsilon^{\hat{\cal{X}}} \bar{\Upsilon}^{\hat{\cal{Y}}} f^i \\ &
		+ 3 N G^{(\hat{4})}_{\hat{\cal{X}}\hat{\cal{Y}},ij}(\underline{t}) \Upsilon^{\hat{\cal{X}}} \bar{\Upsilon}^{\hat{\cal{Y}}} \psi^i \bar{\psi}^j
		- \frac{3}{2} G^{(\hat{4})}_{\hat{\cal{X}}\hat{\cal{Y}},i}(\underline{t}) 
			\Upsilon^{\hat{\cal{X}}} \bar{\Upsilon}^{\hat{\cal{Y}}} ( \psi_0 \psi^i - \bar{\psi}_0 \bar{\psi}^i ) , 
			\eqlabel{L4}
	\eea
	\bea\eqlabel{L14}
	\mathcal{L}^{(1,4)} &= 4 N^{-1} V G^{(1,4)}_{a\bar{b}}(\underline{z}, \underline{\bar{z}}) \dot{z}^a \dot{\bar{z}}^{\bar{b}}
		- 2i V G^{(1,4)}_{a\bar{b}}(\underline{z}, \underline{\bar{z}}) 
			(\kappa^a \dot{\bar{\kappa}}^{\bar{b}} - \dot{\kappa}^a \bar{\kappa}^{\bar{b}}) \\ &
		- 4 N^{-1} V G^{(1,4)}_{a\bar{b}}(\underline{z}, \underline{\bar{z}}) 
			(\psi_0 \kappa^a \dot{\bar{z}}^{\bar{b}} - \bar\psi_0 \bar{\kappa}^{\bar{b}} \dot{z}^a)
		+ 4 N^{-1} V G^{(1,4)}_{a\bar{b}}(\underline{z}, \underline{\bar{z}}) \psi_0 \bar\psi_0 \kappa^a \bar{\kappa}^{\bar{b}} \\ &
		+ 2i V G^{(1,4)}_{a\bar{b}, c}(\underline{z}, \underline{\bar{z}}) \kappa^a \bar{\kappa}^{\bar{b}} \dot{z}^c 
		- 2i V G^{(1,4)}_{a\bar{b}, \bar{c}}(\underline{z}, \underline{\bar{z}}) \kappa^a \bar{\kappa}^{\bar{b}} \dot{\bar{z}}^{\bar{c}} \\ &
		- \oneon{12} N K_i G^{(1,4)}_{a\bar{b}}(\underline{z}, \underline{\bar{z}}) \kappa^a \bar{\kappa}^{\bar{b}} f^i
		- \frac{2}{3} N K_{ij} G^{(1,4)}_{a\bar{b}}(\underline{z}, \underline{\bar{z}}) \kappa^a \bar{\kappa}^{\bar{b}} \psi^i \bar{\psi}^j \\ &
		- \oneon{3!} K_i G^{(1,4)}_{a\bar{b}}(\underline{z}, \underline{\bar{z}}) 
			\psi^i \bar{\kappa}^{\bar{b}} (\dot{z}^a - \oneon{2} \psi_0 \kappa^a)
		+ \oneon{3!} K_i G^{(1,4)}_{a\bar{b}}(\underline{z}, \underline{\bar{z}}) 
			\bar{\psi}^i \kappa^a (\dot{\bar{z}}^{\bar{b}} + \oneon{2} \bar{\psi}_0 \bar{\kappa}^{\bar{b}})\; .
	\eea
We should now compare this Lagrangian with our result obtained from dimensional reduction in the previous section. To do this, we first have to integrate out the auxiliary fields $f^i$ and $H^{\hat{\cal{X}}}$. A quick inspection of their equations of motion derived from \eqrangeref{L11}{L14} shows that they are given by fermion bilinears. Hence, integrating them out only leads to additional four-fermi terms. Since we have not computed four-fermi terms in our reduction from 11 dimensions they are, in fact, irrelevant for our comparison. All other terms, that is purely bosonic terms and terms bilinear in fermions, coincide with our reduction result~\eqref*{S1B}, \eqref*{s11f} and \eqref*{action1_f_pauli}. This shows that \eqref{superspace_action_1d} is indeed the correct superspace action. 

Both the lapse function $N$ and the gravitino $\psi_0$ are non-dynamical and their equations of motion lead to constraints. For the lapse, this constraint implies the vanishing of the Hamiltonian associated with the Lagrangian~\eqrangeref*{L11}{L14} and it reads (after integrating out the $(1,1)$ and $\hat{4}$-form auxiliary fields $f^i$ and $H^{\hat{\cal{X}}}$)
\begin{multline}\eqlabel{mcy5_hamilt_constr}
	   \oneon{4}  G^{(1,1)}_{ij}(\underline{t}) ( \dot{t}^i + 2i \psi^i\psi_0 + 2i \bar{\psi}^i\bar\psi_0) \dot{t}^j 
	+ \oneon{2} G^{(3)}_{\mathcal{PQ}}(\underline{t},\underline{z},\bar{\underline{z}}) (\dot{X}^{\cal{P}}
			+ 2 i \Lambda^{\cal{P}}\psi_0 + 2 i \bar{\Lambda}^{\cal{P}}\bar\psi_0)) \dot{X}^{\cal{Q}} \\
	+ 4 V G^{(1,4)}_{a\bar{b}}(\underline{z}, \underline{\bar{z}}) ( \dot{z}^a \dot{\bar{z}}^{\bar{b}}
			- \psi_0 \kappa^a \dot{\bar{z}}^{\bar{b}} + \bar\psi_0 \bar{\kappa}^{\bar{b}} \dot{z}^a) + (\text{fermi})^4 = 0\;  .
\end{multline}
The equation of motion for $\psi_0$ generates the superpartner of this Hamiltonian constraint and implies the vanishing of the supercurrent.

\subsection{Symmetries of the one-dimensional effective action}\seclabel{symm_1d_eff_action}

Let us now discuss some of the symmetries of the above one-dimensional action. The action~\eqref*{superspace_action_1d} is manifestly invariant under super-worldline reparametrizations $\{\tau,\theta,\bar\theta\}$ $\rightarrow$ $\{\tau^\prime (\tau,\theta,\bar\theta)$, $\theta^\prime (\tau,\theta,\bar\theta)$, $\bar{\theta}^\prime (\tau,\theta,\bar\theta)\}$, which, in particular, includes worldline re\-pa\-ra\-me\-tri\-sations $\tau \rightarrow \tau^\prime (\tau)$ (that is, one-dimensional diffeomorphisms) and local $\susyno=2$ supersymmetry. Note that the super-determinant of the supervielbein $\mathcal{E}$, which transforms as a super-density, is precisely what is needed to cancel off the super-jacobian from the change of $d\tau\, d^2 \theta$, so that $d\tau\, d^2 \theta\,\mathcal{E}$ is an invariant measure.

In particular, the theory is invariant under worldline reparametrizations, $\tau\rightarrow\tau^\prime(\tau)$ which can be regarded as a remnant of the diffeomorphism invariance of the 11-di\-men\-sio\-nal action~\eqref*{S11}. Here, the lapse function $N$ plays the same r\^{o}le as the ``vielbein'' and it transforms as a co-vector under worldline reparametrizations. The transformation properties of the different types of component fields under worldline re\-pa\-ra\-me\-tri\-sa\-tions are summarised in \tabref{WR}.
\begin{table}[t]\begin{center}
\begin{tabular}{|c|l|} \hline
	\emph{Name} & \emph{WR transformation} $\tau \rightarrow \tau^\prime (\tau)$ \\ \hline\hline
	scalar & $z^a \rightarrow z^a {}^\prime (\tau^\prime) = z^a(\tau)$ \\
	co-vector & $N \rightarrow N^\prime (\tau^\prime) = \frac{\partial \tau}{\partial \tau^\prime} N (\tau)$ \\
	spin-$1/2$ & $\kappa^a \rightarrow \kappa^a {}^\prime (\tau^\prime) = \kappa^a (\tau)$ \\
	spin-$3/2$ & $\psi_0 \rightarrow \psi_0^\prime (\tau^\prime) = \frac{\partial \tau}{\partial \tau^\prime} \psi_0 (\tau)$ \\ \hline
\end{tabular}\caption{Worldline reparametrisation (WR) covariance.}\tablabel{WR}\end{center}\end{table}
The bosonic matter fields $t^i$, $X^{\cal{P}}$ and $z^a$ and the bosonic auxiliary fields $f^i$, $g^{\cal{P}}$ and $H^{\hat{\cal{X}}}$ transform as scalars, whereas the fer\-mi\-onic matter fields $\psi^i$, $\Lambda^{\cal{P}}$, $\Upsilon^{\hat{\cal{X}}}$ and $\kappa^a$ transform as spin-$1/2$ fields. Finally, the gravitino $\psi_0$ transforms as a spin-$3/2$ field. 

The 3-form scalars $X^{\cal{P}}$ arise as zero-modes of the M-theory three-form $A$ and, hence, they are axion-like scalars with associated shift transformations acting as
\be\eqlabel{mcy5_PQ_shift_symm}
	X^{\cal{P}} (\tau) \rightarrow {X^{\cal{P}}}^\prime (\tau) = X^{\cal{P}} (\tau) + c^{\cal{P}} \; ,
\ee
where the $c^{\cal{P}}$ are a set of real constants. It is easy to see that the component action~\eqref*{L11}--\eqref*{L14} only depends on $\dot{X}^{\cal{P}}$ but not on $X^{\cal{P}}$ and that, hence, the action is invariant under the above shifts. Also in the 3-form sector, there is a local fermionic symmetry of the form $\delta \Lambda^{\cal{P}} = P_- {}_{\cal{Q}} {}^{\cal{P}} l^{\cal{Q}}$ as discussed around \eqref{3form_symm}.

\section{Flux compactifications}\seclabel{flux_comp}

In \secref{topol_constraint}, we analysed the conditions for unbroken one-dimensional $\susyno=2$ supersymmetry in the presence of non-zero ${\cal O}(\gamma^1)$ $g$-flux and found that it has to satisfy
\be\eqlabel{gflux_KSeq_cond_new0}
	g^{(3,1)} = g^{(1,3)} = 0 \; , \qquad\qquad \tilde{g}^{(2,2)} = 0 \; .
\ee
The second condition can be expressed in a more geometric way by noting that this is precisely the definition of a primitive form as described at the end of \appref{complex_geometry}. Since $h^{0,2}(X)=0$ on a Calabi-Yau five-fold, the $(1,3)$-part $g^{(1,3)}$ of the 4-form flux is automatically primitive since $\tilde{g}^{(1,3)}$ would be a harmonic $(0,2)$-form. We conclude that the entire 4-form flux $g = g^{(1,3)} + g^{(2,2)} + g^{(3,1)}$ must be primitive $g = g_0$ and hence, the two conditions in~\eqref*{gflux_KSeq_cond_new0} are independent in the sense that one of them only restricts the K\"ahler class whereas the other one only restricts the complex structure and the order with which the two are imposed is irrelevant. Furthermore, by applying the general result~\eqref*{Lefschetz1_to_Lefschetz2}, which follows from \eqrangeref{Lefschetz_Laplace_comm}{Lefschetz_self_comm}, the condition of primitivity, $\tilde{g}=0$, is logically equivalent to $g\wedge J^2 = 0$ and thus,~\eqref*{gflux_KSeq_cond_new0} becomes
\be\eqlabel{gflux_KSeq_cond_new}
	g^{(3,1)} = g^{(1,3)} = 0 \; , \qquad\qquad g\wedge J^2 = 0 \; .
\ee
For the particular case of harmonic $(2,2)$-forms, this also follows from \eqref{sigmaJJ} and taking traces thereof. For a harmonic $(1,3)$-form, it follows from the observation that a harmonic $(1,3)$-form wedged with $J^2$ is a harmonic $(3,5)$-form, which, in turn, is Hodge-dual to a harmonic $(0,2)$-form and must hence vanish, since $h^{0,2}(X)=0$.

Our aim in this section is to find explicit Calabi-Yau five-folds that can serve as ${\cal O}(\gamma^1)$ M-theory backgrounds and to compute the next-to-leading order corrections to the one-dimensional effective action. In particular, we expect a scalar potential to arise for at least some of the moduli as a consequence of non-zero flux. The physical contents and implications of the scalar potential will be analysed in more detail in \chref{msp}, where we will also discuss some explicit models as well as some simple classical and quantum solutions.

\subsection{Explicit examples of Calabi-Yau five-fold backgrounds}\seclabel{flux_cy5_examples}

At the purely classical level (that is, at zeroth order in $\gamma$), any Calabi-Yau five-fold in the sense defined in \chref{cy5} is well-suited as compactification background. The situation changes as one goes to ${\cal O}(\gamma^1)$, where the metric is allowed to stay uncorrected and the compactification manifold is still Calabi-Yau, but subjected to a topological constraint and a flux quantisation condition of the form
\be\eqlabel{fluxcond}
	c_4(X) - 12\, g\wedge g = 24\, W \; , \qquad\qquad
	g + \oneon{2} c_2(X) \in H^4(X,\mathbb{Z}) \; ,
\ee
as shown in \secref{topol_constraint}. In addition, if we are interested in supersymmetry preserving backgrounds, $g$-flux must obey $g^{(1,3)} = 0$ and $g\wedge J^2 = 0$, as explained above.

The simplest way to satisfy~\eqref*{fluxcond} is to consider Calabi-Yau five-folds $X$ with vanishing fourth Chern class $c_4(X)=0$. They provide ${\cal O}(\gamma^1)$ M-theory backgrounds without the necessity for any internal flux and membranes to be present. Amongst the CICY five-folds examined in \secref{cicy5}, we did not find any configurations with this property, but since we only scanned the simplest cases with small configuration matrices, this statement is not conclusive and there may well be more complicated configurations with $c_4(X)=0$. To remedy this shortcoming, we constructed the torus quotient $T^{10}/\Z_2^4$ in \secref{torus_quotients}, which is a Calabi-Yau five-fold in the strict sense defined in \secref{cy_def} and whose fourth Chern class indeed vanishes. Hence, the case $c_4(X)=0$ is not entirely pathological, although it remains an open question whether Calabi-Yau five-folds with $c_4(X)=0$ and full $\SU(5)$ holonomy exist.

In the most general case, one must assume that $c_4(X)\neq 0$ and the question then is whether an appropriate combination of internal flux and membranes can be found to satisfy~\eqref*{fluxcond}. We will discuss this question by means of some examples in the form of the simple CICY five-folds presented in \secref{cicy5}.

First, we consider the CICY five-folds that can be defined in a single projective space. There are 11 distinct cases and they were listed in \tabref{tab:cicy1}. Here, we immediately run into the problem that $h^{2,2}(X) = 1 \not> 1 = h^{1,1}(X)$, which implies that there are no primitive $(2,2)$-forms on such manifolds. Indeed, harmonic $(2,2)$-forms are necessarily proportional to $J^2$ with $J$ being the K\"ahler class of the ambient complex projective space. Thus, $g$-flux is given by $g = k J^2$ for some flux parameter $k$. Requiring $g$ to be primitive implies
\be
	g \wedge J^2 = k J^4 = 0 \; ,
\ee
forcing $k=0$, that is $g=0$. In other words, any non-zero internal 4-form flux explicitly breaks supersymmetry for this class of manifolds. Keeping full $\susyno=2$ supersymmetry means restricting to the six cases out of the 11 manifolds of \tabref{tab:cicy1}, where $c_2(X)$ is even and, since $c_4(X)\neq 0$ in all cases, membranes must be present to cancel off the value of the fourth Chern class.
\begin{table}[t]\begin{center}
\begin{tabular}{|l|l|l|l|}
\hline
$[n|q_1\dots q_K]$	&	$J^4/\hat{J}$	&	$c_4(X)/\hat{J}$	&	$W/\hat{J}$	\\\hline\hline
$[7|6\,2]$			&	12			&	5448				&	227			\\\hline
$[7|4\,4]$			&	16			&	3168				&	132			\\\hline
$[8|5\,2\,2]$		&	20			&	4680				&	195			\\\hline
$[8|4\,3\,2]$		&	24			&	3264				&	136			\\\hline
$[9|3\,3\,2\,2]$		&	36			&	2808				&	117			\\\hline
$[11|2\,2\,2\,2\,2\,2]$&	64			&	2496				&	104			\\\hline
\end{tabular}
\caption{The six CICY five-folds in a single projective space for which~\eqref*{fluxcond} can be satisfied in the absence of flux.}
\tablabel{cicy1_fluxcond}\end{center}\end{table}
Due to the factor of 24 in front of the membrane class $W$ in~\eqref*{fluxcond}, it is not automatically guaranteed that the topological constraint can be satisfied for these CICYs. A short calculation reveals, however, that in each of the six cases the fourth Chern class is divisible 24 when expressed in terms of the dual $(4,4)$-form $\hat{J}$ defined in \eqref{def_dual_44_forms} and thus, we have found six examples of five-folds that satisfy~\eqref*{gflux_KSeq_cond_new,fluxcond} with membranes only. Some important properties of these six CICYs are summarised in \tabref{cicy1_fluxcond}.

Even upon relaxing the condition of unbroken supersymmetry, we find that~\eqref*{fluxcond} can not be satisfied with flux alone for the class of CICYs in a single projective space. This is essentially a consequence of $h^{1,1}(X) = h^{2,2}(X) = 1$, which is very restrictive. On the other hand, allowing both flux and membranes to be present,~\eqref*{fluxcond} can frequently be satisfied. For example, let us consider the simplest CICY five-fold, namely the septic in $\CP^6$ with configuration matrix $[6|7]$. We find $J^4 = 7 \hat{J}$ and hence, from \tabref{tab:cicy1}, $c_4([6|7]) = 5733 \hat{J}$ and $c_2([6|7]) = 21 J^2$. If we parametrise $g$-flux and membranes by $g=k J^2$ and $W=l \hat{J}$, respectively, we can write~\eqref*{fluxcond} for the septic as
\be
	(5733 - 84 k^2 - 24 l) \hat{J} = 0 \; , \qquad\qquad \left(k+\frac{21}{2}\right) J^2 \in H^4([6|7],\mathbb{Z}) \; .
\ee
The second condition implies that $k$ must be half-integral $k = (2m+1)/2$ with $m\in\Z$, since $J^2$ is an integral class. Solving for $l$ then yields
\be
	l = 238 - \frac{7}{2} m(m+1) \; ,
\ee
which shows that we have a countably infinite number of choices to satisfy the constraints~\eqref*{fluxcond} for the septic. For example, one may choose $m=7$, which sets $g = 15/2 J^2$ and $W = 42 \hat{J}$.

In order to find explicit examples with non-zero $g$-flux, we need to turn to more complicated constructions that allow for $h^{2,2}(X) > h^{1,1}(X)$. We start with the second example in \tabref{tab:cicy2}, that is
\be\eqlabel{cicy35}
	X\sim\left[\begin{array}{l}2\\4\end{array}\right|\left.\begin{array}{l}3\\5\end{array}\right] .
\ee
Expanding $J = t^1 J_1 + t^2 J_2$ and $g = k_{11} J_1^2 + k_{12} J_1 J_2 + k_{22} J_2^2$, we first compute the volume using~\eqref*{vol_form_cplx} and formul\ae\ of \secref{cicy5}. We find
\be
	V = \oneon{5!} \int_X J^5 = \frac{5}{12} (t^1)^2 (t^2)^3 + \oneon{8} t^1 (t^2)^4
\ee
and thus, $t^i \neq 0$ in order for $X$ to have non-zero volume. With $t^i \neq 0$, the primitivity condition $g\wedge J^2 = 0$ implies $g=0$. However, with membranes alone the topological constraint in~\eqref*{fluxcond} can not be satisfied, because the term in $c_4(X)$ proportional to $\hat{J}^2$ is not divisible by 24. In addition, the flux quantisation condition is not satisfied for $g=0$, since the coefficients in front of $J_1^2$ and $J_1 J_2$ in $c_2(X)$ are odd. This rules out $X$ as a valid M-theory background, if it is to preserve full $\susyno=2$ supersymmetry. Dropping the last assumption, it is easily possible to satisfy~\eqref*{fluxcond}, which becomes
\begin{gather}\begin{split}
  &(3240 - 12 k_{22} (10 k_{12} + 3 k_{22}) - 24 l_1) \hat{J}^1 \\ &\qquad\qquad\qquad
  	+ (3975 - 12 (10 k_{11}k_{22} + 5 k_{12}^2 + 6 k_{12}k_{22}) - 24 l_2) \hat{J}^2 = 0 \; , \\
  &\left(k_{11} + \frac{3}{2}\right) J_1^2 + \left(k_{12} + \frac{15}{2}\right) J_1 J_2 + (k_{22} + 5) J_2^2 \in H^4(X,\mathbb{Z}) \; ,
\eqlabel{cicy35_conditions}\end{split}\end{gather}
using both flux and membranes. Here, the membrane class has been expanded as $W=l_1 \hat{J}^1 + l_2 \hat{J}^2$. An exemplary solution is provided by $(k_{11},k_{12},k_{22}) = (1/2,15/2,0)$ and $(l_1,l_2) = (135,25)$. A solution with only flux and no membranes does not exist, however.

Our next example is the third manifold in \tabref{tab:cicy2}, specified by
\be\eqlabel{cicy44}
	X\sim\left[\begin{array}{l}3\\3\end{array}\right|\left.\begin{array}{l}4\\4\end{array}\right] .
\ee
Using the same expansion, that is $J = t^1 J_1 + t^2 J_2$ and $g = k_{11} J_1^2 + k_{12} J_1 J_2 + k_{22} J_2^2$, we find for the volume
\be
	V = \oneon{5!} \int_X J^5 = \oneon{3} \left(  (t^1)^2 (t^2)^3 +  (t^1)^3 (t^2)^2 \right) .
\ee
Again, $t^i \neq 0$ is needed so that $X$ has non-zero volume and then the primitivity condition $g\wedge J^2 = 0$ forces $g=0$. Since $c_2(X)$ is even and $c_4(X)$ is divisible by 24, the constraints~\eqref*{fluxcond} can be satisfied by membranes only, namely for $W = 150 \hat{J}^1 + 150 \hat{J}^2$. Without the primitivity requirement, both flux and membranes are allowed to be present and~\eqref*{fluxcond} becomes 
\begin{gather}\begin{split}
	&(3600-48k_{12}^2-96k_{11}k_{22}-96k_{12}k_{22} - 24 l_1)\hat{J}^1 \\ &\qquad\qquad\qquad
		+ (3600-96k_{11}k_{12}-48k_{12}^2-96k_{11}k_{22} - 24 l_2)\hat{J}^2 = 0\; , \\
	&(k_{11}+3)J_1^2+(k_{12}+8)J_1J_2+(k_{22}+3)J_2^2 \in H^4(X,\mathbb{Z}) \; ,
\eqlabel{cicy44_conditions}\end{split}\end{gather}
with exemplary solution $(k_{11},k_{12},k_{22}) = (0,1,0)$ and $(l_1,l_2) = (148,148)$. As in the previous case, flux only solutions do not exist.

Finally, we look at an example of a CICY defined in a product of three complex projective spaces. To this end, let us consider the CICY in the last row of \tabref{tab:cicy2}, 
\be\eqlabel{cicy234}
	X\sim\left[\begin{array}{l}1\\2\\3\end{array}\right|\left.\begin{array}{l}2\\3\\4\end{array}\right].
\ee
As usual, we parametrise $J$ and $g$ by $J=t^1 J_1 + t^2 J_2 + t^3 J_3$ and $g=k_{12}J_1J_2+k_{13}J_1J_3+k_{22}J_2^2+k_{23}J_2J_3+k_{33}J_3^2$, respectively. Next, we compute the volume
\be
	V = \oneon{5!} \int_X J^5 = t^1 (t^2)^2 (t^3)^2 + \oneon{2} t^1 t^2 (t^3)^3 + \oneon{6} (t^2)^2 (t^3)^3 \; .
\ee
One may choose $t^1 = 0$, $t^2\neq 0$ and $t^3 \neq 0$ and still have non-zero volume. In this case, the primitivity requirement $g\wedge J^2=0$ is solved by
\be
	g = g_0 = 3 k_{33} x^2 J_2^2 - 2 k_{33} x J_2 J_3 + k_{33} J_3^2 \; ,
\ee
where $x \equiv t^2/t^3$. The flux quantisation condition then becomes
\be
	\left( 3k_{33} x^2 + \frac{3}{2} \right) J_2^2 + (-2k_{33} x + 6) J_2 J_3 + (k_{33} + 3) J_3^2 + 3 J_1 J_2 + 4 J_1 J_3
	\in H^4(X,\mathbb{Z}) \; .
\ee
This is solved by $k_{33} = 4m+2$ and $x=(2n+1)/2$ with $m,n\in\Z$. Finally, we need to check the topological constraint
\begin{multline}
	\left[ 84 - 48 k_{33}^2 x (10x-3) - 24 l_1 \right] \hat{J}^1 + 
	\left[ 114 + 96 k_{33}^2 x - 24 l_2 \right] \hat{J}^2 \\+ 
	\left[ 130 - 240 k_{33}^2 x^2 - 24 l_3 \right] \hat{J}^3 = 0 \; .
\end{multline}
The coefficients are such that this equation can never be satisfied, as can be seen by considering, for example, the term containing $\hat{J}^2$, which leads to $l_2 = 19/4 + 2 (4m+2)^2 (2n+1) \notin \Z$. For completeness, we mention that upon imposing primitivity the case where all $t^i \neq 0$ leads to $g=0$, which is also not compatible with~\eqref*{fluxcond}. Dropping the primitivity requirement, we find that~\eqref*{fluxcond} can be satisfied by flux alone, for example taking $(k_{12}, k_{13}, k_{22}, k_{23}, k_{33})=(1,3,7/2,0,6)$ with $W=0$.

In summary, the topological constraint, the flux quantisation condition and the primitivity requirement together are rather restrictive and at least amongst the simplest CICYs considered in this section they could frequently not be satisfied simultaneously by adding flux but excluding membranes. In the absence of membranes, this leads to a reduction in the number of viable supersymmetry-preserving M-theory backgrounds based on CICY five-folds, although it is conceivable that more complicated CICY configurations can be found that satisfy all requirements and still allow for flux only solutions. The inclusion of membranes alters this situation and supersymmetry-preserving M-theory backgrounds become more abundant amongst the CICY five-folds. The primitivity condition $g\wedge J^2=0$, stemming from the requirement of supersymmetry-pre\-ser\-va\-tion, is already quite restrictive in its own right. After giving up the requirement of supersymmetry-pre\-ser\-va\-tion, it is not too hard to obtain viable M-theory backgrounds based on CICY five-folds by adding flux and membranes as well as membranes only. With some more effort, that is by exploring more complicated examples with larger configuration matrices, solutions with flux only can be found as well.

\subsection{Leading-order corrections to the one-dimensional effective action}\seclabel{action1d_corr}

We have seen that, unless one works with a Calabi-Yau five-fold $X$ satisfying $c_4(X)=0$, flux and/or membranes are required in order to satisfy~\eqref*{fluxcond} and they are expected to contribute to a scalar potential in the one-dimensional effective theory. In this section, we will calculate the leading order contributions to the scalar potential from the ${\cal O}(\beta)$-corrections to the 11-dimensional action~\eqref*{S11}. Given the need for flux and/or membranes in many five-fold compactifications this potential is clearly of great physical significance.

There are three terms in the 11-dimensional theory which can contribute at ${\cal O}(\beta)$ to a scalar potential in the one-dimensional effective theory:\footnote{The fourth term at ${\cal O}(\beta)$, namely the GS-term~\eqref*{sugra11_gs}, has already been dealt with in \secref{what_about_mui}.} The non-topological $R^4$ terms~\eqref*{R4} evaluated on the five-fold background, the kinetic terms $G\wedge\ast G$ for the 4-form field strength if flux is non-zero and the volume term in the membrane action~\eqref*{membraneaction} provided wrapped membranes are present. We will now discuss these terms in turn starting with the $R^4$ one.

Due to its complicated structure, the reduction of this term on a Calabi-Yau five-fold background is not straightforward. Also, this term depends on the unknown four-curvature of the five-fold and the only hope of arriving at an explicit result is that it becomes topological when evaluated on a five-fold background. A fairly tedious, although in principle straightforward calculation shows that this is indeed the case and that it can be expressed in terms of the fourth Chern class, $c_4(X)$, of the five-fold. Explicitly, we find that~\eqref*{R4} reduces to
\be
	\frac{l\beta_1}{4}\int d\tau\, N\, \frac{1}{24}\,c_{4i}(X)t^i\; ,
\ee
where $\beta_1\equiv(2\pi )^4\beta /v^{4/5}$ and we have expanded the fourth Chern class as $c_4(X)=c_{4i}(X)\hat{\omega}^i$ into a basis of harmonic $(4,4)$-forms $\hat{\omega}^i$ dual to the harmonic $(1,1)$-forms $\omega_i$.

Next, we consider the contribution of a membrane wrapping a holomorphic curve $C$ in $X$ with second homology class $W$. Using the explicit parametrisation $X^0=\sigma^0$, $X^\mu =X^\mu (\sigma )$, $X^{\bar{\mu}}=X^{\bar{\mu}}(\bar{\sigma})$, where $\sigma =(\sigma^1+i\sigma^2)/\sqrt{2}$ for the curve $C$, the first term in the membrane action~\eqref*{membraneaction} reduces to
\be
	- \frac{l\beta_1}{4}\int d\tau\, N\, W_it^i\; .
\ee
Here, we have expanded the membrane class as $W=W_i\hat{\omega}^i$ into our basis of harmonic $(4,4)$-forms.

Finally, we need to consider 4-form flux $g$. The ansatz for $g$-flux can be written as
\be\eqlabel{4form_flux_ansatz}
	g = \oneon{2} n^{\cal{X}} O_{\cal{X}} = n^e\sigma_e+(m^x\varpi_x +{\rm c.c.})\; ,
\ee
where $\{ O_{\cal{X}} \}$ with $\mathcal{X,Y},\ldots=1,\ldots,b^4(X)$ is a basis of real harmonic 4-forms, $\{\sigma_e\}$ with $e,f,\ldots = 1,\dots ,h^{2,2}(X)$ is a basis of real harmonic $(2,2)$-forms, $\{\varpi_x\}$ with $x,y,\ldots = 1,\dots ,h^{1,3}(X)$ is a basis of harmonic $(1,3)$-forms and we used the Hodge decomposition to split a real 4-form into $(1,3)$, $(3,1)$ and $(2,2)$ parts. The factor of $1/2$ has been introduced for convenience in view of the flux quantisation condition~\eqref*{gquant}, which demands that $n^{\cal{X}}$ be an even (odd) integer depending on whether $c_2 (X)$ is even (odd). An essential ingredient in the reduction is the 10-dimensional Hodge dual of $g$. From the results in~\eqref{dualforms} we see that this is given by
\be
 \ast g 	= n^e\left(J\wedge\sigma_e-\frac{i}{2}J^2\wedge\tilde{\sigma}_e+\frac{1}{12}\tilde{\tilde{\sigma}}_eJ^3\right) 
 		-  \left(m^xJ\wedge\varpi_x+{\rm c.c.}\right) \; .
\ee
We recall from \secref{cy5_geom} that $\tilde{\sigma}_e$ is a harmonic $(1,1)$-form which is obtained from $\sigma_e$ by a contraction with the inverse metric $g^{\mu\bar{\nu}}$. Likewise, $\tilde{\tilde{\sigma}}_e$ is a scalar on $X$, obtained from $\sigma_e$ by contraction with two inverse metrics. Following the discussion in \secref{cy5_moduli_spaces}, these objects can be written as
\be
	\tilde{\sigma}_e=ik_e^i\omega_i\; ,\qquad \tilde{\tilde{\sigma}}_e=-\frac{5}{\kappa}k_e^i\kappa_i\; ,
\ee
where $k_e^i$ is a set of (moduli-dependent) coefficients. Combining these results, the 4-form kinetic term reduces to
\begin{multline}
	\frac{1}{2\kappa_{11}^2}\int_{\cal M} (-\oneon{2})G\wedge\ast G =
		-\frac{l\beta_1}{4}\int d\tau \frac{N}{2}\left[ n^en^f\left(d_{efi}t^i+\frac{1}{2}k_f^id_{eijk}t^jt^k \right.\right. \\ \left.\left.
		-\frac{5}{12\kappa}k_e^i\kappa_id_{ejkl}t^jt^kt^l\right)-\left(m^x\bar{m}^{\bar{y}}d_{x\bar{y}i}t^i +{\rm c.c.}\right)\right],
\end{multline}
where we have used some of the intersection numbers introduced in \secref{cy5_moduli_spaces}.

We introduce a one-dimensional scalar potential ${\cal U}$ by
\be\eqlabel{SBpot}
	S_{\rm B,pot} = -\frac{l}{4}\int d\tau\, N\, {\cal U} \; .
\ee
This expression should be added to the bosonic action~\eqref*{S1B}. Then, by combining the three contributions above, we find that
\begin{multline}\eqlabel{V1}
 {\cal U} = \beta_1\left[\left(\frac{1}{2}(g\wedge g)_{(2,2)i}-\frac{1}{2}(g\wedge g)_{(1,3)i}+W_i-\frac{1}{24}c_{4i}(X)\right)t^i \right.\\ \left.
 	+\frac{1}{4}n^en^fk_f^i\left({\cal G}^{(1,1)}_{ik}-\frac{25}{12}\frac{\kappa_i\kappa_k}{\kappa^2}\right){\cal G}^{(1,1)kj}d_{ejlm}t^lt^m\right] .
\end{multline}
A few remarks on this result are in order. The first line is linear in the K\"ahler moduli $t^i$ with coefficients which are almost identical to the components of the anomaly condition~\eqref*{intcond}. In fact, only the sign of $(g\wedge g)_{(1,3)}$, the contribution from the $(1,3)$-part of the flux, is opposite to what it is in the anomaly condition~\eqref*{intcond}. After using the anomaly condition~\eqref*{intcond}, the first part of the scalar potential reduces to a term linear in $t^i$ which depends only on $(1,3)$-flux. Recall that $t^i$ are fluctuation variables, which means an explicit term linear in $t^i$ would indicate an instability of the theory and should therefore be absent. In this way, we re-discover the condition $g^{(1,3)}=0$, found in \secref{topol_constraint}, from a one-dimensional perspective.\footnote{Besides $g^{(1,3)}=0$, we also obtained $g\wedge J^2 = 0$ from the Killing spinor equation (see \secref{topol_constraint} and page~\eqpageref*{gflux_KSeq_cond_new0}). The question of how the second condition arises from a one-dimensional perspective will be answered shortly.} The second line in \eqref{V1} only depends on $(2,2)$-flux and can be written in a supersymmetric form deriving from a superpotential, as we will see. 

The Hodge decomposition~\eqref*{4form_flux_ansatz} of 4-form flux into $(1,3)$, $(3,1)$ and $(2,2)$ pieces depends on the complex structure and therefore, the condition $g^{(1,3)}=0$ generally leads to a potential for the complex structure moduli. In other words, the complex structure moduli $z^a$ are only allowed to fluctuate in such a way as to keep the 4-form flux purely of $(2,2)$-type. With the decomposition~\eqref*{31_4_rel3} inserted into \eqref{4form_flux_ansatz}, the condition $g^{(1,3)}=0$ becomes
\be\eqlabel{g13_zero_implicit_14_potential}
	m^x (\underline{z},\underline{\bar{z}}) = n^{\cal{X}} \mathfrak{D}_{\cal{X}} {}^x (\underline{z},\underline{\bar{z}}) = 0 \; , 
	\qquad\qquad (\text{and $\cc$}) \; .
\ee
However, it is not known whether the $\mathfrak{D}_{\cal{X}} {}^x$ and hence the resulting potential for the $z^a$ can be calculated explicitly. It is important to recall that in our analysis of bosonic and fermionic 4-form fields we are restricting to Calabi-Yau five-folds that satisfy \eqref{22_vertical} and, in this case, the potential vanishes, that is the complex structure moduli are restored as flat directions in the moduli space, because for such manifolds the split of a 4-form into a $(2,2)$-piece and a $(1,3)+(3,1)$-piece is complex structure independent. This can also be seen by noting that in this case the condition~\eqref*{g13_zero_implicit_14_potential} turns into the complex structure independent equation $n^{\cal{\hat{X}}} = 0$, with the help of the decomposition~\eqref*{31_4_new_rel3}. Moreover, the $(2,2)$-flux itself, $g^{(2,2)} = \oneon{2} n^{\cal{\tilde{X}}} O_{\cal{\tilde{X}}} = n^e \sigma_e$, becomes a complex structure independent quantity.

This leaves us with the second line in \eqref{V1} and, in order to write this into a more explicit form, we need to compute the coefficients $k_e^i$. This has, in fact, been done in \eqref{kei}. Inserting these results and using eqs.~\eqref*{GtGti} and \eqref*{Gti} we finally find for the scalar potential
\be\eqlabel{UW0}
	{\cal U}=\frac{1}{2}G^{(1,1)ij}{\cal W}_i{\cal W}_j\; ,\qquad {\cal W}_i=\frac{\partial{\cal W}}{\partial t^i} \; ,
\ee
where the ``superpotential'' ${\cal W}$ is given by
\be\eqlabel{superpot}
	{\cal W}(\underline{t}) = \frac{\gamma_1}{3} d_{eijk}n^et^it^jt^k
\ee
and $G^{(1,1)ij}$ is the inverse of the physical $(1,1)$ moduli space metric \eqref*{G11}. Here, we introduced $\gamma_1 \equiv \sqrt{\beta_1}$\newnot{symbol:gamma1}.

The fact that the scalar potential can be written in terms of a superpotential in the usual way suggests it can be obtained as the bosonic part of a superfield expression. This is indeed the case and the term we have to add to the superspace action~\eqref*{superspace_action_1d} is simply
\be\eqlabel{Spot}
	S_{\rm pot} = - \frac{l}{2} \int d\tau \,d^2\theta\,\mathcal{E}\,{\cal W}(\underline{T})\; .
\ee 
Indeed, combining this term with the $(1,1)$ kinetic term in the superspace action~ \eqref*{superspace_action_1d} and working out the bosonic component action using \eqref{L11} one finds the terms
\be
	\frac{l}{2}\int d\tau\, \frac{N}{4} \left( G^{(1,1)}_{ij}f^if^j - 2 f^i{\cal W}_i \right) ,
\ee
which, after integrating out the $(1,1)$ auxiliary fields $f^i = G^{(1,1)ij} \, {\cal W}_j$, reproduce the correct scalar potential. The superpotential~\eqref*{superpot} can be obtained from a Gukov-type formula
\be\eqlabel{superpotGukov}
	{\cal W} (\underline{t}) = \frac{\gamma_1}{3} \int_X g \wedge J^3\; .
\ee
This integral is, in fact, the only topological integral, linear in flux, one can build using the two characteristic forms $J$ and $\Omega$ of the five-fold and $G_{\rm flux}$. In this sense, it is the natural expression for the superpotential. Here, we have explicitly verified by a reduction from 11 dimensions that it gives the correct answer.

So far, we have ignored the fact that $g$ must be primitive in order for the Calabi-Yau five-fold background to preserve $\susyno=2$ supersymmetry (see \secref{topol_constraint} and page~\eqpageref*{gflux_KSeq_cond_new0}). Using the results of \appref{complex_geometry} on primitive forms, we apply the Hard Lefschetz decomposition~\eqref*{Lefschetz_decomp_form} to the general 4-form $g$
\be\eqlabel{Lefschetz_decomp_flux}
	g = g_0 + g_0^{(1,1)} \wedge J + g_0^{(0,0)} \wedge J^2 \; .
\ee
The primitive forms $g_0$ and $g_0^{(1,1)}$ obey $g_0\wedge J^2 = 0$ and $g_0^{(1,1)}\wedge J^4 = 0$, respectively (the $(0,0)$-form $g_0^{(0,0)}$ is automatically primitive). Hence, the first two terms of the decomposition drop out of the superpotential~\eqref*{superpotGukov} and we are left with
\be\eqlabel{superpotGukov_prim}
	{\cal W} (\underline{t}) = 40 \gamma_1 g_0^{(0,0)} V(\underline{t}) \; ,
\ee
where $V(\underline{t})$ is the Calabi-Yau five-fold volume defined in~\eqref*{def_kappa0,vol_form_kappa}. Note that while the left hand side of~\eqref*{Lefschetz_decomp_flux} is independent of the $t^i$, the differential forms on the right hand side are not. In fact, by performing a re-scaling $t^i \to \lambda t^i$, we learn that $g_0$, $g_0^{(1,1)}$ and $g_0^{(0,0)}$ are homogeneous of degree zero, $-1$ and $-2$ in the $t^i$, respectively. Thus, since $V$ scales as $V(\lambda \ul{t}) = \lambda^5 V(\ul{t})$, one concludes that the superpotential is homogeneous of degree three. This is consistent with~\eqref*{superpot,superpotGukov}. From~\eqref*{Lefschetz_decomp_form}, we have
\be
	\text{$g\;\;$ primitive} \qquad \Longleftrightarrow \qquad g_0^{(1,1)} = g_0^{(0,0)} = 0 \; .
\ee
We see that the superpotential~\eqref*{superpotGukov_prim} depends only on the non-primitive components of $g$. In turn, requiring $g$ to be primitive leads to a vanishing superpotential. This can, of course, also be seen more directly and without the Hard Lefschetz decomposition by recalling that the primitivity of $g$ is equivalent to $g\wedge J^2 = 0$ (see~\eqref*{Lefschetz1_to_Lefschetz2}) and the fact that the superpotential is given by the integral of $g\wedge J^3$ over $X$. 

How does this condition arise from a one-dimensional perspective? The answer is the same as for the higher-dimensional context in which we originally discovered it, namely as a supersymmetry preservation condition. To see this, we need to look for vacua with constant bosonic fields and vanishing fermions that preserve the $\susyno=2$ supersymmetry of the one-dimensional theory. Finding such vacua amounts to setting the supersymmetry variations of all fermions to zero and solving the resulting Killing spinor equations, as usual. For the various $2b$ multiplets, the supersymmetry variations of their fermion components vanish automatically for constant bosonic fields and vanishing fermions, as can be seen directly from the results in \secref{sspace_curved}. On the other hand, the supersymmetry variations of the fermions residing in $2a$ multiplets require a bit more care. For the 3-form fermions $\Lambda^{\cal{P}}$, one has from \eqref{sspace_curved_transf_2a,sspace_curved_transf_2a_bar}
\be
	\delta_\epsilon \Lambda^{\cal{P}} = 0 \; , \qquad
	\delta_\epsilon \bar\Lambda^{\cal{P}} = -\oneon{2}\epsilon g^{\cal{P}} = 0 \; ,
\ee
after inserting \eqref{3ferm_constraint_spartner}. For the $(1,1)$ fermions $\psi^i$, the transformations lead to
\be
	\delta_\epsilon\psi^i = 0\; ,\qquad \delta_\epsilon\bar{\psi}^i = -\oneon{2}\epsilon f^i=-\oneon{2}\epsilon G^{(1,1)ij}\, {\cal W}_j \; .
\ee
Hence, constant scalar field vacua which preserve $\susyno=2$ supersymmetry are characterised by the ``F-term'' equations
\be\eqlabel{Ftermeq1}
	{\cal W}_i = 0 \; .
\ee
Upon plugging in~\eqrangeref*{superpotGukov}{Lefschetz_decomp_flux}, the F-term equations translate into
\be\eqlabel{Ftermeq2}
	g_0^{(1,1)j} \kappa_{ij} + g_0^{(0,0)} \kappa_i = 0 \; ,
\ee
where we have expanded $g_0^{(1,1)} = g_0^{(1,1)i} \omega_i$. Note that $g_0^{(1,1)i} \kappa_i = 0$, as a consequence of $g_0^{(1,1)}\wedge J^4 = 0$. Thus, multiplying by $t^i$ makes the first term disappear and we are left with
\be
	g_0^{(0,0)} V(\underline{t}) = 0 \; ,
\ee
which implies $g_0^{(0,0)} = 0$ assuming non-zero volume. With this result and by virtue of $g_0^{(1,1)i} \kappa_i = 0$, we can re-write \eqref{Ftermeq2} as
\be
	g_0^{(1,1)j} G^{(1,1)}_{ij}(\underline{t}) = 0
\ee
and by contracting with the inverse $(1,1)$ metric $G^{(1,1)ij}$, we arrive at $g_0^{(1,1)i} = 0$. The conclusion is that
\be
	\text{$g\;\;$ primitive} \qquad \Longleftrightarrow \qquad {\cal W}_i = 0 \; ,
\ee
which explains how the primitivity condition on $g$ arises from a one-dimensional perspective.

\Eqref{UW0} shows that solutions to the F-term equations are stationary points of the superpotential, although, unlike in four-dimensional ${\cal N}=1$ supergravity, they need not be minima since the $(1,1)$ metric $G^{(1,1)}$ is not positive definite. Another interesting difference to four-dimensional supergravity is that the scalar potential always vanishes for solutions of the F-term equations.

Finally, when $(2,2)$-flux is non-vanishing, another set of bosonic terms arises from the Chern-Simons term $A\wedge G\wedge G$ in the 11-dimensional action~\eqref*{S_CJS_B}. Writing the complete ansatz for the 4-form field strength $G$, including $(2,2)$-flux and zero modes, one has
\be\eqlabel{Gcompl}
 G 	= (2\pi)^2 \gamma \, g^{(2,2)} + \dot{X}^{\cal{P}} d\tau \wedge N_{\cal{P}} 
  	= (2\pi)^2 \gamma \, n^e\sigma_e + \dot{X}^{\cal{P}} d\tau \wedge N_{\cal{P}}\; .
\ee
Here, we recall that $\{N_{\cal{P}}\}$, where $\mathcal{P,Q},\ldots =1,\dots ,b^3(X)$, are a basis of real harmonic 3-forms and $X^{\cal{P}}$ are the associated 3-form zero modes. Inserting this ansatz into the 11-dimensional Chern-Simons term one finds
\be\eqlabel{SBCS}
	S_{\rm B,CS} = - \frac{l}{2} \int d\tau\frac{\gamma_1}{3} d_{\mathcal{PQ}e}n^e\dot{X}^{\cal{P}} X^{\cal{Q}} \; ,
\ee
where $d_{\mathcal{PQ}e} = \int_X N_{\cal{P}} \wedge N_{\cal{Q}} \wedge \sigma_e$. Note that~\eqref*{SBCS} is linear in flux and, hence, appears at order $\gamma = \sqrt{\beta}$. It represents a one-dimensional Chern-Simons term.
%
%

\chapter{M-theory Calabi-Yau Cosmology}\chlabel{msp}

After developing the necessary background knowledge, we spent a great amount of time in the preceding chapter providing a detailed presentation of virtually all aspects of the reduction of M-theory on Calabi-Yau five-folds down to one dimension. In this chapter, we will examine some of the physical contents of the one-dimensional effective action.

At the end of this chapter, we will report on some preliminary results originating from unfinished (and hence as yet unpublished) work in progress on the mini-superspace quantisation of our model.

\section{The bosonic effective action and its classical solutions}\seclabel{class_sol}

In this chapter, we will focus exclusively on the bosonic part of the one-dimensional effective action. To begin with, we summarise our result for the complete bosonic action for later reference. The bosonic action depends on three sets of fields, the real $(1,1)$ moduli $t^i$, the real 3-form moduli $X^{\cal{P}}$ and the complex $(1,4)$ moduli $z^a$. It can be written as a sum of three parts
\be
	S_{\rm B}=S_{\rm B,kin}+S_{\rm B,pot}+S_{\rm B,CS}
\ee
which, from \eqref{S1B,SBpot,UW0,SBCS}, are given by
\begin{align}
	S_{\rm B,kin} &= \frac{l}{2} \int d\tau N^{-1} \left\{ \oneon{4} G_{ij}^{(1,1)}(\underline{t}) \dot{t}^i \dot{t}^j  
		+ \oneon{2} G_{\mathcal{PQ}}^{(3)}(\underline{t},\underline{z},\underline{\bar{z}}) \dot{X}^{\cal{P}} \dot{X}^{\cal{Q}}
		+ 4 V(\underline{t}) G_{a\bar{b}}^{(1,4)} (\underline{z},\underline{\bar{z}}) \dot{z}^a \dot{\bar{z}}^{\bar{b}}  \right\} , 
		\eqlabel{1deff_bos_kin} \\
	S_{\rm B,pot} &= - \frac{l}{4}\int d\tau\, N\, {\cal U} \; , \\
	S_{\rm B,CS} &= - \frac{l}{2} \int d\tau\frac{\gamma_1}{3} d_{\mathcal{PQ}e}n^e\dot{X}^{\cal{P}} X^{\cal{Q}} \; , \eqlabel{1deff_bos_CS}
\end{align}
with scalar potential ${\cal U}$ and superpotential ${\cal W}$
\be\eqlabel{UW}
	{\cal U} = \frac{1}{2}G^{(1,1)ij}{\cal W}_i{\cal W}_j\; ,\qquad {\cal W}(\underline{t})=\frac{\gamma_1}{3}d_{eijk}n^et^it^jt^k\; .
\ee 
The metrics $G^{(1,1)}_{ij}$, $G^{(1,4)}_{a\bar{b}}$ and $G_{\mathcal{PQ}}^{(3)}$ have been defined in~\eqrangeref*{G11}{G3}. The first two parts of this action can be schematically written as
\be\eqlabel{msp_1d_sigma_model_action}
 S_{\rm B,kin}+S_{\rm B,pot}= 
 	\frac{l}{2}\int d\tau\left\{N^{-1}\,G_{IJ}(\underline{\phi})\dot{\phi}^I\dot{\phi}^J - \frac{N}{2} \,{\cal U}(\underline{\phi})\right\} ,
\ee
where we have collectively denoted the various types of fields by $\phi^I=(t^i, X^{\cal{P}}, z^a, \bar{z}^{\bar{b}})$ and $G_{IJ}$ is a block-diagonal metric which contains the above moduli space metrics in the appropriate way. The associated equations of motion then have the general form
\be\eqlabel{eom1d}
	\frac{1}{N}\frac{d}{d\tau}\left(\frac{\dot{\phi}^I}{N}\right) + \Gamma^I_{JK} \frac{\dot{\phi}^J}{N} \frac{\dot{\phi}^K}{N} + 
	\oneon{4} G^{IJ}\frac{\partial{\cal U}}{\partial\phi^J}+C^I=0\; ,
\ee
where $\Gamma^I_{JK}$ is the Christoffel connection associated to $G_{IJ}$ and $C^I$ is the contribution from the Chern-Simons term. Since the Chern-Simons term only depends on $X^{\cal{P}}$, we have $C^i=C^a=C^{\bar{b}}=0$.

When beginning to search for solutions, the first question that arises is whether there are any static solutions, that is, solutions with all $\phi^I=\text{const}$. Since the potential ${\cal U}$ only depends on the $(1,1)$-moduli, it is certainly consistent with the equations of motion~\eqref*{eom1d} to set all other fields to constants. For vacua without $(2,2)$-flux (but possibly with membranes) this can also be done for the $(1,1)$-moduli $t^i$. In this case the scalar potential vanishes identically and the moduli space is completely degenerate.

More generally, we saw in \secref{action1d_corr} that one-dimensional $\susyno=2$ supersymmetry preserving solutions must satisfy the F-term equations ${\cal W}_i = 0$, which lead to flat directions. In that section, we also showed that the F-term equations are equivalent to demanding $g$-flux to be primitive, $g=g_0$. The primitivity condition on $g$-flux was analysed for selected examples of Calabi-Yau five-folds in \secref{flux_cy5_examples} and will not be repeated in full length. Suffice it to say here that in cases where $h^{1,1}(X) = 1$, the only solution to the F-term equations is $t^1=0$. This corresponds to vanishing Calabi-Yau five-fold volume and is thus beyond the regime of validity of the one-dimensional effective theory.

Moving on to cases where $h^{1,1}(X)>1$, we begin with the first example in \tabref{tab:cicy2}, that is
\be\eqlabel{cicy26_chp6}
	X\sim\left[\begin{array}{l}1\\5\end{array}\right|\left.\begin{array}{l}2\\6\end{array}\right] .
\ee
Since $h_0^{2,2}(X) = h^{2,2}(X) - h^{1,1}(X) = 2 - 2 = 0$ (see \eqref{def_prim_Hodge_numbers}), which is a topological property, the condition $g\wedge J^2 = 0$, implies vanishing Calabi-Yau volume if $g\neq 0$ and hence, there is no viable supersymmetric minimum in this case.

The same conclusion also holds true for the second example in \tabref{tab:cicy2}, that is
\be
	X\sim\left[\begin{array}{l}2\\4\end{array}\right|\left.\begin{array}{l}3\\5\end{array}\right] ,
\ee
although the derivation is more complicated. As in \secref{flux_cy5_examples}, we use the expansion $g = k_{11} J_1^2 + k_{12} J_1 J_2 + k_{22} J_2^2$. The superpotential then becomes
\be
	{\cal W} = \frac{4}{3} k_{11} (t^2)^3 + 2 ( 2 k_{12} + k_{22} ) t^1 (t^2)^2 + 4 k_{22} (t^1)^2 t^2 \; .
\ee
Taking into account that $k_{22}$ is integral, whereas $k_{11}$ and $k_{12}$ must be half-integral as demanded by the flux quantisation condition~\eqref*{cicy35_conditions}, it can be shown that ${\cal W}_i=0$ leads to $V=0$ if $g\neq 0$.

The situation improves for the third example in \tabref{tab:cicy2}, that is
\be
	X\sim\left[\begin{array}{l}3\\3\end{array}\right|\left.\begin{array}{l}4\\4\end{array}\right] .
\ee
Again, as in \secref{flux_cy5_examples}, we expand $g$ as $g = k_{11} J_1^2 + k_{12} J_1 J_2 + k_{22} J_2^2$. Then, one finds for the superpotential
\be
	{\cal W} = \frac{4}{3} k_{11} (t^2)^3 + 4( k_{11} + k_{12} ) t^1 (t^2)^2 + \frac{4}{3} k_{22} (t^1)^3 + 4( k_{12} + k_{22} ) (t^1)^2 t^2 \; .
\ee
It is easy to see that setting, for example, $k_{11}=k_{22}=3$ and $k_{12}=-4$, the F-term equations are satisfied along the flat direction $t^1=t^2$. Inspecting \eqref*{cicy44_conditions}, one verifies that this choice also satisfies the topological constraint (with membrane class $W = 130 \hat{J}^1 + 130 \hat{J}^2$) and the flux quantisation condition. Moreover, this flat direction consists of minima of the potential with zero cosmological constant. The existence of such minima is of considerable importance for our M-theory compactifications. Flux-compactifications are often plagued with yielding large potential energies above the compactification scale due to the quantised nature of flux. This is problematic, for it invalidates the low-energy effective theory. We have just seen an example where this problem can be avoided due to a flat direction with vanishing vacuum energy in the two-dimensional K\"ahler moduli space. This means, at least to ${\cal O}(\beta^1)$, self-consistent five-fold compactifications of M-theory with $(2,2)$-flux exist.

In cases where the F-term equations cannot be solved, a few general statements can nonetheless be made, based on the scaling behaviour of the scalar potential. Recall from \secref{action1d_corr} (see below \eqref{superpotGukov_prim}) that the superpotential scales as ${\cal W}(\lambda\ul{t}) \to \lambda^3 {\cal W}(\ul{t})$ under $t^i \to \lambda t^i$. Together with $G^{(1,1)}_{ij}(\lambda\underline{t})=\lambda^3G^{(1,1)}_{ij}(\underline{t})$, it follows that the scalar potential is homogeneous of degree one, that is ${\cal U}(\lambda\underline{t})=\lambda{\cal U}(\underline{t})$. In this context, it is important to note that the metric $G^{(1,1)}_{ij}$ has signature $(-1,+1,\ldots  ,+1)$. Whether the negative direction is ``probed'' by the scalar potential depends on the structure of the superpotential and its derivatives. If it is, the potential contains a piece which is of the form ${\cal U}=-c\tilde{t}$, where $c$ is a positive constant and $\tilde{t}$ points along the negative direction in the K\"ahler moduli space. This indicates an instability which will eventually lead to decompactification. 

Clearly, this is always the case for examples with $h^{1,1}(X)=1$ where the metric is just a negative number. For example, one finds ${\cal U}=-\frac{525}{4}\beta_1 t$ for the septic in $\CP^6$, that is $[6|7]$. As expected, the potential is negative and results in a rapid growth of the volume.

For $h^{1,1}(X)>1$ the picture is less clear and what happens depends on the choice of Calabi-Yau manifold and flux. As an example, we choose the configuration~\eqref*{cicy26_chp6}. Expanding $g = k_{12} J_1 J_2 + k_{22} J_2^2$, one finds for the volume and the superpotential
\be
	V = \oneon{4} t^1 (t^2)^4 + \oneon{60} (t^2)^5 \; ,\qquad 
	{\cal W} = 6 k_{22} t^1 (t^2)^2 + \left(2 k_{12} + \frac{2}{3} k_{22} \right) (t^2)^3 \; .
\ee 
This gives rise to the following scalar potential
\be
	{\cal U} = \frac{\beta_1(15 k_{12} + 2 k_{22})}{6(15t^1+t^2)} \left((3k_{12} - 2 k_{22}) (t^2)^2 - 36 k_{22} t^1 t^2 \right) .
\ee
For the choice $(k_{12}, k_{22}) = (0, 1/2)$, the above potential is negative and such that both $t^i$ will grow. For $(k_{12}, k_{22}) = (1, -1/2)$, on the other hand, the potential is positive. Gradients are such that $t^2$ contracts and, as a result, the total volume eventually approaches zero (while $t^1$ slowly expands).

After presenting a variety of scenarios to illustrate the different classical evolutions that are possible, we will now turn to some of the quantum mechanical aspects of our one-dimensional effective theory, but beforehand we need to introduce the concept of mini-superspace quantisation.

\section[Mini-superspace quantisation in a nutshell]{Mini-superspace quantisation in a nutshell\footnote{This important sub-branch of quantum cosmology is discussed more thoroughly, for example, in refs.~\cite{Kiefer:2004gr,Halliwell:1987eu,Halliwell:1990uy,DEath:1996at}. More recent reviews on quantum geometrodynamics in general can be found in refs.~\cite{Giulini:2006xi,Kiefer:2008bs}. Here, we will merely provide a minimalist account to properly allow this topic to be embedded into this thesis and to present some general results needed later.}}\seclabel{msp_intro} 

This is the second time the word ``superspace'' is introduced in this thesis. There are indeed two entirely different meanings of the word in the physics literature. While nowadays most physicists would probably associate it with supersymmetry (as explained in \chref{superspace}), it had already existed before the invention of supersymmetry in a totally different context: The name was coined by Wheeler in 1968~\cite{Wheeler:1988zr} to denote the configuration space of classical four-dimensional general relativity. The superspace ${\cal S}(\Sigma)$ for a given three-manifold $\Sigma$\newnot{symbol:Sigma} is the space of all three-geometries, that is the space of all three-metrics modded out by the diffeomorphism group: ${\cal S}(\Sigma) \equiv \mathrm{Riem}\;\Sigma/\mathrm{Diff}\; \Sigma$.

\subsection{Geometrodynamics}

During the 1960s, by a program called ``geometrodynamics'' mainly attributed to Wheeler, Einstein's theory of general relativity was reformulated as an initial value problem with constraint- and evolution-equations and with the ultimate goal of unravelling the foundations of quantum gravity. It was hoped that in this way, gravity could be placed on the same footing as the other fundamental forces and the very successful methods of conventional quantum field theory could then be applied. This is expressed in the very name ``(quantum) geometro\emph{dynamics}'', which evidently derives from the two extraordinarily successful predecessors ``(quantum) electro\emph{dynamics}'' and ``(quantum) chromo\emph{dynamics}.''

The starting point of geometrodynamics is the Hamiltonian formulation of general relativity, known as ADM-formalism~\cite{Arnowitt:1962hi}. The four-dimensional line element $ds^2$ is para\-me\-trised as follows
\be\eqlabel{msp_metricansatz}
	ds^2 = g_{MN} dx^M dx^N = - N^2 d\tau^2 + h_{mn} (dx^m + N^m d\tau) (dx^n + N^n d\tau) \; ,
\ee
where $N$ and $N^m$ are called lapse function and shift vector, respectively. The indices are such that $M,N,\ldots=0,1,2,3$ and $m,n,\ldots=1,2,3$. Here, the index labelling conventions are somewhat non-standard as compared to the literature (see, for example, refs.~\cite{Wald:1984rg,Kiefer:2004gr,Giulini:2006xi,Kiefer:2008bs}). This was done to achieve maximal resemblance to eleven dimensions and hence to the forumul\ae\ of the previous chapters. Indeed, this parametrisation closely resembles our 11-dimensional compactification ansatz in~\eqref*{gansatz}. The only difference, besides the index ranges, is the appearance of the shift vector $N^m$, which was absent for Calabi-Yau five-fold compactifications since $h^{1,0}(X)=0$ for a Calabi-Yau five-fold $X$, and the fact that the lapse function and shift vector can depend on $\ul{x}=(x^1,x^2,x^3)$ as well as $\tau$, that is $N=N(\tau,\ul{x})$ and $N^m=N^m(\tau,\ul{x})$.

The three-metric $h_{mn}$, which is an element of superspace, is the intrinsic metric on the spatial hypersurfaces $\tau=\text{const.}$, assuming a global $(3+1)$-foliation\footnote{This requires $\manifold$ to be globally hyperbolic and the global choice $\manifold=\R\times\Sigma$ excludes topology changes.} $\manifold=\R\times\Sigma$ of the four-dimensional space-time manifold $\manifold$. As in \chref{mthy}, the Einstein-Hilbert action reads
\be\eqlabel{msp_EH_orig}
	S_{{\rm EH}} = \oneon{2\kappa^2} \int_\manifold R \ast 1 = \int_\manifold d^4 x\, \lagr \; .
\ee
Plugging in~\eqref*{msp_metricansatz}, one finds (see, for example, ref.~\cite{Wald:1984rg,DEath:1996at})\footnote{The total derivative in~\eqref*{msp_EH_orig} will be ignored in the following, although its presence can have important consequences in cases where, for example, $\Sigma$ is non-compact or not asymptotically flat.}
\be\eqlabel{msp_EH_hamil}
	S_{{\rm EH}} = \oneon{2\kappa^2} \int_{\R\times\Sigma} d\tau d^3 x \left\{ \pi^{mn} \dot{h}_{mn} - \hamil + (\text{total derivative}) \right\} ,
\ee
where the dot denotes the derivative with respect to $\tau$, $\hamil$ is the Hamiltonian density given below and $\pi^{mn}$ is the momentum conjugate to $h_{mn}$, that is $\pi^{mn} \equiv \frac{\delta L}{\delta \dot{h}_{mn}}$. The Hamiltonian density $\hamil$ splits into two parts according to
\bea\eqlabel{msp_hamiltonians}
	\hamil &= N \hamil_0 + N^m \hamil_m \; , \\
	\hamil_0 &= 2\kappa^2 \, \sqrt{h} \left[ \pi^{mn} \pi_{mn} - \oneon{2} (\pi^m {}_m)^2 \right] - \frac{\sqrt{h}}{2\kappa^2} R^{(3)} \; , \\
	\hamil_m &= - 2 \nabla^{(3)}_n \pi_m {}^n \; ,
\eea
where $h\equiv\det h_{mn}$ and, here and in the following, the three-dimensional indices $m,n,\ldots$ are raised and lowered using the three-metric $h_{mn}$. Moreover, $R^{(3)}$ and $\nabla^{(3)}_m$ denote Ricci-scalar and covariant derivative with respect to $h_{mn}$. Note that the conjugate momenta to $N$ and $N^m$ vanish identically and thus, $N$ and $N^m$ act as Lagrange multipliers. Their variations lead to the famous classical Hamiltonian and diffeomorphism constraints of general relativity
\be\eqlabel{msp_class_constraints}
	\hamil_0 = 0 \; , \qquad\qquad \hamil_m = 0 \; ,
\ee
which, in fact, encapsulate the complete dynamics of the theory. 

The full Hamiltonian $H = \int_\Sigma d^3 x\, \hamil = \int_\Sigma d^3 x\, \left\{ \pi^{mn} \dot{h}_{mn} - \lagr \right\}$ takes the form
\be
	H = \int_\Sigma d^3 x \left\{N \hamil_0 + N^m \hamil_m \right\} .
\ee
A very useful concept in this formalism is the extrinsic curvature (or second fundamental form) $K_{mn}$ of the spatial hypersurfaces $\Sigma$ defined as
\be
	K_{mn} \equiv \oneon{2} \pounds_{n(\tau,\ul{x})} h_{mn} \; ,
\ee
where $n$ is the unit normal vector to $\Sigma$ pointing in the future direction and $\pounds_{n(\tau,\ul{x})}$ is the Lie derivative with respect to the vector field $n(\tau,\ul{x})$. The extrinsic curvature can be expressed in terms of the quantities defined above. One finds
\be
	K_{mn} = \oneon{2N} \left( \dot{h}_{mn} - 2 \nabla^{(3)}_{(m} N_{n)} \right) ,
\ee
where $n^M = \oneon{N}(1,-N^m)$ and the expression for the Lie derivative of a symmetric $(0,2)$-tensor~\cite{Wald:1984rg}, $\pounds_X T_{MN} = X^P \nabla_P T_{MN} + 2 (\nabla_{(M} X^P) T_{N)P}$, has been used. 

With the help of the extrinsic curvature, the canonical momenta $\pi^{mn}$ can be written as
\be\eqlabel{msp_momentum_extr_curv}
	\pi^{mn} = \frac{\sqrt{h}}{2\kappa^2} \left( K^{mn} - K\, h^{mn} \right) = \oneon{2\kappa^2} G^{mnpq} K_{pq} \; ,
\ee
where $K\equiv K^m {}_m$ and in the last equality, the famous DeWitt-metric~\cite{DeWitt:1967yk} (a symmetric $6\times 6$ matrix in the space of symmetric index pairs $(mn)$) has been introduced. It is given by
\be
	G^{mnpq} \equiv \frac{\sqrt{h}}{2} \left( h^{mp} h^{nq} + h^{mq} h^{np} - 2 h^{mn} h^{pq} \right) 
\ee
and can be interpreted as a metric on the space $\mathrm{Riem}\;\Sigma$ of all three-metrics on $\Sigma$. It has Lorentzian signature, $(-,+,\ldots,+)$, at each space point $\ul{x}$ and this feature is independent of the signature of the space-time metric $g_{MN}$. Sometimes the DeWitt-metric is also referred to as supermetric (another word re-used in supersymmetry terminology). The zero-component of the Hamiltonian density can then be re-written as follows
\be\eqlabel{msp_H0_dewitt_metric}
	\hamil_0 = 2\kappa^2 \, G_{mnpq} \pi^{mn} \pi^{pq} - \frac{\sqrt{h}}{2\kappa^2} R^{(3)} \; ,
\ee
where the inverse DeWitt-metric is given by 
\be
	G_{mnpq} = \oneon{2\sqrt{h}} \left( h_{mp} h_{nq} + h_{mq} h_{np} - h_{mn} h_{pq} \right) .
\ee
The DeWitt-metric and its inverse satisfy $G^{mnpq} G_{pqrs} = \oneon{2} \left( \delta^m_r \delta^n_s + \delta^m_s \delta^n_r \right)$.

Upon setting $N^m=0$, the Lagrangian density $\lagr$ becomes
\be\eqlabel{lagr_msp_prep}
	2\kappa^2 \, \lagr = \oneon{4} N^{-1} G^{mnpq} \dot{h}_{mn} \dot{h}_{pq} + N\, \sqrt{h}\, R^{(3)} \; ,
\ee
which resembles our one-dimensional sigma-model Lagrangian in~\eqref*{msp_1d_sigma_model_action} quite closely if we also set $N=N(\tau)$ and integrate over $\int_\Sigma d^3 x$. Thus, the geometrodynamics (and mini-superspace) formalism is rather directly applicable to our one-dimensional effective action obtained from Calabi-Yau five-fold reductions of M-theory as we will see in more detail later in \secref{msp_cy5}.

\subsection{Quantum Geometrodynamics}

We will now briefly describe how geometrodynamics can be quantised, at least, at the formal level. We restrict to the canonical quantisation procedure and omit many inherent subtleties. Full details on the path integral quantisation of geometrodynamics can be found in the literature~\cite{Kiefer:2004gr,Halliwell:1990uy,DEath:1996at,Kiefer:2008bs}.

The Poisson bracket between the three-metric $h_{mn}$ and its canonically conjugated momentum $\pi^{mn}$ is given by
\be
	\{ h_{mn} (\ul{x}), \pi^{pq} (\ul{x}^\prime) \} = \delta^p_{(m} \delta^q_{n)} \delta^{(3)} (\ul{x}-\ul{x}^\prime) \; .
\ee
Thus, $h_{mn}$ and $\pi^{mn}$ play the r\^{o}le of the canonical coordinates $x$ and $p$ in ordinary Hamiltonian mechanics. In the same way as $x$ and $p$ in ordinary quantum mechanics, $h_{mn}$ and $\pi^{mn}$ are promoted to operators acting on the superspace-wave-functional $\Psi[h_{mn}(\ul{x})]$ (dubbed ``wave-function of the universe'') in the following way
\bea\eqlabel{msp_field_op_subst}
	\hat{h}_{mn}(\ul{x}) \Psi[h_{mn}(\ul{x})] &= h_{mn}(\ul{x}) \cdot \Psi[h_{mn}(\ul{x})] \; , \\
	\hat{\pi}^{pq}(\ul{x}) \Psi[h_{mn}(\ul{x})] &= -i \frac{\delta}{\delta h_{pq}(\ul{x})} \Psi[h_{mn}(\ul{x})]
\eea
and satisfying the canonical commutation relations
\begin{gather}\begin{split}
	[ \hat{h}_{mn} (\ul{x}), \hat{\pi}^{pq} (\ul{x}^\prime) ] &= i \delta^p_{(m} \delta^q_{n)} \delta^{(3)} (\ul{x}-\ul{x}^\prime) \; , \\
	[ \hat{h}_{mn} (\ul{x}), \hat{h}_{pq} (\ul{x}^\prime) ] &= [ \hat{\pi}^{mn} (\ul{x}), \hat{\pi}^{pq} (\ul{x}^\prime) ] = 0 \; .
\end{split}\end{gather}
With the appearance of these commutation relations, the classical concept of space-time ceases to exist, owing to the uncertainty relation between $h_{mn}$ and $\pi^{mn}$ and the fact that $\pi^{mn}$ is proportional to the extrinsic curvature $K^{mn}$ (see \eqref{msp_momentum_extr_curv}).

Following the Dirac constraint quantisation~\cite{Dirac-LQM}, the classical constraints~\eqref*{msp_class_constraints} are imposed as operator conditions on the wave-functional $\Psi[h_{mn}(\ul{x})]$
\be\eqlabel{msp_quant_constraints}
	\hat{\hamil}_0 \Psi[h_{mn}(\ul{x})] = 0 \; , \qquad\qquad \hat{\hamil}_m \Psi[h_{np}(\ul{x})] = 0 \; ,
\ee
As in the classical case, these equations encapsulate the entire dynamics of the system. With expressions~\eqref*{msp_hamiltonians,msp_H0_dewitt_metric} for the Hamiltonian and the substitution rules in \eqref{msp_field_op_subst}, the quantum constraint equations read as follows
\begin{align}
	\hat{\hamil}_0 \Psi &= \left[ - 2\kappa^2 \, G_{mnpq} \frac{\delta^2}{\delta h_{mn}\delta h_{pq}} 
		- \frac{\sqrt{h}}{2\kappa^2} R^{(3)} \right] \Psi = 0 \; , \eqlabel{WdWeq} \\
	\hat{\hamil}^m \Psi &= 2i \nabla^{(3)}_n \frac{\delta\Psi}{\delta h_{mn}}= 0 \; . \eqlabel{qu_diff_constr}
\end{align}
The first of these equations is the famous Wheeler-DeWitt equation~\cite{DeWitt:1967yk,Wheeler:1988zr}, while the second one is known as quantum diffeomorphism constraints.

These are the central equations of quantum geometrodynamics and there are many interpretational, philosophical and mathematical issues associated with them. Besides primarily technical issues such as operator ordering ambiguities, there are also conceptual issues such as the well-known ``problem of time''~\cite{Isham:1992ms,Isham:1993ji,Butterfield:1998dd} -- one of the prime puzzles of quantum gravity. This is related to the observation that \eqrangeref{WdWeq}{qu_diff_constr} do not contain any reference to the time variable $\tau$. In particular, there is no time derivative-term in \eqref{WdWeq}, which is in stark contrast to the standard Schr\"odinger equation of ordinary quantum mechanics. Therefore, the concept of energy is lacking and the wave-functional $\Psi[h_{mn}(\ul{x})]$ itself is time-independent. Moreover, the interpretation of \eqrangeref{WdWeq}{qu_diff_constr} as evolution equations in time $\tau$ is lost and the only hope is to indirectly regain a derived notion of time from the only dynamical quantity $h_{mn}(\ul{x})$. Ultimately, this is a consequence of general coordinate invariance, which lies at the heart of general relativity.

Leaving all these problems aside and concentrating on the mathematical content, it becomes apparent that \eqrangeref{WdWeq}{qu_diff_constr} are mathematical monstrosities and performing concrete calculations is in general hugely difficult. This is because \eqrangeref{WdWeq}{qu_diff_constr} comprise an infinite number of functional-differential equations (one at each space point $\ul{x}$) in the wave-functional defined over an infinite-dimensional coordinate space (which is superspace $\mathcal{S}(\Sigma)$). This inherent complexity is the main motivation for studying mini-superspace models~\cite{Misner:1969ae}, to which we turn next. The central point of this approach is a specialisation to highly symmetric space-times to reduce the number of degrees of freedom to a finite value and therefore turning a quantum-field-theoretical problem into a more tractable quantum-mechanical problem.

\subsection{Mini-superspace quantisation}

The starting point of mini-superspace models~\cite{Misner:1969ae} is a specialisation of the general metric parametrisation~\eqref*{msp_metricansatz} to the form
\be\eqlabel{msp_reduced_metricansatz}
	ds^2 = - N(\tau)^2 d\tau^2 + h_{mn}(q^I(\tau)) dx^m dx^n \; ,
\ee
which is characterised by the absence of the shift vector, $N^m=0$, and the homogeneity (that is, translational invariance) of $N$ and $h_{mn}$. The degrees of freedom of the three-metric are labelled $q^I$, where $I=1,\ldots,N_f$. Comparing with~\eqref*{gansatz} makes it immediately apparent why this formalism is relevant to our Calabi-Yau five-fold reductions.

Plugging the mini-superspace ansatz~\eqref*{msp_reduced_metricansatz} into the Einstein-Hilbert action and using the intermediate result~\eqref*{lagr_msp_prep} yields the mini-superspace action
\be\eqlabel{msp_action}
	S 	= \oneon{2\kappa^2} \int d\tau\, L 
		= \oneon{2\kappa^2} \int d\tau\left\{N^{-1}\,G_{IJ}(\ul{q})\dot{q}^I \dot{q}^J - \frac{N}{2} \,{\cal U}(\ul{q})\right\} ,
\ee
with the mini-superspace version $G_{IJ}(\ul{q})$ of the DeWitt-metric (sometimes called ``mini-supermetric'')
\be
	G_{IJ}(\ul{q}) \equiv \oneon{4} \int_\Sigma d^3 x\, G^{mnpq} h_{mn,I} h_{pq,J} \; .
\ee
The action~\eqref*{msp_action} is of the same form as the non-linear sigma model~\eqref*{msp_1d_sigma_model_action} obtained by reduction from eleven dimensions. The mini-supermetric $G_{IJ}(\ul{q})$ inherits the Lorentzian signature $(-,+,\ldots,+)$ from the DeWitt-metric~\cite{DeWitt:1967yk}. 
The classical equations of motion derived from this action read as follows
\begin{gather}
	\frac{1}{N}\frac{d}{d\tau}\left(\frac{\dot{q}^I}{N}\right) + \Gamma^I_{JK} \frac{\dot{q}^J}{N} \frac{\dot{q}^K}{N} + 
	\oneon{4} G^{IJ}\frac{\partial{\cal U}}{\partial q^J} = 0 \; , \\
	\frac{2}{N^2} G_{IJ} \dot{q}^I \dot{q}^J + {\cal U} = 0 \; ,
\end{gather}
where $\Gamma^I_{JK} = \oneon{2} G^{IL} \left( G_{LJ,K} + G_{LK,J} - G_{JK,L} \right)$ is the Christoffel symbol associated to $G_{IJ}$. The Hamiltonian is readily found to be
\vspace{\parskip}
\begin{align}
	H &= p_N \dot{N} + p_I \dot{q}^I - L = N H_0 \; , \eqlabel{msp_1d_H} \\
	H_0 &= \oneon{4} G^{IJ} p_I p_J + \oneon{2} {\cal U} \; , \eqlabel{msp_1d_H0}
\end{align}
\vspace{\parskip}\noindent 
where $G^{IJ}$ is the inverse of $G_{IJ}$ and we have used $p_N = \partial L/\partial \dot{N} = 0$ and $p_I = \partial L/\partial \dot{q}^I = 2 N^{-1} G_{IJ} \dot{q}^J$ (with canonical Poisson brackets $\{ q^I, p_J \} = \delta^I_J$).

In order to canonically quantise this system, one first introduces position and momentum operators as usual by
\vspace{\parskip}
\be\eqlabel{msp_qp_ops}
	\hat{q}^I \psi(\ul{q}) = q^I \cdot \psi(\ul{q}) \; , \qquad\qquad
	\hat{p}_I \psi(\ul{q}) = -i \frac{\partial}{\partial q^I} \psi(\ul{q}) \; ,
\ee
\vspace{\parskip}\noindent 
which satisfy the canonical commutation relations
\vspace{\parskip}
\be
	[ \hat{q}^I, \hat{p}_J ] = i \delta^I_J \; , \qquad\qquad [ \hat{q}^I, \hat{q}^J ] = [ \hat{p}_I, \hat{p}_J ] = 0 \; .
\ee
\vspace{\parskip}\noindent 
The next step is to promote $H_0$ to an operator. At this point, however, one encounters a factor ordering ambiguity in the first term in~\eqref*{msp_1d_H0}, since
\vspace{\parskip}
\bea
	&G^{IJ}(\ul{q}) p_I p_J = p_I G^{IJ}(\ul{q}) p_J = p_I p_J G^{IJ}(\ul{q}) \; \text{,$\;$ but} \\
	&G^{IJ}(\ul{q}) \hat{p}_I \hat{p}_J \neq \hat{p}_I G^{IJ}(\ul{q}) \hat{p}_J \neq \hat{p}_I \hat{p}_J G^{IJ}(\ul{q}) \; .
\eea
\vspace{\parskip}\noindent
Several ``natural choices'' to resolve this ambiguity have been proposed~\cite{DeWitt:1967yk,Hawking:1985bk,Louko:1988zb,Misner:1972js,Halliwell:1988wc,Moss:1988wk} and a thorough discussion of this problem is beyond the scope of this section. Here, we simply adopt the common choice (see, for example, refs.~\cite{Wald:1984rg,Kiefer:2004gr,Halliwell:1990uy})
\be
	G^{IJ}(\ul{q}) \hat{p}_I \hat{p}_J \rightarrow 
	- \nabla^2_{{\rm LB}} = - \oneon{\sqrt{-G}} \partial_I \left( \sqrt{-G}\, G^{IJ} \partial_J \right) \; ,
\ee
which has the virtue of covariantising the Wheeler-DeWitt equation. Here, $G\equiv\det G_{IJ}$. The so-called Laplace-Beltrami operator $\nabla^2_{{\rm LB}}$ is the covariant generalisation of the Laplacian with respect to the mini-su\-per\-metric $G_{IJ}$. With this choice of factor ordering, the Wheeler-DeWitt equation of mini-superspace reads
\be\eqlabel{msp_qm_WdW_eq}
	\hat{H}_0 \psi(\ul{q}) = \left[ - \oneon{4} \nabla^2_{{\rm LB}} + \oneon{2} {\cal U}(\ul{q}) \right] \psi(\ul{q}) = 0 \; ,
\ee
which bears a remarkable resemblance to the Klein-Gordon equation.

Although mini-superspace models may be regarded as quantum mechanical ``toy models'' of quantum geometrodynamics, which is a full-fledged quantum field theory, thereby making analytic calculations possible, many of the interpretational and conceptual problems remain. An important question is what the physical interpretation of the wave-function $\psi(\ul{q})$ is. As in ordinary quantum theory, one would like to construct a probability measure, which then allows one to make predictions and there are several competing proposals of such a probability measure.

Perhaps the simplest approach is to precisely mimic quantum mechanics and use the canonical inner product
\be
	\langle \psi_1 , \psi_2 \rangle \equiv \int d^{N_f} q\, \sqrt{-G}\, \psi_1^\ast \psi_2 \; .
\ee
The probability $P(\psi,A)$ of the universe $\psi$ being in the interval $A=(\ul{q},\ul{q}+\delta\ul{q})$ is then given by
\be
	P(\psi,A) = \sqrt{-G}\, |\psi(\ul{q})|^2 \, \delta q^1 \cdots \delta q^{N_f} \; .
\ee
However, this merely yields relative probabilities and turning it into an absolute probability founders on the fact that even simple solutions to \eqref{msp_qm_WdW_eq} are often not normalisable, that is $\langle \psi , \psi \rangle = \infty$.

Another problem is the choice of suitable boundary conditions when solving the Wheeler-DeWitt equation~\eqref*{msp_qm_WdW_eq}. Again several competing proposals exist in the literature. The first one dates back to the early days of quantum geometrodynamics and is called DeWitt's boundary condition~\cite{DeWitt:1967yk}. More recent proposals are the no-boundary condition (or Hartle-Hawking proposal)~\cite{Hartle:1983ai} and the tunnelling condition put forward by Vilenkin~\cite{Vilenkin:1987kf}. A proper discussion of these proposals requires concepts not introduced here (such as path integral quantisation and conserved currents) and we hence refrain from doing so. Instead, we now return to our one-dimensional effective action obtained from our M-theory reduction and take a first look at how to quantise it \`a la mini-superspace.

\section[Mini-superspace meets Calabi-Yau five-folds]{Mini-superspace meets Calabi-Yau five-folds\footnote{A more general treatment of the mini-superspace quantisation of M-theory (for general 10-manifolds, that is not restricted to Calabi-Yau five-folds) can be found in ref.~\cite{Grigorian:2006yva}.}}\seclabel{msp_cy5}

In \secref{class_sol}, we described the bosonic effective action and studied some classical solutions. Ignoring the Chern-Simons term~\eqref*{1deff_bos_CS} led to the action~\eqref*{msp_1d_sigma_model_action}, which is precisely equivalent to the mini-superspace action~\eqref*{msp_action} and the quantisation proceeds exactly as presented in the previous section. If we include the Chern-Simons term, the action schematically reads
\be
	S = \frac{l}{2}\int d\tau\left\{N^{-1}\,G_{IJ}(\underline{\phi})\dot{\phi}^I\dot{\phi}^J - \frac{N}{2} \,{\cal U}(\underline{\phi})
	- H_{IJ} \dot{\phi}^I \phi^J \right\} ,
\ee
where $H_{IJ} = - H_{JI}$ is a constant anti-symmetric matrix composed out of the flux parameters $n^e$
\be
	H_{\mathcal{PQ}} = \frac{\gamma_1}{3} d_{\mathcal{PQ}e} n^e \; , \qquad\qquad H_{IJ} = 0 \; , \quad \text{for $I,J\neq \mathcal{P,Q}$}.
\ee
In other words, only for 3-form indices is $H$ non-zero. In the mini-superspace picture, the metric $G_{IJ}$ corresponds to the mini-supermetric of the superspace of all 10-geometries on the Calabi-Yau five-fold $X$ together with the moduli space of the degrees of freedom originating from the M-theory 3-form. Due to the anti-symmetry of $H_{IJ}$, the Chern-Simons term cannot be written as a total derivative.

The canonically conjugate momenta $p_I$ receive an extra contribution from the Chern-Simons term
\be
	p_I = \frac{\partial L}{\partial \dot{\phi}^I} = \frac{2}{N} G_{IJ} \dot{\phi}^J - H_{IJ} \phi^J \; ,
\ee
which leads to the following expression for the Hamiltonian
\be\eqlabel{msp_1d_eff_full_hamilt}
	H = N H_0 = N \left[ \oneon{4} G^{IJ} p_I p_J + \oneon{2} {\cal U} 
		+ \oneon{2} H_{IJ} \phi^I p^J - \oneon{4} G^{IJ} H_{IK} H_{JL} \phi^K \phi^L \right] ,
\ee
where $p^I \equiv G^{IJ} p_J$. Note that the last term is of the same order in the $\gamma$-expansion as the scalar potential ${\cal U}$. The Hamiltonian constraint $H_0 = 0$, originating from the equation of motion of the lapse $N$, gives rise to the invariance under worldline reparametrizations $\tau\rightarrow\tau^\prime(\tau)$ as discussed in \secref{symm_1d_eff_action}.

Upon promoting $\phi^I$ and $p_I$ to operators following \eqref*{msp_qp_ops}, one encounters, at first sight, another operator ordering ambiguity in the third term in~\eqref*{msp_1d_eff_full_hamilt}
\bea
	H_{IJ} \phi^I p^J = &H_{IJ} G^{JK} \phi^I p_K = H_{IJ} G^{JK} p_K \phi^I = H_{IJ} p_K G^{JK} \phi^I \; \text{,$\;$ but} \\
		&H_{IJ} \hat{G}^{JK} \hat{\phi}^I \hat{p}_K \neq H_{IJ} \hat{G}^{JK} \hat{p}_K \hat{\phi}^I 
		\neq H_{IJ} \hat{p}_K \hat{G}^{JK} \hat{\phi}^I \; .
\eea
A closer inspection reveals, that this is in fact partially resolved thanks to the anti-sym\-metry of $H_{IJ}$, for
\be
	H_{IJ} \hat{G}^{JK} \hat{p}_K \hat{\phi}^I = - i H_{IJ} G^{JK} \partial_K \phi^I = -i\, \tr\, H = 0 \; .
\ee
It remains the case where $\hat{p}_K$ is to the left of $\hat{G}^{JK}$. It is important to recall the particular forms of $G_{IJ}$ and $H_{IJ}$: the metric $G_{IJ}$ is a block-diagonal matrix in the index blocks $I=(i, {\cal P}, a, \bar{b})$ and $H_{IJ}$ is zero unless $I,J=\mathcal{P,Q}$. Thus,
\be
	H_{IJ} \hat{p}_K \hat{G}^{JK} \hat{\phi}^I = H_{\mathcal{PQ}} \hat{p}_K \hat{G}^{\mathcal{Q}K} \hat{\phi}^{\cal{P}} = 
	H_{\mathcal{PQ}} \hat{p}_{\cal{R}} \hat{G}^{\mathcal{QR}} \hat{\phi}^{\cal{P}} = 
	-i H_{\mathcal{PQ}} \partial_{\cal{R}} (G^{\mathcal{QR}}) \phi^{\cal{P}} \; .
\ee
The last expression vanishes since the metric $G_{IJ}$ (and its inverse) only depends on the K\"ahler and complex structure moduli $t^i$ and $z^a$, respectively, as can be seen from~\eqref*{1deff_bos_kin}. This resolves the apparent factor ordering ambiguity and one instead has the universal replacement
\be
	H_{IJ} \phi^I p^J \to H_{IJ} \hat{G}^{JK} \hat{\phi}^I \hat{p}_K = -i H_{IJ} G^{JK} \phi^I \partial_K \; ,
\ee
The Wheeler-DeWitt equation of our effective theory thus reads as follows
\be
	\left[ - \oneon{4} \nabla^2_{{\rm LB}} + \oneon{2} {\cal U}(\ul{\phi})
	- \frac{i}{2} H_{IJ} \phi^I \nabla^J - \oneon{4} H_{IJ} H^I {}_K \phi^J \phi^K \right] \psi(\ul{\phi}) = 0 \; ,
\ee
where $H^I {}_K \equiv G^{IJ} H_{JK}$ and $\nabla^I \psi \equiv G^{IJ} \nabla_J \psi = G^{IJ} \partial_J \psi$, since the wave-function behaves like a scalar in mini-superspace. In this way, the Wheeler-DeWitt equation is written in a manifestly covariant form.

\subsection{An (admittedly pathological) example}

In \secref{cy5_examples}, we presented some explicit examples of Calabi-Yau five-folds. All the CICYs with $q^r_\al>0$ and the torus quotient have the property $h^{1,2}(X)=0$ (and hence $b^3(X)=0$). Therefore, $H_{IJ}=0$ in all our examples and we are left with the original Wheeler-DeWitt equation
\be
	\left[ - \oneon{4} \nabla^2_{{\rm LB}} + \oneon{2} {\cal U}(\ul{\phi}) \right] \psi(\ul{\phi}) = 0 \; .
\ee
As a first example, we consider the CICYs defined in a single projective space since their Hodge diamond is particularly simple (see \tabref{tab:cicy1}). We just have a single $(1,1)$-modulus $t$, which is related to the volume via $V(t) = d\cdot t^5/5!$, where $d$ is a positive integer defined as
\be
	d \equiv \int_X J^5 = \prod_{\al=1}^K q^1_\al
\ee
and $t>0$ so that $V>0$. 

However, even for these simplest of all CICY examples, the complex structure moduli space is highly non-trivial: $h^{1,4}(X)$ is of the order $10^2 - 10^3$. For now, we will therefore concentrate on the special case where the $(1,4)$-moduli are frozen. In other words, we restrict our attention to wave-functions that are constant over the complex structure moduli space, that is $\psi = \psi(t)$.

In this case, the Wheeler-DeWitt equation reduces to a second order ordinary differential equation of the following universal form
\be\eqlabel{WdW_cicy1}
	\psi^{\prime\prime}(t) - \frac{3}{2t} \psi^\prime (t) - (\gamma_1 d k)^2 t^4 \psi(t) = 0 \; ,
\ee
where $(\cdot)^\prime \equiv d/dt$ and $k$ is the flux parameter $g = k J^2$. In the absence of a potential (that is, for $k=0$), the general solution takes the form
\be
	\psi (t) = A t^{\frac{5}{2}} + B \; ,
\ee
with integration constants $A,B\in\R$. This is an example of a non-normalisable wave-function.

\begin{figure}[t]
\centering
\includegraphics[width=10cm]{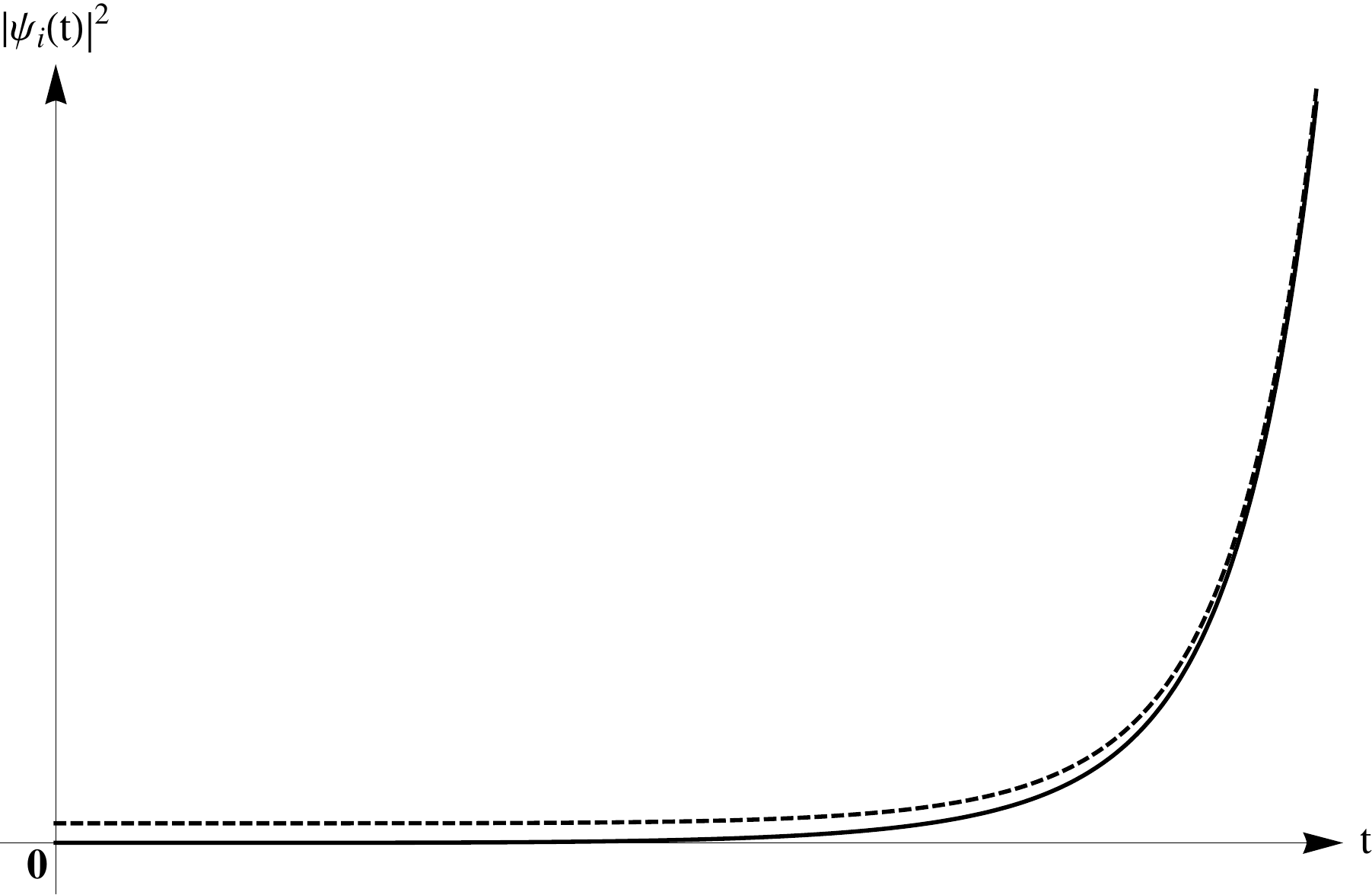}
\caption{Qualitative behaviour of $|\psi_1(t)|^2$ (dashed line) and $|\psi_2(t)|^2$ (solid line).}\figlabel{wf_cicy1}
\end{figure}
Returning to the case $k\neq 0$, we see that the sign of the potential is always negative, ${\cal U} < 0$, and from the experience with ordinary quantum mechanics, one expects an exponential behaviour of $\psi(t)$. Indeed, explicitly solving \eqref{WdW_cicy1} leads to
\bea
	\psi(t) &= \psi_1 (t) + \psi_2 (t) \; , \\
	\psi_1 (t) &= A\, t^{\frac{5}{4}} I_{-\frac{5}{12}} \left(\frac{t^3}{3}\right) , \\
	\psi_2 (t) &= B\, t^{\frac{5}{4}} I_{\frac{5}{12}}  \left(\frac{t^3}{3}\right) .
\eea
where again $A$ and $B$ are arbitrary integration constants and we have set $(\gamma_1 d k)^2 = 1$, without loss of generality. The functions $I_\al (x)$ are the modified Bessel functions of the first kind
\be
	I_\al (x) = \sum_{m=0}^\infty \oneon{m! \Gamma(m+\alpha+1)} {\left({\frac{x}{2}}\right)}^{2m+\alpha} \; ,
\ee
which are solutions of the modified Bessel's differential equation $x^2 y^{\prime\prime}(x) + x y^\prime (x) - \left( x^2 + \al^2 \right) y(x) = 0$. The qualitative behaviour of $|\psi_1(t)|^2$ and $|\psi_2(t)|^2$ is shown in \figref{wf_cicy1}. The wave-functions $\psi_1 (t)$ and $\psi_2 (t)$ differ in their behaviour at the origin, namely
\be
	\psi_1 \to 6^{5/12} (\Gamma(7/12))^{-1} A \sim 1.38 \, A \; , \qquad
	\psi_2 \to 0 \; , \qquad \text{as $t\to 0$.}
\ee
For large $t$, both $\psi_1 (t)$ and $\psi_2 (t)$ diverge exponentially. In other words, a large Calabi-Yau volume is infinitely more probable than a small one. This is in accord with the classical behaviour of decompactification exhibited by this model, which was discussed in \secref{class_sol}.
%
%

\chapter{F-theory on Calabi-Yau Five-Folds}\chlabel{fcy5}

We begin with a few cautionary remarks to save the reader from disappointment when approaching this chapter. Since its inception in 1996, F-theory~\cite{Hull:1995xh,Vafa:1996xn} has developed into a vast subject that is involved both mathematically and conceptually (the latter being mainly due to the ``threat'' to its physical interpretability from the presence of more than one time direction). A thorough study of the topic written in the chapter heading could potentially constitute another thesis in its own right. Here, we will instead take a very narrow view of F-theory and restrict our attention to a particular proposal by Khviengia, Khviengia, L\"u and Pope (KKLP)~\cite{Khviengia:1997rh} of a 12-dimensional Lagrangian that is possibly related to the full F-theory. After briefly presenting this proposal, we will concentrate, for the remainder of this chapter, on the application of our Calabi-Yau five-fold program to this case.

Like in the previous chapter, we can at present merely provide a snapshot of (unfinished and unpublished) work in progress. In this way however, the attention is smoothly drawn towards the concluding remarks in the next chapter, where we end with an outlook and a list of future directions.

\section[The KKLP proposal]{The KKLP proposal\footnote{This is not a misprint and should not be confused with the KKLT proposal (named after Kachru, Kallosh, Linde and Trivedi) in the context of string cosmology.}}

The starting point of the KKLP proposal~\cite{Khviengia:1997rh} is a bosonic field theory in 12 dimensions whose field content comprises the metric $g_{MN}$, a dilaton $\psi$ and Abelian 3- and 4-form gauge fields $A$ and $B$, respectively. With the formul\ae\ of \secref{sugra11}, one counts $54$, $1$, $120$ and $210$ independent on-shell degrees of freedom contained in $g_{MN}$, $\psi$, $A$ and $B$, respectively. This adds up to a total of $54+1+120+210 = 385$ bosonic on-shell degrees of freedom. In the KKLP proposal, it is conjectured that these fields are governed by the following 12-dimensional action~\cite{Khviengia:1997rh}
\be\eqlabel{S12}
	S_{12} = \oneon{2\kappa^2_{12}} \int_\manifold \left\{ R\ast 1 - \oneon{2} d\psi \wedge\ast d\psi
		- \oneon{2} e^{\al\psi} F\wedge\ast F - \oneon{2} e^{\beta\psi} G\wedge\ast G + \gamma B\wedge F\wedge F \right\} ,
\ee
where $\kappa_{12}$ is the anticipated 12-dimensional gravitational constant, $F=dA$ is the 4-form field strength of the 3-form gauge field $A$, $G=dB$ is the 5-form field strength of the 4-form gauge field $B$, $R$ is the Ricci scalar of the 12-dimensional metric $g_{MN}$, $\manifold$ is the space-time manifold and $\al$, $\beta$ and $\gamma$ are \emph{a priori} arbitrary coupling parameters. The equations of motion for $g_{MN}$, $\psi$, $A$ and $B$ derived from this action are respectively given by
\begin{align}
	R_{MN} &= \oneon{2} (\partial_M \psi)(\partial_N \psi) + \oneon{12} e^{\al\psi} F_{M M_2 \ldots M_4} F_N {}^{M_2 \ldots M_4}
		- \oneon{160} e^{\al\psi} g_{MN} F^2 \nonumber\\ &\hspace*{3.5cm}
		+ \oneon{48} e^{\beta\psi} G_{M M_2 \ldots M_5} G_N {}^{M_2 \ldots M_5}
		- \oneon{600} e^{\beta\psi} g_{MN} G^2 \; , \\
	d\ast d\psi &= \frac{\al}{2} e^{\al\psi} F\wedge\ast F + \frac{\beta}{2} e^{\beta\psi} G\wedge\ast G \; , \\
	d\ast \left( e^{\al\psi} F \right) &= \frac{\gamma}{3} G\wedge F \; , \\
	d\ast \left( e^{\beta\psi} G \right) &= - \gamma F\wedge F
\end{align}
written succinctly in differential form language. In the first equation, we have introduced the abbreviations $F^2 \equiv F_{M_1 \ldots M_4} F^{M_1 \ldots M_4}$ and $G^2 \equiv G_{M_1 \ldots M_5} G^{M_1 \ldots M_5}$.

The action~\eqref*{S12} is singled out essentially uniquely by the following two requirements: First, upon dimensional reduction on a circle and in conjunction with a certain consistent truncation, the theory should yield the bosonic part of 11-dimensional supergravity. Second, when dimensionally reduced on a 2-torus, it should be possible to reproduce the bosonic sector of 10-dimensional type IIB supergravity (bar the 5-form self-duality constraint), provided the reduction is, once again, accompanied by a suitable consistent truncation. 

The crucial point here is demanding the existence of the two inequivalent consistent truncations to the field content of 11-dimensional supergravity and type IIB supergravity, respectively. This highly non-trivial requirement poses strong constraints on the possible form of the 12-dimensional theory and fixes the action~\eqref*{S12}. The truncations are necessary because the field content of the 12-dimensional candidate theory is larger than that of the reduced theories (385 versus 128). This enlarged field content, in turn, is necessary to accommodate all fields of the lower dimensional theories and is a consequence of the different field structure of 11-dimensional supergravity and type IIB supergravity.

Even the parameters $\al$, $\beta$ and $\gamma$ are largely fixed by these requirements. In order to reproduce 11-dimensional supergravity, $\gamma$ must be equal to
\be
	\gamma = \frac{\sqrt{3}}{4} \; .
\ee
Moreover, if the theory is to yield the 10-dimensional dilatonic structure with the global classical $\mathrm{SL}(2,\R)$ symmetry of type IIB supergravity, the two remaining parameters $\al$ and $\beta$ need to satisfy
\be
	\al^2 = - \oneon{5} \; , \qquad\qquad \beta^2 = - \frac{4}{5}
\ee
and hence, $\al$ and $\beta$ are imaginary couplings. It should be noted~\cite{Khviengia:1997rh} that $\al$ and $\beta$ can be turned into real couplings by a field re-definition of the 12-dimensional dilaton $\psi \to i \psi$. However, $\psi$ then has the wrong sign for its kinetic term.

Further support for the field content and the form of the action of the 12-dimensional theory is found by oxidising $p$-branes from type IIB and 11-dimensional supergravity to 12 dimensions~\cite{Khviengia:1997rh}. In other words, it is shown how the $p$-brane solutions of type IIB and 11-dimensional supergravity follow from classical solutions of the 12-dimensional theory. However, upon oxidising the M2- and M5-branes from 11 to 12 dimensions, the brane-like structure is lost. That is, the M-theory branes do not descend from F-theory branes in a straightforward way. As for the type IIB branes, they can all be uplifted to 12 dimensions and most of them admit a brane-like interpretation in terms of various configurations of 12-dimensional F-branes ($D=12$ membranes), 3-branes and 6-branes.

After summarising the evidence for the existence of this 12-dimensional field theory, we will now briefly describe the truncations necessary to extract type IIB and 11-dimensional supergravity from it. We begin with the simpler case of 11-dimensional supergravity. The truncation precedes a dimensional reduction on a circle, $\manifold = \manifold^{11} \times S^1$. After splitting the 12-dimensional coordinates into $x^M = (x^m, z)$, where $z$ is the coordinate on the circle, the truncation can be written as follows
\be\eqlabel{KKLP_trunc_sugra11}
	G_{m_1\ldots m_5} = 0\; , \qquad F_{mnp z} = 0 \; , \qquad \mathcal{F} = 0 \; , \qquad \psi = -i \phi \; , \qquad 
	F_{mnpq} = \sqrt{2}\, G_{mnpq z} \; ,
\ee
where $\mathcal{F} = d\mathcal{A}$ is the field strength corresponding to the Kaluza-Klein vector $\mathcal{A}_m = e^{(3/\sqrt{5})\phi} g_{mz}$ and $\phi$ is the Kaluza-Klein scalar $g_{zz} = e^{-(3/\sqrt{5})\phi}$. It is, of course, not obvious that this also constitutes a consistent truncation, but KKLP show that this is indeed the case. The M-theory 3-form $\tilde{A}$ is, up to a rescaling, obtained from the 11-dimensional components of $A$ (or alternatively from $B_{mnp z}$) via $\tilde{A}_{mnp} = \sqrt{3/2}\, A_{mnp} = \sqrt{3}\, B_{mnp z}$. Through the choice $\psi = -i \phi$, the dilaton disappears and one is left with precisely the field content of pure 11-dimensional supergravity.

Though not directly relevant for our purposes, we state, for completeness, the consistent truncation necessary to arrive at type IIB supergravity after the reduction on $\manifold = \manifold^{10} \times T^2$ has been performed. The coordinates are labelled $x^M = (x^m, z_1, z_2)$, with $z_1$ ($z_2$) being the coordinate on the first (second) circle. The truncation then reads as follows
\be\eqlabel{KKLP_trunc_IIB}
	F_{mnpq} = G_{mnpq z_1} = G_{mnpq z_2} = G_{mnp z_1 z_2} = F_{mn z_1 z_2} = \mathcal{F}^{(1)} = \mathcal{F}^{(2)} = 0 \; , \qquad
	\psi = i\varphi \; ,
\ee
where $\mathcal{F}^{(1)}$ ($\mathcal{F}^{(2)}$) is the field strength corresponding to the Kaluza-Klein vector $\mathcal{A}^{(1)}$ ($\mathcal{A}^{(2)}$) of the first (second) circle and $\varphi = \frac{2}{3} \phi_1 + \frac{\sqrt{5}}{3} \phi_2$ with $g_{z_1 z_1} = e^{-(3/\sqrt{5})\phi_1}$ and $g_{z_2 z_2} = e^{-(4/3)\phi_2}$. The type IIB fields (denoted by tilded symbols) are identified with the 12-dimensional fields in the following way: $\tilde{g}_{mn} = g_{mn}$, $\tilde{B}_{mn} = A_{mn z_1}$, $\tilde{\phi} = \frac{\sqrt{5}}{3} \phi_1 - \frac{2}{3} \phi_2$, $\tilde{C}^{(0)} = \mathcal{A}^{(1)}_{z_2}$, $\tilde{C}^{(2)}_{mn} = A_{mn z_2}$ and $\tilde{C}^{(4)}_{mnpq} = B_{mnpq}$. Here, $\tilde{g}_{mn}$, $\tilde{B}$, $\tilde{\phi}$, $\tilde{C}^{(0)}$, $\tilde{C}^{(2)}$ and $\tilde{C}^{(4)}$ label the metric, NS-NS B-field, dilaton and R-R 0-, 2- and 4-form fields, respectively.

\subsection{A word about supersymmetry and the space-time signature}

Any field theoretical study of a 12-dimensional theory that is claimed to be related to string-/M-theory, must of course be accompanied by a discussion of the issues with supersymmetry and the space-time signature. The 12-dimensional Lorentz group $\SO(11,1)$ admits both Majorana and Weyl (but not Majorana-Weyl) spinors as its smallest possible spinor representation. However, both the Majorana and the Weyl representation have 64 real components (see, for example, ref.~\cite{VanProeyen:1999ni}) and the smallest supergravity multiplet (a multiplet with a spin-2 field) gives rise to fields with spin greater than two. It is generally believed that these higher spin fields lead to inconsistencies.\footnote{It has been argued however, that it might be possible to avoid those problems, for example by giving up full 12-dimensional Lorentz invariance, or by restricting to higher spin fields that do not couple to other fields, or by letting the higher spin fields have negative energy.} 

To remedy this situation, it has been suggested to relax the requirement of Lo\-rentzi\-an signature and instead to consider metrics with more general pseudo-Riemannian signature $(-,\ldots,-,+,\ldots,+)$. Indeed, $\SO(10,2)$ possesses a Majorana-Weyl representation with 32 real components, which is exactly the highest number of real supercharges believed to be possible in a consistent supergravity theory. However, the price one has to pay is the presence of two time directions, which potentially means sacrificing (or at least challenging) physical interpretability. For example, with several time directions present, the concept of energy ceases to be well-defined and thus, the action~\eqref*{S12} can no longer be regarded as a low-energy effective field theory of some elusive complete theory. This is one reason why, in a more conservative and pragmatic approach, F-theory is commonly regarded as just a very useful ``book-keeping device'' to organise non-perturbative solutions of type IIB superstring theory. Interpreted in this way, the problem of multiple time directions disappears, since the additional two dimensions of the two-torus are not taken to exist physically and hence, their signature is physically irrelevant.

Returning to the KKLP proposal, one first notices that a detailed discussion of the aforementioned issues is carefully avoided in the original paper~\cite{Khviengia:1997rh}. Fortunately, the general analysis can be carried out in a way that is, to a large degree, independent of the space-time signature and the signature is therefore deliberately left unspecified wherever possible. For example, this can be seen by noting that the action~\eqref*{S12} is manifestly independent of the space-time signature. That is, none of the terms in~\eqref*{S12} changes sign when changing between $(11,1)$- and $(10,2)$-signature. However, regardless of the space-time signature, not all is well when supersymmetry comes into play. On a practical level, it is difficult to see how the 385 bosonic degrees of freedom can be matched with an equal number of fermionic degrees of freedom. It is particularly problematic that the number of bosonic degrees of freedom is odd. A gravitino in $(10,2)$- and $(11,1)$-signature has 144 and 288 on-shell degrees of freedom, respectively. Most intriguingly, $385-1=384= 12 \times 32$, which suggests a possible matching of degrees of freedom upon removal of a bosonic degree of freedom. A natural candidate for removal is the 12-dimensional dilaton $\psi$. However, $\psi$ is a crucial ingredient in the reductions and truncations to type IIB and 11-dimensional supergravity, which, in turn, are the main arguments for the existence of~\eqref*{S12}. The next integer greater than 385 and divisible by 32 is $416 = 13\times 32$, which could be reached by one gravitino and four spin-$1/2$ particles in an $(11,1)$-signature ($416=288+4\times 32$). The excess of $31=416-385$ bosons could be compensated by introducing three additional vectors ($3\times 10$) and a scalar. However, it is not clear how (and if) these extra bosonic fields can be integrated into the KKLP proposal. Besides that, we have merely presented numerology and there are compelling arguments why supergravity models cannot exist in 12 dimensions~\cite{Nahm:1977tg,Castellani:1982ke}.

We will leave this undecided question aside now and instead adopt the approach by KKLP, which is to leave the space-time signature unspecified if possible and keep the discussion general so that it applies to both $(11,1)$ and $(10,2)$. In the remainder of this chapter, we will apply our Calabi-Yau five-fold reduction program to the 12-dimensional field theory proposed by KKLP.

\section{KKLP and Calabi-Yau Five-Folds: From \texorpdfstring{$D=12$}{D=12} to \texorpdfstring{$D=2$}{D=2}}\seclabel{fcy5_KKLP_CY5_12_2}

In this section, we consider the Kaluza-Klein reduction of the 12-dimensional action given by~\eqref*{S12} on a space-time manifold of the form $\manifold = \Sigma \times X$, where $X$ is a Calabi-Yau five-fold and $\Sigma$ is a two-dimensional (pseudo-)Riemannian manifold. This is interesting, because as we will see, $\Sigma$ can be interpreted as a string-worldsheet embedded into the moduli space of Calabi-Yau five-folds.

We will proceed along the lines of the bosonic reduction of M-theory on Calabi-Yau five-folds presented in \secref{bos_red}. In particular, we will employ the same notation for the moduli spaces and follow the same logical order. We thus start by presenting the Kaluza-Klein ansatz for the bosonic fields on $\manifold = \Sigma \times X$.

\subsection{The reduction ansatz}\seclabel{fcy5_bos_red_ansatz}

We begin with the ansatz for the 12-dimensional line element
\be\eqlabel{fcy5_gansatz}
	ds^2 = \gamma_{\al\beta} d\sigma^\al d\sigma^\beta + g_{mn} (t^i,z^a,\bar{z}^{\bar{a}}) dx^m dx^n \; ,
\ee
where $(\sigma^\al, \gamma_{\al\beta})$ and $(x^m, g_{mn})$ are the local coordinates and metrics on $\Sigma$ and $X$, respectively. The indices take the values $\al,\beta,\ldots=1,2$ and $m,n,\ldots = 1,\ldots,10$. The Calabi-Yau metric $g_{mn}$ is Ricci flat and parametrised by the K\"ahler and complex structure moduli denoted $t^i = t^i(\sigma^\al)$ and $z^a=z^a(\sigma^\al)$, respectively. As in the one-dimensional case in \chref{mcy5}, gravity in two-dimensions is non-dynamical, but we will keep $\gamma_{\al\beta}$ as a field, for now. While $g_{mn}$ has purely positive signature, the two-dimensional metric $\gamma_{\al\beta}$ has either $(-,+)$ or $(-,-)$ signature depending on whether one starts with $(11,1)$ or $(10,2)$ signature in 12 dimensions. Note that the most general ansatz does not contain a Kaluza-Klein vector, owing to $h^{1,0}(X)=0$.

The next field is the dilaton, which is a scalar and thus we merely drop the $x^m$-dependence
\be\eqlabel{fcy5_psiansatz}
	\psi (\sigma^\al, x^m) = \psi(\sigma^\al)
\ee
to arrive at the proper Kaluza-Klein ansatz.

Next, we need to consider the form fields. We begin with the 3-form gauge field
\be\eqlabel{fcy5_Aansatz}
	A = X^{\cal{P}} N_{\cal{P}} + A^i \wedge \omega_i \; ,
\ee
with $A^i = A^i_\al d\sigma^\al$ and otherwise using the same notation as in~\eqref*{Aansatz_3forms}. Vectors do not contain any dynamical degrees of freedom in two dimensions and hence $A^i$ can be set to zero right from the outset. Like the 2-metric $\gamma_{\al\beta}$, we will nonetheless keep it for now and set it to zero later.

Finally, the 4-form gauge field $B$ is decomposed as follows
\be\eqlabel{fcy5_Bansatz}
	B = b^{\cal{X}} O_{\cal{X}} + B^{\cal{P}} \wedge N_{\cal{P}} + C^i \wedge \omega_i \; ,
\ee
with $b^4(X)$ real scalars $b^{\cal{X}}$, $b^3(X)$ real vectors $B^{\cal{P}} = B^{\cal{P}}_\al d\sigma^\al$ and $h^{1,1}(X)$ real 2-form fields $C^i = \oneon{2} C^i_{\al\beta} d\sigma^\al \wedge d\sigma^\beta$. The 2-form fields can be dualised to scalars $C^i \sim C^i_{\al\beta} \epsilon^{\al\beta} d^2 \sigma = \tilde{C}^i d^2 \sigma$ and, just as the vectors $A^i$ and $B^p$, are not dynamical. In fact, $C^i \wedge \omega_i$ is closed and thus the $\tilde{C}^i$ play the same r\^ole as the gauge degrees of freedom $\mu_i$ in~\eqref*{Aansatz}.

\subsection{Performing the bosonic reduction}

We will now plug the ansatz into the action~\eqref*{S12} and perform a Kaluza-Klein reduction. Let us begin with the Einstein-Hilbert term. The calculation is largely similar to the one in \secref{bos_red_comp} and will be omitted save for some instructive intermediate results. The first step is calculating the Christoffel symbols for the metric ansatz~\eqref*{fcy5_gansatz}. One finds
\bea
	\Gamma^\al_{\beta\gamma} 	&= \Gamma^{(2)\al}_{\beta\gamma} \; , \qquad
	&\Gamma^\al_{\beta m} 		&= 0 \; , \qquad
	&\Gamma^\al_{mn} 			&= - \oneon{2} \gamma^{\al\beta} g_{mn,\beta} \; , \\
	\Gamma^m_{\al\beta} 		&= 0 \; , \qquad
	&\Gamma^m_{\al n} 		&= \oneon{2} g^{mp} g_{pn,\al} \; , \qquad
	&\Gamma^m_{np} 			&= \tilde{\Gamma}^m_{np} \; ,
\eea
where here and in the following the tilde denotes purely 10-dimensional quantities on $X$. From there, the Riemann tensor is computed. We do not need all components of the Riemann tensor as we are only interested in the Ricci scalar $R = \gamma^{\al\beta} R_{\al\beta} + g^{mn} R_{mn}$ with $R_{MN} = \gamma^{\al\beta} R_{\al M \beta N} + g^{pq} R_{p M q N}$. After exploiting the symmetries of the Riemann tensor, we only need to know three independent components, namely $R_{\al\beta\gamma\delta}$, $R_{\al m \beta n}$ and $R_{mnpq}$. We skip writing down these lengthy expressions and jump directly to the result for the Ricci scalar
\be
	R = R^{(2)} - \oneon{4} \gamma^{\al\beta} g^{mn} g^{pq} \left( g_{mp,\al} g_{nq,\beta} + g_{mn,\al} g_{pq,\beta} \right)
		- \nabla^{(2)}_\al \left( \gamma^{\al\beta} g^{mn} g_{mn,\beta} \right) ,
\ee
where $R^{(2)}$ and $\nabla^{(2)}_\al$ are the Ricci scalar and covariant derivative with respect to the 2-metric $\gamma_{\al\beta}$ and we have dropped terms containing the Ricci tensor of $X$, which vanishes by definition for Calabi-Yau five-folds, that is $\tilde{R}_{mn} = 0$. After integrating by parts, the Einstein-Hilbert term with the ansatz inserted reads as follows
\be
	\int_\manifold R\ast 1 = \int_\Sigma d^2\sigma \sqrt{|\gamma|} \int_X d^{10} x \sqrt{g} \left\{  
		- \oneon{4} \gamma^{\al\beta} g^{mn} g^{pq} \left( g_{mp,\al} g_{nq,\beta} - g_{mn,\al} g_{pq,\beta} \right) + R^{(2)} \right\} ,
\ee
Now, we change to local holomorphic coordinates $z^\mu$ on $X$ and insert \eqref{metric_mode_expansion} (with the dot replaced by $\partial_\al$). The final form of the reduced Einstein-Hilbert term is then found to be
\begin{multline}
	\int_\manifold R\ast 1 = \int_\Sigma d^2\sigma \sqrt{|\gamma|} \left\{ - \oneon{8} G_{ij}^{(1,1)}(\underline{t}) (\partial_\al t^i) (\partial^\al t^i) 
		\right. \\ \left. \vphantom{\oneon{8}} 
		- 2 V(\ul{t}) G_{a\bar{b}}^{(1,4)} (\underline{z},\underline{\bar{z}}) (\partial_\al z^a) (\partial^\al \bar{z}^{\bar{b}}) 
		+ V(\ul{t}) R^{(2)} \right\} ,
\end{multline}
with the physical moduli space metrics as presented in \secref{bos_red_metrics}.

In addition, we need the field strengths $F=dA$ and $G=dB$ for the form field ans\"atze~\eqrangeref*{fcy5_Aansatz}{fcy5_Bansatz} and their Hodge duals
\begin{align}
	F &= dX^{\cal{P}} \wedge N_{\cal{P}} + F^i \wedge\omega_i  \; ,\qquad\qquad
	G = db^{\cal{X}} \wedge O_{\cal{X}} + G^{\cal{P}} \wedge N_{\cal{P}} \; , \\
	\ast F &= - \oneon{2} \cs_{\cal{P}} {}^{\cal{Q}} \left( \ast^{(2)} dX^{\cal{P}} \right) \wedge  N_{\cal{Q}}\wedge J^2
		- \left( \ast^{(2)} F^i \right) \wedge \left( \oneon{3!} J^3 \wedge \omega_i + \frac{i}{4!} \tilde{\omega}_i J^4 \right) , \\
	\ast G &= \left( \ast^{(2)} db^{\cal{X}} \right) \wedge \left( \ast^{(10)} O_{\cal{X}} \right)
		+ \oneon{2} \cs_{\cal{P}} {}^{\cal{Q}} \left( \ast^{(2)} G^{\cal{P}} \right) \wedge  N_{\cal{Q}}\wedge J^2 \; ,
\end{align}
where we have introduced the two-dimensional field strengths $F^i \equiv dA^i = \oneon{2} \partial_\al A^i_\beta d\sigma^\al \wedge d\sigma^\beta$ and $G^{\cal{P}} = d B^{\cal{P}}$. To derive the expressions for $\ast F$ and $\ast G$, we have used the results~\eqref*{dualforms,3form_hodgestar}. Where no confusion is possible, we will henceforth denote the two-dimensional Hodge star $\ast^{(2)}$ simply by $\ast$.

Inserting the ans\"atze~\eqrangeref*{fcy5_gansatz}{fcy5_Bansatz} together with the last four equations into the action~\eqref*{S12} and integrating over the Calabi-Yau five-fold, one arrives at the following two-dimensional effective action
\begin{multline}\eqlabel{S2D}
	S_2 = \frac{T}{2} \int_\Sigma \left\{  - \oneon{8} G_{ij}^{(1,1)}(\ul{t}) dt^i \wedge\ast dt^i
		- 2 V(\ul{t}) G_{a\bar{b}}^{(1,4)} (\ul{z},\ul{\bar{z}}) dz^a \wedge\ast d\bar{z}^{\bar{b}}
		- \oneon{2} V(\ul{t}) d\psi \wedge\ast d\psi \right. \\ \left.
		- \oneon{2} e^{\al\psi} \left[ G_{\mathcal{PQ}}^{(3)}(\ul{t},\ul{z},\ul{\bar{z}}) dX^{\cal{P}} \wedge\ast dX^{\cal{Q}}  
		+ 2 V(\ul{t}) \mathcal{G}_{ij}^{(1,1)}(\ul{t}) F^i \wedge\ast F^j \right] \right. \\ \left.
		- \oneon{2} e^{\beta\psi} \left[ G_{\mathcal{XY}}^{(4)}(\ul{t}) db^{\cal{X}} \wedge\ast db^{\cal{Y}} 
		+ G_{\mathcal{PQ}}^{(3)}(\ul{t},\ul{z},\ul{\bar{z}}) G^{\cal{P}} \wedge\ast G^{\cal{Q}} \right] \right. \\ \left. \vphantom{\oneon{8}}
		+ V(\ul{t}) R^{(2)}\ast 1
	\right\} + S_{2,{\rm CS}} \; ,
\end{multline}
where $T\equiv v^{4/5}/\kappa^2_{12}$ is the ``string tension\footnote{The power of the coordinate volume $v = \int_X d^{10} x$ has been inserted into the definition of $T$ so that it has the correct mass dimension for a tension, namely $(\text{mass})^2$ (or $(\text{length})^{-2}$). Compensating powers of $v$ also need to be inserted into the definitions of the moduli space metrics but are suppressed in order to keep equations concise. If needed, the powers of $v$ can easily be reconstructed by demanding objects to take on their correct mass dimensions.}'', the moduli space metrics $G_{ij}^{(1,1)}$, $G_{a\bar{b}}^{(1,4)}$, $G_{\mathcal{PQ}}^{(3)}$ and $\mathcal{G}_{ij}^{(1,1)}$ have been defined in~\eqrangeref*{G11def}{G3def} and~\eqref*{metric11}, respectively, and the only new object is the 4-form moduli space metric defined as
\be
	G_{\mathcal{XY}}^{(4)}(\ul{t}) \equiv \int_X O_{\cal{X}} \wedge\ast O_{\cal{Y}} \; .
\ee
The Chern-Simons contribution $S_{2,{\rm CS}}$ is explicitly given by
\be\eqlabel{S2CS}
	S_{2,{\rm CS}} = - \frac{T}{2} \gamma d_{\mathcal{XPQ}} \int_\Sigma b^{\cal{X}} \wedge dX^{\cal{P}} \wedge dX^{\cal{Q}} \; ,
\ee
with intersection number $d_{\mathcal{XPQ}} \equiv \int_X O_{\cal{X}} \wedge N_{\cal{P}} \wedge N_{\cal{Q}}$. By employing differential form language in~\eqref*{S2D}, we have arrived at an action that is the same for both $(-,+)$ and $(-,-)$ signature. 

Assuming $(-,+)$ signature, all but two kinetic terms have the appropriate sign such as to yield a positive energy expression. The two exceptions are the $\psi$ kinetic term and another one in the $(1,1)$-sector corresponding to the negative eigenvalue direction of the $(1,1)$-metric. To see the former explicitly, perform a simultaneous rescaling $\psi\to i\psi$, $\al\to -i\al$ and $\beta\to -i \beta$. The only change in the action~\eqref*{S2D} induced by this rescaling is a flipped sign in front of the dilaton kinetic term, while all other terms remain unchanged in form but are now manifestly real. Thus, the only terms with a negative contribution to the total energy stem from the dilaton kinetic term and, by the same arguments as in \chref{msp}, from the single negative direction in the $(1,1)$-sector.

\subsection{Strings on moduli spaces}

The action~\eqref*{S2D} still contains terms involving the (non-dynamical) gauge degrees of freedom $A^i$ and $B^{\cal{P}}$. It is however consistent with their equations of motion
\be
	d\ast \left( e^{\al\psi} V \mathcal{G}_{ij}^{(1,1)} F^j \right) = 0 \; , \qquad\qquad
	d\ast \left( e^{\beta\psi} G_{\mathcal{PQ}}^{(3)} G^{\cal{Q}} \right) = 0
\ee
to set them to zero and we will do so henceforth. By combining the other fields into one large vector $\phi^I = (t^i, z^a, \bar{z}^{\bar{b}}, \psi, X^{\cal{P}}, b^{\cal{X}})$, the remaining terms in the action~\eqref*{S2D} can be written in a concise way as
\be\eqlabel{S2string}
	S_2 = - \frac{T}{2} \int_\Sigma d^2 \sigma \sqrt{|\gamma|} \left\{ \gamma^{\al\beta} G_{IJ} (\ul{\phi}) \partial_\al \phi^I \partial_\beta \phi^J
		- R^{(2)} (\gamma) V(\ul{\phi}) \right\} + S_{2,{\rm CS}} \; ,
\ee
with a block-diagonal metric $G_{IJ}$ comprising the individual moduli space metrics in the appropriate way. This two-dimensional sigma-model is a generalisation of the Polyakov action (also known as Brink-DiVecchia-Howe-Deser-Zumino action) of a bosonic string moving in a target space with embedding coordinates $\phi^I(\ul{\sigma})$. In our case, the target space is the $n$-dimensional moduli space of deformations around the Calabi-Yau five-fold $X$, where $n$ can be read off from the $I$-index range
\be\eqlabel{fvy5_string_target_space_dim}
	n = h^{1,1}(X) + 2 h^{1,4}(X) + 2 h^{1,2}(X) + 2 h^{1,3}(X) + h^{2,2}(X) + 1
\ee
and the target space metric $G_{IJ}$ has $(n-2,2)$-signature irrespective of the signature of the 12-dimensional space-time manifold $\manifold$. The two negative directions correspond to the worldsheet dilaton $\psi$ and the negative eigenvalue in the $(1,1)$-metric as explained above. Intriguingly, the Calabi-Yau volume $V$ plays the r\^ole of the target-space dilaton and its expectation value $\langle V\rangle$ is related to the string coupling constant $g_s$ via $g_s = e^{\langle V\rangle}$ (or $g_o^2 = e^{\langle V\rangle}$, for open strings) in the string interpretation of the sigma-model~\eqref*{S2string}.

The classical equations of motion following from~\eqref*{S2string} are
\begin{gather}
	\nabla_\al \nabla^\al \phi^I + \Gamma^I_{JK} \partial_\al \phi^J \partial^\al \phi^K + \oneon{2} R^{(2)} G^{IJ} \frac{\partial V}{\partial \phi^J}
		+ C^I = 0 \; , \\
	G^{(2)}_{\al\beta} = V^{-1} G_{IJ} \partial_\al \phi^I \partial_\beta \phi^J 
		- \oneon{2} \gamma_{\al\beta} V^{-1} G_{IJ} \partial_\gamma \phi^I \partial^\gamma \phi^J  \; ,
\end{gather}
where $\Gamma^I_{JK}$ is the Christoffel symbol associated with $G_{IJ}$, $C^I$ is the contribution from the Chern-Simons term~\eqref*{S2CS} and $G^{(2)}_{\al\beta}$ is the two-dimensional Einstein tensor. Recall, that gravity is non-dynamical in two-dimensions and may be gauged away. After gauge fixing the worldsheet metric, the two-dimensional Einstein tensor $G^{(2)}_{\al\beta}$ vanishes and Einstein's equation turns into the constraint of the vanishing of the worldsheet energy-momentum tensor
\be
	T_{\al\beta} = V^{-1} G_{IJ} \partial_\al \phi^I \partial_\beta \phi^J 
			- \oneon{2} \gamma_{\al\beta} V^{-1} G_{IJ} \partial_\gamma \phi^I \partial^\gamma \phi^J = 0 \; .
\ee
This is analogous to the Hamiltonian constraint~\eqref*{mcy5_hamilt_constr} in \chref{mcy5}. 

After choosing conformal gauge $\gamma_{\al\beta} = \eta_{\al\beta}$ (with $\eta_{\al\beta} = \mathrm{diag}(-1,+1)$ or $\eta_{\al\beta} = \mathrm{diag}(-1,-1)$) and changing to holomorphic coordinates $z = \sigma^1 + i \sigma^2$, the gauge fixed action is simply given by
\be
	S_2 = \frac{T}{2} \int_\Sigma d^2 z\, G_{IJ} (\ul{\phi}) 
			\left( \partial \phi^I \partial \phi^J + \bar\partial \phi^I \bar\partial \phi^J \right) + S_{2,{\rm CS}} \; ,
\ee
for $\eta_{\al\beta} = \mathrm{diag}(-1,+1)$ and
\be
	S_2 = T \int_\Sigma d^2 z\, G_{IJ} (\ul{\phi}) \partial \phi^I \bar\partial \phi^J + S_{2,{\rm CS}} \; ,
\ee
for $\eta_{\al\beta} = \mathrm{diag}(-1,-1)$. Here, $d^2 z = dz d\bar{z} \equiv 2 d\sigma^1 d\sigma^2$, $\partial = \partial/\partial z$ and $\bar\partial = \partial/\partial\bar{z}$.

The action~\eqref*{S2string} has the well-known symmetries of bosonic string theory, namely target-space Poincar\'e invariance, worldsheet reparametrisation (or two-dimensional diffeomorphism) invariance and local Weyl (or conformal) invariance. To check that the Chern-Simons term respects conformal invariance, recall the conformal weights, denoted $\mathrm{wt}(\cdot)$, of various objects in the theory
\be
	\mathrm{wt} (\sigma^\al) = -1 \; , \qquad \mathrm{wt} (\partial_\al) = 1 \; , \qquad \mathrm{wt} (\gamma_{\al\beta}) = 2 \; , \qquad 
	\mathrm{wt} (\phi^I) = 0 \; .
\ee
It is then straightforward to see that $\mathrm{wt} (d) = 0$ and $\mathrm{wt} (S_{2,{\rm CS}}) = 0$, which means the Chern-Simons term is indeed conformally invariant. In addition to the standard symmetries of the bosonic string, the two axion-like scalars $X^{\cal{P}}$ and $b^{\cal{X}}$ originating from the 12-dimensional form fields $A$ and $B$ possess a Peccei-Quinn shift symmetry
\begin{align}
	X^{\cal{P}} (\ul{\sigma}) &\rightarrow {X^{\cal{P}}}^\prime (\ul{\sigma}) = X^{\cal{P}} (\ul{\sigma}) + c^{\cal{P}} \; , \eqlabel{2D_PQ_3form} \\
	b^{\cal{X}} (\ul{\sigma}) &\rightarrow {b^{\cal{X}}}^\prime (\ul{\sigma}) = b^{\cal{X}} (\ul{\sigma}) + c^{\cal{X}} \; , \eqlabel{2D_PQ_4form}
\end{align}
where the $c^{\cal{P}}$ and $c^{\cal{X}}$ are sets of real constants. This is in full analogy to the case of M-theory on Calabi-Yau Five-Folds (see \secref{symm_1d_eff_action}). The invariance of the first two terms in the action~\eqref*{S2string} is a consequence of the fact that $G_{IJ}$ and $V$ are independent of $X^{\cal{P}}$ and $b^{\cal{X}}$ and those fields only occur differentiated there. The Chern-Simons term~\eqref*{S2CS} varies into a total derivative under the Peccei-Quinn transformations~\eqrangeref*{2D_PQ_3form}{2D_PQ_4form}.

We end this section with a curious observation. It is well-known that quantum consistency (namely, the absence of a conformal anomaly) requires the target-space of the bosonic string to be 26-dimensional. In our case, the string moves in the moduli space of deformations around the Calabi-Yau five-fold $X$ and the dimension of this target space was given in \eqref{fvy5_string_target_space_dim} in terms of the Hodge numbers of $X$. Thus, in order for our dimensional reduction to be consistent at the quantum level, it seems that we must restrict to those Calabi-Yau five-folds $X$ which satisfy
\be
	n = h^{1,1}(X) + 2 h^{1,4}(X) + 2 h^{1,2}(X) + 2 h^{1,3}(X) + h^{2,2}(X) + 1 = 26 \; .
\ee
This places constraints on the possible topology of $X$. Intriguingly enough, amongst all the explicit examples of Calabi-Yau five-folds studied in this thesis, there is precisely one that satisfies this constraint. It is the torus quotient $(T^2)^5/\mathbb{Z}_2^4$ introduced in \secref{torus_quotients}. With the result~\eqrangeref*{torus_quotient_hpq1}{torus_quotient_hpq2} for the Hodge numbers of $(T^2)^5/\mathbb{Z}_2^4$, one verifies
\be
	n = 5 + 2\cdot 5 + 2\cdot 0 + 2\cdot 0 + 10 + 1 = 26 \; .
\ee
By the same argument, the simplest CICYs with small configuration matrices studied in \secref{cicy5} are ruled out on the sole ground that $h^{1,4}(X)$ is of the order $10^2$ or even $10^3$.

\section{Connection with M-theory on Calabi-Yau Five-Folds}

To end this chapter, we would like to mention how the two-dimensional action~\eqref*{S2string} is related to the one-dimensional effective theory obtained from reducing M-theory on Calabi-Yau Five-Folds. In the original KKLP proposal, it was shown how 11-dimensional supergravity emerges from the 12-dimensional action~\eqref*{S12} by a reduction on $S^1$ in conjunction with the consistent truncation stated in~\eqref*{KKLP_trunc_sugra11}. We will now apply the same steps to the case where $\manifold = \Sigma\times X$. In other words, the total space is further decomposed as $\manifold = \R \times S^1\times X$.

We may start with the reduction on a circle from 12 to 11 dimensions and directly adopt the result given in ref.~\cite{Khviengia:1997rh}
\begin{multline}\eqlabel{S11_KKLP}
	S_{11}^\prime = \oneon{2\kappa^2_{11}} \int \left\{ R\ast 1 - \oneon{2} d\psi \wedge\ast d\psi - \oneon{2} d\phi \wedge\ast d\phi
		- \oneon{2} e^{-(1/\sqrt{5})\phi + \al\psi} F_4\wedge\ast F_4 \right. \\ \left.
		- \oneon{2} e^{(7/3\sqrt{5})\phi + \al\psi} F_3\wedge\ast F_3 
		- \oneon{2} e^{-(4/3\sqrt{5})\phi + \beta\psi} G_5\wedge\ast G_5 \right. \\ \left.
		- \oneon{2} e^{(2/\sqrt{5})\phi + \beta\psi} G_4\wedge\ast G_4
		- \oneon{2} e^{-(10/3\sqrt{5})\phi} \mathcal{F} \wedge\ast \mathcal{F} \right\} + S_{11,{\rm CS}}^\prime \; ,
\end{multline}
where $\kappa_{11}^2 = \kappa_{12}^2 / (2\pi)$, $F=F_4 + F_3 \wedge dz = dA_3 + dA_2 \wedge dz$, $G=G_5 + G_4 \wedge dz = dB_4 + dB_3\wedge dz$, $\mathcal{F} = d\mathcal{A}$ and
\be\eqlabel{KKLP_gansatz_12_11}
	ds^2_{12} = e^{\phi/(3\sqrt{5})} ds^2_{11} + e^{-(3/\sqrt{5})\phi} (dz + \mathcal{A})^2 \; ,
\ee
with $z$ the coordinate on the circle. The Chern-Simons terms, including the corrections to the field strengths $F_4$ and $G_5$, are given by
\begin{multline}\eqlabel{S11_KKLP_CS}
	S_{11,{\rm CS}}^\prime = \oneon{2\kappa^2_{11}} \int \left\{ \gamma B_3 \wedge F_4 \wedge F_4 
		+ 2\gamma B_4 \wedge F_3 \wedge F_4 \vphantom{e^{-(1/\sqrt{5})\phi + \al\psi}}\right. \\ \left. 
		+ e^{-(1/\sqrt{5})\phi + \al\psi} F_3\wedge\mathcal{A} \wedge\ast F_4 
		+ e^{-(4/3\sqrt{5})\phi + \beta\psi} G_4\wedge\mathcal{A}\wedge\ast G_5 \right\} .
\end{multline}
The next step is our Calabi-Yau five-fold reduction. For the metric, we choose the same ansatz as in \secref{bos_red_ansatz} (see~\eqref*{gansatz})
\be
	ds^2_{11} = -\oneon{4}N(\tau )^2 d\tau^2 + g_{mn} (t^i,z^a,\bar{z}^{\bar{a}}) dx^m dx^n \; .
\ee
The ans\"atze for the form fields are chosen as follows (cf.~\eqref*{Aansatz_3forms})
\be
	A_3 = \tilde{X}^{\cal{P}} N_{\cal{P}} \; , \qquad A_2 = A^i \omega_i \; , \qquad B_4 = b^{\cal{X}} O_{\cal{X}} \; , \qquad
	B_3 = \tilde{Y}^{\cal{P}} N_{\cal{P}} \; , \qquad \mathcal{A} = 0 \; .
\ee
Note that $\mathcal{A}$ does not contribute any dynamical degrees of freedom to the one-dimensio\-nal action, since $h^{1,0}(X) = 0$ and one-dimensional vectors are non-dynamical.

Plugging this reduction ansatz into the action~\eqref*{S11_KKLP} and using the results from \secref{bos_red_comp} leads to the following one-dimensional action
\begin{multline}\eqlabel{S1_KKLP}
	S_1^\prime = \frac{l}{2} \int d\tau N^{-1} \left\{ \oneon{4} G_{ij}^{(1,1)}(\ul{t}) \dot{t}^i \dot{t}^j 
		+ 4 V(\ul{t}) G_{a\bar{b}}^{(1,4)} (\ul{z},\ul{\bar{z}}) \dot{z}^a \dot{\bar{z}}^{\bar{b}} 
		+ V \left( \dot{\psi}^2 + \dot{\phi}^2 \right) \right. \\ \left.
		+ \oneon{2} e^{-(1/\sqrt{5})\phi + \al\psi} 
			G_{\mathcal{PQ}}^{(3)}(\ul{t},\underline{z},\ul{\bar{z}}) \dot{\tilde{X}}^{\cal{P}} \dot{\tilde{X}}^{\cal{Q}} 
		+ \oneon{2} e^{(2/\sqrt{5})\phi + \beta\psi} 
			G_{\mathcal{PQ}}^{(3)}(\ul{t},\underline{z},\ul{\bar{z}}) \dot{\tilde{Y}}^{\cal{P}} \dot{\tilde{Y}}^{\cal{Q}} \right. \\ \left.
		+ 2 e^{(7/3\sqrt{5})\phi + \al\psi} V(\ul{t}) \mathcal{G}_{ij}^{(1,1)}(\ul{t}) \dot{A}^i \dot{A}^j
		+  e^{-(4/3\sqrt{5})\phi + \beta\psi} G_{\mathcal{XY}}^{(4)}(\ul{t}) \dot{b}^{\cal{X}} \dot{b}^{\cal{Y}} \vphantom{\oneon{4}} \right\} .
\end{multline}
Note that none of the Chern-Simons terms in~\eqref*{S11_KKLP_CS} contributes to the one-dimensional action.

Of course, one could also have arrived at the same one-dimensional action by reducing first on the Calabi-Yau five-fold $X$ to two dimensions and then one the circle. In this case, the metric ansatz~\eqref*{KKLP_gansatz_12_11} is re-grouped to read
\be
	ds^2_{12} = \gamma_{\al\beta} d\sigma^\al d\sigma^\beta + e^{\phi/(3\sqrt{5})} g_{mn} (t^i,z^a,\bar{z}^{\bar{a}}) dx^m dx^n \; ,
\ee
which amounts to a Weyl rescaling of the Calabi-Yau metric, as compared to the standard reduction ansatz~\eqref*{fcy5_gansatz}. The form field ans\"atze are as in \secref{fcy5_bos_red_ansatz}. This ansatz leads to a modified two-dimensional action, denoted $S_2^\prime$, which contains a kinetic term for $\phi$ and powers of $e^\phi$ inserted as pre-factors into the terms in the action~\eqref*{S2D}. The reduction to one dimension is then performed using the following ansatz for the 2-metric
\vspace{\parskip}
\be
	ds^2 	= \gamma_{\al\beta} d\sigma^\al d\sigma^\beta 
			= -\oneon{4} e^{\phi/(3\sqrt{5})} N (\tau )^2 d\tau^2 + e^{-(3/\sqrt{5})\phi} dz^2
\ee
\vspace{\parskip}\noindent 
and for the other fields in~\eqref*{S2D}, we merely drop the $z$-dependence. The result of that reduction is our modified one-dimensional action~\eqref*{S1_KKLP} (after some relabelling of fields), as expected.

Now, we will turn to the question of how the action~\eqref*{S1_KKLP} can be truncated consistently to yield the one-dimensional effective theory from our M-theory on Calabi-Yau five-folds reduction in \chref{mcy5}. We will be guided by the original KKLP proposal and use the truncation~\eqref*{KKLP_trunc_sugra11} with our reduction ansatz inserted. This first leads to
\vspace{\parskip}
\be
	\dot{A}^i = \dot{b}^{\cal{X}} = 0 \; .
\ee
\vspace{\parskip}\noindent 
This is consistent with their respective equations of motion, which read
\begin{align}
	\frac{d}{d\tau} \left( N^{-1} e^{(7/3\sqrt{5})\phi + \al\psi} V(\ul{t}) \mathcal{G}_{ij}^{(1,1)}(\ul{t}) \dot{A}^j \right) &= 0 \; , \\
	\frac{d}{d\tau} \left( N^{-1} e^{-(4/3\sqrt{5})\phi + \beta\psi} G_{\mathcal{XY}}^{(4)}(\ul{t}) \dot{b}^{\cal{Y}} \right) &= 0 \; .
\end{align}
Following ref.~\cite{Khviengia:1997rh}, we now choose $\al = i/\sqrt{5}$ and $\beta = - 2i/\sqrt{5}$ and assemble $\psi$ and $\phi$ into a single complex field $w$ according to
\be
	w = \oneon{\sqrt{2}} (-\phi + i \psi) \; , \qquad\qquad \bar{w} = \oneon{\sqrt{2}} (-\phi - i \psi) \; .
\ee
After setting $\dot{A}^i = \dot{b}^{\cal{X}} = 0$ and introducing the new field $w$, the action~\eqref*{S1_KKLP} reduces to
\begin{multline}\eqlabel{S1_KKLP_trunc1}
	S_1^\prime = \frac{l}{2} \int d\tau N^{-1} \left\{ \oneon{4} G_{ij}^{(1,1)}(\ul{t}) \dot{t}^i \dot{t}^j 
		+ 4 V(\ul{t}) G_{a\bar{b}}^{(1,4)} (\ul{z},\ul{\bar{z}}) \dot{z}^a \dot{\bar{z}}^{\bar{b}} 
		+ 2 V \dot{w} \dot{\bar{w}} \right. \\ \left.
		+ \oneon{2} e^{\sqrt{2/5}\, w} 
			G_{\mathcal{PQ}}^{(3)}(\ul{t},\underline{z},\ul{\bar{z}}) \dot{\tilde{X}}^{\cal{P}} \dot{\tilde{X}}^{\cal{Q}} 
		+ \oneon{2} e^{- 2 \sqrt{2/5}\, w} 
			G_{\mathcal{PQ}}^{(3)}(\ul{t},\underline{z},\ul{\bar{z}}) \dot{\tilde{Y}}^{\cal{P}} \dot{\tilde{Y}}^{\cal{Q}} \vphantom{\oneon{4}} \right\} 
\end{multline}
and the equations of motion for $w$ and $\bar{w}$ can be read off directly
\begin{align}
	\frac{d}{d\tau} \left( N^{-1} V(\ul{t}) \dot{w} \right) &= 0 \; , \\
	\frac{d}{d\tau} \left( N^{-1} V(\ul{t}) \dot{\bar{w}} \right) &= 
		\oneon{4} \sqrt{\frac{2}{5}} \left( 
			e^{\sqrt{2/5}\, w} G_{\mathcal{PQ}}^{(3)}(\ul{t},\underline{z},\ul{\bar{z}}) \dot{\tilde{X}}^{\cal{P}} \dot{\tilde{X}}^{\cal{Q}} - 2
			e^{- 2 \sqrt{2/5}\, w} G_{\mathcal{PQ}}^{(3)}(\ul{t},\underline{z},\ul{\bar{z}}) \dot{\tilde{Y}}^{\cal{P}} \dot{\tilde{Y}}^{\cal{Q}} \right) ,
\end{align}
which shows that it is consistent to set $w=\bar{w}=0$, provided
\be
	\dot{\tilde{X}}^{\cal{P}} = \sqrt{2}\, \dot{\tilde{Y}}^{\cal{P}} \; .
\ee
With this identification and $w=\bar{w}=0$, the action can be written as
\be
	S_1 = \frac{l}{2} \int d\tau N^{-1} \left\{ \oneon{4} G_{ij}^{(1,1)}(\ul{t}) \dot{t}^i \dot{t}^j 
		+ 4 V(\ul{t}) G_{a\bar{b}}^{(1,4)} (\ul{z},\ul{\bar{z}}) \dot{z}^a \dot{\bar{z}}^{\bar{b}} 
		+ \oneon{2} G_{\mathcal{PQ}}^{(3)}(\ul{t},\underline{z},\ul{\bar{z}}) \dot{X}^{\cal{P}} \dot{X}^{\cal{Q}} \right\} ,
\ee
where $X^{\cal{P}} = (\sqrt{3/2}) \tilde{X}^{\cal{P}} = \sqrt{3}\, \tilde{Y}^{\cal{P}}$. This is precisely the one-dimensional effective action~\eqref*{S1B} obtained from the reduction of M-theory on Calabi-Yau five-folds. Note that in order for this identification to work, we had to reduce both M-theory and F-theory on the same Calabi-Yau five-fold $X$. Hence, this does not give rise to mirror symmetry for Calabi-Yau five-folds.

To summarise, we illustrate the result of this section with the help of the following commutative diagram
\be\begin{CD}
	S_{12} 		@>S^1>> 	S_{11}^\prime	@>\text{trunc.}>> 	S_{\rm CJS,B} \\
	@VVCY5V				@VVCY5V					@VVCY5V	\\
	S_2^\prime	@>S^1>> 	S_1^\prime	@>\text{trunc.}>>	S_1
\end{CD}\ee
We showed in this section that one may follow around the arrows in this diagram in an arbitrary order. Thus, the two-dimensional string action~\eqref*{S2D} obtained from reducing the KKLP action~\eqref*{S12} contains the one-dimensional effective action~\eqref*{S1B} that originated from the M-theory reduction as a special case.
%
%

\chapter{Conclusion}\chlabel{conclusions}

\section{Summary}

This thesis explored Calabi-Yau five-folds as compactification manifolds for physical theories. These manifolds are ten-dimensional and therefore, compactifications on such spaces can only be performed for theories originally defined in more than ten dimensions. This uniquely singles out M-theory and F-theory as the realm of Calabi-Yau five-fold compactifications. Both of these possibilities have been studied in this thesis at least at the field theoretical level. The focus, however, has clearly been on the M-theory setting.

In \chref{mthy,cy5}, we developed the necessary physical (\chref{mthy}) and mathematical (\chref{cy5}) background knowledge. More precisely, we reviewed some aspects of M-theory and 11-dimensional supergravity in \chref{mthy}, mostly focusing on those elements that are directly relevant to later parts of the thesis. \Chref{cy5} introduced the mathematics of Calabi-Yau manifolds in general and briefly detoured into the history of how Calabi-Yau spaces entered the physics literature. The second part of this chapter concentrated on special properties of five-folds and on constructing explicit examples of such manifolds. The latter turned out to be important, since M-theory -- as we learned in \chref{mcy5} -- places constraints on the possible Calabi-Yau five-fold backgrounds. In such a situation, explicit examples that satisfy these constraints serve as an existence proof of such compactification models or -- in other words -- are a way of re-assuring that one is not dealing with an empty set.

\Chref{superspace} was devoted to another preparatory topic, namely to one-dimensional $\susyno=2$ superspace. While this is in general not a new subject, we found that we needed some generalisations of the results contained in the literature. In order to develop a solid base for our application to M-theory, we opted for a systematic exposition of one-dimensional $\susyno=2$ flat and curved superspace.

The great amount of time we spent developing the background material in the first review-like chapters was rewarded in \chref{mcy5}, where the reduction of M-theory on Calabi-Yau five-folds is presented in its entirety. We started off by exploring the implications of a Calabi-Yau five-fold background. In other words, what properties must such a background have in order to be consistent with the 11-dimensional equations of motion? We found that the M-theory corrected equations of motion impose a topological constraint forcing the fourth Chern class of the Calabi-Yau five-fold to be proportional to the square of the internal fluxes and to the membrane classes on the manifold. In the absence of flux and membranes, this constraint allows only those five-folds as valid compactification backgrounds that have vanishing fourth Chern class.

We also analysed the Killing spinor equation for a Calabi-Yau five-fold background. In the quest to obtain supersymmetric solutions, we found that they can be satisfied by constraining the 4-form flux to be primitive and purely of $(2,2)$-type. This, in turn, leads to a lifting of the otherwise totally flat moduli space, since the K\"ahler and complex structure moduli are now forced to evolve only in such a way as to ensure the chosen 4-form flux obeys the aforementioned restrictions at all times (or otherwise the solution is non-supersymmetric). 

Before studying the flux compactifications in more detail, we presented the lowest-order Kaluza-Klein reduction of 11-dimensional supergravity on Calabi-Yau five-folds. Due to the peculiar features of one-dimensional supersymmetry, the usual procedure of performing only the bosonic reduction and then guessing the supersymmetric completion is not guaranteed to work in our case. Therefore, we also dimensionally reduced the fermionic part up to the two-fermi level and indeed we found purely fermionic multiplets, which we would have missed had we only reduced the bosonic side. We then used the results from \chref{superspace} to write the one-dimensional effective action in superspace. In this way, we could summarise the effective theory in a concise way and make all its symmetries manifest. Moreover, it served as a check that the lower-dimensional theory indeed possesses $\susyno=2$ supersymmetry.

In the last part of \chref{mcy5}, we studied flux compactifications. A first concern was establishing the existence of viable examples of compactification manifolds. At this point, the explicit constructions of \chref{cy5} were put to use. We showed how the various constraints can be satisfied for some five-folds, while in other cases the constraints rule out certain examples. After re-assuring ourselves that we were not dealing with an empty set of compactification manifolds, we moved on to calculating the corrections from flux and higher-order terms in the 11-dimensional action to the one-dimensional effective theory. In accord with the general expectation, the inclusion of flux leads to a scalar potential in the one-dimensional theory. After finding the potential, we showed how it can be obtained from a Gukov-type superpotential. This derivation is in parts similar to the case of Calabi-Yau four-folds. However, differences arise when one attempts to incorporate the restrictions stemming from the Killing spinor equation into the lower-dimensional effective theories. The condition of primitivity of the 4-form flux can in both cases be implemented by requiring the superpotential to be independent of the K\"ahler moduli. In our case, this actually leads to a vanishing scalar potential. The second condition -- namely the vanishing of the $(1,3)$-part of the 4-form flux -- could only be included as an implicit constraint equation. Ultimately, this is related to the fact that our scalar potential only depends on the K\"ahler moduli and not on the complex structure moduli, which, in turn, is a consequence of a sign difference between the expressions for the Hodge duals of $(2,2)$- and $(1,3)$-forms. On the other hand, in the four-fold case, a second -- complex structure moduli dependent -- superpotential arises and imposing its independence of the complex structure moduli ensures the absence of the $(1,3)$-part of the 4-form flux.

In the remaining two chapters, we provided an overview of some ongoing (and largely unpublished) work in progress. \Chref{msp} is concerned with connecting the one-dimensional effective theory with physics, focusing on a possible cosmological interpretation. In the first part, we studied some classical solutions of the bosonic part of the one-dimensional effective action, again putting to use the explicit examples of Calabi-Yau five-folds disclosed in \chref{cy5}. Since it is not clear whether one is eventually only interested in supersymmetric solutions (after all, the world around us is non-supersymmetric), we also included some non-supersymmetric examples.

Leaving the realm of classical field theory, one first notices that our one-dimensional action is actually a mini-superspace model. To illustrate this fact, we briefly reviewed some prominent topics in quantum gravity research, namely (quantum) geometrodynamics and mini-superspace quantisation. We then applied mini-superspace techniques to quantise our one-dimensional sigma model and obtained a Wheeler-DeWitt equation, which is a generalisation of the usual Wheeler-DeWitt equation of mini-superspace models. The extra corrections originate from the Chern-Simons term in the one-dimensional action. We showed that despite the corrections, the Wheeler-DeWitt equation can be written in a manifestly covariant way and an apparent additional factor ordering ambiguity is resolved owing to the particular properties of the Chern-Simons term. Just to give a taste, the model was quantised for a simple (yet unrealistic) class of examples, namely those with only a single K\"ahler modulus. The solution is merely formal as it is exponentially diverging leading to decompactification and thus to an invalidation of our approximations. 

The last chapter looked beyond M-theory and eleven dimensions. Indeed, in \chref{fcy5}, we applied our Calabi-Yau five-folds reduction program to a particular proposal by Khviengia, Khviengia, L\"u and Pope (KKLP) of a 12-dimensional field theory that is possibly related to F-theory. We began by summarising the KKLP proposal and discussing some of the issues with 12-dimensional theories. Finally, we dimensionally reduced the KKLP action on a Calabi-Yau five-fold from 12 to two dimensions. The result is a two-dimensional sigma-model coupled to two-dimensional gravity, which can be interpreted as a bosonic string moving in the moduli space of Calabi-Yau five-folds. It is known from ordinary string theory that quantum consistency requires the target space to be 26-dimensional. In our case, the target space is the moduli space and its dimension is determined by the Hodge numbers of the Calabi-Yau manifold. Hence, quantum consistency turns into a topological constraint on the possible Calabi-Yau backgrounds and with the torus quotient $T^{10}/\Z_2^4$, we found an example which satisfies this constraint.

A guiding principle of KKLP was the condition that 11-dimensional supergravity (and type IIB supergravity) be embeddable into the 12-dimensional action. It was therefore interesting to see how (and if) this embedding can be performed at the level of the lower-dimensional theories obtained from Calabi-Yau five-fold reductions. The way 11-dimensional supergravity emerges from the KKLP action is by a dimensional reduction on a circle accompanied by a consistent truncation of the fields. A similar process was anticipated in our lower-dimensional scenario. To check this, we first reduced the KKLP action from 12 to one dimension on a Calabi-Yau five-fold times a circle (this can actually be done in two equivalent ways, since $\nfold{5}\times S^1 \cong S^1 \times\nfold{5}$). Then -- led by the original KKLP case -- we found a way to consistently truncate our one-dimensional action such as to land on the effective theory obtained from reducing M-theory.

\section{Outlook and future directions}

The possible applications of Calabi-Yau five-folds in M- and F-theory have by far not been exhausted. By investing considerable efforts into developing the generalities of Calabi-Yau five-fold compactifications, we have paved the way for a wide variety of different directions in which to pursue and extend this research programme further.

A very interesting question within the context of ``moduli space cosmology'' is whether it is possible to find explicit models in which three spatial dimensions (that is, three directions inside the compact Calabi-Yau five-fold) dynamically grow large at late times thereby offering a novel way of explaining why our universe has three spatial dimensions. First steps towards this admittedly ambitious aim were made in \chref{msp}. However, the Wheeler-DeWitt equation from the mini-superspace quantisation of our non-linear sigma models is in general a highly-coupled, non-linear, higher-order partial differential equation, which means resorting to numerical methods save for the simplest cases.

Before turning to such a notoriously difficult question regarding the dynamics, it is perhaps instructive to first study a much simpler kinematical question: Assuming one is in a region of the five-fold moduli space where three directions are large, what properties does an effective decompactification limit to four dimensions have? For example, how much supersymmetry does it possess and can one learn something about the cosmological constant (perhaps along the lines of ref.~\cite{Witten:1995rz})? A simple example that can presumably be worked out rather explicitly is the torus quotient $T^{10}/\Z_2^4$ with a $T^{10}=T^3 \times T^7$ split, where $T^3$ is large compared to $T^7$ and such that the full manifold is still complex, of course. It is conceivable (by supersymmetry arguments) that such decompactifications to four dimensions could be linked to M-theory on $G_2$-manifolds.

Returning to the question regarding the dynamics, it would also be interesting to investigate into the important question of ``naturalness.'' In other words, how abundant are models that exhibit a $3+7$ split at late times in the entire Calabi-Yau five-fold landscape? Finally, another possible approach to making contact with four dimensions, is to study the application of brane gas cosmology to our Calabi-Yau five-fold reductions of M-theory. 

In a somewhat different direction goes the full-fledged treatment of general five-fold flux compactifications, which first and foremost amounts to studying the 11-dimensional Killing spinor equation for Calabi-Yau five-fold backgrounds (and including higher-order corrections) in the language of $G$-structures (here, $G=\SU(5)$). This also involves studying higher-order corrections to the lowest order Calabi-Yau five-fold background, in a way that is more general than the analysis we begun in \chref{mcy5}. For example, one may wish to allow for a warping of the full 11-dimensional metric or one may study deviations away from genuine Calabi-Yau manifolds.

Finally, the possible generalisation to F-theory that was started in \chref{fcy5} is far from a complete story. Instead of restricting to field theoretical methods \`a la KKLP, it would be interesting to study full F-theory on Calabi-Yau five-folds. This involves considering Calabi-Yaus with toric fibrations to allow for the embedding of the F-theory torus. Returning to KKLP, there are a number of question that can be asked about the two-dimensional sigma-model action obtained from our reduction. For example, can it be completed supersymmetrically in such a way as lead to our one-dimensional effective theory from M-theory on five-folds upon reduction on a circle plus a consistent truncation \`a la KKLP and does this have implications for the elusive fermionic side of the 12-dimensional theory? Does the Calabi-Yau five-fold reduction favour a particular space-time signature and is there a relation to four dimensions?

Several of the problems mentioned in this section are currently under investigation.

%
%

\chapter*{Acknowledgements}
\vspace*{-8mm}
I would like to take this opportunity and express the greatest possible thanks to my supervisor Kelly Stelle for his unlimited support, guidance, encouragement and patience and for being so generous with his time. Our frequent meetings were not only interesting and enjoyable, but they also greatly helped improving my understanding of theoretical physics. 

I wish to thank Andre Lukas for our fruitful collaboration and for allowing me to participate in the Calabi-Yau five-fold research project. Throughout my time as graduate student, I benefited from his support as well as from his patience and efforts in sharing his knowledge of physics and mathematics with me and I am very grateful for that.

I have made many friends in the theoretical physics community through various ways -- be it as office mates, as fellow conference participants, as travel companions or as mountain climbers -- and I would like to extend my sincere thanks to all of them.

The Theoretical Physics Group and the Geometry and String Theory Programme at the Institute for Mathematical Sciences have been my main ``intellectual homes'' for the past four years. It is a pleasure to thank my friends and colleagues for creating such an enjoyable atmosphere. Special thanks to Chris Hull and Daniel Waldram for support.

Also, I am grateful to Philip Candelas and Paul Howe for having kindly agreed to examine this thesis.

I have had the pleasure to be able to attend many interesting conferences, schools and workshops. This was possible thanks to generous travel funding from the Institute for Mathematical Sciences, the Imperial College Trust, the C R Barber Trust Fund of the Institute of Physics, the Research Student Conference Fund of the Institute of Physics, the EU RTN Network Forces-Universe and the European Superstring Theory Network.

I gratefully acknowledge the award of a postgraduate studentship by the Institute for Mathematical Sciences, Imperial College London. During the academic year 2007/08, I was given the chance to visit the Albert Einstein Institute (AEI) in Golm to which I express my gratitude for hospitality and generous financial support. At the AEI, I would like to thank, in particular, Hermann Nicolai and Stefan Theisen.

Finally, it is my friends and family to whom I am indebted. Above all else, it is my parents whom I profoundly thank for their encouragement and for their unwavering emotional, intellectual and not least financial support during the ups and downs of the past years.

\part*{Appendix}

\cleardoublepage
\thispagestyle{empty}
\vspace*{\stretch{1}}
\begin{quotation}
``Let no one ignorant of Mathematics enter here.''
\begin{flushright}
--- Inscription above the doorway of \textsc{Plato}'s Academy
\end{flushright}
\end{quotation}
\vspace*{\stretch{3}}

\appendix
%
%

\chapter{Notation and Conventions}\applabel{conventions}

\section{Symbols, Generalities and Space-Time Signatures}\applabel{indices_etal}

A complete list of symbols was given at the beginning of this thesis on pages~\pageref{LoS}-\pageref{LoSend} and is to be consulted first for any questions regarding the meaning of a symbol, a notation or an otherwise unknown expression. Beyond this list, some explanatory remarks on the general philosophy of our conventions are in order. 

First of all, a word on index choices: Indices are by default curved indices and we refer to their corresponding tangent space indices by underlining the same set of letters. Despite this convention, a large number of different index sets need to be used and the restricted size of the alphabet inevitably leads to a re-using of the same letters to serve as index labels in different contexts. It is hoped that this does not lead to confusion and a strong effort is made to always (re-)define the meaning and range of indices before their first (re-)use. The list of symbols on pages~\pageref{LoS}-\pageref{LoSend} provides information on all global index choices. The scope of index labels not listed there is restricted to the section in which they are defined.

Moreover, Einstein's summation convention is implicitly used throughout the entire thesis. This means any index that appears twice in an expression is summed over unless otherwise stated. Also, we set $c=\hbar=1$ everywhere. In the physics-context, the number of space-time dimensions is invariably denoted by $D$. On the other hand, we often use $n$ to denote the dimension of a manifold in the mathematical chapters and appendices. In addition, we always choose the ``mainly plus'' notation $\eta_{\ul{MN}} = \mathrm{diag}(-,+,+,\ldots,+)$ for manifolds equipped with a Lorentzian signature metric. In other words, all spatial directions are positive, whereas the time direction is singled out as being the negative direction. This generalises smoothly to the two-time signatures in the F-theory context, where the 10 spatial directions are still positive while the two time directions are negative, $(-,-,+,+,\ldots,+)$.

Symmetrisations and anti-symmetrisations over sets of indices are carried out with ``weight one.'' In other words
\newnot{symbol:ind_asymm}\newnot{symbol:ind_symm}
\be\eqlabel{sym_antisym_conv}
	\omega_{(\mu_1 \ldots \mu_n)} \equiv \frac{1}{n!} (\omega_{\mu_1 \ldots \mu_n} +   \text{perm.}) \; , \qquad\;\;
	\omega_{[\mu_1 \ldots \mu_n]} \equiv \frac{1}{n!} (\omega_{\mu_1 \ldots \mu_n} \pm \text{perm.}) \; ,
\ee
and indices enclosed in $|\ldots |$\newnot{symbol:ind_no_asymm} are not to be included in the (anti)-symmetrisation. A useful formula is the expansion of the anti-symmetrisation $[\ldots]$ in terms of $\mu_1$
\be\eqlabel{expand_anti_symm_bracket}
	\omega_{[\mu_1 \ldots \mu_n]} = \frac{1}{n} 
	\sum_{i=1}^n (-1)^{i-1} \omega_{[\mu_2\ldots\mu_i |\mu_1| \mu_{i+1}\ldots\mu_n]} \; ,
\ee
with special case
\be\eqlabel{expand_anti_symm_bracket_special}
	\omega_{[\mu_1 \ldots \mu_n]} = 
	\frac{1}{n} \omega_{\mu_1\ldots\mu_n} + 
	(-1)^{n-1} \frac{n-1}{n} \omega_{[\mu_2\ldots\mu_n]\mu_1} 
\ee
applicable when $\omega$ is totally anti-symmetric in the last $(n-1)$ indices.

\section{Spinor and \texorpdfstring{$\gamma$}{gamma}-matrix conventions in various dimensions}\applabel{spinorconv}

We now turn to our spinor conventions and start in 11 dimensions. The 11-dimensional ($32 \times 32$-component) gamma matrices $\Gamma^{\underline{M}}$ satisfy the Clifford algebra
\be\eqlabel{cliff11}
	\{ \Gamma^{\underline{M}}, \Gamma^{\underline{N}} \} = 2 \eta^{\underline{M}\underline{N}} \id_{32\times 32} ,
\ee
where $\eta_{\underline{M}\underline{N}} = \diag (-1,+1,\ldots,+1)$. Curved gamma matrices $\Gamma^M$ are constructed by contracting with an inverse vielbein $\Gamma^M = e_{\underline{N}}^M \Gamma^{\underline{N}}$. Dirac spinors $\Psi$ in $11$ dimensions have $32$ complex components and are anti-commuting objects. We are working in the Majorana representation in which the gamma matrices are real, $(\Gamma^{\underline{M}})^\ast = \Gamma^{\underline{M}}$, and all spatial gamma matrices are symmetric, $(\Gamma^{\underline{m}})^\transp = \Gamma^{\underline{m}}$, whereas the timelike gamma matrix is anti-symmetric, $(\Gamma^{\underline{0}})^\transp = - \Gamma^{\underline{0}}$. These properties combine into the following important formul\ae:
\be\eqlabel{gamma_11d_dagger}
	(\Gamma^{\underline{M}})^\dagger = \Gamma^{\underline{0}} \Gamma^{\underline{M}} \Gamma^{\underline{0}} \qquad\text{and}\qquad
	(\Gamma^{\underline{M}})^\transp = \Gamma^{\underline{0}} \Gamma^{\underline{M}} \Gamma^{\underline{0}} \; .
\ee
We define the Dirac conjugation matrix $D$ and charge conjugation matrix $C$ by
\be\eqlabel{gamma11_D_and_C}
	(\Gamma^{\underline{M}})^\dagger = - D \Gamma^{\underline{M}} D^{-1} \qquad\text{and}\qquad
	(\Gamma^{\underline{M}})^\transp = - C \Gamma^{\underline{M}} C^{-1} \; ,
\ee
which are satisfied by choosing $D = C = i \Gamma^\UZ$. From \eqref{gamma_11d_dagger}, one can infer useful expressions for the transpose (and hermitian conjugate) of anti-symmetric products of gamma matrices $\Gamma^{\ul{M}_1 \ldots \ul{M}_n} \equiv \Gamma^{[\ul{M}_1} \cdots \Gamma^{\ul{M}_n]}$, namely
\begin{align}
	(\Gamma^{\ul{M}\ul{N}})^\transp 				&= 
		\Gamma^\UZ \Gamma^{\ul{M}\ul{N}} 					\Gamma^\UZ \; , \eqlabel{gamma_11d_dagger_mult2} \\
	(\Gamma^{\ul{M}\ul{N}\ul{P}})^\transp 			&= -
		\Gamma^\UZ \Gamma^{\ul{M}\ul{N}\ul{P}} 				\Gamma^\UZ \; , \eqlabel{gamma_11d_dagger_mult3} \\
	(\Gamma^{\ul{M}\ul{N}\ul{P}\ul{Q}})^\transp 		&= -
		\Gamma^\UZ \Gamma^{\ul{M}\ul{N}\ul{P}\ul{Q}} 			\Gamma^\UZ \; , \eqlabel{gamma_11d_dagger_mult4} \\	
	(\Gamma^{\ul{M}\ul{N}\ul{P}\ul{Q}\ul{R}})^\transp 	&= 	
		\Gamma^\UZ \Gamma^{\ul{M}\ul{N}\ul{P}\ul{Q}\ul{R}} 		\Gamma^\UZ \; , \eqlabel{gamma_11d_dagger_mult5} \\
	(\Gamma^{\ul{M}\ul{N}\ul{P}\ul{Q}\ul{R}\ul{S}})^\transp	&= 	
		\Gamma^\UZ \Gamma^{\ul{M}\ul{N}\ul{P}\ul{Q}\ul{R}\ul{S}} 	\Gamma^\UZ \; , \eqlabel{gamma_11d_dagger_mult6}
\end{align}
with the same formul\ae\ holding true for the hermitian conjugate. The Dirac and Majorana conjugates of an 11-dimensional Dirac spinor $\Psi$ are given by $\bar\Psi \equiv \Psi^\dagger D = i \Psi^\dagger \Gamma^\UZ$\newnot{symbol:dirac_conjug} and $\Psi^c \equiv \Psi^\transp C = i \Psi^\transp \Gamma^\UZ$, respectively. The Majorana condition
\be\eqlabel{majorana_cond_gen}
	\Psi^c = \bar\Psi \; , 
\ee
thus amounts to a Majorana spinor $\Psi_{\rm Maj.}$ having real components, that is
\be\eqlabel{majorana_cond_11}
	\Psi_{\rm Maj.}^\ast = \Psi_{\rm Maj.} \; .
\ee
All 11-dimensional spinors appearing in this thesis are taken to be Majorana and therefore satisfy \eqref{majorana_cond_11}.

The 11-dimensional gamma matrix identities used in this thesis can be succinctly summarised in the following formula (see ref.~\cite{Miemiec:2005ry,Naito:1986cr}; for an outline of the proof, see ref.~\cite{Bergshoeff:1982az})
\be\eqlabel{gamma11_prod_id}
	\Gamma^{\ul{M}_i\ldots \ul{M}_1}\Gamma_{\ul{N}_1\ldots \ul{N}_j} = \sum\limits_{k=0}^{{\rm min}(i,j)} k!
	\binom{i}{k} \binom{j}{k}
	\delta^{[\ul{M}_1}_{[\ul{N}_1}\cdots\delta^{\ul{M}_k}_{\ul{N}_k}
	\Gamma^{\ul{M}_i\ldots \ul{M}_{k+1}]}{}_{\ul{N}_{k+1}\ldots \ul{N}_j]} \; ,
\ee
which is valid for $i+j \leq 11$. The case $i+j > 11$ can be reduced to the previous case by means of the following important duality relation~\cite{Miemiec:2005ry,Naito:1986cr,Bergshoeff:1982az}
\be\eqlabel{gamma11_dual}
	\Gamma^{\ul{M}_1 \ldots \ul{M}_i} = 
		\frac{(-1)^{(i+1)(i-2)/2}}{(11-i)!} \varepsilon^{\ul{M}_1 \ldots \ul{M}_{11}} \Gamma_{\ul{M}_{i+1} \ldots \ul{M}_{11}} \; ,
\ee
where the totally anti-symmetric $\varepsilon$-symbol is such that $\varepsilon^{\ul{0} \ldots \ul{10}} = - \varepsilon_{\ul{0} \ldots \ul{10}} = - 1$ (see \appref{forms}).

The 10-dimensional Euclidean gamma matrices $\gamma^{\underline{m}}$ satisfy the Clifford algebra
\be
	\{ \gamma^{\underline{m}}, \gamma^{\underline{n}} \} = 2 \delta^{\underline{m}\underline{n}} \id_{32\times 32}\,  .
	\eqlabel{Cliff10}
\ee
In accordance with our 11-dimensional conventions they are chosen to be real matrices and are, hence, also symmetric. The ten dimensional chirality operator $\gamma^{(11)}$ is given by
\be
	\gamma^{(11)} = i \gamma^{\underline{1}} \cdots \gamma^{\underline{10}}  \eqlabel{gamma11},
\ee
and it satisfies the relations $(\gamma^{(11)})^2 = \id_{32\times 32}$, $(\gamma^{(11)})^\ast = - \gamma^{(11)}$, $(\gamma^{(11)})^T = - \gamma^{(11)}$ and $\{\gamma^{(11)} , \gamma^{\underline{m}} \}$ $=$ $0$. Ten-dimensional Dirac spinors $\eta$ are $32$-dimensional complex, as in $11$ dimensions, and are taken to be commuting. Positive (negative) chirality spinors $\eta$ ($\eta^\star$) are then defined by $\gamma^{(11)}\eta = \eta$ ($\gamma^{(11)}\eta^\star = -\eta^\star$). Written in complex coordinates the anti-commutation relations for the gamma matrices read
\be\eqlabel{cy5_hol_cliff_alg}
	\{ \gamma^{\underline{\mu}}, \gamma^{\bar{\underline{\nu}}} \} = 2\, \delta^{\underline{\mu}\bar{\underline{\nu}}} \id_{32\times 32} , \qquad\qquad 
	\{ \gamma^{\underline{\mu}},\gamma^{\underline{\nu}} \} = \{ \gamma^{\bar{{\underline{\mu}}}}, \gamma^{\bar{\underline{\nu}}} \} = 0\; .
\ee
As usual, the gamma matrices in complex coordinates can be interpreted as creation and annihilation operators. If one defines a ``ground state'' $\eta$ by
\be
	\gamma^{\bar{\mu}}\eta = 0 \; ,
\ee
then $\eta$ has positive and $\eta^\star$ negative chirality. The other spinor states are obtained by acting with up to five creation operators $\gamma^\mu$ on $\eta$.

In one dimension, there is only a single gamma matrix $\gamma$ (a $1\times 1$ matrix) which we take to be~$\gamma=-i$. This choice complies with the one-dimensional Clifford algebra $\{\gamma,\gamma\} = - 2$, which has the correct sign since our single dimension $\tau$ is time-like. One-dimensional Dirac spinors $\psi$ are complex one-component anti-commuting objects and we often denote their complex conjugate by $\bar{\psi} = (\psi)^\ast$. Complex conjugation of a product of two anti-commuting objects is defined to be
\be
	(\psi_1 \psi_2)^\ast \equiv \bar{\psi}_2 \bar{\psi}_1 \; .
\ee
Note the change of order on the right hand side, which resembles the operation of Hermitian conjugation in higher dimensions.

Spinorial differentiation and Berezin integration are the same operations and satisfy the relations
\bea
	&\partial_\theta \theta = 1 , \;\; \partial_\theta \bar\theta = 0 , \;\; \partial_{\bar\theta} \theta = 0 , \;\; \partial_{\bar\theta} \bar\theta = 1 , \\
	&\partial_\theta^2 = 0, \;\; \partial_{\bar\theta}^2 = 0, \;\; \{ \partial_\theta, \partial_{\bar\theta} \} = 0 \; ,
\eea
where $\partial_\theta \equiv \partial/\partial\theta$ and $\partial_{\bar\theta} \equiv \partial/\partial\bar\theta = - (\partial_\theta)^\ast$. The rules for Berezin integration can be read off by replacing $\partial_\theta \rightarrow \int d\theta$ and $\partial_{\bar\theta} \rightarrow \int d\bar\theta$. We also abbreviate $d^2 \theta \equiv d\theta d\bar\theta$ so that
\be
	\int d^2\theta\, \theta\bar\theta = -1 \; .
\ee

The relation between $11$-, $10$- and one-dimensional gamma matrices is summarised by the decomposition
\be\eqlabel{gamma_matrix_dim_red_split}
	\Gamma^\UZ = (- i) \otimes \gamma^{(11)} , \qquad \Gamma^{\underline{m}} = \id_{1\times 1} \otimes \gamma^{\underline{m}} ,
\ee
where the tensor product between a complex number and a $32\times 32$ matrix has been introduced solely to make contact with similar formul\ae\ for compactifications to more than one dimension. As can be checked quickly, the matrices~\eqref*{gamma_matrix_dim_red_split} indeed satisfy the 11-dimensional anti-commutation relations~\eqref*{cliff11,gamma_11d_dagger}, provided the $\gamma^{\underline{m}}$ satisfy the 10-dimensional anti-commutation relations~\eqref*{Cliff10}. Dirac spinors $\Psi$ in 11-dimensions can be written as (linear combinations of) tensor products of the form $\psi\otimes\eta$, where $\psi$ and $\eta$ are one- and $10$-dimensional spinors, respectively. An 11-dimensional Majorana spinor $\Psi_{\rm Maj.}$ can be decomposed as
\be
	\Psi_{\rm Maj.} = \psi\otimes\eta + \bar{\psi}\otimes\eta^\star\; .
\ee

\section{Differential forms, cohomology and (pseudo-)Rieman\-ni\-an geometry}\applabel{forms}

We choose conventions -- largely following ref.~\cite{Nakahara:1990th} -- such that a $p$-form $\omega$ on a differentiable manifold $\manifold$ can be written in local coordinates $x^\mu$ as follows
\be
	\omega \equiv \oneon{p!}\omega_{\mu_1 \ldots \mu_p} dx^{\mu_1} \wedge \cdots \wedge dx^{\mu_p} \; .
\ee
The components $\omega_{\mu_1 \ldots \mu_p}$ are by definition totally anti-symmetric in all indices, that is $\omega_{\mu_1 \ldots \mu_p} = \omega_{[\mu_1 \ldots \mu_p]}$. The combinatorial pre-factor ensures, in particular, that for $p=D$ we have $\omega = \omega_{0 \ldots d} dx^0 \wedge \cdots \wedge dx^d$, where $D$ denotes the dimension of $\manifold$, $d\equiv D-1$ and Lorentzian signature is assumed. For Euclidean signature, the corresponding relation is $\omega = \omega_{1 \ldots D} dx^1 \wedge \cdots \wedge dx^D$.

A $p$-form $\omega$ and a $q$-form $\eta$ can be combined to yield a $(p+q)$-form by means of the wedge-product $\wedge$ defined as
\begin{gather}
\begin{split}
	\omega\wedge\eta &= \oneon{(p+q)!} (\omega\wedge\eta)_{\mu_1\ldots\mu_{p+q}} 
		dx^{\mu_1} \wedge \cdots \wedge dx^{\mu_{p+q}} \\ &\equiv
		\oneon{p!q!} \omega_{\mu_1\ldots\mu_p}\eta_{\mu_{p+1}\ldots\mu_{p+q}} dx^{\mu_1} \wedge \cdots \wedge dx^{\mu_{p+q}} \; ,
\eqlabel{def_wedge_product}\end{split}
\end{gather}
where $(\omega\wedge\eta)_{\mu_1\ldots\mu_{p+q}} = \frac{(p+q)!}{p!q!}
\omega_{[\mu_1\ldots\mu_p}\eta_{\mu_{p+1}\ldots\mu_{p+q}]}$. The exterior derivative $d$ maps the $p$-form $\omega$ into the $(p+1)$-form $d\omega \equiv \partial\wedge\omega$. In local coordinates, it is given by
\be
	d\omega = \oneon{p!} \partial_\mu \omega_{\mu_1 \ldots \mu_p}
		dx^\mu \wedge dx^{\mu_1} \wedge \cdots \wedge dx^{\mu_p} \; .
\ee
The exterior derivative is nilpotent $d^2 = 0$ and satisfies the generalised Leibniz rule
\be
	d(\omega\wedge\eta) = d\omega\wedge\eta + (-1)^{\deg \omega} \omega\wedge d\eta \; ,
\ee
where $\deg\omega \equiv p$ for a $p$-form $\omega$.

We now turn to Riemannian manifolds (or pseudo-Riemannian manifolds for the case of Lorentzian metrics). With the help of the metric, the Hodge-star operation $\ast$, which linearly maps the $p$-form $\omega$ into the $(D-p)$-form $\ast\omega$, is defined as follows
\be
	\ast\omega \equiv \frac{1}{p!(D-p)!} 
		\omega^{\mu_1 \ldots \mu_p} \epsilon_{\mu_1 \ldots \mu_p\mu_{p+1} \ldots \mu_D}
		dx^{\mu_{p+1}} \wedge \cdots \wedge dx^{\mu_D} \; ,
\ee
where $\epsilon_{\mu_1 \ldots \mu_D}$ is the curved totally-antisymmetric $\epsilon$-tensor in $D$ dimensions. 

Starting with flat space, we adopt the convention $\varepsilon_{0\ldots d}\equiv +1$ ($\varepsilon_{1\ldots D}\equiv +1$ for Euclidean signature) and define the $\varepsilon$-symbol with upper indices as $\varepsilon^{\mu_1\ldots \mu_D} \equiv (-1)^s \, \varepsilon_{\mu_1\ldots \mu_D}$, where $s$ corresponds to the number of minus signs in the signature of the metric. In flat space, $\varepsilon$ is a numerical invariant. The curved space equivalent is the totally anti-symmetric tensor density $\epsilon$ related to $\varepsilon$ by
\be
	\epsilon_{\mu_1\ldots\mu_D} \equiv \sqrt{|g|} \varepsilon_{\mu_1\ldots\mu_D} \; ,
\ee
where $g \equiv \det g_{\mu\nu}$. The indices on $\epsilon$ are raised and lowered as usual, using the metric $g_{\mu\nu}$, and it follows that
\be
	\epsilon^{\mu_1\ldots\mu_D} = \oneon{\sqrt{|g|}} \varepsilon^{\mu_1\ldots\mu_D}
															= (-1)^s \oneon{\sqrt{|g|}} \varepsilon_{\mu_1\ldots\mu_D} \; .
\ee
The fundamental $\varepsilon$-symbol identity in $D$ dimensions is (see, for example, ref.~\cite{Wald:1984rg} pp. 432):
\be\eqlabel{epsilon_identity1}
	\varepsilon_{\mu_1\ldots\mu_n \nu_1\ldots\nu_p} \varepsilon_{\mu_1\ldots\mu_n \rho_1\ldots\rho_p}
		= n!\,p!\, \delta_{[\nu_1|\rho_1|} \delta_{\nu_2 |\rho_2|} \cdots 
			\delta_{\nu_{p-1} |\rho_{p-1}|} \delta_{\nu_p]\rho_p} \; ,
\ee
valid for all $n$ and $p$ with $n+p=D$. The curved space version is given by
\be\eqlabel{epsilon_identity2}
	\epsilon^{\mu_1\ldots\mu_n \nu_1\ldots\nu_p} \epsilon_{\mu_1\ldots\mu_n \rho_1\ldots\rho_p}
		= (-1)^s \, n!\,p!\, \delta^{\nu_1 \ldots \nu_p}_{\rho_1 \ldots \rho_p} \; ,
\ee
valid for all $n$ and $p$ with $n+p=D$ and the generalised Kronecker-$\delta$ is defined as
$\delta^{\nu_1 \ldots \nu_p}_{\rho_1 \ldots \rho_p} \equiv
 \delta^{[\nu_1}_{[\rho_1} \cdots \delta^{\nu_p]}_{\rho_p]} = 
 \delta^{\nu_1}_{[\rho_1} \cdots \delta^{\nu_p}_{\rho_p]} =
 \delta^{[\nu_1}_{\rho_1} \cdots \delta^{\nu_p]}_{\rho_p}$\newnot{symbol:asymm_kronecker}.

With the help of the $\epsilon$-tensor identity, one verifies the Hodge duality relation
\be\eqlabel{dbl_hodge_star}
	\ast\ast\omega = (-1)^{s+(D+1)\deg\omega} \omega \; .
\ee
The Hodge-star induces an inner product $(\cdot,\cdot)$ on the space of $p$-forms. It is defined as
\be\eqlabel{forms_inner_product}
	\left(\omega,\eta\right) \equiv \int_\manifold \omega\wedge\ast\eta 
		= \oneon{p!} \int_\manifold \omega_{\mu_1\ldots\mu_p}\eta^{\mu_1\ldots\mu_p}\sqrt{|g|}d^D x 
\ee
for two $p$-forms $\omega$ and $\eta$. Via the inner product, the defining property of the adjoint $d^\dagger$ of the exterior derivative $d$ can be stated as follows
\be
	\left(d\omega, \eta\right) = \left(\omega, d^\dagger \eta\right)
\ee
for all $(p-1)$-forms $\omega$ and all $p$-forms $\eta$. The adjoint -- or codifferential -- maps $p$-forms into $(p-1)$-forms and, using $\int d(\omega\wedge\ast\eta) = 0$ and~\eqref*{dbl_hodge_star}, one arrives at the explicit expression $d^\dagger = (-1)^{D(p+1)+s+1} \ast d \ast$ when acting on $p$-forms. In local coordinates, one finds
\be
	d^\dagger \omega = - \oneon{(p-1)!} \nabla^{\mu_1} \omega_{\mu_1\ldots\mu_p} dx^{\mu_2} \wedge \cdots \wedge dx^{\mu_p} \; .
\ee
Like $d$, the codifferential is nilpotent, $d^\dagger {}^2 = \ast d \ast\ast d \propto \ast d^2 \ast = 0$. A further differential operator -- the Laplacian $\Delta$ -- is defined by
\be\eqlabel{def_laplacian}
	\Delta \equiv (d+d^\dagger)^2 = dd^\dagger + d^\dagger d \; .
\ee
It maps $p$-forms into $p$-forms and satisfies
\be
	\Big[ \Delta \ast - (-1)^s \ast \Delta \Big] \omega = 0 \; .
\ee

A $p$-form $\omega$ is called closed (co-closed) if it satisfies $d\omega=0$ ($d^\dagger \omega=0$) and exact (co-exact) if there exists a $(p\begin{smallmatrix}-\vspace{-2pt}\\(+)\end{smallmatrix}1)$-form $\eta$ such that $\omega = d\eta$ ($\omega = d^\dagger \eta$), globally. The space of  closed (exact) $p$-forms on $\manifold$ is denoted by $Z^p (\manifold)$ ($B^p (\manifold)$). The $p$-th de Rham cohomology group $H^p (\manifold)$ is defined by $H^p (\manifold) \equiv Z^p (\manifold) / B^p (\manifold)$ and the dimension of $H^p (\manifold)$ is usually referred to as the $p$-th Betti number, $b^p (\manifold) \equiv \dim H^p (\manifold)$. A differential form $\omega$ satisfying $\Delta\omega=0$ is called harmonic. Harmonicity is equivalent to being both closed and co-closed. Hodge's theorem states that on a compact orientable Riemannian manifold $\manifold$, the space of harmonic $p$-forms ${\rm Harm}^p (\manifold)$ is isomorphic to $H^p (\manifold)$.

Finally, the invariant volume element $\Omega_\manifold$ is defined as 
\be\eqlabel{volume_form}
	\Omega_\manifold \equiv\begin{cases} 	\sqrt{|g|} \, dx^0 \wedge \cdots \wedge dx^d & \qquad\text{for Lorentzian manifolds} \; ,\\
									\sqrt{|g|} \, dx^1 \wedge \cdots \wedge dx^D & \qquad\text{for Euclidean manifolds} \; . \end{cases}
\ee
When integrated over the entire manifold $\manifold$, it measures the volume $V$ of $\manifold$
\be\eqlabel{def_volume}
	V \equiv {\rm vol}(\manifold) = \int_\manifold \Omega_\manifold \; .
\ee
By virtue of the Hodge star,~\eqref*{volume_form} can be re-expressed in a coordinate-invariant fashion
\be
	\Omega_\manifold = \ast 1
\ee
upon using $dx^{\mu_1} \wedge \cdots \wedge dx^{\mu_D}=(-1)^s \sqrt{|g|}\epsilon^{\mu_1\ldots\mu_D} dx^0 \wedge \cdots \wedge dx^d$. The invariant volume element $\Omega_\manifold$ is an example of a top form, that is a form whose degree is equal to the number of dimensions of the manifold it is defined on.

\section{Complex geometry and K\"ahler manifolds}\applabel{complex_geometry}

In this section, we lay down our conventions, which largely follow refs.~\cite{Nakahara:1990th,Candelas:1987is}, and summarise some mathematical results on differential forms on complex manifolds and on complex geometry. The \emph{complex} dimension of a complex manifold $\manifold$ is denoted by $n \equiv \dim_\C \manifold$ and we will consider Hermitian metrics with Euclidean signature on $\manifold$.

On a complex manifold, the $2n$ real coordinates $x^m$, where $m=1,\ldots,2n$\newnot{symbol:10indEucl}, can locally be re-arranged into $n$ holomorphic coordinates $z^\mu$, where $\mu=1,\ldots,n$\newnot{symbol:10indEuclhol}, and $n$ anti-holomorphic coordinates $\bar{z}^{\bar\mu}$, where $\bar\mu=\bar{1},\ldots,\bar{n}$\newnot{symbol:10indEuclahol}, via
\be\eqlabel{real_complex_coords}
	z^\mu = \oneon{\sqrt{2}}\left(x^\mu+i\, x^{\mu + n}\right) \; , \qquad 
	\bar{z}^{\bar\mu} =\oneon{\sqrt{2}}\left(x^{\bar{\mu}}-i\, x^{\bar{\mu} +n}\right) \; .
\ee
This amounts to fixing the complex structure, a globally-defined smooth tensor ${\cal J}$ of rank $(1,1)$ with defining properties\footnote{In this section, the notations $(\cdot)^\ast$ and $\overline{(\cdot)}$ are used interchangeably to denote complex conjugation.}\newnot{symbol:complex_conjug} ${\cal J}^2 = -\id$ and ${\cal J}^\ast = {\cal J}$, to
\be
	{\cal J}_\mu {}^\nu = i \delta_\mu {}^\nu \; , \qquad\qquad 
	{\cal J}_{\bar\mu} {}^{\bar\nu} = - i \delta_{\bar\mu} {}^{\bar\nu} \; .
\ee
The complexification~\eqref*{real_complex_coords} applies more generally to contravariant vectors. Higher rank tensors are complexified by transforming each index separately according to eqs.~\eqref*{real_complex_coords} and are split into $(p,q)$ index types by means of the projection operators $P_\pm \equiv (\id \mp i {\cal J})/2$, which obey $(P_\pm)^2 = P_\pm$, $P_\pm P_\mp = 0$, $(P_\pm)^\ast = P_\mp$ and $P_+ + P_- = \id$. That is, a $(p,q)$-tensor $T$ satisfies
\be\eqlabel{real_complex_pq_tensor}
	(P_+)^p (P_-)^q T = T \; .
\ee
Covariant vectors $v_\mu$ are complexified according to
\be\eqlabel{real_complex_covariant_vectors}
	v^\C_\mu = \oneon{\sqrt{2}} \left( v^\R_\mu - i v^\R_{\mu + n} \right) \; , \qquad
	v^\C_{\bar\mu} = \oneon{\sqrt{2}} \left( v^\R_{\bar\mu} + i v^\R_{\bar\mu + n} \right) \; .
\ee
The choice of signs and numerical factors in~\eqref*{real_complex_coords,real_complex_covariant_vectors} ensures that index contractions are converted with weight one
\be\eqlabel{real_complex_tensor_contractions_conversion}
	T^{m_1\ldots m_i} U_{m_1 \ldots m_i} = \sum_{j=0}^i
	T^{\mu_1 \ldots \mu_j \bar{\nu}_1 \ldots \bar{\nu}_{i-j}} U_{\mu_1 \ldots \mu_j \bar{\nu}_1 \ldots \bar{\nu}_{i-j}} \; .
\ee
The metric $g_{mn}$ is complexified using~\eqref*{real_complex_covariant_vectors} for each index. The complexified metric is denoted by $G_{\mu\bar\nu}$. By assumption, it is a Hermitian, ${\cal J}_\mu {}^\rho {\cal J}_{\bar\nu} {}^{\bar\gamma} G_{\rho\bar\gamma} = G_{\mu\bar\nu}$, $(0,2)$-tensor with $(1,1)$ index structure. Note that
\begin{gather}
	ds^2 = g_{mn} dx^m dx^n = 2 G_{\mu\bar\nu} dz^\mu d\bar{z}^{\bar\nu} \; , \eqlabel{real_complex_ds_conversion} \\
	g^{mn} g_{mn} = 2 \, G^{\mu\bar\nu} G_{\mu\bar\nu} = 2n \; , \\
	\sqrt{|g|} = G \; ,
\end{gather}
where $g\equiv\det g_{mn}$ and $G\equiv\det G_{\mu\bar\nu}$.

After these preliminary remarks, we now turn to differential forms on complex manifolds. An $r$-form can be split according to~\eqref*{real_complex_pq_tensor} into $(p,q)$-forms, with $p+q=r$. In our conventions, a $(p,q)$-form $\omega$ is defined in local holomorphic coordinates as follows
\be\eqlabel{complex_form_expansion}
	\omega \equiv \oneon{p!q!} \omega_{\mu_1 \ldots \nu_p\bar{\nu}_1 \ldots \bar{\nu}_q} 
		dz^{\mu_1} \wedge \cdots \wedge dz^{\mu_p} \wedge 
		d\bar{z}^{\bar{\nu}_1} \wedge \cdots \wedge d\bar{z}^{\bar{\nu}_q} \; .
\ee
The components of $\omega$ are separately anti-symmetric in the holomorphic and anti-ho\-lo\-mor\-phic indices, $\omega_{\mu_1 \ldots \nu_p\bar{\nu}_1 \ldots \bar{\nu}_q} = \omega_{[\mu_1 \ldots \mu_p]\bar{\nu}_1 \ldots \bar{\nu}_q} = \omega_{\mu_1 \ldots \mu_p[\bar{\nu}_1 \ldots \bar{\nu}_q]}$. With this definition and the original expression~\eqref*{def_wedge_product} for the wedge product, one arrives at
\begin{multline}
	\omega\wedge\eta = \oneon{p_{\omega}!q_{\omega}!p_{\eta}!q_{\eta}!}
	\omega_{\mu_1 \ldots \mu_{p_{\omega}}\bar{\nu}_1 \ldots \bar{\nu}_{q_{\omega}}}
	\eta_{\rho_1 \ldots \rho_{p_{\eta}}
				\bar{\sigma}_1 \ldots \bar{\sigma}_{q_{\eta}}} \\ \shoveright{
	dz^{\mu_1} \wedge \cdots \wedge dz^{\mu_{p_{\omega}}} \wedge 
	d\bar{z}^{\bar{\nu}_1} \wedge \cdots \wedge d\bar{z}^{\bar{\nu}_{q_{\omega}}} \wedge } \\ 
	dz^{\rho_1} \wedge \cdots \wedge dz^{\rho_{p_{\eta}}} \wedge 
	d\bar{z}^{\bar{\sigma}_1} \wedge \cdots \wedge d\bar{z}^{\bar{\sigma}_{q_{\eta}}} \; ,
\end{multline}
for the wedge-product of a $(p_{\omega},q_{\omega})$-form $\omega$ and a $(p_{\eta},q_{\eta})$-form $\eta$. The wedge-product satisfies $\overline{\omega\wedge\eta} = \overline{\omega}\wedge\overline{\eta}$. The Hodge-star $\ast\omega$ of a $(p,q)$-form $\omega$ is an $(n-q,n-p)$-form and in local holomorphic coordinates, it is given by
\begin{multline}
	\ast\omega = \frac{i^n (-1)^{\frac{n(n-1)}{2}+np}}{p!(n-p)!q!(n-q)!} 
		\omega^{\bar{\mu}_1 \ldots \bar{\mu}_p \nu_1 \ldots \nu_q} 
		\epsilon_{\bar{\mu}_1 \ldots \bar{\mu}_n} 	
		\epsilon_{\nu_1 \ldots \nu_n} \\
		dz^{\nu_{q+1}} \wedge \cdots \wedge dz^{\nu_n} \wedge
		d\bar{z}^{\bar{\mu}_{p+1}} \wedge \cdots \wedge d\bar{z}^{\bar{\mu}_n} \; .
\end{multline}
Note that $\epsilon_{m_1 \ldots m_{2n}}$ becomes $\epsilon_{\mu_1 \ldots \mu_n} \epsilon_{\bar{\nu}_1 \ldots \bar{\nu}_n}$ when changing from real to complex (holomorphic) coordinates on a Hermitian manifold. The $\epsilon$-tensor conventions are the same as in \appref{forms}. The Hodge-star satisfies
\begin{align}
	\overline{\ast\omega}	&= 	\ast\bar{\omega} \; , \\
	\ast\ast\omega 			&=	(-1)^{p+q} \omega \; .
\end{align}
The inner product is inherited from~\eqref*{forms_inner_product}. For two $(p,q)$-forms $\omega$ and $\eta$, it reads
\be
	\left(\omega,\eta\right) 	= \int_\manifold \omega\wedge\bar{\ast}\eta 
						= \frac{i^n (-1)^{\frac{n(n-1)}{2}}}{p!q!}\int_\manifold
							\omega_{\mu_1 \ldots \mu_p\bar{\nu}_1 \ldots \bar{\nu}_q}
							\overline{\eta}^{\mu_1 \ldots \mu_p\bar{\nu}_1 \ldots \bar{\nu}_q} \, G \, d^{2n} z \; ,
\ee
where $\bar{\ast}\eta \equiv \overline{\ast\eta} = \ast\bar{\eta}$ and $d^{2n} z \equiv dz^1 \wedge\cdots\wedge dz^n\wedge d\bar{z}^{\bar{1}} \wedge\cdots\wedge d\bar{z}^{\bar{n}}$. The expression $\left(\omega,\eta\right)$ is hermitian, $\overline{\left(\omega,\eta\right)}=\left(\eta,\omega\right)$, as is indeed required for an inner product.

From the complex structure ${\cal J}$ and the Hermitian metric $G_{\mu\bar\nu}$, one can build a $(1,1)$-form $J_{\mu\bar\nu} \equiv {\cal J}_\mu {}^\rho G_{\rho\bar\nu}$ called the K\"ahler -- or fundamental -- form. In local holomorphic coordinates, one has
\be
	J = i G_{\mu\bar\nu} dz^{\mu}\wedge d\bar{z}^{\bar\nu} \; .
\ee
It is straightforward to verify that $J$ is real, $J^\ast = J$, and that it is related to the invariant volume element $\Omega_\manifold = \ast 1$ via 
\begin{gather}\begin{split}\eqlabel{vol_form_cplx}
	\frac{J^n}{n!} 	&= i^n (-1)^{\frac{n(n-1)}{2}} \, G \, d^{2n}z \\
				&= \sqrt{|g|} \, d^{2n}x = \ast 1 \; ,
\end{split}\end{gather}
where we have used $d^{2n}z = (-i)^n (-1)^{\frac{n(n-1)}{2}} d^{2n}x$, which follows from~\eqref*{real_complex_coords}. Upon integration over $\manifold$,~\eqref*{vol_form_cplx} becomes Wirtinger's theorem. A Hermitian manifold whose K\"ahler form is closed, $dJ=0$, is called K\"ahler manifold. Closedness of $J$ also implies co-closedness, $d^\dagger J = 0$, and is hence logically equivalent to $J$ being harmonic, $\Delta J = 0$.

In analogy to the local coordinates, the exterior derivative $d$ can be decomposed into holomorphic and anti-holomorphic parts, $d=\partial + \bar\partial$, called Dolbeault operators. They map a $(p,q)$-form $\omega$ into a $(p+1,q)$- and a $(p,q+1)$-form, respectively. In local coordinates, they are given by
\begin{align}
	\eqlabel{dolbeault1}
	\partial \omega &= \oneon{p!q!} \partial_{\mu_1} \omega_{\mu_2 \ldots \mu_{p+1} \bar{\nu}_1 \ldots \bar{\nu}_q}
		dz^{\mu_1} \wedge\cdots\wedge dz^{\mu_{p+1}} \wedge d\bar{z}^{\bar{\nu}_1} \wedge\cdots\wedge d\bar{z}^{\bar{\nu}_q} \; , \\
	\eqlabel{dolbeault2}
	\bar\partial \omega &= \frac{(-1)^p}{p!q!} \partial_{\bar{\nu}_1} \omega_{\mu_1 \ldots \mu_p \bar{\nu}_2 \ldots \bar{\nu}_{q+1}}
		dz^{\mu_1} \wedge\cdots\wedge dz^{\mu_p} \wedge d\bar{z}^{\bar{\nu}_1} \wedge\cdots\wedge d\bar{z}^{\bar{\nu}_{q+1}} \; .
\end{align}
They share with $d$ the property of being nilpotent, $d^2 = \partial^2 = {\bar\partial}^2 = 0$. Furthermore, they satisfy $(\partial)^\ast = \bar\partial$ and $\{ \partial, \bar\partial \} = 0$.

Similarly, the codifferential $d^\dagger$ decomposes into holomorphic and anti-holomorphic parts, $d^\dagger = \partial^\dagger + \bar\partial^\dagger$, where $\partial^\dagger \equiv - \ast \bar\partial \ast$ and $\bar\partial^\dagger \equiv - \ast \partial \ast$. The local coordinate versions are
\begin{align}
	\eqlabel{codolbeault1}
	\partial^\dagger \omega &= 
	\frac{(-1)^{q+1}}{(p-1)!q!} \nabla^{\mu_1} \omega_{\mu_1\ldots\mu_p\bar{\nu}_1\ldots\bar{\nu}_q}
	dz^{\mu_2} \wedge\cdots\wedge dz^{\mu_p} \wedge d\bar{z}^{\bar{\nu}_1} \wedge\cdots\wedge d\bar{z}^{\bar{\nu}_q} \; , \\
	\eqlabel{codolbeault2}
	\bar\partial^\dagger \omega &= 
	\frac{(-1)^{p+1}}{p!(q-1)!} \nabla^{\bar{\nu}_1} \omega_{\mu_1\ldots\mu_p\bar{\nu}_1\ldots\bar{\nu}_q}
	dz^{\mu_1} \wedge\cdots\wedge dz^{\mu_p} \wedge d\bar{z}^{\bar{\nu}_2} \wedge\cdots\wedge d\bar{z}^{\bar{\nu}_q} \; .
\end{align}
They map $(p,q)$-forms into $(p-1,q)$- and $(p,q-1)$-forms, respectively and satisfy $\partial^\dagger {}^2 = \bar\partial^\dagger {}^2 = 0$, $(\partial^\dagger)^\ast = \bar\partial^\dagger$ and $\{ \partial^\dagger, \bar\partial^\dagger \} = 0$. In analogy to~\eqref*{def_laplacian}, one defines Laplacians corresponding to $\partial$ and $\bar\partial$ via
\begin{gather}
	\Delta_\partial \equiv (\partial + \partial^\dagger)^2 = \partial \partial^\dagger + \partial^\dagger \partial \; , \\
	\Delta_{\bar\partial} \equiv (\bar\partial + \bar\partial^\dagger)^2 = \bar\partial \bar\partial^\dagger + \bar\partial^\dagger \bar\partial \; ,
\end{gather}
which map $(p,q)$-forms into $(p,q)$-forms and satisfy
\be\eqlabel{cplx_lapl_props}
	\bar\ast \Delta_{\bar\partial} = \Delta_{\bar\partial} \bar\ast \; , \qquad
	\ast \Delta_{\bar\partial} = \Delta_{\partial}\!\ast \; .
\ee

The space of $\bar\partial$-closed $(p,q)$-forms is called $(p,q)$-cocycle and is denoted by $Z^{p,q}_{\bar\partial} (\manifold)$, whereas the space of $\bar\partial$-exact $(p,q)$-forms is called $(p,q)$-coboundary and is denoted by $B^{p,q}_{\bar\partial} (\manifold)$. The quotient of the two spaces is called the $(p,q)$-th $\bar\partial$-cohomology group, $H^{p,q}_{\bar\partial} (\manifold) \equiv Z^{p,q}_{\bar\partial} (\manifold) / B^{p,q}_{\bar\partial} (\manifold)$. The Hodge numbers $h^{p,q} (\manifold)$ are defined as the complex dimension of the $(p,q)$-th $\bar\partial$-cohomology group, $h^{p,q} (\manifold) \equiv \dim_\C H^{p,q}_{\bar\partial} (\manifold)$. A $(p,q)$-form $\omega$ satisfying $\Delta_{\bar\partial}\omega=0$ is called $\bar\partial$-harmonic and the space of $\bar\partial$-harmonic $(p,q)$-forms is denoted by ${\rm Harm}^{p,q}_{\bar\partial} (\manifold)$. The form $\omega$ is $\bar\partial$-harmonic if and only if it is both $\bar\partial$-closed and $\bar\partial$-co-closed. Hodge's theorem carries over to the case of complex manifolds, where it becomes ${\rm Harm}^{p,q}_{\bar\partial} (\manifold) \cong H^{p,q}_{\bar\partial} (\manifold)$. 

An important result is the fact that on a K\"ahler manifold the three different Laplacians $\Delta$, $\Delta_{\partial}$ and $\Delta_{\bar{\partial}}$ coincide, 
\be\eqlabel{kahler_laplacian}
	\Delta = 2 \Delta_{\partial} = 2 \Delta_{\bar{\partial}} \; ,
\ee
which establishes a connection between the Hodge and Betti numbers, namely $b^r (\manifold) = \sum_{p+q=r} h^{p,q} (\manifold)$.

Finally, we introduce a concept that proves useful in the context of physics applications. Namely, a natural map $\tilde{(\cdot)}$\newnot{symbol:tmap} from the space of $(p,q)$-forms to the space of $(p-1,q-1)$-forms defined by contraction with the inverse Hermitian metric $G^{\mu\bar\nu}$
\be\eqlabel{def_tilde_map}
		\tilde{\omega} \equiv \oneon{(p-1)!(q-1)!}
		{\omega_{\mu_1\ldots\mu_p \bar{\nu}_1\ldots\bar{\nu}_{q-1}}}^{\mu_p} dz^{\mu_1}\wedge \cdots \wedge 
		dz^{\mu_{p-1}} \wedge d\bar{z}^{\bar{\nu}_1} \wedge \cdots \wedge d\bar{z}^{\bar{\nu}_{q-1}}
\ee
for $p > 0$ and $q > 0$. Using~\eqref*{kahler_laplacian} and the covariant constancy of the metric, $\nabla_\mu G_{\nu\bar\rho} = \nabla_{\bar\mu} G_{\nu\bar\rho} = 0$, one can prove that on a K\"ahler manifold $\tilde\omega$ is harmonic if $\omega$ is. To show this, the following two intermediate results are useful
\begin{align}
	\bar\partial \tilde\omega 			&= - \widetilde{(\bar\partial \omega)} \; , \eqlabel{dtildeomega1} \\
	\bar\partial^\dagger \tilde\omega 	&= - \widetilde{(\bar\partial^\dagger \omega)} \; . \eqlabel{dtildeomega2}
\end{align}
They are obtained by direct computations in local coordinates, using~\eqrangeref*{dolbeault1}{codolbeault2} and the covariant constancy of the metric. The proof of harmonicity is immediately completed by recalling that $\Delta_{\bar\partial} \omega = 0$ if and only if $\bar\partial \omega = \bar\partial^\dagger \omega = 0$.

Note that $\tilde{J} = n\, i$ and more generally, if $\omega$ is a harmonic $(1,1)$-form, then $\tilde\omega$ is a harmonic $(0,0)$-form, that is a constant. To determine this constant, we first introduce a map $\kappa$ from a set of $n$ $(1,1)$-forms $\{ \omega_i \}_{i=1,\ldots.n}$ to $\C$ via
\be\eqlabel{def_kappa_map}
	\kappa (\omega_1, \ldots, \omega_n) \equiv \int_\manifold \bigwedge_{i=1}^n \omega_i \; .
\ee
The map $\kappa$ is related to the invariant volume element $\Omega_\manifold$ in the following way
\be\eqlabel{vol_form_kappa}
	\int_\manifold \Omega_\manifold = \frac{\kappa(J,\ldots,J)}{n!} \; ,
\ee
where \eqref{vol_form_cplx} has been used. A direct computation in local coordinates then yields
\be\eqlabel{kappa00}
	\kappa (\omega,J,\ldots,J) = (n-1)! (-i) \, \tilde\omega \int_\manifold \Omega_\manifold \; ,
\ee
where the fact that $\tilde\omega$ is a constant, if $\omega$ is a harmonic $(1,1)$-form, has been used to take it out of the integral. \Eqref{vol_form_kappa,kappa00} determine the constant $\tilde\omega$ to be
\be\eqlabel{tilde_11}
	\tilde\omega = n\, i\, \frac{\kappa(\omega,J,\ldots,J)}{\kappa(J,\ldots,J)} \; .
\ee

Beyond the usefulness in the foregoing calculation, there is, in fact, a rich area of mathematics related to the $\tilde{(\cdot)}$-map. This area is intimately linked to the representation theory of $\mathrm{SL}(2,\C)$. To make contact with the mathematical literature (see, for example, refs.~\cite{GH1978,Wells1980,Huybrechts2004,PetersSteenbrink2008}), we introduce the so-called Lefschetz operator $L$ and its dual $\Lambda=L^\dag$ by
\bea
	&L : \;\; A^{p,q} (\manifold) \rightarrow A^{p+1,q+1} (\manifold), \quad L\omega \equiv \omega\wedge J \; , \\
	&\Lambda : \;\; A^{p,q} (\manifold) \rightarrow A^{p-1,q-1} (\manifold), \quad 
		\Lambda\omega \equiv \begin{cases} (-1)^q \tilde\omega& \text{if $p,q\geq 1$,}\\ 0& \text{otherwise,} \end{cases}
\eea
where $A^{p,q} (\manifold)$ is the vector space of $(p,q)$-forms on $\manifold$. As is customary for linear maps, the parentheses around the arguments of $L$ and $\Lambda$ are omitted. The results we list in the following are valid for compact K\"ahler manifolds $\manifold$ of complex dimension $n$ and can be found, for example, in the textbooks~\cite{GH1978,Wells1980,Huybrechts2004,PetersSteenbrink2008}. We have already shown above that $\Lambda$ preserves the harmonicity of a differential form. The same is also true for $L$. This is succinctly summarised by
\be\eqlabel{Lefschetz_Laplace_comm}
	[L, \Delta] = [\Lambda, \Delta] = 0 \; .
\ee
Another useful relation is
\be\eqlabel{Lefschetz_self_comm}
	[L, \Lambda] = p+q - n \; ,
\ee
valid for all $\omega \in A^{p,q} (\manifold)$. The lengthy proof, which can for example be found in refs.~\cite{GH1978,Wells1980}, is done in a local coordinate patch chosen such that the metric is flat in this patch. \Eqrangeref{Lefschetz_Laplace_comm}{Lefschetz_self_comm} are a subset of the so-called Hodge identities, which summarise relations among the various operators defined on K\"ahler manifolds. Owing to \eqref{Lefschetz_Laplace_comm}, $L$ and $\Lambda$ naturally map between the cohomology groups $H^{p,q} (\manifold)$. In particular, one defines the primitive $(p,q)$-cohomology $H_0^{p,q} (\manifold) \subset H^{p,q} (\manifold)$ as the space of harmonic $(p,q)$-forms $\omega$ satisfying $\Lambda\omega=0$\newnot{symb:prim_form}. In other words
\be
	H_0^{p,q} (\manifold) \equiv (\mathrm{ker}\,\Lambda) \cap H^{p,q} (\manifold) \; .
\ee
In addition, the complex dimension of $H_0^{p,q} (\manifold)$, denoted by $h_0^{p,q} (\manifold)$, is related to the Hodge numbers via
\be\eqlabel{def_prim_Hodge_numbers}
	h_0^{p,q} (\manifold) = h^{p,q} (\manifold) - h^{p-1,q-1} (\manifold) \; ,
\ee
for $p+q \leq n$. The formula~\eqref*{def_prim_Hodge_numbers} shows that $h_0^{p,q} (\manifold)$ is a topological invariant, which is surprising given that $H_0^{p,q} (\manifold)$ is defined with respect to the K\"ahler class $J$. It can be shown that
\be\eqlabel{Lefschetz1_to_Lefschetz2}
	\Lambda\omega = 0 \quad \Longleftrightarrow \quad L^{n-p-q+1} \omega = 0
\ee
and hence $H_0^{p,q} (\manifold) = (\mathrm{ker}\,L^{n-p-q+1}) \cap H^{p,q} (\manifold)$. An explicit example of a primitive $(p,q)$-form is given by the local coordinate expression $dz^1 \wedge\cdots\wedge dz^p \wedge d\bar{z}^{\bar{p}+\bar{1}} \wedge\cdots\wedge d\bar{z}^{\bar{p}+\bar{q}}$ (see ref.~\cite{PetersSteenbrink2008}). An important result is the so-called Hard-Lefschetz decomposition, which asserts that the cohomology groups can be decomposed into a sum of primitive pieces as follows
\be\eqlabel{Lefschetz_decomp}
	H^{p,q} (\manifold) = \bigoplus_{0\leq i\leq \frac{p+q}{2}} L^{i} H_0^{p-i,q-i} (\manifold) \; .
\ee
More explicitly, when this is applied to a harmonic $(p,q)$-form $\omega^{(p,q)}$, one learns that $\omega^{(p,q)}$ can be decomposed according to
\be\eqlabel{Lefschetz_decomp_form}
	\omega^{(p,q)} = \omega_0^{(p,q)} + \omega_0^{(p-1,q-1)} \wedge J + \omega_0^{(p-2,q-2)} \wedge J^2 + \ldots \; ,
\ee
where $\omega_0^{(p,q)} \in H_0^{p,q} (\manifold)$ and $\omega^{(p,q)}$ is primitive precisely if $\omega^{(p,q)} = \omega_0^{(p,q)}$. The Hodge-dual of a primitive $(p,q)$-form $\omega_0$ is given by
\be
	\ast \omega_0 = c\, \omega_0 \wedge J^{n-p-q} \; ,
\ee
where the complex coefficient $c$ can be read off from~\eqref*{dualforms} and together with~\eqref*{Lefschetz1_to_Lefschetz2}, it follows that $L\ast\omega_0 = (\ast\omega_0)\wedge J = 0$.

\chapter{Detailed calculations and proofs}\applabel{long_calcs}

Details of some longer calculations and proofs, which were omitted in the main text to avoid distraction, are collected together in this appendix. These details are not crucial for the main text and if anything, they would have threatened to interrupt the flow of the presentation. Nonetheless, the interested reader might find these additional details useful, especially when a closer exposition to the content matter is desired, for example through checking and reproducing some of the results presented in this thesis. Therefore, instead of omitting these calculations and proofs altogether, a compromise was found by relegating them to this appendix.

\section{Outline of the proof of local supersymmetry of the CJS action}\applabel{proof_CJS_action_susy}

\Secref{sugra11} constitutes a short review of some of the key elements of 11-dimensional supergravity found by Cremmer, Julia and Scherk (CJS) in 1978~\cite{Cremmer:1978km}. The explicit proof of supersymmetry of the CJS action is not found in the recent literature very frequently. Hence, we decided to present some details of the proof in this appendix. A check of supersymmetry will also reassure us that there are no mistakes regarding signs or numerical factors in the formul\ae\ we present in \secref{sugra11}. This is relevant for later chapters where the supersymmetry of the 11-dimensional theory is used to draw conclusions about properties of the lower dimensional effective theories obtained by dimensional reduction. 

We will consider bosonic terms and terms quadratic in fermions at most. That means, in particular, 4-fermi terms are ignored and we will deal with the action as written in \eqref{S_CJS_B,S_CJS_F1} and the supersymmetry transformations~\eqref*{sugra11_susy_transf_g,sugra11_susy_transf_A,sugra11_susy_transf_gravitino1}. For more detailed calculations, we refer to refs.~\cite{Miemiec:2005ry,Naito:1986cr}. In contrast to \secref{sugra11}, we drop the superscript $(11)$ on the infinitesimal 11-dimensional supersymmetry parameter $\epsilon^{(11)}$ and abbreviate it simply to $\epsilon$ for ease of notation.

Before actually applying a supersymmetry variation on the action, we need two auxiliary results derived in most textbooks on general relativity (see for example refs.~\cite{Wald:1984rg,Ortin:2004ms}):
\begin{align}
	\delta \sqrt{-g} &= \oneon{2} \sqrt{-g} g^{MN} \delta g_{MN} \; , \eqlabel{delta_sqrt_of_g} \\
	\delta R &= R_{MN} \delta g^{MN} + (\text{total divergence}) \; .
\end{align}
By varying both sides of $\delta_M {}^N = g_{MP} g^{PN}$, one learns
\be
	\delta g^{MN} = - g^{MP} g^{NQ} \delta g_{PQ} \overset{\delta \rightarrow \delta_\epsilon}{=} - 2 \bar\epsilon \Gamma^{(M} \Psi^{N)} \; .
\ee
Recall that the 4-form field strength $G$ is, at least locally, given by $G=dA$ or in components $G_{MNPQ} = 4 \partial_{\left[M\right.} A_{\left. NPQ\right]}$. For its variations, one learns
\begin{gather}
	\delta G_{MNPQ} = 4 \partial_{\left[M\right.} \delta A_{\left. NPQ\right]} 
		\overset{\delta \rightarrow \delta_\epsilon}{=} - 12 \nabla_{\left[M\right.} (\bar\epsilon \Gamma_{NP} \Psi_{\left. Q\right]}) \; , \\
	\delta G^2 = 2 G^{MNPQ} \delta G_{MNPQ} + 4 G_M {}^{PQR} G_{NPQR} \delta g^{MN} \; .
\end{gather}
where $G^2 \equiv G^{MNPQ} G_{MNPQ}$. Combining the previous two equations, one obtains
\begin{gather}
	\begin{split}
	\delta_\epsilon \int_\manifold d^{11}x \sqrt{-g} \, G^2
		= \int_\manifold d^{11}x \sqrt{-g} \left\{ G^2 \bar\epsilon \Gamma^M \Psi_M 
		+ 4! (\nabla_M G^{MNPQ}) \bar\epsilon \Gamma_{NP} \Psi_Q \right. \\ \left.
		- 8 G_M {}^{PQR} G_{NPQR} \bar\epsilon \Gamma^M \Psi^N \right\} , \end{split} \\
	\delta_\epsilon \int_\manifold G\wedge G\wedge A 
		= \frac{3}{2 (4!)^2} \int_\manifold d^{11}x \sqrt{-g} \, \epsilon^{M_1\ldots M_{11}} 
		G_{M_1\ldots M_4} G_{M_5\ldots M_8} \bar\epsilon \Gamma_{M_9 M_{10}} \Psi_{M_{11}}
\end{gather}
after integrating by parts and using the Bianchi identity $dG=0$.

We are now ready to compute the supersymmetry variation of the bosonic action~\eqref*{S_CJS_B}:
\begin{multline}\eqlabel{susy_trans_S_CJS_B}
	\delta_\epsilon S_{\rm CJS,B} = - \oneon{2\kappa^2_{11}} \int_\manifold d^{11}x \sqrt{-g} \left\{
	  2 \bar\epsilon \Gamma^M \Psi^N \left[ G_{MN} - \oneon{12} G_{MPQR} G_N {}^{PQR} + \oneon{96} g_{MN} G^2 \right] \right. \\ \left. + 
	  \oneon{2} \bar\epsilon \Gamma_{NP} \Psi_Q \left[ \nabla_M G^{MNPQ} + 
	  \oneon{2 (4!)^2} G_{M_1\ldots M_4} G_{M_5\ldots M_8} \epsilon^{M_1\ldots M_8 NPQ} \right] \right\} ,
\end{multline}
deploying the Einstein tensor $G_{MN} \equiv R_{MN} - \oneon{2} g_{MN} R$. Note that the two square brackets, when set to zero, correspond to the equations of motion for the metric and 3-form, respectively (cf. \eqref{sugra11_geom,sugra11_Geom_class_loc_coords}).

Next, we turn to the supersymmetry variation of the fermionic action~\eqref*{S_CJS_F1}. Since we are working at the $(\text{fermi})^2$-level, we only need to keep the supersymmetry variations of the gravitino, for the supersymmetry variations of the metric and 3-form solely contribute 4-fermi terms when applied to the fermionic action. In a first step, we vary the gravitino in the fermionic action~\eqref*{S_CJS_F1} without inserting the explicit expression for the variation:
\be\eqlabel{susy_trans_S_CJS_F1}
	\delta_\epsilon S_{\rm CJS,F} = - \oneon{2\kappa^2_{11}} \int_\manifold d^{11}x \sqrt{-g} \left\{
		\overline{(\delta_\epsilon \Psi_M)} \Phi_{(1)}^M + \bar\Phi_{(2)}^M (\delta_\epsilon \Psi_M) + (\text{fermi})^4 \right\} ,
\ee
where
\begin{align}
	\Phi_{(1)}^M &= \Gamma^{MNP} D_N(\omega) \Psi_P + \oneon{96} \left( \Gamma^{MNPQRS} \Psi_S + 
		12 g^{MN} \Gamma^{PQ} \Psi^R \right) G_{NPQR} \; , \eqlabel{LambdaM_1} \\
	\bar\Phi_{(2)}^M &= \bar{\Psi}_P \Gamma^{PNM} D_N(\omega) + \oneon{96} \left( \bar{\Psi}_N \Gamma^{NPQRSM} + 
		12 \bar{\Psi}^P \Gamma^{QR} g^{MS} \right) G_{PQRS} \; . \eqlabel{LambdaM_2}
\end{align}
The first step is to show that the first two terms in \eqref{susy_trans_S_CJS_F1} are equal to each other. To this end, note first of all that the first term in \eqref{LambdaM_2} contains a covariant derivative acting to the right. Integrating the first term in \eqref{LambdaM_2} by parts yields
\be
	\bar\Phi_{(2)}^M = - \overline{(D_N \Psi_P)} \Gamma^{PNM} + \oneon{96} \left( \bar{\Psi}_N \Gamma^{NPQRSM} + 
		12 \bar{\Psi}^P \Gamma^{QR} g^{MS} \right) G_{PQRS} \; ,
\ee
where the covariant constancy of the metric $\nabla_M g_{NP} = 0$ and vielbein $\nabla_M e_N^{\underline{P}} = 0$ has been used to push the covariant derivative through $\sqrt{-g}$ and the curved gamma matrices (the notation $\nabla_M$ and $D_M$ is used interchangeably in this section). The expression $\overline{D_N \Psi_P}$ is explicitly given by
\be\eqlabel{dirac_conj_cov_deriv}
	\overline{D_N \Psi_P} = i (D_N \Psi_P)^\transp \Gamma^\UZ = i \Psi_P^\transp \overset{\leftarrow}{D} {}_N^\transp \Gamma^\UZ \; ,
\ee
where
\be
	\overset{\leftarrow}{D} {}_N^\transp = \overset{\leftarrow}{\partial}_N + 
		\oneon{4} {\omega_N}^{\underline{Q}\underline{R}} \Gamma^\UZ \Gamma_{\underline{Q}\underline{R}} \Gamma^\UZ \; .
\ee
and we have used \eqref{gamma_11d_dagger_mult2}. \Eqref{dirac_conj_cov_deriv} together with \eqrangeref{gamma_11d_dagger_mult2}{gamma_11d_dagger_mult6} can be utilised to compute the transpose of $\bar\Phi_{(2)}^M$. One finds
\be
	(\bar\Phi_{(2)}^M)^\transp = - i \Gamma^\UZ \Phi_{(1)}^M \; ,
\ee
where \eqref{majorana_cond_11} has been used. A spinor bilinear $\bar\Psi \Phi$ is a scalar in spinor space and hence invariant under transposition $\bar\Psi \Phi = (\bar\Psi \Phi)^\transp = - \Phi^\transp \bar\Psi^\transp$ (the minus sign on the right hand side stems from the anti-commuting nature of the spinors). Thus, the second term in \eqref{susy_trans_S_CJS_F1} becomes
\be
	\bar\Phi_{(2)}^M (\delta_\epsilon \Psi_M) = [\bar\Phi_{(2)}^M (\delta_\epsilon \Psi_M)]^\transp = 
	- (\delta_\epsilon \Psi_M)^\transp (\bar\Phi_{(2)}^M)^\transp = \overline{(\delta_\epsilon \Psi_M)} \Phi_{(1)}^M \; .
\ee
We have established the equivalence of the first two terms in \eqref{susy_trans_S_CJS_F1}. The supersymmetry variation of the fermionic action~\eqref*{S_CJS_F1} can now be summarised as follows
\be\eqlabel{susy_trans_S_CJS_F1_2}
	\delta_\epsilon S_{\rm CJS,F} = - \oneon{2\kappa^2_{11}} \int_\manifold d^{11}x \sqrt{-g} \left\{
		2 \overline{(\delta_\epsilon \Psi_M)} \Phi_{(1)}^M + (\text{fermi})^4 \right\} .
\ee
The full expression reads
\vspace{\parskip}
\begin{multline}
	\overline{(\delta_\epsilon \Psi_M)} \Phi_{(1)}^M = \left[ 2 \overline{D_M \epsilon} -
		\oneon{144} \bar\epsilon (\Gamma_M {}^{NPQR} + 8 \delta_M^N \Gamma^{PQR}) G_{NPQR} \right] \\ \times
		\left[ \Gamma^{MST} D_S(\omega) \Psi_T + \oneon{96} \left( \Gamma^{MSTUVW} \Psi_W + 
		12 g^{MS} \Gamma^{TU} \Psi^V \right) G_{STUV} \right] .
\end{multline}
\vspace{\parskip}\noindent 
The first term in the first square bracket is integrated by parts to take the covariant derivative off the supersymmetry parameter $\epsilon$. After multiplying out the two square brackets, the resulting terms may be split into four groups
\vspace{\parskip}
\begin{align}
	&\overline{(\delta_\epsilon \Psi_M)} \Phi_{(1)}^M = X_1 + \ldots + X_4 \; , \\
	&X_1 = - 2 \bar\epsilon \Gamma^{MNP} D_M D_N \Psi_P \; , \eqlabel{X1} \\ \begin{split}
	&X_2 = - \oneon{144} \bar\epsilon (3 \Gamma^{NPQRST} + 36 g^{NS} g^{PT} \Gamma^{QR}
		+ \Gamma_M {}^{NPQR} \Gamma^{MST} \\ & \hspace{160pt} + 8 \Gamma^{PQR}\Gamma^{NST}) (D_S \Psi_T) G_{NPQR} \; , 
		\eqlabel{X2} \end{split} \\
	&X_3 = - \oneon{4} \bar\epsilon \Gamma_{NP} \Psi_Q \nabla_M G^{MNPQ} \; , \eqlabel{X3} \\ \begin{split}
	&X_4 = - \oneon{(4!)^3} \bar\epsilon \left( \Gamma_M {}^{NPQR} \Gamma^{MSTUVW} + 
		12 \Gamma^{SNPQR} \Gamma^{TU} g^{VW} \right. \\ & \qquad + \left.
		8 \Gamma^{PQR} \Gamma^{NSTUVW} + 
		96 \Gamma^{PQR} \Gamma^{TU} g^{NS} g^{VW} \right) \Psi_W G_{NPQR} G_{STUV} \; . \eqlabel{X4} \end{split}
\end{align}
\vspace{\parskip}\noindent 
One term in $X_3$ has already been dropped by virtue of the Bianchi identity $dG=0$.

$X_1$ can be simplified by using $2 D_{\left[M\right.} D_{\left. N\right]} = [D_M, D_N ] = \oneon{4} R_{MN} {}^{PQ} \Gamma_{PQ}$, which yields
\be
	X_1 = - \oneon{4} \bar\epsilon \Gamma^{MNP} \Gamma_{QR} R_{MN} {}^{QR} \Psi_P \; .
\ee
The product $\Gamma^{MNP} \Gamma_{QR}$ can be re-written by means of \eqref{gamma11_prod_id}. This results in
\be
	X_1 = - \bar\epsilon \Gamma^M \Psi^N G_{MN} \; ,
\ee
where we have used $R_{MN} = R^P {}_{MPN}$, $R_{M\left[NPQ\right]} = 0$, $R_{MNPQ} = - R_{NMPQ} = - R_{MNQP}$ and $R_{MNPQ} = R_{PQMN}$.

Now, we turn to $X_2$. For the third and fourth term in $X_2$, we use again \eqref{gamma11_prod_id} to obtain
\begin{align}
	\begin{split}
	\Gamma_M {}^{NPQR} \Gamma^{MST} (D_S \Psi_T) G_{NPQR} &= (5 \Gamma^{NPQRST} + 48 g^{N\left[S\right.} \Gamma^{\left. T\right] PQR} 
		\\ &\qquad - 84 g^{NS} g^{PT} \Gamma^{QR} ) (D_S \Psi_T) G_{NPQR} \; , \end{split} \\ \begin{split}
	8 \Gamma^{PQR}\Gamma^{NST} (D_S \Psi_T) G_{NPQR} &= (- 8 \Gamma^{NPQRST} - 48 g^{N\left[S\right.} \Gamma^{\left. T\right] PQR} 
		\\ &\qquad + 48 g^{NS} g^{PT} \Gamma^{QR} ) (D_S \Psi_T) G_{NPQR} \; . \end{split}
\end{align}
After putting this back into the expression~\eqref*{X2} for $X_2$, one concludes
\be
	X_2 = 0 \; .
\ee

$X_3$ as stated in \eqref{X3} can not be transformed any further and is already written in its final form. As for $X_4$, the four terms in parenthesis can be re-written by means of \eqref{gamma11_prod_id}. The result reads as follows
\begin{gather}
	\begin{split}
		&\Gamma_M {}^{NPQR} \Gamma^{MSTUVW} \Psi_W G_{NPQR} G_{STUV} = (2 \Gamma^{NPQRSTUVW} \\ & \quad - 
		288 \Gamma^{NPSTW} g^{QU} g^{RV} + 144 \Gamma^W g^{NS} g^{PT} g^{QU} g^{RV}) \Psi_W G_{NPQR} G_{STUV} \\ & \quad + 
		(12 \Gamma^{NPQRSTU} - 192 \Gamma^{NPQST} g^{RU} - 720 \Gamma^{NPS} g^{QT} g^{RU} \\ & \quad +
		576 \Gamma^N g^{PS} g^{QT} g^{RU}) \Psi^V G_{NPQR} G_{STUV} \; , \eqlabel{X_4_1}
	\end{split} \\
	\begin{split}
	12 \Gamma^{SNPQR} \Gamma^{TU} \Psi^V G_{NPQR} G_{STUV} &= 
		(12 \Gamma^{NPQRSTU} + 96 \Gamma^{NPQST} g^{RU} \\ &- 144 \Gamma^{NPS} g^{QT} g^{RU} ) \Psi^V G_{NPQR} G_{STUV} \; ,
	\eqlabel{X_4_2} \end{split} \\
	\begin{split}
	8 \Gamma^{PQR} &\Gamma^{NSTUVW} \Psi_W G_{NPQR} G_{STUV} = (-8 \Gamma^{NPQRSTUVW} \\ & +
	288 \Gamma^{NPSTW} g^{QU} g^{RV} ) \Psi_W G_{NPQR} G_{STUV} + ( -
	24 \Gamma^{NPQRSTU} \\ & + 192 \Gamma^{NPQST} g^{RU} + 288 \Gamma^{NPS} g^{QT} g^{RU} ) \Psi^V G_{NPQR} G_{STUV} \; ,
	\eqlabel{X_4_3} \end{split} \\
	\begin{split}
	96 \Gamma^{PQR} \Gamma^{TU} g^{NS} \Psi^V G_{NPQR} G_{STUV} &= (-96 \Gamma^{NPQST} g^{RU} +
	576 \Gamma^{NPS} g^{QT} g^{RU} \\ &+ 576 \Gamma^N g^{PS} g^{QT} g^{RU} ) \Psi^V G_{NPQR} G_{STUV} \; . \eqlabel{X_4_4}
	\end{split}
\end{gather}
The first term in \eqref{X_4_1,X_4_3} contains $\Gamma^{M_1 \ldots M_9}$, which is dualised to $\Gamma^{N_1 N_2}$ by means of the dualisation relation stated in \eqref{gamma11_dual}.

Finally, collecting all terms and plugging them back into \eqref{X4}, one finds for $X_4$
\begin{multline}
	X_4 = \bar\epsilon \Gamma^M \Psi^N \left[ \oneon{12} G_{MPQR} G_N {}^{PQR} - \oneon{96} g_{MN} G^2 \right] \\
		- \oneon{8 (4!)^2} \bar\epsilon \Gamma_{NP} \Psi_Q G_{M_1 \ldots M_4} G_{M_5 \ldots M_8} \epsilon^{M_1 \ldots M_8 NPQ} \; ,
\end{multline}
and for $S_{\rm CJS,F}$
\be\eqlabel{susy_trans_S_CJS_F1_3}
	\delta_\epsilon S_{\rm CJS,F} = - \oneon{2\kappa^2_{11}} \int_\manifold d^{11}x \sqrt{-g} \left\{
		2 (X_1 + \ldots X_4) + (\text{fermi})^4 \right\} = - \delta_\epsilon S_{\rm CJS,B} \; ,
\ee
with $\delta_\epsilon S_{\rm CJS,B}$ as stated in \eqref{susy_trans_S_CJS_B}. The full 11-dimensional CJS action~\eqref*{S_CJS} is given by $S_{\rm CJS} = S_{\rm CJS,B}+S_{\rm CJS,F}$. From \eqref{susy_trans_S_CJS_F1_3}, we therefore learn that
\be
	\delta_\epsilon S_{\rm CJS} = \delta_\epsilon S_{\rm CJS,B} + \delta_\epsilon S_{\rm CJS,F} = 0 \; ,
\ee
which constitutes an \emph{off-shell} symmetry since the equations of motion have not been invoked. This completes the verification of the vanishing of the supersymmetry variations of the 11-dimensional CJS action~\eqref*{S_CJS} up to the $(\text{fermi})^2$-level.

\section{Chern classes of CICY five-folds}\applabel{CICY_chern_classes}

In this short appendix, we provide some details of how to obtain the formul\ae~\eqrangeref*{c1}{c5} for the Chern classes of CICY five-folds used in \secref{cicy5}. For a complete intersection of hypersurfaces in products of complex projective spaces described by a configuration matrix $[\mathbf{n}|\mathbf{q}]$ as defined in~\eqref*{conf2}, the total Chern class $c([\mathbf{n}|\mathbf{q}])$ is given by the succinct formula~\cite{Hubsch:1992,Fre:1995bc}
\be\eqlabel{CICY_total_chern_class}
	c([\mathbf{n}|\mathbf{q}]) = \frac{ \prod_{r=1}^m (1+J_r)^{n_r+1} } { \prod_{\al=1}^K (1+\sum_{s=1}^m q_\al^s J_s) } \; .
\ee
This is to be regarded as a formal expression that should be expanded into a power series in the K\"ahler forms $J_r$. The $(p,p)$-part of that power series then yields the correct expression for the $p$-th Chern class $c_p([\mathbf{n}|\mathbf{q}])$. The power series automatically truncates after finitely many terms since any $(l,l)$-form with $l > \dim_\C [\mathbf{n}|\mathbf{q}] = \sum_{r=1}^m n_r - K$ vanishes on $[\mathbf{n}|\mathbf{q}]$. 

To obtain the explicit expression for the power series, one performs a multi-variable Taylor expansion about the origin. The following result on the multi-variable Taylor expansion of a function $f(J_1,\ldots,J_m)$ of $m$ variables $J_r$ will be useful
\be
	f(J_1,\ldots,J_m) = \sum_{i_1=0}^{n_1} \cdots \sum_{i_m=0}^{n_m} \left(\frac{\partial^{i_1}}{\partial K_1^{i_1}} \cdots 
		\frac{\partial^{i_m}}{\partial K_m^{i_m}} \frac{f(K_1,\ldots,K_m)}{i_1!\cdots i_m!} \right)\Bigg|_{\mathbf{K} = 0} J_1^{i_1}\cdots J_m^{i_m} \; ,
\ee
where we have already made use of the fact that $J^{n_r+1}_r = 0$ to truncate the sums and we have specialised to an expansion about the origin by setting $\mathbf{K} = 0$.

Regarding the right hand side of \eqref{CICY_total_chern_class} as a function $f(J_1,\ldots,J_m)$ of $m$ variables $J_r$, applying the Taylor expansion formula and collecting terms of equal degrees such that $c([\mathbf{n}|\mathbf{q}]) = 1 + \sum_{p=1}^{\dim_\C [\mathbf{n}|\mathbf{q}]} c_p([\mathbf{n}|\mathbf{q}])$, one has
\be
	c_p ([\mathbf{n}|\mathbf{q}]) = \sum_{r_1, \ldots, r_p=1}^m 
		\frac{\partial^p c([\mathbf{n}|\mathbf{q}])}{\partial J_{r_1}\cdots \partial J_{r_p}} \Bigg|_{\mathbf{J} = 0} J_{r_1} \cdots J_{r_p}
\ee
for $p=1,\ldots,\dim_\C [\mathbf{n}|\mathbf{q}]$. For the first derivative of \eqref{CICY_total_chern_class}, one finds
\be\eqlabel{c_prime1}
	\frac{\partial c([\mathbf{n}||\mathbf{q}])}{\partial J_r} = c([\mathbf{n}||\mathbf{q}])
	\left[ (n_r+1)(1+J_r)^{-1} - \sum_{\al=1}^K q_\al^r (1+\sum_{s=1}^m q_\al^s J_s)^{-1} \right] \; ,
\ee
which is of the symbolic form $f^\prime (x) = f(x)g(x)$ with $g(x)$ representing the expression in square brackets in the equation above. Symbolically, the $n$-th derivative $f^{(n)} (x)$ is thus given by a variant of Fa\`a di Bruno's formula (the generalisation of the chain rule to higher derivatives)
\be
	f^{(n)}(x) = f(x) \left[ \sum_{(k_1,\ldots,k_n)\in T_n} \frac{n!}{k_1! \cdots k_n!} 
		\prod_{m=1, k_m\geq 1}^{n} \left( \frac{g^{(m-1)}(x)}{m!} \right)^{k_m} \right] ,
\ee
where the sum runs over the set $T_n$ of all $n$-tuples $(k_1,\ldots,k_n)$ of non-negative integers satisfying the constraint $1 k_1 + 2 k_2 + \cdots + n k_n = n$. The number of terms in the above sum is equal to the partition number $p(n)$, that is the number of possible partitions of $n$. Here, we just need the cases $n=2,\ldots,5$, which read as follows
\begin{align}
	f^{(2)} &= f [g^2 + g^\prime] \; , \\
	f^{(3)} &= f [g^3 + 3 g g^\prime + g^{(2)}] \; , \\
	f^{(4)} &= f [g^4 + 6 g^2 g^\prime + 3(g^\prime)^2 + 4g g^{(2)} + g^{(3)}] \; , \\
	f^{(5)} &= f [g^5 + 10 g^3 g^\prime + 10 g^2 (g^{(2)})^2 + 15 g (g^\prime)^2 + 10 g^\prime g^{(2)} + 5 g g^{(3)} + g^{(4)}] \; .
\end{align}
Putting everything together, using the above formul\ae\ and noting that $c([\mathbf{n}|\mathbf{q}])|_{\mathbf{J} = 0} = 1$, we find for general configurations (not necessarily Calabi-Yau)
\begin{align}
	c_1^r &\equiv \frac{\partial c([\mathbf{n}||\mathbf{q}])}{\partial J_r}\Bigg|_{\mathbf{J} = 0} = 
		\left[n_r+1-\sum_{\alpha=1}^Kq^r_\alpha\right] , \eqlabel{app_c1} \\
	c_2^{rs} &\equiv \oneon{2} \frac{\partial^2 c([\mathbf{n}||\mathbf{q}])}{\partial J_r \partial J_s}\Bigg|_{\mathbf{J} = 0} = 
		\oneon{2}\left[-(n_r+1)\delta^{rs} + \sum_{\al=1}^K q_\al^r q_\al^s + c_1^r c_1^s \right] , \eqlabel{app_c2} \\
	c_3^{rst} &\equiv \oneon{3!} \frac{\partial^3 c([\mathbf{n}||\mathbf{q}])}{\partial J_r \partial J_s \partial J_t}\Bigg|_{\mathbf{J} = 0} = 
		\oneon{3}\left[(n_r+1)\delta^{rst} - \sum_{\al=1}^K q_\al^r q_\al^s q_\al^t + 3 c_2^{\left(rs\right.} c_1^{\left. t\right)} - c_1^r c_1^s c_1^t \right] , \eqlabel{app_c3} \\
	c_4^{rstu} &\equiv \oneon{4!} \frac{\partial^4 c([\mathbf{n}||\mathbf{q}])}{\partial J_r \cdots \partial J_u}\Bigg|_{\mathbf{J} = 0} = 
		\oneon{4}\left[ -(n_r+1)\delta^{rstu} + \sum_{\al=1}^K q_\al^r q_\al^s q_\al^t q_\al^u + 2 c_2^{\left(rs\right.} c_2^{\left. tu\right)} 
		\right. \nonumber \\ & \left. \qquad\qquad + 4 c_3^{\left(rst\right.} c_1^{\left. u\right)} - 4 c_2^{\left(rs\right.} c_1^t c_1^{\left. u\right)} 
		+ c_1^r c_1^s c_1^t c_1^u \vphantom{\sum_{\al=1}^K} \right] , \eqlabel{app_c4} \\
	c_5^{rstuv} &\equiv \oneon{5!} \frac{\partial^5 c([\mathbf{n}||\mathbf{q}])}{\partial J_r \cdots \partial J_v}\Bigg|_{\mathbf{J} = 0} = 
		\oneon{5}\left[ (n_r+1)\delta^{r \ldots v} - \sum_{\al=1}^K q_\al^{r} \cdots q_\al^{v} 
		+ 5 c_3^{\left(rst\right.} c_2^{\left. uv\right)} \right. \nonumber \\ & \left. \qquad\qquad + 5 c_4^{\left(rstu\right.} c_1^{\left. v\right)} 
		- 5 c_2^{\left(rs\right.} c_2^{tu} c_1^{\left. v\right)} - 5 c_3^{\left(rst\right.} c_1^u c_1^{\left. v\right)} 
		+ 5 c_2^{\left(rs\right.} c_1^t c_1^u c_1^{\left. v\right)} - c_1^r \cdots c_1^v \vphantom{\sum_{\al=1}^K}\right] . \eqlabel{app_c5}
\end{align}
The Calabi-Yau condition $c_1([\mathbf{n}|\mathbf{q}])=0$ translates into a set of algebraic constraints $c_1^r=0$, that is
\be\eqlabel{app_cycondition}
	\sum_{\al=1}^K q^r_\al = n_r+1
\ee
for all $r = 1,\ldots, m$. The numerous terms in~\eqrangeref*{app_c1}{app_c5} involving the coefficients $c_1^r$ thus vanish for CICYs leading to a significant simplification of the expressions above and we arrive at the formul\ae~\eqrangeref*{c1}{c5} quoted in the main text.

Finally, we turn to some inequalities involving Chern classes. From the Calabi-Yau condition~\eqref*{app_cycondition} and the fact that the $q_\al^r$ are non-negative integers, we obtain
\be\eqlabel{cicy_config_ineqs1}
	\sum_{\al=1}^K (q_\al^r)^j \geq n_r+1 \; ,
\ee
valid for all $r = 1,\ldots, m$ and all positive integers $j=1,2,\ldots$. A linear constraint in a single $\CP^n$ produces $\CP^{n-1}$ and is hence redundant. Such a linear constraint would satisfy $\sum_{r=1}^m q_\al^r = 1$ for some $\al$ and should be excluded to avoid over-counting. We therefore require~\cite{Hubsch:1992}
\be\eqlabel{cicy_config_ineqs2}
	\sum_{r=1}^m q_\al^r \geq 2
\ee
for all $\al=1,\ldots,K$. In the following, we distinguish between two different cases and consider them separately.

In the first case, at least one of the $q_\al^r$ is greater than one, for example as in $[6|7]$. Let us call such an element $q_\beta^s$, that is $q_\beta^s \geq 2$ for some $\beta$ and $s$. Then, for $r=s$, the left hand side of the inequality~\eqref*{cicy_config_ineqs1} is genuinely larger than the right hand side and thus
\begin{align}
	c_2^{ss} 		&= \oneon{2}\left[\sum_{\al=1}^K (q_\al^s)^2 - (n_s+1) \right] > 0 \; , \\
	c_3^{sss}		&= \oneon{3}\left[(n_s+1) - \sum_{\al=1}^K (q_\al^s)^3 \right] < 0 \; , \\
	c_4^{ssss}	&= \oneon{4}\left[\sum_{\al=1}^K (q_\al^s)^4 - (n_s+1) + 2 (c_2^{ss})^2 \right] > 0 \; , \\
	c_5^{sssss} 	&= \oneon{5}\left[(n_s+1) - \sum_{\al=1}^K (q_\al^s)^5 + 5 c_3^{sss} c_2^{ss} \right] < 0 \; .
\end{align}
The second case constitutes configurations where all $q_\al^r$ are either zero or one. For example, $\left[\begin{smallmatrix} 4 \\ 4 \\ 2\end{smallmatrix}\Big|\begin{smallmatrix} 1 & 1 & 1 & 1 & 1 \\ 1 & 1 & 1 & 1 & 1 \\ 1 & 1 & 1 & 0 & 0\end{smallmatrix}\right]$. In this case,~\eqref*{cicy_config_ineqs1} becomes an equality and hence, $c_2^{rr} = c_3^{rrr} = c_4^{rrrr} = c_5^{rrrrr} = 0$ for all $r$, while~\eqref*{cicy_config_ineqs2} guarantees that $m \geq 2$. Hence, we choose two different values, $r\neq s$, for $r$ and $s$ and then find
\begin{align}
	c_2^{rs} 		&= 	\oneon{2} \sum_{\al=1}^K q_\al^r q_\al^s 	= 	\oneon{2} \min(n_r+1,n_s+1) > 0 \; , \\
	c_3^{rss}		&= - 	\oneon{3} \sum_{\al=1}^K q_\al^r (q_\al^s)^2	= - 	\oneon{3} \min(n_r+1,n_s+1) < 0 \; , \\
	c_4^{rsss} 	&= 	\oneon{4} \sum_{\al=1}^K q_\al^r (q_\al^s)^3	=	\oneon{4}	\min(n_r+1,n_s+1) > 0 \; , \\
	c_5^{rssss}	&= -	\oneon{5} \sum_{\al=1}^K q_\al^r (q_\al^s)^4	= - 	\oneon{5}	\min(n_r+1,n_s+1) < 0 \; .
\end{align}
This shows that for every configuration at least some of the coefficients $c_2^{rs}$, $c_3^{rst}$, $c_4^{rstu}$ and $c_5^{rstuv}$ are non-zero and of definite sign.

Of course, this is not sufficient to conclude that the Chern classes themselves are non-zero, since it is in general configuration-dependent whether the coefficients considered above dot into combinations of $J_r$s that vanish or not. Instead, using~\eqref*{intersec,mudef}, it is easy to see that $\kappa(J_r,J_s,J_t,J_u,J_v) \geq 0$ and hence
\begin{gather}
	\hat{c}_{4r} = c_4^{stuv} \kappa(J_r,J_s,J_t,J_u,J_v) \geq 0 \; , \\
	\eta ([\mathbf{n}|\mathbf{q}]) = \int_{[\mathbf{n}|\mathbf{q}]} c_5([\mathbf{n}|\mathbf{q}]) \leq 0 \; .
\end{gather}
The following configuration is an example of a complete intersection manifold with vanishing Euler number
\be
	\left[\begin{array}{c}2\\5\end{array}\right|\left.\begin{array}{cc}3&0\\0&6\end{array}\right] .
\ee
For its fourth Chern class, we find $c_4 = 7830 \hat{J}^1$. However, this configuration can be written as $[2|3]\times [5|6]$, which means it decomposes into $T^2 \times \nfold{4}$ and is therefore not a Calabi-Yau five-fold in the strict sense defined in \chref{cy5}.

\section{\texorpdfstring{$X_8$}{X8} and the relationship between Chern classes and Pontrjagin classes}\applabel{cp_classes}

As mentioned in \secref{sugra_mthy_corr}, one of the ${\cal O}(\beta)$-corrections to the 11-dimensional supergravity action~\eqref*{S_CJS} is the famous Green-Schwarz term~\eqref*{sugra11_gs}~\cite{Duff:1995wd}. It consists of a wedge product between the 3-form gauge field $A$ and an object denoted $X_8$, which is a quartic polynomial in the curvature two-form $\curvtwofrm^\ul{M} {}_\ul{N} \equiv \oneon{2} R^\ul{M} {}_{\ul{N}PQ} dx^P \wedge dx^Q$
\be\eqlabel{def_x8_orig}
	X_8 = \oneon{(2\pi)^4} \left[ - \oneon{768} (\tr\, \curvtwofrm^2)^2 + \oneon{192} \tr\, \curvtwofrm^4 \right] .
\ee
In~\eqref*{def_x8}, it was claimed without proof that $X_8$ can be written in terms of the first and second Pontrjagin classes, which are topological invariants of the 11-dimensional space-time manifold $\manifold$. This fact is important in deriving the topological constraint~\eqref*{intcond} in \secref{topol_constraint} and the lack of proof is remedied now.

Before discussing Pontrjagin classes, we begin with a preliminary exercise. Let us consider the expression $\det{(\id+A)}$ for a complex diagonalisable $n\times n$ matrix $A$ and denote its eigenvalues by $\{x_i\}_{i=1,\ldots,n}$. Thus, $\det A = x_1 \cdots x_n$ and the expression $\det{(\id+A)}$ reads as follows
\be\eqlabel{det1A}
	\det{(\id+A)} = (1 + x_1)\cdots (1 + x_n) \; .
\ee
The next step is to multiply out the right hand side and collect together terms of the same degree in the $x_i$. One finds
\be\eqlabel{symm_fnct}
	\det{(\id+A)} = \prod_{i=1}^n (1 + x_i) = 1 + S_1(\ul{x}) + \ldots + S_n(\ul{x}) \; ,
\ee
where $S_k (\ul{x})$ is the $k$-th elementary symmetric function in the $x_i$ defined by $S_k (\ul{x}) \equiv \sum_{i_1 < \ldots < i_k} x_{i_1} \cdots x_{i_k}$. In particular, $S_1(\ul{x}) = \sum_{i=1}^n x_i$, $S_2 (\ul{x}) = \sum_{i<j} x_i x_j$ and $S_n (\ul{x}) = x_1 \cdots x_n$ and we see immediately that $S_1 (\ul{x}) = \tr\, A$ and $S_n (\ul{x}) = \det A$. For the remaining $S_k (\ul{x})$, more complicated expressions in terms of $A$ emerge
\begin{align}
	S_1 (\ul{x}) &= \tr\, A \; , \eqlabel{S1} \\
	S_2 (\ul{x}) &= \oneon{2} \left[(\tr\, A)^2 - \tr\, A^2\right] , \eqlabel{S2} \\
	S_3 (\ul{x}) &= \oneon{3!} \left[2\,\tr\, A^3 - 3(\tr\, A^2)\tr\, A + (\tr\, A)^3\right] , \eqlabel{S3} \\
	S_4 (\ul{x}) &= \oneon{4!}\left[-6\,\tr\, A^4 + 3(\tr\, A^2)^2 + (\tr\, A)^4 + 8(\tr\, A^3)\tr\, A - 6(\tr\, A)^2\tr\, A^2\right] , \eqlabel{S4} \\
	&\;\;\vdots \nonumber \\
	S_k (\ul{x}) &= \det A \; . \eqlabel{Sk}
\end{align}
These formul\ae\ can be most straightforwardly checked in a reverse manner, that is by starting from the expressions on the right hand side and applying the so-called multinomial theorem, which reads
\be
	(x_1 + \ldots + x_n)^k = \sum_{i_1 + \ldots + i_n = k} \binom{k}{i_1, \ldots, i_n} x_1^{i_1} \cdots x_n^{i_n}
\ee
and the multinomial coefficients are defined by $\binom{k}{i_1, \ldots, i_n} \equiv k! / (i_1! \cdots i_n!)$.

The total Pontrjagin class $p(E)$ of a real vector bundle $E$ over an $n$-dimensional manifold $\manifold$ equipped with a connection $\omega$ is defined by 
\be\eqlabel{Pontrjagin_class_defn}
	p(E) \equiv \det\left(\id + \oneon{2\pi}\Omega \right) = 1 \oplus p_1(E) \oplus p_2(E) \oplus \cdots \; ,
\ee	
where $\Omega = d\omega + \omega\wedge\omega$ is the curvature two-form of $E$ with respect to the connection $\omega$. Viewed as a matrix, $\Omega$ is anti-symmetric, $\Omega^\transp = -\Omega$, and not diagonalisable by an element of $\GL(k,\R)$, but by one of $\GL(k,\C)$, where $k\equiv\dim E$. The result is
\vspace{\parskip}
\be
	\Omega \stackrel{\GL(k,\C)}{\longrightarrow} \diag(ix_1,-ix_1,ix_2,-ix_2,\ldots)
\ee
\vspace{\parskip}\noindent 
and $\Omega^{2i} \stackrel{\GL(k,\C)}{\longrightarrow} (-1)^i \diag(x_1^{2i},x_1^{2i},x_2^{2i},x_2^{2i},\ldots)$. Hence, $\tr\,\Omega^{2i} = 2 (-1)^i \sum_{j=1}^{\lfloor k/2\rfloor} x_j^{2i}$ and the total Pontrjagin class can be expanded as follows
\vspace{\parskip}
\be
	\det\left(\id+\oneon{2\pi}\Omega\right) = \det\left(\id-\oneon{2\pi}\Omega\right) = 
	\prod_{i=1}^{\lfloor k/2\rfloor} \left(1 + \oneon{(2\pi)^2} x_i^2 \right) \; ,
\ee
\vspace{\parskip}\noindent 
where the first equality is a consequence of $\Omega^\transp = -\Omega$ and $\lfloor k/2\rfloor$\newnot{symbol:floor} denotes the largest integer less than or equal to $k/2$. The $p_i(E)$ in the expansion~\eqref*{Pontrjagin_class_defn} are the analogue of the elementary symmetric functions $S_k (\ul{x})$ in~\eqref{symm_fnct} and we read off $(2\pi)^{2i} p_i(E) = \sum_{j_1 < \ldots < j_i} x_{j_1}^2 \cdots x_{j_i}^2$. By the same procedure as described above for the $S_k (\ul{x})$, one finds expressions for the $p_i(E)$ as polynomials of degree $2i$ in the curvature two-form
\vspace{\parskip}
\begin{align}
	p_1(E) &= - \oneon{2}  \oneon{(2\pi)^2} \tr\, \Omega^2 \; , \eqlabel{p1} \\
	p_2(E) &=   \oneon{8}  \oneon{(2\pi)^4} \left[(\tr\, \Omega^2)^2 - 2\, \tr\, \Omega^4\right] , \eqlabel{p2} \\
	p_3(E) &=   \oneon{48} \oneon{(2\pi)^6} \left[-(\tr\,\Omega^2)^3 + 6\,\tr\,\Omega^2 \,\tr\,\Omega^4 - 8\,\tr\,\Omega^6 \right] , \eqlabel{p3} \\
		&\;\;\vdots \nonumber \\
	p_{\lfloor k/2\rfloor}(E) &= \oneon{(2\pi)^{2\lfloor k/2\rfloor}} \det \Omega \eqlabel{pk} 
\end{align}
\vspace{\parskip}\noindent 
and $[p_i(E)] \in H^{4i}(\manifold,\Z)$. Also, $p_i(E) = 0$ for $2i>k$ or $4i>n$. The Pontrjagin classes $p_i(\manifold)$ of the manifold $\manifold$ are defined with respect to its tangent bundle $T\manifold$ and curvature $\curvtwofrm$ as $p_i(\manifold) \equiv p_i(T\manifold)$. With the help of~\eqref*{def_x8_orig,p1,p2}, one straightforwardly verifies the expression~\eqref*{def_x8} given for $X_8$ in \secref{sugra_mthy_corr}.

Next, we turn to complex vector bundles, where an analogue of the Pontrjagin classes, namely the Chern classes, exist. Let $E$ be a complex vector bundle over an $n$-dimensional manifold $\manifold$ equipped with a connection $\omega$. The total Chern class of $E$ is then defined as
\be
	c(E) \equiv \det{(\id + \frac{i}{2\pi}\Omega)} = 1 \oplus c_1(E) \oplus c_2(E) \oplus \cdots \; .
\ee
The (matrix-valued) curvature two-form $\Omega$ can be diagonalised by an element of $\GL(k,\C)$, where $k\equiv\dim_\C E$. If we denote its eigenvalues by $\{x_i\}_{i=1,\ldots,k}$, we can directly apply \eqrangeref{det1A}{symm_fnct} to re-write $c(E)$ as
\be
	\det{(\id + \frac{i}{2\pi}\Omega)} = \prod_{i=1}^k (1 + \frac{i}{2\pi} x_i)
\ee
and use the expressions~\eqrangeref*{S1}{Sk} for the elementary symmetric functions $S_k (\ul{x})$ to find the $c_i(E)$ as polynomials of degree $i$ in $\Omega$
\begin{align}
	c_1(E) &= \frac{i}{2\pi}\tr\, \Omega \; , \eqlabel{c1_curv} \\
	c_2(E) &= \oneon{8\pi^2} \left[\tr\, \Omega^2 - (\tr\, \Omega)^2\right] , \eqlabel{c2_curv} \\
	c_3(E) &= \frac{i}{48\pi^3} \left[-2\,\tr\, \Omega^3 + 3(\tr\, \Omega^2)\tr\, \Omega - (\tr\, \Omega)^3\right] , \eqlabel{c3_curv} \\
	c_4(E) &= \oneon{384\pi^4}\left[-6\,\tr\, \Omega^4 + 3(\tr\, \Omega^2)^2 + (\tr\, \Omega)^4 
				+ 8(\tr\, \Omega^3)\tr\, \Omega - 6(\tr\, \Omega)^2\tr\, \Omega^2\right] , \eqlabel{c4_curv} \\
		&\;\;\vdots \nonumber \\
	c_k(E) &= \left(\frac{i}{2\pi}\right)^k \det \Omega \; . \eqlabel{ck_curv}
\end{align}
Note that $[c_i(E)] \in H^{2i}(\manifold,\Z)$ and $c_i(E) = 0$ for $i>k$ or $2i>n$. The Chern classes $c_i(\manifold)$ of a complex manifold $\manifold$ are defined with respect to its holomorphic tangent bundle $T\manifold$ as $c_i(\manifold) \equiv c_i(T\manifold)$.

Neglecting the complex structure, a complex vector bundle $E$ with $\dim_\C E=k$ can also be viewed as a real vector bundle with $\dim_\R E=2k$, which is why one expects a relationship between the Chern and Pontrjagin classes of $E$. In order to find this relationship, we need the following embedding of $\U(k)$ into $\SO(2k)$ (see, for example, ref.~\cite{GSW})
\be
	F:\;\; \U(k) \rightarrow \SO(2k), \qquad F(U)\equiv\left(
	\begin{array}{cc}
		U_R & U_I \\
		-U_I & U_R
	\end{array}\right),
\ee
where $U=U_R +i U_I$ and both $U_R$ and $U_I$ are $k\times k$ matrices with real entries. With the explicit form of $F$ and a bit of algebra, one may verify 
\be\eqlabel{u_so_tr}
	\tr_\R\, T^{2k} \equiv \tr_\R\, F(T^{2k}) = 2\, \tr\, T^{2k} \; , \qquad\qquad \forall\, k>0 \; ,
\ee
where $T$ is an element in the fundamental representation of the Lie algebra $\mathfrak{u}(k)$ corresponding to $\U(k)$ and we have introduced $\tr_\R$ to denote the trace taken in the fundamental representation of the Lie algebra $\mathfrak{so}(2k)$ corresponding to $\SO(2k)$ and induced by the map $F$. Recalling that the curvature two-form $\Omega$ is Lie algebra-valued (with the Lie algebra corresponding to the structure group of the vector bundle $E$), we compute the Pontrjagin classes of the complex vector bundle $E$
\begin{align}
	p_1(E) &= - \oneon{2}  \oneon{(2\pi)^2} \tr_\R\, \Omega^2
					= - 					 \oneon{(2\pi)^2} \tr\, \Omega^2 \; , \eqlabel{pontr_compl1} \\
	p_2(E) &=   \oneon{8}  \oneon{(2\pi)^4} \left[(\tr_\R\, \Omega^2)^2 - 2\, \tr_\R\, \Omega^4\right]
					=   \oneon{2}  \oneon{(2\pi)^4} \left[(\tr\, \Omega^2)^2 - \tr\, \Omega^4\right] , \eqlabel{pontr_compl2} \\
	p_3(E) &=   \oneon{48} \oneon{(2\pi)^6} \left[-(\tr_\R\,\Omega^2)^3 + 6\,\tr_\R\,\Omega^2 \,\tr_\R\,\Omega^4 
						- 8\,\tr_\R\,\Omega^6 \right] \nonumber \\
				 &=   \oneon{6}  \oneon{(2\pi)^6} \left[-(\tr\,\Omega^2)^3 + 3\,\tr\,\Omega^2 \,\tr\,\Omega^4 
				 		- 2\,\tr\,\Omega^6 \right] , \eqlabel{pontr_compl3} \\
		&\;\;\vdots \nonumber
\end{align}
from \eqrangeref{p1}{p3} and~\eqref*{u_so_tr}. Plugging \eqrangeref{c1_curv}{c4_curv} into \eqrangeref{pontr_compl1}{pontr_compl2} directly implies
\begin{align}
	p_1 &= c_1^2 - 2 c_2 \; , \eqlabel{pontrj_chern1} \\
	p_2 &= c_2^2 - 2 c_1 c_3 + 2 c_4 \; . \eqlabel{pontrj_chern2} \\
			&\;\;\vdots \nonumber 
\end{align}
This represents a specialisation of the general formula~\cite{Milnor:1974,Hubsch:1992}
\be
	p_i = c_i^2 + 2 \sum_{j=0}^{i-1} (-1)^{i+j} c_j c_{2i-j} \; ,
\ee
with $c_0 = 1$.

Finally, we quote without proof~\cite{Milnor:1974,Nakahara:1990th}: $p(E\oplus F) = p(E) \wedge p(F)$. For $\R\times X$, this implies
\be
	p(\R\times X) = p(T\R\oplus TX) = p(T\R) \wedge p(TX) = p(X) \; ,
\ee
where we have used $p(\R)=1$ to arrive at the last equality. This means $p_i(\R\times X) = p_i(X)$ and for $X$ a Calabi-Yau five-fold (that is, with $c_1(X)=0$), one finds
\be
	X_8(\R \times X) = \oneon{48} \left(\left(\frac{p_1}{2}\right)^2 - p_2\right) = - \oneon{24} c_4(X)
\ee
after using \eqrangeref{pontrj_chern1}{pontrj_chern2}. This proves \eqref{X8res}, which was crucial in deriving the topological constraint present in Calabi-Yau five-fold compactifications of M-theory considered in \secref{topol_constraint}.

\section{Details of the fermionic reduction: An example}\applabel{ferm_red_11_example}

In this section, we will take a closer look at how the results of \secref{ferm_red_perform} are obtained. To illustrate the techniques used to calculate the fermionic part of the one-dimensional effective action, we specialise to a particular term and follow it through each step of the calculation.

As exemplary term, we choose the kinetic term of the $(1,1)$ fermions $\psi^i$. In other words, we will now show explicitly how the first term in~\eqref*{s11f} arises from the fermionic reduction. In the ansatz~\eqrangeref*{ansatz_gravitino_0}{ansatz_gravitino_1} for the 11-dimensional gravitino, we may drop all but the $(1,1)$-part
\be\eqlabel{ansatz_gravitino_11}
	\Psi_0 = 0 \; , \qquad
	\Psi_{\bar{\mu}} = \psi^i(\tau) \otimes (\omega_{i,\nu\bar{\mu}} \gamma^\nu \eta) \; , \qquad
	\Psi_{\mu} = - \bar{\psi}^i(\tau) \otimes (\omega_{i,\mu\bar{\nu}} \gamma^{\bar\nu} \eta^\star) \; .
\ee
The minus sign in the last expression follows from $\Psi_{\mu} = (\Psi_{\bar{\mu}})^\ast$, which descends from the 11-dimensional Majorana condition, and $(\omega_{i,\mu\bar\nu})^\ast = - \omega_{i,\nu\bar\mu}$, which is the component version of the reality condition $(\omega_i)^\ast = \omega_i$ of the $(1,1)$-forms.

The starting point for the fermionic reduction is the fermionic part of the CJS action~\eqref*{S_CJS_F1}. Here, we only need to consider the 11-dimensional Rarita-Schwinger action, that is the first term in~\eqref*{S_CJS_F1}, for none of the other terms contain derivatives of fermions. In the remainder of this section, the ellipsis symbol $(\ldots)$ indicates fermionic terms not capable of contributing to the first term in~\eqref*{s11f}. We can then write~\eqref*{S_CJS_F1} as
\bea\eqlabel{11_ex_action1}
	S_{\rm CJS,F} &= - \oneon{2\kappa^2_{11}} \int d^{11}x \sqrt{-g} \, \bar{\Psi}_M \Gamma^{MNP} D_N(\omega) \Psi_P + \ldots \\
	 &= - \oneon{2\kappa^2_{11}} \int d\tau d^{10}x \frac{N}{2} \sqrt{g^{(10)}} \, i \Psi^\dag_m \Gamma^\UZ \Gamma^{m0n} \dot{\Psi}_n + \ldots \; ,
\eea
where in the second line, we used~\eqref*{gansatz},~\eqrangeref*{covder1}{covder3} and the definition of the Dirac conjugate and we restricted to the index structure of~\eqref*{ansatz_gravitino_11}. The Clifford algebra~\eqref*{cliff11} allows us to simplify the gamma matrix expression to
\be
	\Gamma^\UZ \Gamma^{m0n} = - \Gamma^\UZ \Gamma^0 \Gamma^{mn} 
	= - e_\UZ {}^0 (\Gamma^\UZ)^2 \Gamma^{mn} = - 2 N^{-1} \gamma^{mn} \; .
\ee
Re-inserting this result into~\eqref*{11_ex_action1} yields
\be\eqlabel{11_ex_action2}
	S_{\rm CJS,F} = - \oneon{2\kappa^2_{11}} \int d\tau d^{10}x \sqrt{g^{(10)}} \, (-i) \Psi^\dag_m \gamma^{mn} \dot{\Psi}_n + \ldots \; .
\ee
The next step is to change to holomorphic coordinates\footnote{The minus sign in front of $X_i^\ast$ is a consequence of the anti-commuting nature of $\Psi_M$. It renders $\Psi^\dag_m \gamma^{mn} \dot{\Psi}_n$ imaginary and, together with the additional factor of $i$ in~\eqref*{11_ex_action2}, it ensures that the action $S_{\rm CJS,F}$ is real.}
\be\eqlabel{11_ex_RS1}
	\Psi^\dag_m \gamma^{mn} \dot{\Psi}_n = 
		\Psi^\dag_\mu \gamma^{\mu\nu} \dot{\Psi}_\nu + \Psi^\dag_\mu \gamma^{\mu\bar\nu} \dot{\Psi}_{\bar\nu} - \cc = 
		X_1 + X_2 - X_1^\ast - X_2^\ast \; ,
\ee
and to insert the ansatz~\eqref*{ansatz_gravitino_11}
\begin{align}
	X_1	&= \Psi^\dag_\mu \gamma^{\mu\nu} \dot{\Psi}_\nu
		= \eta^\dag \gamma^{\bar\rho} \gamma^{\mu\nu} \gamma^{\bar\sigma} \eta^\star \,
			\omega_{i,\mu\bar\rho} \omega_{j,\nu\bar\sigma} \bar{\psi}^i \dot{\bar\psi}^j \; , \\
	X_2	&= \Psi^\dag_\mu \gamma^{\mu\bar\nu} \dot{\Psi}_{\bar\nu}
		= - \eta^\dag \gamma^{\bar\rho} \gamma^{\mu\bar\nu} \gamma^\sigma \eta \,
			\omega_{i,\mu\bar\rho} \omega_{j,\sigma\bar\nu} \bar{\psi}^i \dot{\psi}^j \; .
\end{align}
The 10-dimensional spinor bilinears can be computed explicitly by means of the gamma matrix algebra~\eqref*{cy5_hol_cliff_alg} together with the annihilation condition~\eqref*{eta_annihil}. One finds
\begin{align}
	X_1	&= - 2 \eta^\dag \eta^\star \, 
		(\omega_{i,\mu} {}^\nu \omega_{j,\nu} {}^\mu - \tilde{\omega}_i \tilde{\omega}_j) \bar{\psi}^i \dot{\bar\psi}^j = 0 \; , \\
	X_2	&= - 2 (\omega_{i,\mu} {}^\nu \omega_{j,\nu} {}^\mu - 2 \tilde{\omega}_i \tilde{\omega}_j) \dot{\psi}^i \bar{\psi}^j \; ,
\end{align}
where, in the second line, the normalisation $\eta^\dag \eta = 1$ (see \eqref{etanorm}) was used and the one-dimensional spinors were re-ordered and re-labelled at the cost of a minus sign. The first line vanishes since $\eta^\dag \eta^\star = 0$ (see \eqref{etaTeta_is_zero}). Thus, \eqref{11_ex_RS1} reads
\be
	\Psi^\dag_m \gamma^{mn} \dot{\Psi}_n = (\psi^i \dot{\bar{\psi}}^j - \dot{\psi}^i \bar{\psi}^j) 
		(2 \omega_{i,\mu} {}^\nu \omega_{j,\nu} {}^\mu - 4 \tilde{\omega}_i \tilde{\omega}_j) \; ,
\ee
and the action~\eqref*{11_ex_action2} becomes
\be
	S_{\rm CJS,F} = - \oneon{2\kappa^2_{11}} \int d\tau \, \frac{i}{2} (\psi^i \dot{\bar{\psi}}^j - \dot{\psi}^i \bar{\psi}^j)
		\left[ -4 \int_X d^{10}x \sqrt{g^{(10)}} \, (\omega_{i,\mu} {}^\nu \omega_{j,\nu} {}^\mu - 2 \tilde{\omega}_i \tilde{\omega}_j)\right]
		+ \ldots \; .
\ee
With the help of the formul\ae\ of \appref{complex_geometry}, it is straightforward to see that the expression in square brackets is equal to the physical $(1,1)$-metric $G^{(1,1)}_{ij}$ defined in~\eqref*{G11def}. We thereby arrive at the final result
\be
	S_{\rm CJS,F} = - \frac{l}{2} \int d\tau \, \frac{i}{2} G^{(1,1)}_{ij}(\underline{t}) (\psi^i \dot{\bar{\psi}}^j - \dot{\psi}^i \bar{\psi}^j)
		+ \ldots \; ,
\ee
where $l = v/\kappa^2_{11}$. We thus precisely reproduced the first term in~\eqref*{s11f}, as claimed. All other terms in the fermionic part of the one-dimensional effective action as presented in \secref{ferm_red_perform} follow from calculations that are similar in spirit to the one picked out here as an example.

\bibliographystyle{utphys}
\bibliography{thesis,mcy5v2,phd_main_reading_db}

\end{document}